\documentclass[reprint,aps,prx,twocolumn,groupedaddress, floatfix,nofootinbib]{revtex4-2}

\usepackage{graphicx}
\usepackage{bm}
\usepackage[utf8]{inputenc}
\usepackage{amsmath,amsfonts,calc}
\usepackage{comment}
\usepackage[normalem]{ulem}
\usepackage[usenames,dvipsnames]{xcolor}
\usepackage{graphicx}
\usepackage{scalerel}
\usepackage{mathtools}
\usepackage{oplotsymbl}
\usepackage{appendix}
\usepackage[colorlinks, linkcolor = Green, citecolor = Blue, bookmarksdepth=2, linktocpage=true, hypertexnames=true]{hyperref}

\usepackage[capitalize]{cleveref}
\usepackage{tikz}
\usepackage{scrextend}
\usepackage{tabularx,booktabs}
\usepackage{longtable}

\usepackage{stackengine}

\usepackage{multirow}
\usepackage{xcolor}
\DeclareFontFamily{U}{matha}{\hyphenchar\font45}
\DeclareFontShape{U}{matha}{m}{n}{
      <5> <6> <7> <8> <9> <10> gen * matha
      <10.95> matha10 <12> <14.4> <17.28> <20.74> <24.88> matha12
      }{}
\DeclareSymbolFont{matha}{U}{matha}{m}{n}
\newcommand{\op}[1]{\hat{#1}} 
\newcommand{\opd}[1]{\hat{#1}^\dagger} 
\newcommand{\moyal}[2]{\{\!\!\{#1, #2\}\!\!\}} 
\newcommand{\comm}[2]{[#1, #2]} 
\newcommand{\cu}[1]{\mathcal{#1}} 
\newcommand{\cl}[1]{\tilde{#1}} 
\newcommand{\clu}[1]{\tilde{\cu #1}} 

\DeclareMathSymbol{\varstar}{2}{matha}{"0F}

\newcommand{\tikzpic}[2]{
 \!\!\begin{tikzpicture}[baseline=#1 pt]
    \node[anchor = north](image) 
    at (0,0) {\includegraphics{#2}};
    \end{tikzpicture}\!\!}

\newcommand{\suppresstocsubsection}[1]{%
  \refstepcounter{subsection}%
  \subsectionmark{#1}%
  \addtocontents{toc}{}%
  \begin{center}
    \textbf{\thesubsection\quad#1}
  \end{center}}

\makeatletter
\def\l@subsubsection#1#2{}

\AtBeginDocument{\let\oldcontentsline\contentsline}

\newcommand{\apptoccontentsline}[4]{\oldcontentsline{#1}{\hspace{18pt} Appendix \!\!\!\!\!#2\vspace{-5pt}}{#3}{#4}}

\newcommand{\apptoc}{\addtocontents{toc}{\let\protect\contentsline\protect\apptoccontentsline}}

\newcommand{\reftoccontentsline}[4]{\oldcontentsline{#1}{#2}{#3}{#4}}
\newcommand{\reftoc}{\addtocontents{toc}{\let\protect\contentsline\protect\reftoccontentsline}}

\newcommand{\acktoccontentsline}[4]{\oldcontentsline{#1}{#2\vspace{-5pt}}{#3}{#4}}
\newcommand{\acktoc}{\addtocontents{toc}{\let\protect\contentsline\protect\acktoccontentsline}}

\makeatother
\date{\today}

\begin{document}
\title{\texorpdfstring{A diagrammatic method to compute the \\effective Hamiltonian of driven nonlinear oscillators}{}}
\author{Xu Xiao}
\thanks{These two authors contributed equally}
\email{xu.xiao@yale.edu, jaya.venkat@yale.edu}
\affiliation{Department of Applied Physics and Physics, Yale University, New Haven, CT 06520, USA}
\author{Jayameenakshi Venkatraman}
\thanks{These two authors contributed equally}
\email{xu.xiao@yale.edu, jaya.venkat@yale.edu}
\affiliation{Department of Applied Physics and Physics, Yale University, New Haven, CT 06520, USA}
\author{Rodrigo G. Corti\~nas}
\affiliation{Department of Applied Physics and Physics, Yale University, New Haven, CT 06520, USA}
\author{Shoumik Chowdhury}
\thanks{present address: Massachusetts Institute of Technology, Cambridge, MA 02139}
\affiliation{Department of Applied Physics and Physics, Yale University, New Haven, CT 06520, USA}
\author{Michel H. Devoret}
\email{michel.devoret@yale.edu}
\affiliation{Department of Applied Physics and Physics, Yale University, New Haven, CT 06520, USA}
\date{\today}
\begin{abstract}
In this work, we present a new diagrammatic method for computing the effective Hamiltonian of driven nonlinear oscillators. At the heart of our method is a self-consistent perturbation expansion developed in phase space, which establishes a direct correspondence between the diagram and algebra. Each diagram corresponds to a Hamiltonian term, the prefactor of which, like those in Feynman diagrams, involves a simple counting of topologically equivalent diagrams. Leveraging the algorithmic simplicity of our diagrammatic method, we provide a readily available computer program that generates the effective Hamiltonian to arbitrary order. We show the consistency of our schemes with existing perturbation methods such as the Schrieffer-Wolff method. Furthermore, we recover the classical harmonic balance scheme from our result in the limit of $\hbar\rightarrow0$. Our method contributes to the understanding of dynamic control within quantum systems and achieves precision essential for advancing future quantum information processors. To demonstrate its value and versatility, we analyze five examples from the field of superconducting circuits. These include an experimental proposal for the Hamiltonian stabilization of a three-legged Schr\"odinger cat, modeling of energy renormalization phenomena in superconducting circuits experiments, a comprehensive characterization of multiphoton resonances in a driven transmon, a proposal for an inductively shunted transmon circuit, and a characterization of classical ultra-subharmonic bifurcation in driven oscillators.  Lastly, we benchmark the performance of our method by comparing it with experimental data and exact Floquet numerical diagonalization.
\end{abstract}

\maketitle

\tableofcontents

\section{Introduction and Motivation}\label{sec:intro}

The nonlinear oscillator driven by a sinusoidally time-varying force, whose archetype is the driven pendulum, is a rich playground of diverse dynamical behaviors that include bifurcation and chaos \cite{dykman2012,guckenheimer2013}. Latest advances in quantum electrodynamics with Josephson circuits (cQED) transpose century-old, classical nonlinear dynamical phenomena into a new quantum regime. This results from the property of the Josephson junction to behave at dilution refrigerator temperatures as an extremely fast non-dissipative pendulum-like electromagnetic oscillator. Quantum-limited amplification \cite{siddiqi2004,manucharyan2007,clerk2010,frattini2018} for high-fidelity readout \cite{krantz2016}, fast parametric gates for bosonic codes \cite{gao2018,wustmann2019}, and the generation and stabilization of non-classical states like Schr\"odinger cat states \cite{grimm2020} and Gottesman-Kitaev-Preskill (GKP) states \cite{campagne2020} are examples of processes taking place in this new quantum regime. 

The Hamiltonian of the generic driven nonlinear oscillator can be written as
\begin{align}\label{eq:nl-osc-H}
\begin{split}
    \frac{\op{\mathcal{H}}(t)}{\hbar} = \omega_o\op a^\dagger \op a &+ \sum_{m\ge3}\frac{g_m}{m}(\op a + \op a^\dagger)^m \\
    &- i\Omega_d (\op a - \op a^\dagger)\cos\omega_d t,
\end{split}
\end{align}
where $\omega_o$ and $g_m \ll \omega_o$ are the natural frequency and $m$-th rank nonlinearity coefficient of the oscillator, $\op a$ is the bosonic annihilation operator, and the driving force (thereafter shortened to ``drive'') is specified by its amplitude $\Omega_d$ and frequency $\omega_d$. In the context of cQED, the Hamiltonian \cref{eq:nl-osc-H} represents a charge-driven Josephson circuit with a single degree of freedom, whose hardware design determines the nonlinear constants $g_m$.\footnote{In certain devices, and to a certain extent, these coefficients are tunable \textit{in situ} by the magnetic flux threading the loops of the circuit.} 

The Hamiltonian \cref{eq:nl-osc-H} gives rise to a variety of dynamical processes with no static counterpart. Their relative strengths can be adjusted \textit{in situ} by the drive amplitude $\Omega_d$ and frequency $\omega_d$. Remarkably, in the perturbative regime, to be defined rigorously below, these processes can be described by a time-independent effective Hamiltonian. We now illustrate this claim by describing two simple examples whose exact calculation will be treated in the rest of the work.

\begin{figure}
\centering
    \includegraphics{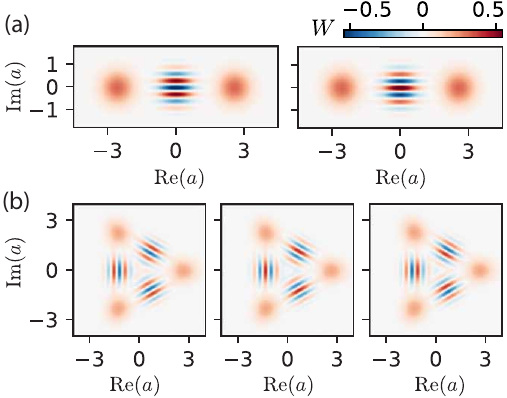}
    \caption{
    Wigner functions $W$ of the ground state manifold for a driven nonlinear oscillator with nonlinearities $g_3/\omega_o = 0.04, g_4/\omega_o = 0.003$ under different driving conditions. They are obtained by numerical diagonalization. Panel (a) corresponds to \cref{eq:Kerr-cat-RWA} with $\omega_d/\omega_o = 1.996$ and $\Omega_d/\omega_o = 0.2$.  The ground states are two perfectly degenerate Schr\"odinger cat states. Panel (b) corresponds to \cref{eq:3-legged-cat-H} with $\omega_d/\omega_o = 1.481$ and $\Omega_d/\omega_o = 0.7$. The ground states are three-fold nearly degenerate comprising three-legged cat-like states.}
    \label{fig:cat-wigner}
\end{figure}

\textit{Example I: the Kerr-cat qubit Hamiltonian.} We first consider the case where we limit the range of nonlinearity rank to $g_3$ and $g_4$ and the effective Hamiltonian terms to first order in these coefficients. This situation can be obtained in practice with a SNAIL circuit \cite{frattini2017} operating with non-zero external magnetic flux. When the drive frequency $\omega_d$ is tuned in the vicinity of $2\omega_o$, the system undergoes a period-doubling bifurcation. In the quantum regime, the effective ground state of the system will be doubly quasi-degenerate and exhibit a Schr\"odinger cat manifold (see Figure \ref{fig:cat-wigner} (a) and \cite{puri2017,grimm2020}). Under a frame transformation amounting to $\op a \rightarrow \op a e^{-i\frac{\omega_d}{2}t} + \xi  e^{-i\omega_dt}+\mathcal{O}(1/\omega_d)$ with $\xi \approx \frac{4i\Omega_d}{3\omega_d}$,\footnote{Here $\Omega_d$ is taken to be $\mathcal{O}(\omega_d)$ and thus $\xi$ to be $\mathcal{O}(1)$.} we arrive at an effective time-independent Hamiltonian
\begin{align}\label{eq:Kerr-cat-RWA}
\begin{split}
\frac{\op K_{\frac{\omega_d}{2}}}{\hbar} = \Delta \op a^\dagger \op a &+ \frac{3g_4}{2}\op a^{\dagger 2} \op a^2 \\
&+ g_3\xi^* \op a^{\dagger 2} + g_3\xi\op a^2 + \mathcal{O}\Big(\frac{1}{\omega_d}\Big)
\end{split}
\end{align}
where $\Delta = \omega_o - \frac{\omega_d}{2} + 3g_4  + 6g_4|\xi|^2$. The subscript in the Hamiltonian \cref{eq:Kerr-cat-RWA} indicates the rotating frame to which the oscillator is transformed. This level of approximation is often called the rotating wave approximation (RWA). Inside the $\mathcal O(1/\omega_d)$ will be obtained later in this article. The Hamiltonian \cref{eq:Kerr-cat-RWA} has received lately theoretical and experimental attention in the context of quantum information processing \cite{puri2017,grimm2020}. 

The emergence of the effective time-independent Hamiltonian \cref{eq:Kerr-cat-RWA} from the time-dependent \cref{eq:nl-osc-H} can be understood from diagrams that have been previously proposed \cite{Mundhada2017,frattini2018,hillman21}. In this way, the static effective squeezing term $g_3\xi^* \op a^{\dagger2}$ in \cref{eq:Kerr-cat-RWA} corresponds to the diagram
\begin{subequations}\label{eq:Kerr-cat-RWA-diagram}
\begin{align}
    \tikzpic{-27}{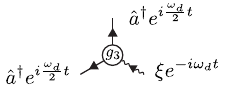}\;.
\end{align}
This diagram should be read in the usual ``Feynman diagram" manner: each diagrammatic element corresponds to a factor that is multiplied with other factors to give the term corresponding to the whole. The diagram illustrates an harmonic mixing process: a 3-wave interaction, which stems from the 3rd rank nonlinearity of the oscillator, takes in a drive excitation $\tikzpic{-10}{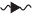}$ at frequency $\omega_d$ and generates two resonant excitations of the oscillator $\tikzpic{-10}{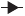}$, which are at frequencies in the vicinity of the oscillator's natural frequency, i.e., $\omega_d/2 \approx \omega_o$.  The corresponding term has a net zero frequency oscillation and thus belongs to the effective time-independent Hamiltonian. 

Similarly, the Kerr nonlinearity $\frac{3g_4}{2}\op a^{\dagger 2} \op a^2 $ in \cref{eq:Kerr-cat-RWA} can be represented by the diagram
\begin{align}
    \tikzpic{-40}{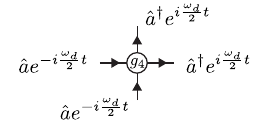},
\end{align}
\end{subequations}
in which a 4-wave interaction, which stems from the 4th rank nonlinearity of the oscillator,  intakes two resonant excitations and produces two resonant excitations of the oscillator. 

\textit{Example II: three-legged cat Hamiltonian.} Our second example consists of another type of bifurcation that requires more involved diagrams. When the same hardware described above is submitted to a drive with frequency $\omega_d$ in the vicinity of  $3\omega_o/2$, the system undergoes a period-tripling bifurcation. The effective ground state becomes triply quasi-degenerate and exhibits a three-legged Schr\"odinger cat manifold (see \cref{fig:cat-wigner} (b)). Under a frame transformation amounting to $\op a \rightarrow \op a e^{-i\frac{2}{3}\omega_d t} + \xi  e^{-i\omega_dt}+\mathcal{O}(1/\omega_d)$ where $\xi \approx \frac{9i\Omega_d}{5\omega_d}$, \cref{eq:nl-osc-H} transforms into another time-independent Hamiltonian capturing this effective dynamics
\begin{align}
\begin{split}
    \frac{\op K_{\frac{2}{3}\omega_d}}{\hbar} = \Delta\op a^\dagger \op a &+ \Big(\frac{3g_4}{2}-\frac{5g_3^2}{\omega_d}\Big) \op a^{\dagger 2} \op a^2 \\
    &+ \Big(\frac{195g_3^3}{4\omega_d^2}-\frac{165g_3g_4}{8\omega_d}\Big)\xi^{2} \op a^{\dagger 3}
    + \text{{h.c.}} \\&+ \mathcal O\left(\frac{1}{\omega_d^3}\right),
\end{split}\label{eq:3-legged-cat-H}
\end{align}
where $\Delta = \omega_o - \frac{2\omega_d}{3} + 3g_4-\frac{10g_3^2}{\omega_d}+(6g_4-\frac{180g_3^2}{7\omega_d})|\xi|^2 + \mathcal O(\frac{1}{\omega_d^2})$. The threefold symmetry emerges from the ``beyond RWA'' term $\xi^2\op a^{\dagger 3}$, which, in this condition, is resonant. The three-legged cats emerging from Hamiltonian in the form of \cref{eq:3-legged-cat-H} have also received theoretical and experimental attention recently for quantum information processing \cite{zhang2017,lorch2019,zhang2019_2}.

From the diagrammatic heuristics, the beyond RWA terms in \cref{eq:3-legged-cat-H} can be understood as cascaded diagrams. For example, the term $(\frac{195g_3^3}{4\omega_d^2}-\frac{165g_3g_4}{8\omega_d})\xi^2\op a^{\dagger 3}$ corresponds to the diagrams
\begin{align}\label{eq:3-legged-cat-diagram}
     \tikzpic{-35}{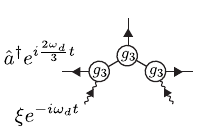}\quad  \tikzpic{-23}{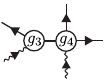}\quad
\end{align}
in which three 3-wave ``mixing vertices'', nicknamed ``mixers'' in the following, are cascaded together to form the first diagram and a 3-wave and a 4-wave mixers are cascaded together to form the second. Each of the diagrams intakes two drive excitations and produces three resonant excitations of the oscillator. However, it is not easy to see how the negative sign in $-\frac{165g_3g_4}{8\omega_d}$ emerges from the diagrams above. This problem points to the limitations of the heuristic diagrams, which, at this stage, have only an interpretational role instead of a calculational one. 

\textit{Beyond heuristic diagrams.} Can the heuristic diagrams that we have presented in these two examples be formalized as components of a systematic procedure?
In this article, we answer positively to this question and construct a stand-alone diagrammatic method to efficiently compute effective static Hamiltonian of driven nonlinear oscillators. At the level of pen and paper, our formalization of the diagrammatic heuristics leads to the complete list of Hamiltonian terms at any order. Their prefactors involve the simple counting of topologically equivalent diagrams. Because the analytical work becomes tedious as order in perturbation is increased, we also provide a ready-to-use computer program to generate the full expansion to any order \cite{qhb2022}. We validate the approach by benchmarking our results against numerical simulation and experimental data. We also use it to explain and predict novel phenomena in the context of cQED.

The rest of the article is divided into two main parts: First, in \cref{sec:system,sec:averaging,sec:diagrams-pert}, we formally introduce the diagrammatic formulation. In \cref{sec:discussion,sec:examples} we discuss the implications of this new method and apply it to analyze particular systems in cQED. Each of these sections and the subsections within are written in a modular manner so that the readers can selectively engage with the topics. Specifically, in \cref{sec:system} we discuss the Hamiltonian of a generic driven nonlinear oscillator and the required frame transformations to prepare it for the perturbative analysis. In \cref{sec:averaging}, we derive our diagrammatic expansion from a novel quantum averaging perturbation method named as quantum harmonic balance (QHB). This method is developed in phase space that puts classical and quantum dynamics on equal footing while highlighting their difference arising from  order-$\hbar$ corrections to the classical Poisson bracket. By carrying out QHB iteratively to lowest orders, we derive the rules for operating and evaluating diagrams from those of an algebra.  In \cref{sec:diagrams-pert}, we present the diagrammatic method in an axiomatic and non-iterative form, which provides a succinct and compact recipe. In \cref{sec:discussion}, we discuss the general form of the effective Hamiltonian, the extension of the diagrams to multi-mode and multi-tone nonlinear oscillators, its extension to open quantum systems described by the effective Lindbladian,  the necessary conditions and limits of choosing a specific rotating frame to construct the diagrams, and the relation between our diagrammatic method and other Floquet methods. In \cref{sec:3lcat} we use diagrams to compute the effective Hamiltonian governing the three-legged Schr\"odinger cat as an illustrative example. In \cref{sec:renormalization}, we compute the energy renormalization of a multi-mode superconducting circuit, which consists of a cavity coupled to a transmon, and show good agreement with experimental data. In \cref{sec:mpnr}, we provide a systematic characterization of multiphoton resonance in Josephson circuits, which limits most of the readout and pumping schemes in cQED. These processes lacked an analytical description until our work. We validate these description with the exact Floquet numerical diagonalization and comment on the relationship between the descriptions in terms of an effective static Hamiltonian and the quasienergies provided by Floquet diagonalization. In \cref{sec:IST}, we propose a circuit, the inductively-shunted transmon (IST), to mitigate the multiphoton resonance. In \cref{sec:USH}, we characterize the general ultra-subharmonic bifurcation process that occurs in a driven nonlinear oscillator, which has important implications in the design of bifurcation-based amplifiers, Schr\"odinger-cat logical qubits, and readout schemes. In \cref{sec:conclusion} we conclude our work and discuss future directions.  

\section{The System of Interest}
\label{sec:system}
In this section, we formally introduce the system of interest, its inherent perturbative structure, and the frame transformations to prepare it for the perturbation analysis. We write the general Hamiltonian of a driven quantum oscillator as
\begin{align}
\label{eq:op-H-q-p}
    \op{\mathcal H}(t)= \frac{\op{p}^2}{2\mu} + \op U(\op q) + f \cos{(\omega_d t)} \op q,
\end{align}
where $\mu$ is the mass of the oscillator, $\op q$ and $\op p$ are the conjugate position and momentum coordinates satisfying $[\op q, \op p ] = i \hbar$. The function $\op U(\op q)$ is the nonlinear potential of the oscillator, which we assume to have the form of
\begin{align*}
    \op U(\op q) = \sum_{m \ge 2} c_m \op{q}^m
\end{align*}
when Taylor expanded around the potential minimum at $q =0$. Here the coefficients $c_2= \frac{1}{2} \mu \omega_o^2$ and $c_{m > 2}$ define the harmonic and anharmonic parts of the oscillator potential respectively, where $\omega_o$ is the small-oscillation frequency of the oscillator. The parameters $f$ and $\omega_d$ in \cref{eq:op-H-q-p} are respectively the strength and frequency of the external drive.\footnote{Unlike in \cref{eq:nl-osc-H}, \cref{eq:op-H-q-p} features a drive that couples to the position degree of freedom, which is more natural for a driven mechanical oscillator. This choice does not affect qualitatively our treatment and we address any change arising from the momentum-coupling usually found in superconducting circuits in the following footnote.}

The Hamiltonian \cref{eq:op-H-q-p} can be re-expressed in a bosonic basis as 
\begin{align}\label{eq:op-H-a-ad}
\begin{split}
    \op{\mathcal H}(t) = &\omega_o\op a^\dagger \op a + \sum_{m\ge3}\frac{g_m}{m}(\op a+\op a^\dagger)^m \\
    &\quad+ \Omega_d (\op a+\op a^\dagger)\cos(\omega_d t),
\end{split}
\end{align}
where we have defined the scaled bosonic coordinates $\op a, \op a^\dagger$ with $\op a = \frac{\sqrt{\hbar}}{2}\left(\frac {\op q}{q_{\mathrm{zps}}}  + i \frac{\op p}{p_{\mathrm{zps}}} \right)$, and $q_{\mathrm{zps}} = \sqrt{\hbar/2 \mu \omega_o}$ and $p_{\mathrm{zps}} = \sqrt{\hbar \mu \omega_o/2}$ are the zero-point-spreads of the corresponding coordinates. Here, we impose the commutator $[\op{a}, \opd{a}] = \hbar$ to explicitly track the quantum corrections to the classical driven nonlinear dynamics. Rescaling the coordinates, such that $[\op{a}, \opd{a}] = 1$, yields the standard bosonic ladder operators used in \cref{eq:nl-osc-H}.  The coefficient $g_m  = \frac{m c_m \hbar\omega_o}{4 c_2 \hbar^{m/2}}  q_{\rm{zps}}^{m-2}$ in \cref{eq:op-H-a-ad} is the $m$-th rank nonlinearity of the oscillator and $\Omega_d = fq_\text{zps}/\sqrt{\lambda}$ is the drive amplitude.

We further perform a frame transformation amounting to $\op a \rightarrow  \op a e^{-i\omega_o^\prime t} + \xi  e^{-i\omega_dt}$, where $\xi =\frac{ \Omega_d\omega_o}{\omega_o^2-\omega_d^2}$ (see footnote\footnote{When the drive couples to the system through the momentum degree of freedom $\op p$ as in \cref{eq:nl-osc-H}, we have $\Omega_d = \sqrt{\lambda}fp_\text{zps}$ and  $\xi = \frac{i\Omega_d \omega_d}{\omega_d^2 - \omega_o^2}$ in \cref{eq:H-tran}, with no further change in the rest of the treatment.} and \cref{app:frame} for details), to bring the pertinent nonlinear process into focus. The transformed Hamiltonian reads
\begin{align}\label{eq:H-tran}
\begin{split}
    \op H(t) = \delta\op a^\dagger \op a+  \sum_{\substack{m \ge 3 }} \frac{g_m}{m}  &(\hat{a} e^{- i \omega_o^\prime t} + \hat{a}^{\dagger} e^{i \omega_o^\prime t} \\
    & \;+ \xi e^{-i \omega_d t} +  \xi^* e^{i \omega_d t} )^m,
\end{split}\raisetag{0.4\baselineskip}
\end{align}
where $\delta = \omega_o - \omega_o^\prime$. We note that the choice of $\omega_o^\prime$ contains a priori knowledge about the rotating frame in which the nonlinear process of interest occurs. For example, in the Kerr-cat and three-legged cat systems in \cref{sec:intro}, $\omega_o^\prime$ is respectively taken to be $\omega_d/2$ and $2\omega_d/3$, corresponding to the dominant ultra-subharmonics generated from the drive in a period-doubling and a period-tripling bifurcation. This relevant frame, which is discussed in more detail in \cref{subsec:frame}, is usually some $p/q$-th ultra-subharmonic of the drive that is near-resonant with the oscillator, i.e. $\delta = \omega_o - p \omega_d/q \ll \omega_o\sim\omega_d$ for some integers $q, p$. For now, we assume there is only one such frame containing nontrivial nonlinear processes. The treatment for multiple nontrivial nonlinear processes coexisting will be discussed later in the text.

The Hamiltonian \cref{eq:H-tran} constitutes the starting point of the perturbative analysis underlying the diagrammatic method. Our perturbative parameters are $\delta$ and $g_m$'s. We also demand the oscillator to be in the weakly excited quantum regime, which translates into the nonlinearities obeying the perturbative structure: \begin{align}\label{eq:pert-condition}
\frac{\hbar^{\frac{m}{2}}}{\hbar\omega_o}|\xi| g_{m+1}\lesssim\frac{\hbar^{\frac{m}{2}}}{\hbar\omega_o}\sqrt{\langle\op a^\dagger \op a\rangle}g_{m+1}\ll\frac{\hbar^{\frac{m}{2}}}{\hbar\omega_o} g_{m}\ll 1,
\end{align}
where $\langle\op a^\dagger \op a\rangle$ is the expectation value of the pertinent quantum states over the operator $\op a^\dagger \op a$.
Note that with non-zero $g_m$'s, only the Fock states below a certain excitation number satisfy the above perturbative condition and our treatment is restricted to them. Recalling the definition of $g_m\sim \frac{m c_m \hbar\omega_o}{4 c_2 \hbar^{m/2}}  q_{\rm{zps}}^{m-2}$, it is then convenient to introduce the notation $g_m = \mathcal O(q_\text{zps}^{m-2})$ and use $q_\text{zps}$ to count the perturbative order of terms in \cref{eq:H-tran} and the subsequent perturbative expansion. For detuning $\delta\ll \omega_o$ and effective drive strength $\xi =\frac{ \Omega_d\omega_o}{\omega_d^2-\omega_o^2}$, we assign the perturbative order $\mathcal O (q_\text{zps})$ and $\mathcal O (q_\text{zps}^0)$, respectively. We also remark that in the case of multiple modes or drives, one still obtains a Hamiltonian in the transformed frame resembling \cref{eq:H-tran}, but with more participants in the multinomial expansion. The diagrammatic rule derived for the single-mode problem can be extended to the multi-mode or multi-tone cases, which we discuss in \cref{subsec:multi}. 

Note that with non-zero $g_m$'s, only the Fock states below a certain excitation number satisfy the above perturbative condition and our treatment is restricted to them. Recalling the definition of $g_m\sim \frac{m c_m \hbar\omega_o}{4 c_2 \hbar^{m/2}}  q_{\rm{zps}}^{m-2}$, it is then convenient to introduce the notation $g_m = \mathcal O(q_\text{zps}^{m-2})$ and use $q_\text{zps}$ to count the perturbative order of terms in \cref{eq:H-tran} and the subsequent perturbative expansion. For detuning $\delta\ll \omega_o$ and effective drive strength $\xi =\frac{ \Omega_d\omega_o}{\omega_d^2-\omega_o^2}$, we assign the perturbative order $\mathcal O (q_\text{zps})$ and $\mathcal O (q_\text{zps}^0)$, respectively. We also remark that in the case of multiple modes or drives, one still obtains a Hamiltonian in the transformed frame resembling \cref{eq:H-tran}, but with more participants in the multinomial expansion. The diagrammatic rule derived for the single-mode problem can be extended to the multi-mode or multi-tone cases, which we discuss in \cref{subsec:multi}. 

\section{Quantum Harmonic Balance
Approach to the Diagrammatic
Perturbation Method}
\label{sec:averaging}

\subsection{Self-consistent perturbation expansion}

\subsubsection{proposing a canonical transformation}
We are now ready to introduce the perturbation method underlying our diagrammatic method, which we refer to as the quantum harmonic balance (QHB), because of its analogy to the harmonic balance method in classical nonlinear dynamics \cite{nayfeh1995}. The starting point of the QHB is the Heisenberg equation of motion for the operator $\op{a}$ submitted to the Hamiltonian $\hat{H}$ in \cref{eq:H-tran}:
\begin{align}
\label{eq:a}
    d_t\op a = -\frac{1}{i \hbar} [\op{H}, \op{a}].
\end{align}
Under the drive, the evolution of $\op a$ exhibits distinct time-scales: a \textit{slow} evolving dynamics on the time-scale of $\delta$ and $g_m/\hbar^{\frac{m}{2}-1}$, and a \textit{fast} oscillating one comparable to the drive oscillation at frequency $\omega_d$. In many driven systems, the nonlinear processes of interest happen on a relatively slow time-scale\footnote{For instance, the Rabi oscillation frequency between two nearby levels in a quantum nonlinear system is usually much lower than the Larmor frequency inducing the transition.} while the fast dynamics is considered a secondary ``micromotion''. While this separation of time-scales provides simplification in describing the driven nonlinear dynamics, it is important to recognize that the slow dynamics and the micromotion couple to each other through the oscillator's nonlinearity. Therefore, the goal here --- as in high frequency perturbation theories in general \cite{venkatraman2021} --- is to seek a frame transformation $\hat a\rightarrow\hat{\mathcal A}$ placing the relevant dynamics at the forefront. In this sought-after frame, the dynamics of the transformed operators $\hat{\mathcal A}, \hat{\mathcal{A}^\dagger}$ should be generated by a time-independent hermitian ``Kamiltonian" \cite{goldstein2002} $\op K$ through the equation of motion:
\begin{align}\label{eq:K-eom}
    d_t \op{\cu{A}} &=-\frac{1}{i\hbar}\comm{\op K(\op{\cu A}, \op{\cu A}^\dagger)}{\op{\cu{A}}},
\end{align}
and we call $\op K$ as the static effective Hamiltonian. 

To find this effective frame, we assume that the original frame is only perturbatively different from the sought-after one by a time-independent quantity  $\op\sigma=\op\sigma(\op{\cu{A}}, \opd{\cu{A}})$ and a purely oscillating quantity $\op\mu=\op\mu(\op{\cu{A}}, \opd{\cu{A}}, t)$, i.e. 
\begin{align}\label{eq:QHB-ansatz}
    \op a = \op{ \cu A} + \op\sigma(\op{\cu A}, \op{\cu A}^\dagger) + \op\mu(\op{\cu A}, \op{\cu A}^\dagger,t).
\end{align}
A non-zero periodic function $f$ is deemed purely oscillating if its time-average is zero, i.e. $\int_0^T dt f = 0$ for $T$ being the period of the function $f$.  \cref{eq:QHB-ansatz} should be understood as an additive representation of the frame transformation that generalizes the widely-employed  averaging methods \cite{krylov1937,bogoliubov1961,landau1976,rahav2003,mirrahimi2015} that address rapidly-driven systems. In addition, to enforce that this frame transformation is canonical, we further express \cref{eq:QHB-ansatz} as an exponential map:
\begin{align}\label{eq:exponential-map}
\begin{split}
\op a &= e^{L_{\op S}}\op{\cu A}\\
& = \op{\cu A} + \frac{1}{i\hbar}[\op S, \op{\cu A}] + \frac{1}{2!(i\hbar)^2}[\op S, [\op S, \op{\cu A}]] + \cdots
\end{split}
\end{align}
where $\hat S = \hat S(\op{\cu A},\op{\cu A}^\dagger,t)$ is assumed to be a \textit{hermitian} oscillating function generating the frame transformation. The super-operator  $L_{\op S}\,\tikzpic{-10}{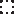} = \frac{1}{i\hbar}[\op S, \tikzpic{-10}{figsv6_small/notation/dash_box.pdf}\,]$ is the Hilbert space Lie derivative with respective to $\hat S$. This exponential map is also a unitary transformation, i.e. $e^{L_{\op S}}\op{\cu A} = e^{\hat S/i\hbar}\op{\cu A} e^{-\hat S/i\hbar}$, and its equivalence to \cref{eq:exponential-map} can be obtained via the Baker-Campbell-Hausdorff formula. For clarity in our discussion, we opt for the super-operator notation $L_{\hat S}$ for the generator of the transformation in \cref{eq:exponential-map}.

We note that the canonical constraint \cref{eq:exponential-map} is a theoretical improvement we introduce in this work over the conventional averaging methods \cite{krylov1937,bogoliubov1961,landau1976,rahav2003,mirrahimi2015} which often suffer from the issue of non-canonicity \cite{grozdanov1988}. We will elaborate this problem and its remedy in more detail in \cref{app:averaging}. In addition, we also remark that there exists an alternative class of high-frequency perturbation techniques, such as the Schrieffer-Wolff expansion \cite{kuchment1993,eckardt2015,venkatraman2021} and the Floquet-Magnus expansion \cite{casas2001}, that construct the exponential map directly through finding $\op S$ in \cref{eq:exponential-map}. As we will subsequently demonstrate in \cref{subsec:other-pert} and rigorously prove in \cref{app:canonicity}, the perturbation method we develop in this work aligns with the Schrieffer-Wolff expansion and provides deeper insights into the dynamics of driven nonlinear systems, along with computational speed-up. 

With this frame transformation defined by \cref{eq:QHB-ansatz}, \cref{eq:a} can be rewritten as
\begin{align}
\label{eq:A-eom}
  \frac{1}{i\hbar}\comm{\op K}{\op{\cu{A}}} - \partial_t \op \mu + \frac{1}{i \hbar} \comm{\op K}{\op\sigma+\op \mu} = \frac{1}{i \hbar} \comm{\op H}{\op a}\Big|_{\!\!\!\!\!\! \substack{\op{a}\; =\; \op{\cu A} + \op\sigma +\op{\mu} \\ \;\;\;\;\; \opd{a} =\; \opd{\cu A} +\opd\sigma + \opd{\mu}}},
\end{align}
where, on the left-hand side of the equation, we have used the Heisenberg equation of motion for $\op {\cu A}, \op\sigma$, and $\op \mu$. The task now is to perturbatively find $\op K, \op\sigma$, and $\op \mu$ that satisfy the \cref{eq:A-eom} at each order of the perturbative parameter.

\subsubsection{phase-space representation of quantum mechanics}
Prior to performing the explicit computation, it is important to note that our perturbative analysis will be developed in phase space rather than Hilbert space. To facilitate this approach, we introduce the Husimi transform $\mathfrak H$, a modification of Wigner transform \cite{curtright2013,hillery1984,puri2001}, that transforms physical quantities in the Hilbert space to the Husimi Q phase space. Comparing with the Hilbert space formulation that operates via matrix algebra, the phase-space formulation of quantum mechanics operates by taking partial derivatives of phase-space functions and thus lends itself more naturally to a diagrammatic representation that features quantum corrections to classical results.  Moreover, the Husimi Q representation is chosen, among other isomorphic representations of the phase-space formulation, because it privileges the normal-ordering of operators in Hilbert space, a convenient choice for studying quantum nonlinear dynamics \cite{puri2001}. We emphasize that developing a diagrammatic method in Hilbert space is also feasible using bosonic operator identities. Yet many diagrammatic properties and the relationship between quantum and classical regimes would then be obscured.

In the Husimi space, \cref{eq:A-eom} transforms under $\mathfrak H$ as:
\begin{align}\label{eq:eom-phase}
\begin{split}
    \moyal{ \cl K}{ {\clu A}}_{\cu {\cl A}, \cu {\cl A}^*}  - \partial_t {\cl \mu} &+ \moyal{ \cl K}{ {\cl\sigma+\cl \mu}}_{\cu {\cl A}, \cu {\cl A}^*} \\
    &\qquad\qquad= \moyal{ \cl H}{ \cl a}\Big|_{\!\!\!\! \substack{\!\!\!\!{\cl a}\; =\; {\cu {\cl A}} + {\cl\sigma+\cl \eta} \\ \;\;\; {\cl a}^* =\; \cu {\cl A}^* + {\cl\sigma}^*+ {\cl \eta}^*}},
    \end{split}
\end{align}
with the Husimi transform $\mathfrak H$ obeying the properties 
\begin{align}\label{eq:Husimi-transform}
\begin{split}
    \mathfrak H (\hat{f}) &= \tilde{f}, \qquad
    \mathfrak H (\hat{f} \hat{g})  = \tilde{f} \varstar \tilde g, \\
    \mathfrak H \left(\frac{1}{i \hbar}[\op f ,\op g ]\right) &=  \moyal{\cl f}{\cl g} = \frac{1}{i\hbar}(\tilde f \varstar \tilde g - \tilde g \varstar \tilde f),
\end{split}
\end{align}
where Hilbert-space operators $\op f$, $\op g$ are transformed to phase-space functions $\cl f$ and $\cl g$. We have here introduced the symbol $\moyal{\tikzpic{-9}{figsv6_small/notation/dash_box.pdf}\,}{\tikzpic{-9}{figsv6_small/notation/dash_box.pdf}}$ standing for the Husimi bracket (a modification of Moyal's bracket \cite{curtright2013}). Crucial to this scheme, the bidifferential operator $\varstar$ (a modification of Groenwold's star-product \cite{curtright2013, hillery1984}) is the underlying product of this phase-space algebra and is defined as
\begin{align}
\label{eq:six-star}
\begin{split}
    \cl f \varstar \cl g &= \cl f \exp \left(\hbar\overleftarrow{\partial}_{\cl a} \overrightarrow \partial_{\cl {a}^*}\right) \cl g \\
    &= \sum_{k\ge0} \frac{\hbar^k}{k!}\cl f\,(\overleftarrow{\partial}_{\cl a} \overrightarrow{\partial}_{\cl a^*})^k \cl g \\
    &= \cl f\cl g + \hbar\partial_{\cl a} \cl f\partial_{{\cl a}^*}\cl g+\cdots.
\end{split}
\end{align}
When we subscript a Husimi bracket, for example as $\moyal{\tikzpic{-9}{figsv6_small/notation/dash_box.pdf}\,}{\tikzpic{-9}{figsv6_small/notation/dash_box.pdf}}_{\cu {\cl A}, \cu {\cl A}^*}$, it is evaluated over the subscript phase space complex coordinates. As one can easily verify, from a normal-ordered operator $\hat f$, the phase-space function $\cl f = \mathfrak H(\op f)$ is obtained readily by replacing its arguments $\op a$ and $\op a^\dagger$ by (now commuting) arguments which are complex scalars $\cl {a}$ and $\cl a^*$. Conversely, the operator $\op f = \mathfrak H^{-1}(\cl f)$ can be obtained from $\cl f$ by simply moving the $\cl a^{*}$ factors in $\cl f$ to the left of $\cl a$ and replacing the phase-space coordinates by their respective Hilbert space operators, with no other change.

To leading order in $\hbar$, the star product defined by \cref{eq:six-star} is just the regular multiplication, and at the same order, the Husimi bracket is just the Poisson bracket that generates classical dynamics. Thanks to the continuous deformation between the phase-space formulation of quantum mechanics and classical mechanics, controlled by the single real parameter $\hbar$ \cite{curtright2013}, all of the analysis we develop next can be applied to a classical driven nonlinear oscillator by taking the limit $\hbar\rightarrow0$.

\subsubsection{self-consistent equations underlying the perturbative method}
We now plug $\tilde H$, the phase-space function associated with \cref{eq:H-tran}, into the \cref{eq:eom-phase} and rewrite the latter as
\begin{widetext}
\begin{align}\label{eq:eom-phase-expand}
\begin{split}
     \partial_{\!\clu A^*}\cl K \!+ i\partial_t \cl\mu -i\moyal{\cl K}{\cl\sigma+\cl\mu}_{\cl{\cu A}, \cl{\cu A}^*}=&\sum_{\substack{m \ge 3}}\!\!e^{i\omega_o^\prime t} g_m  \Big((\cl{\cu A}+\cl\sigma+\cl\mu) e^{- i \omega_o^\prime t}+ (\cl{\cu A}^* \!+ \cl\sigma^*+ \cl\mu^*) e^{i \omega_o^\prime  t}
     +\xi e^{-i \omega_d t} +  \xi^* e^{i \omega_d t} \Big)^{m-1}_\varstar  \\[5pt]
     &\;+ \delta(\clu A +\cl\sigma+\cl\mu).
\end{split}\raisetag{1.5\baselineskip}
\end{align}
\end{widetext}
In writing the right-hand side of \cref{eq:eom-phase-expand}, we have employed the chain rule $\partial_{\clu A^*}(\alpha \clu A + \beta \clu A^*)_{\varstar}^m = m\beta (\alpha \clu A + \beta \clu A^*)^{m-1}_{\varstar}$ for constants $\alpha$ and $\beta$ (see \cref{app:star-chain} for a proof), and the multinomial expansion with subscript $\varstar$ means that any two terms in the expansion are associated with a star product, which here and in the rest of the paper is evaluated over $\clu A$ and $\clu A^*$. 

With \cref{eq:eom-phase-expand}, analyzing the dynamics of the oscillator translates to solving the equation for the unknowns $\cl K$, $\cl\sigma$, and $\cl\mu$ subject to the constraints that (1) $\cl K$ and $\cl\sigma$ are time-independent (2) $\cl\mu$ is purely oscillating, and (3) the transformation $\cl a\rightarrow \clu A$ is canonical, i.e.
\begin{align}\label{eq:exponential-map-phase}
\clu A+\cl\sigma+\cl\mu = e^{L_{\cl S}}\clu A,
\end{align}
with a real function $\cl S = \cl S(\clu A, \clu A^*, t)$. However, solving \cref{eq:eom-phase-expand} directly is challenging due to the intricate interdependency between $\cl K$, $\cl\sigma$, and $\cl\eta$. A critical strategy we adopt involves a set of change of variables:
\begin{subequations}\label{eq:composite-var}
\begin{align}\label{eq:composite-var-M}
&\cl M = i\partial_t\cl\mu -i\moyal{K}{\cl\mu}-\delta\cl\mu,\\\label{eq:composite-var-B}
&\clu B = \clu A + \cl\sigma,\\\label{eq:composite-var-Gamma}
&\cl\Gamma =-i\moyal{\cl K}{\clu B}= \partial_{\clu A^*}\cl K - i\moyal{\cl K}{\clu B-\clu A},
\end{align}
\end{subequations}
where $\cl B$ and $\cl \Gamma$ are time-independent, and $\cl M$ is purely oscillating by construction. These changes of variables are inspired by the diagrammatic formulation yet to be introduced. For the moment, they should be understood as to decompose \cref{eq:eom-phase-expand} into a set of coupled equations each of simpler form and clearer physical meaning. 

Specifically, with the new set of variables, \cref{eq:eom-phase-expand} can be written in a more succinct form as:
\begin{align}\label{eq:eom-dressed}
\begin{split}
\cl\Gamma + \cl M &= \sum_{\substack{m \ge 3}}\!\!e^{i\omega_o^\prime t} g_m  \Big((\cl{\cu B}+\cl\mu) e^{- i \omega_o^\prime t} \!+ (\cl{\cu B}^* \!+ \cl\mu^*) e^{i \omega_o^\prime  t}\\[-5pt]
     &\qquad\qquad+\xi e^{-i \omega_d t} +  \xi^* e^{i \omega_d t} \Big)^{m-1}_\varstar + \delta\clu B
\end{split}\raisetag{1.0\baselineskip}
\end{align}
By further imposing that $\cl \Gamma$ is time-independent and $\cl M$ is purely oscillating, we have
\begin{subequations}\label{eq:self-consistent}
\begin{align}\label{eq:self-consistent-Gamma}
\begin{split}
\cl\Gamma &= \text{Sta}\Bigg(\sum_{\substack{m \ge 3}}\!\!e^{i\omega_o^\prime t} g_m  \Big((\cl{\cu B}+\cl\mu) e^{- i \omega_o^\prime t} \!+ (\cl{\cu B}^* \!+ \cl\mu^*) e^{i \omega_o^\prime  t}\\[-5pt]
     &\qquad\qquad\qquad+\xi e^{-i \omega_d t} +  \xi^* e^{i \omega_d t} \Big)^{m-1}_\varstar + \delta\clu B\Bigg)
\end{split}\raisetag{1.5\baselineskip}\\\label{eq:self-consistent-M}
\begin{split}
\cl M &= \text{Rot}\Bigg(\sum_{\substack{m \ge 3}}\!\!e^{i\omega_o^\prime t} g_m  \Big((\cl{\cu B}+\cl\mu) e^{- i \omega_o^\prime t} \!+ (\cl{\cu B}^* \!+ \cl\mu^*) e^{i \omega_o^\prime  t}\\[-5pt]
     &\qquad\qquad\qquad+\xi e^{-i \omega_d t} +  \xi^* e^{i \omega_d t} \Big)^{m-1}_\varstar\Bigg)
\end{split}\raisetag{1.5\baselineskip},
\end{align}
where the operators $\text{Sta}(\cl f) = \int_0^T\!dt \,\cl f$ and $\text{Rot}(\cl f)=\cl f-\text{Sta}(\cl f)$ respectively extract the static and rotating components of a given periodic function $\cl f$ with period $T$. From the canonicity of the transformation enforced by \cref{eq:exponential-map-phase} and from \cref{eq:composite-var-M,eq:composite-var-B,eq:composite-var-Gamma}, we can also obtain 
\begin{align}
\label{eq:self-consistent-mu}
\cl\mu &= -i\!\int\! dt \,\cl M + \int dt \big(-i\delta\mu + \moyal{\cl K}{\cl \mu}\big)\\
\label{eq:self-consistent-B}
\clu B &= \clu A + \Big(\text{Sta}(e^{L_{\cl S}}\clu A)-\clu A\Big)\\\label{eq:self-consistent-S}
i\cl S &= \!\int \!\!d\clu A^*\, \cl\mu + \!\int \!\!d\clu A^* \Big(\!-\!\text{Rot}(e^{L_{\cl S}}\clu A) +i\moyal{\clu A}{i\cl S}\Big)\\
\label{eq:self-consistent-K}
\cl K &= \!\int \!d\clu A^*\,\cl\Gamma + \int \!d\clu A^*\,i\moyal{\cl K}{\clu B-\clu A}
\end{align}
\end{subequations}
where \cref{eq:self-consistent-S} represents the inversion of the relation $\cl\mu = \text{Rot}(e^{L_{\hat S}}\clu A)$, thereby defining $\cl S$ self-consistently.

\cref{eq:self-consistent-Gamma,eq:self-consistent-M,eq:self-consistent-mu,eq:self-consistent-K,eq:self-consistent-B,eq:self-consistent-S} form a set of self-consistent equations, each having the structure
\[x = f(\vec y) + g(x,\vec y),\] 
where $x \in \{\cl\Gamma, \cl M, \cl\mu, \clu B, \cl S, \cl K\}$ is an unknown of the problem, and $\vec y$ represents the other unknowns. The term $f(\vec y)$ provides the leading-order contribution in $x$, while $g(x,\vec y)$ contains higher-order corrections that depend self-consistently on $x$ itself.\footnote{For \cref{eq:self-consistent-Gamma,eq:self-consistent-M}, their corresponding $g(x,\vec y)$'s are zero.} To systematically solve these equations order by order, we introduce a perturbative expansion in terms of the small parameter $q_\text{zps}$:
\begin{align}\label{eq:pert-structure}
\begin{split}
    &\cl\Gamma = \sum_{n>0} \cl\Gamma^{(n)},\;\cl M = \sum_{n>0} \cl M^{(n)}, \; \cl\mu = \sum_{n>0} \cl\mu^{(n)}, \\
    &\cl K = \sum_{n>0} \cl K^{(n)},\; \clu B = \clu A + \sum_{n> 0} \clu B^{(n)}, \;\cl S = \sum_{n>0} \cl S^{(n)},
\end{split}
\end{align}
where $\clu A$ is considered the zeroth-order term. Within this perturbative framework, one can substitute these expansions into the original self-consistent equations and determine each correction $\cl\Gamma^{(n)}, \cl M^{(n)}, \cl\mu^{(n)}, \cl K^{(n)}, \clu B^{(n)}, \cl S^{(n)}$ iteratively, ensuring a systematic approach to solving the full nonlinear problem. Notably, \cref{eq:self-consistent} is structurally akin to the Dyson equation in quantum electrodynamics, which defines the Green’s function of a particle in a self-consistent manner and serves as a foundation for deriving Feynman diagrams \cite{dyson1949,weinberg1995}. Analogously, \cref{eq:self-consistent} can be regarded as a stand-alone perturbative framework from which our diagrammatic method will be constructed.

In the remainder of this section, we will implement the self-consistent equations in \cref{eq:self-consistent} using diagrammatic representation that clarifies their relevance to physical processes. Later, in \cref{sec:diagrams-pert}, we will introduce a stand-alone diagrammatic approach that does not depend on \cref{eq:self-consistent}. It is also important to remark that, while our current focus is on driven bosonic systems, the self-consistent framework presented herein is versatile and can be adapted to other perturbative techniques. In \cref{app:RS}, we demonstrate the universal applicability of this approach by reformulating the well-established Rayleigh-Schr\"odinger perturbation theory as operator-valued self-consistent equations. This reformulation not only allows for more insightful physical interpretations but also bridges it with another widely-used method, the Brillouin-Wigner perturbative theory.

\subsection[Diagrammatic representation]{Introducing the diagrammatic representation}\label{subsec:diagram-prep}

To introduce the diagrammatic representation, we first translate \cref{eq:self-consistent-Gamma,eq:self-consistent-M}, each with an additional global phase $e^{-i\omega_o^\prime t}$, into the diagrams as follows:
\begin{subequations}\label{eq:sc-diagram}
\begin{align}\label{eq:sc-diagram-Gamma}
\begin{split}
    \tikzpic{-13}{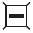}\!&=
    \!\!\sum_{m\ge3}\!\! \tikzpic{-11}{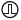}\tikzpic{-14}{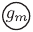}\Bigg(
    \tikzpic{-21}{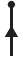}+\! \tikzpic{-24}{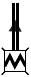}
    \!+\tikzpic{-21}{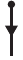}+\! \tikzpic{-24}{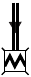}\!+\!
    \tikzpic{-21}{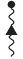}
    +\tikzpic{-21}{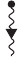}\Bigg)^{\!\!m-1}_{\!\!\scaleto{\varstar}{5pt}}\!\!+\!\tikzpic{-30}{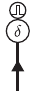}
\end{split}\\
\begin{split}\label{eq:sc-diagram-M}
    \tikzpic{-13}{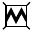}\!&=
    \!\!\sum_{m\ge3}\!\! \tikzpic{-11}{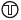}\tikzpic{-14}{figsv6_small/notation/gm.pdf}\Bigg(
    \tikzpic{-21}{figsv6_small/notation/B.pdf}+\! \tikzpic{-24}{figsv6_small/notation/eta.pdf}
    \!+\tikzpic{-21}{figsv6_small/notation/Bs.pdf}+\! \tikzpic{-24}{figsv6_small/notation/etas.pdf}\!+\!
    \tikzpic{-21}{figsv6_small/notation/xi.pdf}
    +\tikzpic{-21}{figsv6_small/notation/xis.pdf}\Bigg)^{\!\!m-1}_{\!\!\scaleto{\varstar}{5pt}},
\end{split}
\end{align}
where each diagrammatic element corresponds to an algebraic expression in \cref{eq:self-consistent-Gamma,eq:self-consistent-M}. The correspondence is:
\begin{align*}
\begin{split}
&\tikzpic{-13}{figsv6_small/notation/Gamma.pdf}= \cl \Gamma e^{-i\omega_o^\prime t}; \;\,\tikzpic{-13}{figsv6_small/notation/M.pdf}= \cl M e^{-i\omega_o^\prime t}; \;\,
 \tikzpic{-14}{figsv6_small/notation/gm.pdf}= g_m;\;\,\tikzpic{-14}{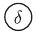}= \delta;\\[5pt]
 &\tikzpic{-11}{figsv6_small/notation/band_pass.pdf}= e^{-i\omega_o^\prime t}\,\text{Sta}\,(e^{i\omega_o^\prime t}\,\tikzpic{-9}{figsv6_small/notation/dash_box.pdf}\,);\quad\tikzpic{-11}{figsv6_small/notation/band_stop.pdf}= e^{-i\omega_o^\prime t}\,\text{Rot}\,(e^{i\omega_o^\prime t}\,\tikzpic{-9}{figsv6_small/notation/dash_box.pdf}\,);\\[3pt]
 &\tikzpic{-20}{figsv6_small/notation/B.pdf}\;,
 \tikzpic{-20}{figsv6_small/notation/Bs.pdf}= \clu Be^{-i\omega_o^\prime t}, \clu B^*e^{i\omega_o^\prime t};\quad
\tikzpic{-23}{figsv6_small/notation/eta.pdf},
 \tikzpic{-23}{figsv6_small/notation/etas.pdf}= \cl\mu e^{-i\omega_o^\prime t}, \cl\mu^*e^{i\omega_o^\prime t};\\[-4pt]
 &\tikzpic{-20}{figsv6_small/notation/xi.pdf}\;,
 \tikzpic{-20}{figsv6_small/notation/xis.pdf}= \xi e^{-i\omega_dt}, \xi^*e^{i\omega_dt}.
\end{split}\raisetag{0.8\baselineskip}
\end{align*}

The diagrams, while initially inspired by heuristic principles introduced in \cref{sec:intro}, are elaborated by rigorous algebraic construction. Specifically, each diagram is invariant by rotation of the page, for example,
\begin{align*}
\begin{split}
\tikzpic{-18}{figsv6_small/notation/B.pdf}=
\tikzpic{-18}{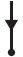}=
\tikzpic{-11}{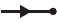}= 
\tikzpic{-11}{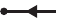}= \clu Be^{-i\omega_o^\prime t}.
\end{split}
\end{align*}
Note that the black dot $\bullet$ in  \!\tikzpic{-9}{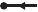} (and other similar diagrams) is employed to unambiguously define the arrow's traveling direction. Moreover, a diagram consisting of multiple components has to be interpreted as the product of the algebraic expressions corresponding to its constituent diagrams. For example, the last diagram in \cref{eq:sc-diagram-Gamma} corresponds to:
\begin{align*}
    \tikzpic{-30}{figsv6_small/notation/delta_B.pdf} \;
    = \;\,e^{-i \omega_o^\prime t}\text{Sta}\left(e^{i \omega_o^\prime t}\delta(\clu B e^{-i \omega_o^\prime t})\right)=\;\delta\clu B e^{-i \omega_o^\prime t}.
\end{align*}
Here, the black dot $\bullet$ in \!\tikzpic{-9}{figsv6_small/notation/B_small.pdf} is not shown since the direction of the arrow can be inferred from the position of \!\tikzpic{-11}{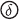}. The diagram above showcases two types of multiplications underlying composite diagrams. The conjunction between \!\tikzpic{-11}{figsv6_small/notation/delta_small.pdf} and \!\tikzpic{-9}{figsv6_small/notation/B_small.pdf} represents a standard commutative multiplication, while the conjunction between \!\tikzpic{-11}{figsv6_small/notation/band_pass.pdf} and \!\tikzpic{-11}{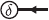} represents a non-commutative multiplication. In the second case,  \!\tikzpic{-11}{figsv6_small/notation/delta_B_no_bp_small.pdf} ($\delta\clu B e^{-i \omega_o^\prime t}$) is taken as the argument for \!\tikzpic{-11}{figsv6_small/notation/band_pass.pdf} ($e^{-i\omega_o^\prime t}\text{Sta}\,(e^{i\omega_o^\prime t}\,\tikzpic{-9}{figsv6_small/notation/dash_box.pdf}\,)$) represented by the placeholder \tikzpic{-9}{figsv6_small/notation/dash_box.pdf}\,. We also remark that  $e^{-i\omega_o^\prime t}\text{Sta}\,(e^{i\omega_o^\prime t}\,\tikzpic{-9}{figsv6_small/notation/dash_box.pdf}\,)$ selectively retains the Fourier component with the phase $e^{-i\omega_o^\prime t}$ in \tikzpic{-9}{figsv6_small/notation/dash_box.pdf}\,. This motivates introduction of the non-commutative diagram \!\tikzpic{-11}{figsv6_small/notation/band_pass.pdf} resembling a band-pass filter that selectively passes through the frequency $\omega_o^\prime$. Conversely, the diagram \tikzpic{-11}{figsv6_small/notation/band_stop.pdf} ($e^{-i\omega_o^\prime t}\text{Rot}\,(e^{i\omega_o^\prime t}\,\tikzpic{-9}{figsv6_small/notation/dash_box.pdf}\,)$) in \cref{eq:sc-diagram-M} acts like a band-stop filter retaining all Fourier components in \tikzpic{-9}{figsv6_small/notation/dash_box.pdf}\, except for those with phase $e^{-i\omega_o^\prime t}$.  

\cref{eq:sc-diagram-Gamma,eq:sc-diagram-M} illustrates a key frequency mixing process within our diagrammatic method. The multinomial expansion in the right-hand side of \cref{eq:sc-diagram-Gamma,eq:sc-diagram-M} can be understood as frequency mixing factors. Each mixing factor comprises of an $m$-wave mixer \!\tikzpic{-11}{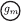} of strength $g_m$, receiving $m-1$ inputs chosen from the items inside the parenthesis.\footnote{ The second term in the right-hand side of \cref{eq:sc-diagram-Gamma} is a 2-wave mixer of strength $\delta$ and intakes one input \!\tikzpic{-9}{figsv6_small/notation/B_small.pdf}.} Each element in the parenthesis is interpreted as some type of \textit{excitation} ---  \!\!\tikzpic{-9}{figsv6_small/notation/B_small.pdf} ($\clu Be^{-i\omega_o^\prime t}$) and \!\!\tikzpic{-9}{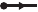} ($\clu B^*e^{i\omega_o^\prime t}$) represent the \textit{dressed resonant excitations} at frequency $\omega_o^\prime$ resonant with the oscillator,  \!\!\tikzpic{-9}{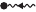} ($\cl\xi e^{-i\omega_d t}$) and \!\!\tikzpic{-9}{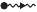} ($\cl\xi^* e^{i\omega_d t}$) represent \textit{drive excitations} at the drive frequency $\omega_d$, and \!\tikzpic{-11}{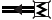} ($\cl\mu e^{-i\omega_o^\prime t}$) and \!\tikzpic{-11}{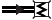} ($\cl\mu^*e^{i\omega_o^\prime t}$) represent \textit{dressed off-resonant excitations} with various frequency components. The term ``off-resonant excitation" is appropriate because $\cl\mu$ is purely rotating, meaning that $\cl\mu e^{-i\omega_o^\prime t}$ lacks any Fourier component resonant with the oscillator. By applying the frequency conservation law, the output frequency of each mixer is then determined and subsequently ``filtered" by either \tikzpic{-11}{figsv6_small/notation/band_pass.pdf} or \tikzpic{-11}{figsv6_small/notation/band_stop.pdf}. The outputs at frequency $\omega_o^\prime$ or otherwise contribute to \tikzpic{-10.5}{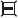} ($\cl\Gamma e^{-i\omega_o^\prime t}$) or \tikzpic{-11}{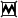} ($\cl M e^{-i\omega_o^\prime t}$) on the left-hand side of the equations, making $\cl\Gamma$ and $\cl M$ time-independent and purely rotating, respectively, as anticipated. As diagrammatically suggested, \tikzpic{-11}{figsv6_small/notation/M_small.pdf}  will in turn determine the dressed off-resonant excitation \!\tikzpic{-11}{figsv6_small/notation/eta_small.pdf} (their relation will be discussed shortly), which will ``feedback'' to the mixing processes on the right-hand side of \cref{eq:sc-diagram-Gamma,eq:sc-diagram-M}. This ``feedback link'', stemming from the self-consistent nature of \cref{eq:self-consistent}, will result in cascaded mixers or \textit{mixing networks} that constitute \tikzpic{-10.5}{figsv6_small/notation/Gamma_small.pdf} and \tikzpic{-11}{figsv6_small/notation/M_small.pdf}. We will elaborate more on this in \cref{subsec:order-1,subsec:bare-virtual}.

Similarly to Feynman diagrams, in our diagrammatic method, a dressed excitation is represented as a bare excitation plus higher order correction. The dressing of these excitations is formally defined by the diagrammatic representation of \cref{eq:sc-diagram-mu,eq:sc-diagram-B,eq:sc-diagram-S,eq:sc-diagram-K}, which we will now briefly discuss. Specifically, with an additional global phase $e^{-i\omega_o^\prime t}$, \cref{eq:self-consistent-mu} is diagrammatically represented as:
\begin{align}\label{eq:sc-diagram-mu}
\tikzpic{-23}{figsv6_small/notation/eta.pdf} &=\; \tikzpic{-23}{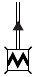}
    \;+\;\tikzpic{-33}{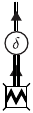}
    \;+\;  \tikzpic{-33}{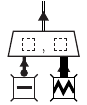},
\end{align}
where we have introduced several new diagrams:
\begin{align*}
&\tikzpic{-15}{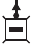}= \cl K;\quad\tikzpic{-20}{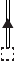}= -ie^{-i\omega_o^\prime t}\!\int dt\,e^{i\omega_o^\prime t}\;\tikzpic{-9}{figsv6_small/notation/dash_box.pdf}\,;\\
&\tikzpic{-13}{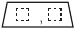}=i\moyal{\,\tikzpic{-9}{figsv6_small/notation/dash_box.pdf}\,}{\,\tikzpic{-9}{figsv6_small/notation/dash_box.pdf}\,}.
\end{align*}
The diagram connected to a \tikzpic{-9}{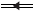} ($-ie^{-i\omega_o^\prime t}\!\int dt\,e^{i\omega_o^\prime t}$\;\tikzpic{-9}{figsv6_small/notation/dash_box.pdf}\,) represents the argument for the placeholder \tikzpic{-9}{figsv6_small/notation/dash_box.pdf}\,. For instance, the first term on the right-hand side of \cref{eq:sc-diagram-mu} corresponds to
\begin{align*}
    \tikzpic{-23}{figsv6_small/notation/eta_bare.pdf} = -ie^{-i\omega_o^\prime t}\!\int dt\,e^{i\omega_o^\prime t} (\cl M e^{-i\omega_o^\prime t}).
\end{align*}
Similarly, the dashed boxes \tikzpic{-9}{figsv6_small/notation/dash_box.pdf} in the Husimi box \tikzpic{-11}{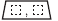} ($i\moyal{\,\tikzpic{-9}{figsv6_small/notation/dash_box.pdf}\,}{ \tikzpic{-9}{figsv6_small/notation/dash_box.pdf}\,}$) should be understood as placeholders, each taking the diagrams under \tikzpic{-11}{figsv6_small/notation/husimi_small.pdf} as the first and second arguments of the Husimi bracket, respectively. For example, 
\begin{align*}
\tikzpic{-26}{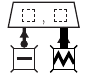} = i\moyal{\cl K}{\cl\mu e^{-i\omega_o^\prime t}}.
\end{align*}

Written in the form of \cref{eq:sc-diagram-mu}, the leading order contribution of the dressed off-resonant excitation \tikzpic{-11}{figsv6_small/notation/eta_small.pdf} is \tikzpic{-11}{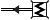}, interpreted as the \textit{bare off-resonant excitation}. Diagrammatically, both \tikzpic{-11}{figsv6_small/notation/eta_small.pdf} and \tikzpic{-11}{figsv6_small/notation/eta_bare_small.pdf} are generated by the mixing networks in \!\tikzpic{-11}{figsv6_small/notation/M_small.pdf}, appended with either a \tikzpic{-9}{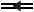} or a \tikzpic{-9}{figsv6_small/notation/prop_small.pdf}, respectively. Therefore, \tikzpic{-9}{figsv6_small/notation/prop_dressed_small.pdf} and \tikzpic{-9}{figsv6_small/notation/prop_small.pdf} are designated as the \textit{dressed propagator} and \textit{bare propagator} of an off-resonant excitation. Their physical significance and evaluation rules will be further detailed in \cref{subsec:bare-virtual,subsec:dressing}. 

Moreover, \cref{eq:self-consistent-K,eq:self-consistent-B,eq:self-consistent-S} can also be translated into diagrammatic form as follows:
\begin{align}\label{eq:sc-diagram-B}
\tikzpic{-21}{figsv6_small/notation/B.pdf} \;&=  \; \tikzpic{-11}{figsv6_small/notation/band_pass.pdf}\exp\Bigg(\tikzpic{-23}{figsv6_small/canonicity/Husimi_zeta.pdf}\Bigg)\tikzpic{-21}{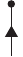}\\
\tikzpic{-13}{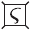} &= \;\tikzpic{-23}{figsv6_small/notation/eta.pdf} \;-\; \tikzpic{-11}{figsv6_small/notation/band_stop.pdf}\exp\Bigg(\tikzpic{-23}{figsv6_small/canonicity/Husimi_zeta.pdf}\Bigg)\tikzpic{-21}{figsv6_small/notation/A.pdf}\;+\;\tikzpic{-28}{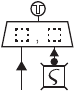}\label{eq:sc-diagram-S}\\
\label{eq:sc-diagram-K}
    \tikzpic{-18}{figsv6_small/notation/K.pdf} \;&=\;\tikzpic{-18}{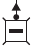} \;+\; \tikzpic{-36}{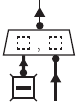} \;-\; \tikzpic{-36}{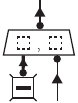}
\end{align}
\end{subequations}
In these diagrams, \cref{eq:sc-diagram-B} corresponds to \cref{eq:self-consistent-B} with an added phase $e^{-i\omega_o^\prime t}$, while \cref{eq:sc-diagram-S} is derived from \cref{eq:self-consistent-S} with a global phase $e^{-i\omega_o^\prime t}$ and a derivative operator $\partial_{\clu A^*}$ applied to both sides. These diagrammatic equations introduce several new diagrams and new rules:
\begin{align}\label{eq:diagram-dict-3}
\begin{split}
  &\tikzpic{-20}{figsv6_small/notation/A.pdf}\;,
 \tikzpic{-20}{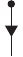}= \clu Ae^{-i\omega_o^\prime t}, \clu A^*e^{i\omega_o^\prime t};\quad
\tikzpic{-14}{figsv6_small/notation/zeta.pdf} = ie^{-i\omega_o^\prime t}\partial_{\clu A^*}\cl S ;\\
&\tikzpic{-15}{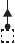}= \int d\clu A^* e^{i\omega_o^\prime t}\;\tikzpic{-9}{figsv6_small/notation/dash_box.pdf}\,.
\end{split}
\end{align}

\cref{eq:sc-diagram-B,eq:sc-diagram-S} provides a self-consistent definition of dressed oscillator excitation \!\!\tikzpic{-9}{figsv6_small/notation/B_small.pdf}~($\clu Be^{-i\omega_o^\prime t}$) with \!\tikzpic{-10}{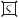}~($ie^{-i\omega_o^\prime t}\partial_{\clu A^*}\cl S$) serving as an intermediate variable. To leading order, the dressed oscillator excitation corresponds to \!\!\tikzpic{-9}{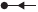}~($\clu Ae^{-i\omega_o^\prime t}$), which we thus interpret as the \textit{bare resonant excitation}. In \cref{subsec:dressing-osc}, we will detail the computation of the dressed resonant excitation. 

Lastly, \cref{eq:sc-diagram-K} establishes a self-consistent relationship for the effective Hamiltonian \!\!\tikzpic{-10}{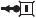}, with its leading order contribution being \!\!\tikzpic{-10}{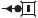} as shown on the right-hand side. Diagrammatically, the unbold diagram \!\!\tikzpic{-10}{figsv6_small/notation/K_bare_small_h.pdf} is similar to the bold diagram \!\!\tikzpic{-10}{figsv6_small/notation/K_small_h.pdf} but replaces the dressed resonant excitation \!\!\tikzpic{-9}{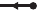} with the bare one \!\!\tikzpic{-9}{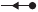}. Algebraically,
\tikzpic{-11}{figsv6_small/notation/K_bare_small_h.pdf} corresponds to
\begin{align}\label{eq:bare-K-integral}
    \tikzpic{-18}{figsv6_small/notation/K_bare.pdf} = \int d\clu A^* e^{i\omega_o^\prime t}(e^{-i\omega_o^\prime t}\cl\Gamma).
\end{align}
Here, when a bare resonant excitation  \!\!\tikzpic{-9}{figsv6_small/notation/As_small_r.pdf} ($\clu A^*e^{i\omega_o^\prime t}$) is appended to the output of a mixing network, which is \!\!\tikzpic{-10.5}{figsv6_small/notation/Gamma_small.pdf} ($e^{-i\omega_o^\prime t}\cl\Gamma$) in the diagram above, the composite diagram is evaluated as an integration of the mixing network with respect to $\clu A^*e^{i\omega_o^\prime t}$ (c.f. the last term in \cref{eq:diagram-dict-3}). Analogously, we can reinterpret \!\!\tikzpic{-10}{figsv6_small/notation/K_small_h.pdf} as ``integrating" over \!\!\tikzpic{-10.5}{figsv6_small/notation/Gamma_small.pdf} with respect to the dressed resonant excitation \!\!\tikzpic{-9}{figsv6_small/notation/Bs_small_r.pdf} ($\clu B^*e^{i\omega_o^\prime t}$), where this modified integration action is formally defined by \cref{eq:sc-diagram-K}. In  \cref{subsec:dressing-osc}, we will discuss the physical significance and evaluation rules of this equation in more detail.

Up to now,  we have translated the algebraic self-consistent equations \cref{eq:self-consistent} at the core of our perturbative method into their diagrammatic form \cref{eq:sc-diagram}. These diagrammatic equations depict nonlinear mixing processes involving dressed excitations at various frequencies, while further elaborating how each dressed quantity relates to its bare counterparts. In the remainder of this section, we will exploit the diagrammatic representations by doing one perturbative iteration step of the self-consistent equations \cref{eq:sc-diagram}. Specifically, in \cref{subsec:order-1,subsec:bare-virtual}, we will explore the structure of the mixing networks generated from these processes through bare diagrams that only involve bare excitations. In \cref{subsec:dressing,subsec:dressing-osc}, we will detail the dressing of bare excitations and their implications.

\subsection{Enter the bare diagrams: order 1}
\label{subsec:order-1}
In the next two subsections, we will solve \cref{eq:sc-diagram} to the leading orders, which only involve the bare form of each type of excitation. The resulting diagram is the bare diagram, which is illustrative to study as it resembles the structures of the dressed diagram representing the full solution of \cref{eq:sc-diagram}. To start, we consider \cref{eq:sc-diagram-Gamma,eq:sc-diagram-M} at the first order, which reads 
\begin{subequations}\label{eq:eom-diagram-order1}
\begin{align}\label{eq:eom-diagram-order1-a}
    \tikzpic{-13}{figsv6_small/notation/Gamma.pdf}\strut^{(1)}\!=\;&
    \tikzpic{-11}{figsv6_small/notation/band_pass.pdf}\tikzpic{-14}{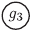}\Bigg(\,
    \tikzpic{-21}{figsv6_small/notation/A.pdf}+\tikzpic{-21}{figsv6_small/notation/As.pdf}
    +\tikzpic{-21}{figsv6_small/notation/xi.pdf}+\tikzpic{-21}{figsv6_small/notation/xis.pdf}\,\Bigg)^2_{\!\scaleto{\varstar}{5pt}}+\;\tikzpic{-30}{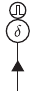},\\
    \tikzpic{-13}{figsv6_small/notation/M.pdf}\strut^{(1)}\!=\;&
    \tikzpic{-11}{figsv6_small/notation/band_stop.pdf}\tikzpic{-14}{figsv6_small/notation/g3.pdf}\Bigg(\,
    \tikzpic{-21}{figsv6_small/notation/A.pdf}+\tikzpic{-21}{figsv6_small/notation/As.pdf}
    +\tikzpic{-21}{figsv6_small/notation/xi.pdf}+\tikzpic{-21}{figsv6_small/notation/xis.pdf}\,\Bigg)^2_{\!\scaleto{\varstar}{5pt}}\label{eq:eom-diagram-order1-b}.
\end{align}
\end{subequations}
To understand the order of the above equations, recall that \tikzpic{-11}{figsv6_small/notation/gm_small.pdf} $= g_m$ is of order ${m-2}$ in the perturbative parameter $q_{\mathrm{zps}}$,  \!\tikzpic{-11}{figsv6_small/notation/delta_small.pdf} $= \delta$ is of order 1, and the drive excitations \tikzpic{-9}{figsv6_small/notation/xi_small.pdf} ($\xi e^{-i\omega_d t}$) \tikzpic{-9}{figsv6_small/notation/xis_small.pdf} ($\xi e^{i\omega_d t}$) are of order 0. The resonant excitations in \cref{eq:eom-diagram-order1} are taken in their bare forms \tikzpic{-9}{figsv6_small/notation/A_small.pdf} ($\clu A e^{-i\omega_o^\prime t}$) and \tikzpic{-9}{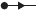} ($\clu A^* e^{i\omega_o^\prime t}$) introduced in \cref{eq:sc-diagram-B}, which are also of order 0. The off-resonant excitations \tikzpic{-11}{figsv6_small/notation/eta_small.pdf} ($\cl\mu e^{-i\omega_o^\prime t}$) and \tikzpic{-11}{figsv6_small/notation/etas_small.pdf} ($\cl\mu^* e^{i\omega_o^\prime t}$) are not included because they are of order 1 (c.f. \cref{eq:pert-structure}). 

In \cref{eq:eom-diagram-order1}, we first note that the last diagram in \cref{eq:eom-diagram-order1-a} is evaluated as $\delta\cl Ae^{-i\omega_o^\prime t}$, which is a direct consequence of the rotating frame at $\omega_o^\prime $, featured in \cref{eq:H-tran}, being detuned from the frequency of the oscillator $\omega_o =\omega_o^\prime + \delta$. This diagram represents a linear process that exists even in the absence of any nonlinearity $g_m$. For the nonlinear part, i.e. the binomial expansions in the right-hand of \cref{eq:eom-diagram-order1-a,eq:eom-diagram-order1-b}, we diagrammatically carry it out by picking two elements within the parenthesis and appending each as an input to the 3-wave mixer in an ordered manner. Specifically, the binomial expansion yields 16 diagrams:
\begin{align}
\label{eq:order-1-sum}
\begin{split}
    &\tikzpic{-20}{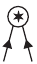}\!\!+\!\!       \tikzpic{-20}{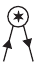}\!\!+\!\!     \tikzpic{-20}{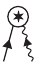}\!\!+\!\!     \tikzpic{-20}{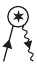}\!\!+\!\!     \tikzpic{-20}{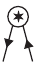}\!\!+\!\!     \tikzpic{-20}{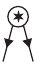}\!\!+\!\!
   \tikzpic{-20}{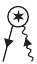}\!\!+\!\!       \tikzpic{-20}{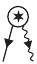}\\
   &\!\!+\!\!\tikzpic{-20}{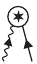}\!\!+\!\!     \tikzpic{-20}{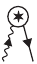}\!\!+\!\!   \tikzpic{-20}{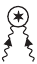}\!\!+\!\!
   \tikzpic{-20}{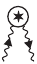}\!\!+\!\!       \tikzpic{-20}{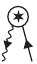}\!\!+\!\!  \tikzpic{-20}{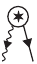}\!\!+\!\!     \tikzpic{-20}{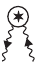}\!\!+\!\!     \tikzpic{-20}{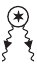},
\end{split}
\end{align} 
where $\varstar$ in each 3-wave mixer indicates the product among the underlying inputs; we have also suppressed both the black dot $\bullet$ associated with each arrow in \cref{eq:eom-diagram-order1} and the mixer label $g_3$. These simplifications introduce no ambivalence to the diagrams as the direction of the inputs can now be inferred from the position of the mixer and the rank of nonlinearity can be inferred from the number of inputs. As we have previously introduced, the types of the inputs are indicated by the associated arrows:\tikzpic{-10}{figsv6_small/notation/straight_arrow_small.pdf} represents the resonant excitation of the oscillator at frequency $\omega_o^\prime$ and\tikzpic{-10}{figsv6_small/notation/wiggly_arrow_small.pdf} represents the drive excitation at frequency $\omega_d$; an excitation traveling into/away from the mixer adds/subtracts the corresponding frequency from the mixer. The output, which will leave at the top of each mixer, is at the frequency $\omega_\text{out}$ that can be read off from the diagram directly by imposing frequency conservation at the mixer element; for example, the output frequencies of the first two diagrams in \cref{eq:order-1-sum} are $2\omega_o^\prime$ and $0$, respectively. 

Each bare diagram in \cref{eq:order-1-sum} is evaluated by multiplying the factor(s) associated with its constituent elements. Importantly, because $\varstar$ is a non-commutative product, we choose to read off the diagram counterclockwise starting from the output. With these rules, \cref{eq:order-1-sum} is evaluated as
\begin{align}
\label{eq:order-1-sum-al}
\begin{split}
    &\quad g_3 e^{-i 2\omega_o^\prime t} \clu A \varstar \clu A + g_3 \clu A  \varstar \clu A^* + g_3 e^{-i (\omega_d+\omega_o^\prime) t} \clu A  \varstar \xi\\
    &+ g_3 e^{i (-\omega_o^\prime+\omega_d)  t} \clu A \varstar \xi^* + g_3\clu A^* \varstar \clu A  + g_3 e^{i 2\omega_o^\prime  t} \clu A^* \varstar \clu A^*\quad
    \\&+ g_3 e^{i (\omega_o^\prime - \omega_d) t} \clu A^*  \varstar\xi + g_3 e^{i ( \omega_o^\prime + \omega_d) t} \clu A^* \varstar \xi^* 
    \\ & + g_3 e^{-i(\omega_o^\prime+\omega_d) t} \xi  \varstar \clu A + g_3 e^{i (\omega_o^\prime - \omega_d) t} \xi \varstar \clu A^*
    \\&+ g_3 e^{-i 2\omega_d t} \xi  \varstar\xi + g_3 \xi  \varstar \xi^* + g_3 e^{i (-\omega_o^\prime+\omega_d) t} \xi^* \varstar \clu A \\
    &+ g_3 e^{i (\omega_o^\prime+\omega_d ) t} \xi^* \varstar \clu A^*  + g_3 \xi^* \varstar\xi + g_3 e^{i  2\omega_d t} \xi^* \varstar \xi^*,
\end{split}\raisetag{4\baselineskip}
\end{align}
and the phase of each term above is simply $e^{-i\omega_\text{out} t}$. 

The output of each diagram in \cref{eq:order-1-sum} will then go through the ``band-pass filter'' \tikzpic{-11}{figsv6_small/notation/band_pass.pdf} or the ``band-stop filter'' \tikzpic{-11}{figsv6_small/notation/band_stop.pdf} in \cref{eq:eom-diagram-order1-a,eq:eom-diagram-order1-b}, respectively. These two filters select the outputs at frequency $\omega_\text{out} = \omega_o^\prime$ and those not, which respectively contribute to \!\tikzpic{-10.5}{figsv6_small/notation/Gamma_small.pdf} and \!\tikzpic{-10}{figsv6_small/notation/M_small.pdf}. Consequently, $\!\tikzpic{-10.5}{figsv6_small/notation/Gamma_small.pdf}^{\,(1)}$ is determined according to:
\begin{align}\label{eq:order1-gamma}
\tikzpic{-18}{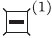} = 
    \begin{cases}
    \tikzpic{-22}{figsv6_small/order1/term7.pdf} + \tikzpic{-22}{figsv6_small/order1/term10.pdf} + \;\tikzpic{-20}{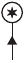} & \omega_o^\prime = \omega_d/2 \\
    \tikzpic{-22}{figsv6_small/order1/term11.pdf} + \;\tikzpic{-20}{figsv6_small/notation/delta_A_abridged.pdf}& \omega_o^\prime = 2\omega_d \\
    \,\,\,\;\tikzpic{-23}{figsv6_small/notation/delta_A_abridged.pdf} & \text{otherwise},
    \end{cases}
\end{align}
where the last term in each case above is just the last term in \cref{eq:eom-diagram-order1-a} and for consistency of style we have suppressed the $\delta$ symbol inside this diagram and replaced it with $\varstar$. Moreover, in the diagrams in \cref{eq:order1-gamma} we have also suppressed \tikzpic{-11}{figsv6_small/notation/band_pass.pdf} for simplicity, since it contributes a trivial factor of 1 with the given driving condition. 

With \!\!\tikzpic{-10.5}{figsv6_small/notation/Gamma_small.pdf}$\,^{(1)}$ determined in \cref{eq:order1-gamma}, \!\tikzpic{-11}{figsv6_small/notation/M_small.pdf}$^{\,(1)}$ consequently corresponds to the remaining diagrams in \cref{eq:order-1-sum}, and thus \cref{eq:sc-diagram-Gamma,eq:sc-diagram-M} are solved at order 1. The effective Hamiltonian \!\tikzpic{-11}{figsv6_small/notation/K_small_h.pdf}$^{\,(1)}$ ($\cl K^{(1)}$) can then be constructed from \!\!\tikzpic{-10.5}{figsv6_small/notation/Gamma_small.pdf}$\,^{(1)}$ ($\cl\Gamma^{(1)}$) using \cref{eq:sc-diagram-K} while only considering the leading order term, i.e. the first one \!\tikzpic{-11}{figsv6_small/notation/K_bare_small_h.pdf} ($\int d\clu A^*e^{i\omega_o^\prime t}\cl\Gamma$) on the right-hand side of the equation. Specifically, we have
\begin{align}
\label{eq:K-1}
\tikzpic{-22}{figsv6_small/notation/K.pdf}^{(1)} = 
    \begin{cases}
    \tikzpic{-24}{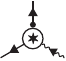} + \tikzpic{-24}{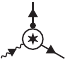}+ \tikzpic{-24}{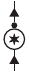} & \omega_o^\prime = \omega_d/2 \\
    \tikzpic{-24}{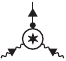} + \tikzpic{-24}{figsv6_small/notation/K_delta_abridged.pdf}& \omega_o^\prime = 2\omega_d \\[15pt]
    \,\,\,\tikzpic{-25}{figsv6_small/notation/K_delta_abridged.pdf} & \text{otherwise}.
    \end{cases}
\end{align}
As illustrated in \cref{eq:bare-K-integral}, the \!\!\tikzpic{-9}{figsv6_small/notation/As_small_r.pdf} ($\clu A^*e^{i\omega_o^\prime t}$) appending to the outcome of a mixer is interpreted as integration over $\clu A^*e^{i\omega_o^\prime t}$. Therefore, \cref{eq:K-1} is evaluated as:
\begin{align}\label{eq:K-1-algebra}
\cl K^{(1)} = 
    \begin{cases}
    g_3\xi \clu A^{*2} + g_3\xi^*\clu A^2 + \delta \clu A^*\clu A & \omega_o^\prime = \omega_d/2 \\
    g_3\xi^2\clu A^* + g_3\xi^{*2}\clu A + \delta \clu A^*\clu A  & \omega_o^\prime = 2\omega_d \\
    \,\,\, \delta \clu A^*\clu A  & \text{otherwise},
    \end{cases}
\end{align}
where we have explicitly carried out the Husimi product $\varstar$ in the relevant terms in \cref{eq:K-1} according to its definition in \cref{eq:Husimi-transform}. We also remark that one can alternatively construct the effective Hamiltonian through the EOM over $\clu A^*$ (instead of that over $\clu A$ as in \cref{eq:eom-phase-expand}) and obtain the corresponding diagram $\tikzpic{-11}{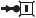} (\cl K^*)$. Since the effective Hamiltonian is real, we have $\tikzpic{-11}{figsv6_small/notation/Ks_small_h.pdf}=\tikzpic{-11}{figsv6_small/notation/K_small_h.pdf}$. 

We have so far introduced a set of diagrammatic notations and operations to solve the equation of motion \cref{eq:eom-phase-expand} up to leading order. Before discussing the second order, we will take a detour and present an alternative representation of the diagrams discussed thus far. This new representation explicitly displays the classical and quantum components of a diagram by expanding the mixing vertices \tikzpic{-11}{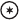} in the diagram. Therefore, we refer to this new type of diagrams as the \textit{expanded diagrams}, whereas the ones with \tikzpic{-11}{figsv6_small/notation/star_small.pdf} as the mixing vertices, e.g. those in \cref{eq:order-1-sum,eq:order1-gamma,eq:K-1}, are referred to as the \textit{unexpanded diagrams}. 

To derive the expanded diagrams, we first recall that the inputs of a mixing vertex \tikzpic{-11}{figsv6_small/notation/star_small.pdf} are associated by Husimi product $\varstar$ defined as
\begin{align}
\label{eq:six-star-A}
\begin{split}
    \cl f \varstar \cl g = \sum_{k\ge0} \frac{\hbar^k}{k!}\cl f\,(\overleftarrow{\partial}_{\!\!\clu A} \overrightarrow{\partial}_{\!\!\clu A^*})^k \cl g
\end{split}
\end{align}
for some generic phase-space functions $\cl f$ and $\cl g$. The classical component of the Husimi product, i.e. the $k=0$ term in \cref{eq:six-star-A}, is just an ordinary commutative product. The quantum components, i.e. those terms with $k>0$, involve $k$ number of $\hbar\overleftarrow{\partial}_{\!\!\clu A} \overrightarrow{\partial}_{\!\!\clu A^*}$ bidifferential operators and introduce the non-commutativity. 

With these observations, we can similarly expand each diagram in \cref{eq:order-1-sum} into a classical diagram and quantum diagrams. For example, the first two diagrams in \cref{eq:order-1-sum} can be expanded as
\begin{align}
\begin{split}
    \tikzpic{-20}{figsv6_small/order1/term1.pdf} = \underbrace{\tikzpic{-20}{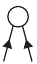}\vphantom{\tikzpic{-20}{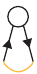}}}_{\substack{\text{classical}}}, \qquad
    \tikzpic{-20}{figsv6_small/order1/term2.pdf} =
    \underbrace{\tikzpic{-20}{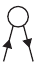}\vphantom{\tikzpic{-20}{figsv6_small/order1/term2_qm.pdf}}}_{\substack{\text{classical}}} \quad+\!\!\!\! \underbrace{\tikzpic{-20}{figsv6_small/order1/term2_qm.pdf}}_{\substack{\text{quantum}}}.
\end{split}
\label{eq:unabridged-order1}
\end{align}
Here, we construct the classical diagram simply by replacing the mixing vertex \tikzpic{-11}{figsv6_small/notation/star_small.pdf} in an unexpanded diagram with \tikzpic{-11}{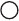} to indicate that its inputs are associated with a regular commutative product. To construct the quantum diagrams, we diagrammatically interpret each factor $\hbar\overleftarrow{\partial}_{\!\!\clu A} \overrightarrow{\partial}_{\!\!\clu A^*}$ in \cref{eq:six-star-A} as forming a \textit{quantum bond}, i.e. the orange bond in \cref{eq:unabridged-order1}, between two inputs in a classical diagram. The non-commutativity of $\hbar\overleftarrow{\partial}_{\!\!\clu A} \overrightarrow{\partial}_{\!\!\clu A^*}$ is reflected by the directionality of the quantum bond --- the arrow to the left of the bond represents a resonant excitation traveling into the mixer and is associated with $\clu A$, while the arrow to the right represents a resonant excitation traveling away and is associated with $\clu A^*$. It is easy to verify that, at $k$-th order in $\hbar$, the combinatorial factor resulting from \cref{eq:six-star-A} equals to the number of conﬁgurations with $k$ quantum bonds formed. In the two examples in \cref{eq:unabridged-order1}, no quantum bond can be formed in the first classical diagram, and thus its associated quantum diagram is absent. In the second classical diagram, there is only one possible configuration for forming such a bond. Later in the text, we will examine more complicated examples involving more than two resonant excitations in a diagram. Since the Husimi product is associative, this diagrammatic interpretation of \cref{eq:six-star-A} is also applicable in those more involved cases.

The expanded diagrams are evaluated the same way as the unexpanded ones but with ordinary multiplication as the underlying product. In addition, each quantum bond is associated with a factor of $\hbar$ and each pair of connected arrows associated with a factor of $1$. For example, \cref{eq:unabridged-order1} is evaluated as 
\begin{align}
\begin{split}
        &g_3 e^{-i \omega_o^\prime t} \clu A \varstar \clu A =  g_3 e^{-i \omega_o^\prime t} \clu A^2, \\
    &g_3 e^{i \omega_o^\prime t} \clu A  \varstar \clu A^* =  g_3 e^{i \omega_o^\prime t}\clu A^*\clu A + \hbar g_3 e^{i \omega_o^\prime t}.
    \end{split}\label{eq:unabridged-order1-algebra}
\end{align}
Note that in the case of a commutative product such as the terms in the right-hand side of \cref{eq:unabridged-order1-algebra}, we have chosen to write the $\clu A^*$-factors to the left of the $\clu A$-factors, so that the Hilbert space representation can be readily recovered by replacing $\clu A$ with $\op{\cu A}$, and $\clu A^*$ with $\op{\cu A^\dagger}$.

Lastly, we would like to make two remarks about the expanded diagram. First, the specific order of the inputs of each vertex in an expanded diagram is not relevant since their underlying product is a commutative ordinary product. Therefore, when evaluating all the unexpanded diagrams in the effective Hamiltonian, one can simply evaluate the \textit{unordered} expanded diagram, which is the class of topologically equivalent ordered expanded diagrams that we have treated thus far, and multiply the result with an additional factor of the total number of topologically equivalent diagrams associated with the unordered diagram. Second, we note that if one chooses to develop expanded diagrams in a different phase-space representation of quantum mechanics (such as the Wigner representation with Groenewold star product ($\star$) \cite{moyal1949,zachos2005} as the underlying product), the classical diagrams remain unchanged, whereas the quantum diagrams and the evaluation rules for their quantum bonds must be adapted accordingly.

\subsection{Enter the off-resonant excitations: order 2}
\label{subsec:bare-virtual}
We now return to the task of solving \cref{eq:sc-diagram} iteratively. At order two, \cref{eq:sc-diagram-Gamma,eq:sc-diagram-M} read
\begin{subequations}\label{eq:eom-diagram-order2}
\begin{align}\label{eq:eom-diagram-order2-a}
\begin{split}
    \tikzpic{-13}{figsv6_small/notation/Gamma.pdf}\strut^{(1)}\!=\;&
    \tikzpic{-11}{figsv6_small/notation/band_pass.pdf}\tikzpic{-14}{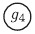}\Bigg(
    \tikzpic{-21}{figsv6_small/notation/A.pdf}+\tikzpic{-21}{figsv6_small/notation/As.pdf}
    +\tikzpic{-21}{figsv6_small/notation/xi.pdf}+\tikzpic{-21}{figsv6_small/notation/xis.pdf}\Bigg)^3_{\!\scaleto{\varstar}{5pt}}\\ 
    &+ \tikzpic{-11}{figsv6_small/notation/band_pass.pdf}\tikzpic{-14}{figsv6_small/notation/g3.pdf}\Bigg(
    \tikzpic{-21}{figsv6_small/notation/A.pdf}+\!\!\tikzpic{-23}{figsv6_small/notation/eta_bare.pdf}
    \!\!+\tikzpic{-21}{figsv6_small/notation/As.pdf}+\!\! \tikzpic{-23}{figsv6_small/notation/eta_bare.pdf}\!+\!
    \tikzpic{-21}{figsv6_small/notation/xi.pdf}
    +\tikzpic{-21}{figsv6_small/notation/xis.pdf}\Bigg)^{\!2}_{\!\scaleto{\varstar}{5pt}}\!\rule{0pt}{22pt}^{(1)},
\end{split}\\\label{eq:eom-diagram-order2-b}
\begin{split}
    \tikzpic{-13}{figsv6_small/notation/M.pdf}\strut^{(1)}\!=\;&
    \tikzpic{-11}{figsv6_small/notation/band_stop.pdf}\tikzpic{-14}{figsv6_small/notation/g4.pdf}\Bigg(
    \tikzpic{-21}{figsv6_small/notation/A.pdf}+\tikzpic{-21}{figsv6_small/notation/As.pdf}
    +\tikzpic{-21}{figsv6_small/notation/xi.pdf}+\tikzpic{-21}{figsv6_small/notation/xis.pdf}\Bigg)^3_{\!\scaleto{\varstar}{5pt}}\\ 
    &+ \tikzpic{-11}{figsv6_small/notation/band_stop.pdf}\tikzpic{-14}{figsv6_small/notation/g3.pdf}\Bigg(
    \tikzpic{-21}{figsv6_small/notation/A.pdf}+\!\!\tikzpic{-23}{figsv6_small/notation/eta_bare.pdf}
    \!\!+\tikzpic{-21}{figsv6_small/notation/As.pdf}+\!\! \tikzpic{-23}{figsv6_small/notation/eta_bare.pdf}\!+\!
    \tikzpic{-21}{figsv6_small/notation/xi.pdf}
    +\tikzpic{-21}{figsv6_small/notation/xis.pdf}\Bigg)^{\!2}_{\!\scaleto{\varstar}{5pt}}\!\rule{0pt}{22pt}^{(1)},
\end{split}
\end{align}
\end{subequations}
where for the resonant and off-resonant excitation only their bare constituents are relevant at this order.\footnote{For the dressed resonant excitation, the correction to its bare counterpart is $\mathcal O(q_\text{zps}^2)$ and thus irrelavent at this order. We will discuss this in more detail in \cref{subsec:dressing-osc}.} In the first multinomial expansion in the right-hand side of \cref{eq:eom-diagram-order2-a,eq:eom-diagram-order2-b}, a 4-wave mixer \!\tikzpic{-11}{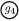} mediates the mixing between three inputs selected from \tikzpic{-9}{figsv6_small/notation/A_small}, \tikzpic{-9}{figsv6_small/notation/As_small}, \tikzpic{-9}{figsv6_small/notation/xi_small.pdf}, and \tikzpic{-9}{figsv6_small/notation/xis_small}. The multinomial expansion results in a sum of $4^3=64$ bare diagrams, each represented in an unexpanded form. We evaluate these diagrams using the same approach as discussed in \cref{subsec:order-1}, and therefore, do not repeat the process here.

In the second multinomial expansion, a 3-wave mixer \tikzpic{-11}{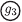} mediates the mixing between two inputs chosen from the six terms inside the parentheses. These terms consist of two bare resonant excitations and two drive excitations, which we have discussed, and the off-resonant excitations \!\tikzpic{-11}{figsv6_small/notation/eta_bare_small.pdf} and \!\tikzpic{-11}{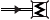}, which we will elaborate shortly. Furthermore, the superscript $(1)$ on the parentheses of the second multinomial expansion selects only first-order terms resulting from the expansion. Thus, the entire mixing process, which involves another factor $g_3=\mathcal O(q_\text{zps})$, is at second order in $q_\text{zps}$.

Let us now consider the off-resonant excitations participating in the 3-wave mixing process in \cref{eq:eom-diagram-order2}. For the diagram \!\tikzpic{-11}{figsv6_small/notation/eta_bare_small.pdf}, it should be understood as being constructed from  \tikzpic{-11}{figsv6_small/notation/M_small.pdf} ($\cl M$) and a bare propagator \tikzpic{-9}{figsv6_small/notation/prop_small.pdf}~($-ie^{-i\omega_o^\prime t}\!\int dt\,e^{i\omega_o^\prime t}$\;\tikzpic{-9}{figsv6_small/notation/dash_box.pdf}\,), which is defined in \cref{eq:sc-diagram-mu} and the discussion below it. Considering the mixing processes constituting  \tikzpic{-11}{figsv6_small/notation/M_small.pdf}~($\cl M$) each corresponds to an oscillating function with phase $e^{-i\omega_\text{out}^\prime t}$ (e.g. \cref{eq:order-1-sum,eq:order-1-sum-al}), the bare propagator associated with each participating process is thus of the form $1/(\omega_\text{out}-\omega_o^\prime)$. This definition justifies the notation of the propagator; intuitively, the propagator is weaker when the off-resonant excitation is more detuned from  the oscillator. With this definition, the first two diagrams in \cref{eq:order-1-sum}, for example,  would yield the off-resonant excitation at order 1 as
\begin{align}
\label{eq:order-2-eta}
\begin{split}
\tikzpic{-32}{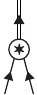} = \frac{g_3e^{-2i\omega_o^\prime t}}{2\omega_o^\prime - \omega_o^\prime} \clu A\varstar \clu A\,, \quad \tikzpic{-32}{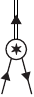} = \frac{g_3}{0 - \omega_o^\prime} \clu A\varstar \clu A^*.
\end{split}
\end{align}
We reiterate that a propagator always remains finite by construction: any diagram with $\omega_{\rm{out}} = \omega_o^\prime$ is collected by\tikzpic{-10.5}{figsv6_small/notation/Gamma_small.pdf}$^{(1)}$ and does not propagate into another mixing process as an input.

Given a diagram constituting \!\tikzpic{-10}{figsv6_small/notation/eta_bare_small}, there exists a corresponding one constituting \!\tikzpic{-10}{figsv6_small/notation/etas_bare_small} whose algebraic expression is just the complex conjugate of the former. Diagrammatically, the complex conjugate expression is formed and evaluated in the same way as the direct expression with modifications stipulating that: (1) the travelling direction of all arrows are inverted, and (2) the inverted inputs of each mixer are reflected about the line drawn through the output of the mixer. These rules apply to constructing the complex conjugate of any diagram. For example, the complex conjugate of \cref{eq:order-2-eta} is
\begin{align}
\label{eq:order-2-etas}
\begin{split}
\tikzpic{-32}{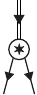} = \frac{g_3e^{2i\omega_o^\prime t}}{2\omega_o^\prime - \omega_o^\prime} \clu A^*\!\varstar \clu A^*,\;\tikzpic{-32}{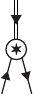} = \frac{g_3}{0 - \omega_o^\prime} \clu A\varstar \clu A^*
\end{split}
\end{align}
and they should be understood as the off-resonant excitation traveling in the opposite direction compared with those in \cref{eq:order-2-eta}. 

At order 1, each off-resonant diagram (at frequency $\omega_\text{out}\ne\omega_o^\prime$) in \cref{eq:order-1-sum} yields an input constituting \!\tikzpic{-10}{figsv6_small/notation/eta_bare_small} and also the corresponding conjugate input constituting  \!\tikzpic{-10}{figsv6_small/notation/etas_bare_small}. Once all of them are determined with the above procedure, one is ready to explicitly draw the mixing processes resulting from the $g_3$ multinomial expansion in \cref{eq:eom-diagram-order2}. For example, a mixing process involving the first diagram in \cref{eq:order-2-etas} as an input is evaluated as
\begin{align}\label{eq:order2-term1}
\begin{split}
    \tikzpic{-20}{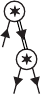} \!\!&= g_3(\clu Ae^{-i\omega_o^\prime t})\!\varstar\!\Big(\frac{1}{2\omega_o^\prime -\omega_o^\prime }\times\\[-25pt]
    &\qquad\qquad\qquad\qquad\qquad g_3(\clu A^*e^{i\omega_o^\prime t})\!\varstar\!(\clu A^*e^{i\omega_o^\prime t})\Big) \\[0pt]
    &=\frac{g_3^2e^{i\omega_o^\prime t}}{\omega_o^\prime}\clu A\varstar \clu A^*\varstar \clu A^*.
\end{split}\raisetag{1\baselineskip}
\end{align}
The diagram in \cref{eq:order2-term1} should be understood as a graphical representation of cascaded mixers, whose output is that of the 3-wave mixer at the top. Such a diagram is evaluated in the same way as those shown in order one and starting from the output of the diagram; this is illustrated by the first line in right-hand side of \cref{eq:order2-term1}.

Similarly to \cref{eq:unabridged-order1}, a bare diagram involving off-resonant excitations can also be represented in the expanded form. Specifically, following the same rules discussed in \cref{subsec:order-1}, \cref{eq:order2-term1} can be rewritten as 
\begin{align}\label{eq:order2-term1-unabridged}
    \tikzpic{-30}{figsv6_small/order2/term1.pdf} =
    \tikzpic{-30}{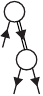} +
    \tikzpic{-30}{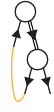} +
    \tikzpic{-30}{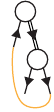}, 
\end{align}
and with each diagram evaluated explicitly the above equation reads
\begin{align}
\begin{split}
    \frac{g_3^2e^{i\omega_o^\prime  t}}{\omega_o^\prime}\clu A\varstar \clu A^*\varstar \clu A^* = & \frac{g_3^2e^{i\omega_o^\prime  t}}{\omega_o^\prime }\clu A^{*2}\clu A 
    +  \frac{\hbar g_3^2e^{i\omega_o^\prime t}}{\omega_o^\prime}\clu A^*\\
    &+  \frac{\hbar g_3^2e^{i\omega_o^\prime t}}{\omega_o^\prime}\clu A^*.
\end{split}\raisetag{1\baselineskip}
\end{align}

Up to now, we have introduced the bare diagrams by iteratively solving \cref{eq:sc-diagram} at the first two orders. At each iteration, the multinomial expansions in the right-hand side of \cref{eq:sc-diagram-Gamma,eq:sc-diagram-M} generate mixing diagrams, among which the ones at frequency $\omega_o^\prime$ and those not contribute to \!\tikzpic{-10.5}{figsv6_small/notation/Gamma_small.pdf} ($\cl\Gamma$) and \tikzpic{-10}{figsv6_small/notation/M_small.pdf} ($\cl M$) in the left-hand side, respectively. With \!\tikzpic{-10.5}{figsv6_small/notation/Gamma_small.pdf} ($\cl\Gamma$) determined, the effective Hamiltonian \tikzpic{-10}{figsv6_small/notation/K_small_h.pdf} ($\cl K$) can be computed via \cref{eq:sc-diagram-K}, which to leading order is \tikzpic{-10}{figsv6_small/notation/K_bare_small_h.pdf} ($\int d\clu A^*e^{i\omega_o^\prime t}\cl\Gamma$). Meanwhile, the diagram \tikzpic{-10}{figsv6_small/notation/M_small.pdf} ($\cl M$) at this order will produce off-resonant excitations following \cref{eq:sc-diagram-mu}, which to leading order are the bare ones \!\tikzpic{-10}{figsv6_small/notation/eta_bare_small}~($e^{-i\omega_o^\prime t}\int dt e^{i\omega_o^\prime t}\cl M$) and \!\tikzpic{-10}{figsv6_small/notation/etas_bare_small}~($e^{i\omega_o^\prime t}\int dt e^{-i\omega_o^\prime t}\cl M^*$).  

As the iteration proceeds, the off-resonant excitations at lower order feeds back to the mixing processes at higher order, and the generated diagrams are of the same type as those in \cref{eq:eom-diagram-order1,eq:order2-term1} but involving more and more cascaded mixers. In the notation of graph theory, these bare diagrams in the unexpanded form\footnote{When an diagram is expressed in its expanded form, e.g. as in the right-hand side of \cref{eq:order2-term1-unabridged}, the underlying graph of a classical diagram is a rooted tree while that of a quantum diagram contains one or more loops.} share a simple graphical structure of  \textit{rooted trees} (see \cref{app:gloss} for the exact definition). The root vertex in the tree is the output of the entire diagram, the internal vertices are the mixers, the external edges are drive excitations or bare resonant excitations, and the internal edges are the propagators of off-resonant excitations. 

It is important to recognize that the dressed diagrams, corresponding to the full solution of \cref{eq:sc-diagram}, share the same structure as the bare diagrams but with all bare quantities replaced but their dressed counterparts. These dressed quantities are determined by \cref{eq:sc-diagram-mu,eq:sc-diagram-B,eq:sc-diagram-S,eq:sc-diagram-K}. In the rest of the section, we will elaborate on the evaluation rule and the physical significance of these dressed quantities.

\subsection{Dressed off-resonant excitation}
\label{subsec:dressing}

In this subsection, we discuss the dressed off-resonant excitation, represented as \!\tikzpic{-11}{figsv6_small/notation/eta_small}, along with its corresponding dressed propagator, depicted by \tikzpic{-9}{figsv6_small/notation/prop_dressed_small.pdf}. These concepts are formally defined through the self-consistent equation presented in \cref{eq:sc-diagram-mu}:
\begin{align*}
\tikzpic{-23}{figsv6_small/notation/eta.pdf} &=\; \tikzpic{-23}{figsv6_small/notation/eta_bare.pdf}
    \;+\;\tikzpic{-33}{figsv6_small/dressed_off/eta_dressing2.pdf}
    \;+\;  \tikzpic{-33}{figsv6_small/dressed_off/eta_dressing1_K.pdf},
\end{align*}
which to the leading order is just the bare off-resonant excitation \!\tikzpic{-11}{figsv6_small/notation/eta_bare_small} elaborated in \cref{subsec:bare-virtual}. To shed lights on the physical meaning of the dressed off-resonant excitation, we first rewrite the self-consistent equation \cref{eq:sc-diagram-mu} into two alternative forms. The first one is an infinite series:
\begin{align}
    \tikzpic{-25}{figsv6_small/notation/eta.pdf}  &= \tikzpic{-25}{figsv6_small/notation/eta_bare.pdf}+\tikzpic{-35}{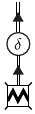}
    \,+\,  \tikzpic{-35}{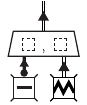}
    \;+  \;\tikzpic{-45}{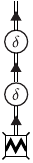}
    +\;\tikzpic{-45}{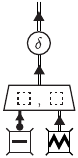}\nonumber\\[-8pt]
    &\quad+\quad\tikzpic{-45}{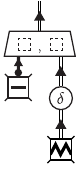}
    \quad+\quad\tikzpic{-45}{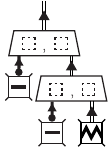} \quad+\quad\cdots,\label{eq:eta-dressed-3}\raisetag{\baselineskip}
\end{align}
achieved by expanding  \cref{eq:sc-diagram-mu} in a self-consistent manner, i.e. successively substituting each \!\tikzpic{-11}{figsv6_small/notation/eta_small} component appearing in the right-hand of \cref{eq:sc-diagram-mu} with all the tree diagrams in the right-hand. Further, by identifying that \cref{eq:eta-dressed-3} is a series of the same form as the Taylor series of $1/(1-x)$, we can succinctly express \cref{eq:eta-dressed-3} as
\begin{align}
    \tikzpic{-25}{figsv6_small/notation/eta.pdf}&=\frac{1}{\;\tikzpic{-20}{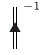} - \tikzpic{-14}{figsv6_small/notation/delta.pdf} - \;\tikzpic{-23}{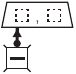}\;} \,\,\tikzpic{-15}{figsv6_small/notation/M.pdf}.\label{eq:eta-dressed-4}
\end{align}

\cref{eq:eta-dressed-3,eq:eta-dressed-4} each defines the dressed off-resonant excitation in a form that is useful for evaluating and interpreting this physical quantity, respectively. We first focus on \cref{eq:eta-dressed-4} to discuss the physical meaning of dressing an off-resonant excitation and will discuss its evaluation later. In particular, from \cref{subsec:order-1}, we note that \tikzpic{-11}{figsv6_small/notation/M_small.pdf} corresponds to an oscillating function containing various Fourier components
\begin{align}\label{eq:seed-network}
\begin{split}
    &\tikzpic{-14}{figsv6_small/notation/M.pdf} = \cl M (\clu A, \clu A^*, t) =  \sum_k \cl M_k (\clu A, \clu A^*, t),\\
    &\cl M_k (\clu A, \clu A^*, t) = \cl f_k(\clu A, \clu A^*)e^{-i\omega_{\text{out}, k}t}
\end{split}
\end{align}
where $\cl f_k(\clu A, \clu A^*)$ is a polynomial function of the phase-space coordinates with no explicit time dependence and $\omega_{\text{out}, k}\ne\omega_o^\prime$ is the frequency of the off-resonant excitation output from the corresponding diagram (c.f. \cref{eq:order-2-eta}). Besides the fast-oscillating phase $e^{-i\omega_{\text{out},k} t}$ in \cref{eq:seed-network}, it is also important to recognize that $\cl f (\clu A, \clu A^*)$ contains an implicit time dependence from the slow dynamics of $\clu A$ and $\clu A^*$ generated by the effective Hamiltonian $\cl K$. In particular, the equation of motion for \tikzpic{-11}{figsv6_small/notation/M_small.pdf}~($\cl M$) in diagrammatic and algebraic forms, respectively, reads
\begin{subequations}\label{eq:seed-network-eom}
\begin{align}
    d_t \tikzpic{-15}{figsv6_small/notation/M.pdf} &= \partial_t \tikzpic{-15}{figsv6_small/notation/M.pdf} - \moyal{\tikzpic{-21}{figsv6_small/notation/K.pdf}}{\tikzpic{-15}{figsv6_small/notation/M.pdf}}\\
    \begin{split}
        \sum_kd_t\cl M_k &= \sum_k\left(\partial_t \cl M_k  - \moyal{\cl K}{\cl M_k}\right)\\
        &= \sum_k\left(-i\omega_{\text{out},k}\cl M_k  - \moyal{\cl K}{\cl M_k}\right).
    \end{split}
\end{align}
\end{subequations}
From the above, it is clear that, heuristically,\footnote{This is a heuristically expression because dividing operation is not formally defined in the phase space.} $\omega_{\text{out},k} =i \partial_t \cl M_k/ \cl M_k$
corresponds to the bare frequency of the off-resonant excitation stemming from its explicit time dependence. It is therefore natural to define a ``dressed frequency'' imitating the same form of the bare one as $\tilde{\omega}_{\text{out},k} = i d_t \cl M_k/ \cl M_k$, which corresponds to the ``true'' frequency of the off-resonant excitation output by the constituent diagram in \!\tikzpic{-11}{figsv6_small/notation/M_small.pdf}. The frequency dressing is $\delta \omega_{\mathrm{out},k} = \tilde{\omega}_{\text{out},k} - {\omega}_{\text{out},k} =  - i\moyal{\cl K}{\cl M_k}/\cl M_k$, whose diagrammatic representation, remarkably, is just the last term in the denominator of \cref{eq:eta-dressed-4}. We remind the reader that the first term $($\!\!\tikzpic{-9}{figsv6_small/notation/prop_small.pdf}$)^{-1}$ in the denominator is evaluated as $\omega_\text{out}-\omega_o^\prime$, the detuning between the bare off-resonant and resonant excitations, while the latter is itself detuned from the oscillator's natural oscillation frequency by $\delta = \omega_o-\omega_o^\prime$. Therefore the full denominator in \cref{eq:eta-dressed-4} simply corresponds to $\tilde{\omega}_\text{out} - \omega_o$, the detuning between the dressed off-resonant excitation and the natural frequency of the oscillator, and the inversion of this dressed detuning defines the dressed propagator.\footnote{We also note that one can further simplify \cref{eq:eta-dressed-4} by redefining the bare off-resonant excitation as \tikzpic{-8}{figsv6_small/notation/prop_small.pdf} $:=1/(\omega_\text{out}-\omega_o)$ and then removing \tikzpic{-10}{figsv6_small/notation/delta_small.pdf}. Consequently \cref{eq:sc-diagram-mu,eq:eta-dressed-3} are also simplified since any term involving \tikzpic{-10}{figsv6_small/notation/delta_small.pdf} in the equations is removed. In the rest of this section, we continue to follow the definition in \cref{eq:eta-dressed-4} as it is.\label{fn:dressed-prop}} 

We have so far introduced the definition of dressed propagator in three different forms ---  as a self-consistent relation in \cref{eq:sc-diagram-mu}, as an infinite series in \cref{eq:eta-dressed-3}, and as a simple fraction in \cref{eq:eta-dressed-4}. Akin to Feynman diagrams, evaluating a dressed diagram involves expanding each dressed propagator in the diagram as the series in \cref{eq:eta-dressed-3} and further evaluating each resulting diagram individually. We evaluate the first diagram \!\tikzpic{-11}{figsv6_small/notation/eta_bare_small.pdf} in the series following the procedure discussed in \cref{subsec:bare-virtual} when the diagram constituting \tikzpic{-11}{figsv6_small/notation/M_small.pdf} is specified. We then evaluate the second diagram and the fourth diagram by multiplying the expression corresponding to the first diagram with one or two extra factors of $\delta/(\omega_\text{out} - \omega_o^\prime )$, each contributed by a diagrammatic component \tikzpic{-11}{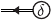}. For the rest of diagrams explicitly displayed in \cref{eq:eta-dressed-3}, each of them contains the component \tikzpic{-11}{figsv6_small/notation/husimi_small.pdf}$=i\moyal{\,\tikzpic{-9}{figsv6_small/notation/dash_box.pdf}\,}{\,\tikzpic{-9}{figsv6_small/notation/dash_box.pdf}\,}$. Evaluating this type of diagram involves non-associative diagrammatic operations, stemming from the Husimi bracket, which we now discuss. 

We take the third term in the series in \cref{eq:eta-dressed-3} as an example. This term is evaluated as another series of diagrams obtained by specifying the particular diagram constituting  \tikzpic{-10}{figsv6_small/notation/M_small.pdf} and writing out each diagram constituting \!\tikzpic{-10}{figsv6_small/notation/K_small_h.pdf}. A term in such a series, for example, reads
\begin{align}\label{eq:eta-dressing-husimi}
    \tikzpic{-50}{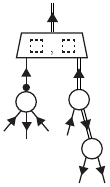},
\end{align}
where for the diagram constituting \tikzpic{-11}{figsv6_small/notation/M_small.pdf} we have chosen the classical diagram in \cref{eq:order2-term1} and for \!\tikzpic{-11}{figsv6_small/notation/K_small_h.pdf} we have chosen a term resulting from the first multinomial expansion in \cref{eq:eom-diagram-order2}. With \cref{eq:eta-dressing-husimi} written as it is, the algebraic expression associated with the left and right subdiagrams under \tikzpic{-11}{figsv6_small/notation/husimi_small.pdf}$=i\moyal{\,\tikzpic{-9}{figsv6_small/notation/dash_box.pdf}\,}{\,\tikzpic{-9}{figsv6_small/notation/dash_box.pdf}\,}$ should be understood as the first and second arguments of the Husimi bracket. Therefore, \cref{eq:eta-dressing-husimi} is evaluated as 
\begin{align}\label{eq:eta-dressing-value}
\begin{split}
\frac{i}{-2\omega_o^\prime}&\moyal{\frac{g_4}{2}\clu A^{*2}\clu A^2}{\frac{g_3^2e^{i\omega_o^\prime t}}{-2\omega_o^{\prime2}}\clu A^{*2} \clu A}\\
&\qquad\qquad=\,\frac{g_3^2g_4}{4\omega_o^{\prime3}}(\clu A^{*3}\clu A^2+\hbar\clu A^{*2}\clu A)e^{i\omega_o^\prime t}.
\end{split}
\end{align}
We note that the non-associativity of the Husimi bracket manifests in the rule that each of the two subdiagrams below a \tikzpic{-11}{figsv6_small/notation/husimi_small.pdf} must be treated as a whole. If a subdiagram itself contains another \tikzpic{-11}{figsv6_small/notation/husimi_small.pdf} the \tikzpic{-11}{figsv6_small/notation/husimi_small.pdf} within the subdiagram should be evaluated first, and then the entire subdiagram is treated as an argument of the upper \tikzpic{-11}{figsv6_small/notation/husimi_small.pdf}. For example, the last term of \cref{eq:eta-dressed-3} is evaluated as 
\begin{align*}
\tikzpic{-45}{figsv6_small/dressed_off/eta_dressing3.pdf} =  \int dt\moyal{\cl K}{\int dt\moyal{\cl K}{\int dt \cl M e^{-i\omega_o^\prime t}}}.
\end{align*}

The evaluation rules associated with \tikzpic{-11}{figsv6_small/notation/husimi_small.pdf}$=i\moyal{\,\tikzpic{-9}{figsv6_small/notation/dash_box.pdf}\,}{\,\tikzpic{-9}{figsv6_small/notation/dash_box.pdf}\,}$ elaborated above enable us to systematically evaluate any diagram containing dressed propagators. In the remainder of this subsection, we introduce the expanded representation of the diagrams involving \tikzpic{-11}{figsv6_small/notation/husimi_small.pdf}. Similar to \cref{eq:unabridged-order1}, this representation explicitly exhibits the classical and quantum components of a dressed propagator by expanding \tikzpic{-11}{figsv6_small/notation/husimi_small.pdf} within it. Furthermore, this representation replaces the complex algebraic operations underlying the Husimi bracket with simple diagrammatic components.

We use the diagram in \cref{eq:eta-dressed-3} as an example to elaborate on this expanded representation. By noting that 
\begin{align}\label{eq:ihusimi}
i\moyal{\cl f}{\cl g} = \frac{1}{\hbar}\cl f\varstar\cl g-\frac{1}{\hbar}\cl g\varstar\cl f,
\end{align}
we first rewrite the diagram in \cref{eq:eta-dressed-3} in a way resembling \cref{eq:ihusimi} as
\begin{align}\label{eq:eta-dressing_abridged}
\tikzpic{-43}{figsv6_small/dressed_off/term_husimi.pdf} =\quad\tikzpic{-40}{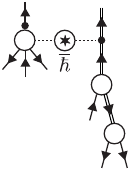} \;- \quad\tikzpic{-40}{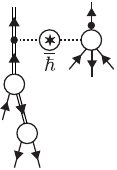}.\raisetag{1.5\baselineskip}
\end{align}
Each diagram in the right-hand side of \cref{eq:eta-dressing_abridged} consists of two subdiagrams, which are essentially the same as those appearing under \tikzpic{-11}{figsv6_small/notation/husimi_small.pdf} on the left-hand side, but with the order of the subdiagrams permuted. It is important to note that the left subdiagram under \tikzpic{-11}{figsv6_small/notation/husimi_small.pdf} contributes to \!\tikzpic{-11}{figsv6_small/notation/K_small_h.pdf}~$(\cl K)$. In the first diagram in the right-hand side of \cref{eq:eta-dressing_abridged}, this contribution is written as its conjugate, which contributes to\!\tikzpic{-11}{figsv6_small/notation/Ks_small_h.pdf}~$(\cl K^*)$. This substitution is made for the convenience of later diagrammatic operations and is valid since $\cl K = \cl K^*$.

For each of the diagrams in the right-hand side of \cref{eq:eta-dressing_abridged}, its two subdiagrams are related by a dashed line decorated by $\tikzpic{-11}{figsv6_small/notation/star_small.pdf}/\hbar$ resembling the $\varstar$ and the $1/\hbar$ factor \cref{eq:ihusimi}. From the definition of star product in \cref{eq:six-star}, we know that  $\tikzpic{-11}{figsv6_small/notation/star_small.pdf}/\hbar$ is algebraically associated with the operation
\begin{align}\label{eq:star-hbar}
\tikzpic{-11}{figsv6_small/notation/star_small.pdf}/\hbar = \sum_{s\ge0} \frac{\hbar^s}{(s+1)s!}(\overleftarrow{\partial}_{\!\!\!\clu A} \overrightarrow{\partial}_{\!\!\!\clu A^*})^{s+1},
\end{align}
where $s$ indexes the order of $\hbar$ in the expansion.\footnote{The series does not contain $\hbar^{-1}$ order term because, at this order, the two terms in the right-hand side of \cref{eq:eta-dressing_abridged} cancel out. In other words, algebraically, the Husimi bracket $\moyal{\cl f}{\cl g} = (\cl f\cl g - \cl g\cl f)(i\hbar)^{-1} + \{\cl f, \cl g\}\hbar^0 + \mathcal O(\hbar)$ is zero at order $\hbar^{-1}$  
while its classical contribution is a Poisson bracket.}
Similar to \cref{eq:six-star-A,eq:unabridged-order1}, each factor of $\overleftarrow\partial_{\!\!\clu A}\overrightarrow\partial_{\!\!\clu A^*}$ in \cref{eq:star-hbar}  can be translated to a diagrammatic operation of forming a bond between the two subdiagrams related by $\tikzpic{-11}{figsv6_small/notation/star_small.pdf}/\hbar$ --- the $\overleftarrow\partial_{\!\!\clu A}$ factor translates to consuming an arrow associated with $\clu A$ in the left subdiagram, the $\overrightarrow\partial_{\!\!\clu A^*}$ factor translates to consuming another arrow associated with $\clu A^*$ in the right, and a directional bond if formed between the two consumed arrows. For the term $\hbar^s(\overleftarrow\partial_{\!\!\clu A}\overrightarrow\partial_{\!\!\clu A^*})^{s+1}$, we then interpret it as forming one classical bond and $s$ number of quantum bonds. With this, \cref{eq:eta-dressing_abridged} can be expanded as the following diagrams, by writing out all unique configurations of forming such bonds
\begin{align}\label{eq:eta-dressing_unabridged}
\begin{split}
    &\quad\tikzpic{-40}{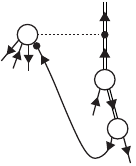}  \quad+\quad
    \tikzpic{-40}{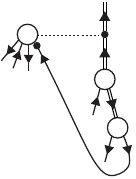}\\[-10pt]
    &+\;
    \tikzpic{-40}{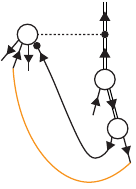} \!+\;\;
    \tikzpic{-40}{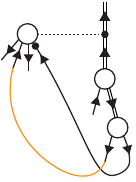}-\;
    \tikzpic{-40}{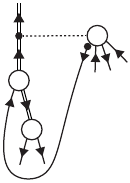}.
\end{split}\raisetag{1.5\baselineskip}
\end{align}
The first two diagrams above are expanded from the first one in the right-hand side of \cref{eq:eta-dressing_abridged}. They correspond to the two configurations of forming one classical bond and zero quantum bond $(s=0)$ between the left and right subdiagrams. Comparing with the one in \cref{eq:eta-dressing_abridged}, in these two diagrams we have rotated the output of the left subdiagram for visual simplicity without changing its relative ordering with respect to the inputs. Moreover, by choice, the output of this subdiagram connects to the classical bond. On top of the first two, the third and fourth diagrams in \cref{eq:eta-dressing_unabridged} correspond to the configurations with one extra quantum bond (orange). Note that each bond stemming from $\overleftarrow\partial_{\!\!\clu A}\overrightarrow\partial_{\!\!\clu A^*}$ is directional, i.e. the associated arrow always points from the right subdiagram to the left one, therefore in these specific diagrams no more one quantum bond can be formed. Similarly, the last diagram in \cref{eq:eta-dressing_unabridged} corresponds to the only way of forming a bond in the last diagram in \cref{eq:eta-dressing_abridged}, i.e. between the output of the right subdiagram constituting \tikzpic{-11}{figsv6_small/notation/K_small_h.pdf} ($\cl K$) to an input in the left subdiagram. 

A diagram like those in \cref{eq:eta-dressing_unabridged} can be evaluated with the same rules associated with the bare diagrams but with two additional ones: (1) the two resonant excitations connected by the classical bond between the two subdiagrams contributes only a factor of $1$ (instead of $\clu A^*\clu A$), and (2) the dashed line contributes another factor $1/(s+1)$ where $s$ is the number of orange bonds drawn \textit{between} the subdiagrams. 
As a result, the five diagrams in \cref{eq:eta-dressing_unabridged} are respectively evaluated as 
\begin{align*}
&\,\,\,\frac{1}{-2\omega_o^\prime}g_4\clu A^{*2}\!\clu A\,\frac{g_3^2e^{i\omega_o^\prime t}}{-2\omega_o^{\prime2}}\clu A^*\!\clu A + \frac{1}{-2\omega_o^\prime}g_4\clu A^{*2}\!\clu A\,\frac{g_3^2e^{i\omega_o^\prime t}}{-2\omega_o^{\prime2}}\clu A^*\!\clu A\\
&\,\,\,+\frac{\hbar}{-2\omega_o^\prime}g_4\clu A^{*2}\,\frac{g_3^2e^{i\omega_o^\prime t}}{-2\omega_o^{\prime2}}\clu A + \frac{\hbar}{-2\omega_o^\prime}g_4\clu A^{*2}\,\frac{g_3^2e^{i\omega_o^\prime t}}{-2\omega_o^{\prime2}}\clu A \\
&\,\,\,- \frac{1}{-2\omega_o^\prime}\times\frac{g_3^2e^{i\omega_o^\prime t}}{-2\omega_o^{\prime2}}\clu A^{*2} g_4\clu A^{*}\!\clu A^2\,\\
&=\frac{g_3^2g_4}{4\omega_o^{\prime3}}(\clu A^{*3}\clu A^2+\hbar\clu A^{*2}\clu A)e^{i\omega_o^\prime t},
\end{align*}
which equals to \cref{eq:eta-dressing-value} obtained from evaluating \cref{eq:eta-dressing-husimi} directly. 

\subsection{Dressed resonant excitation}\label{subsec:dressing-osc}

We have now discussed the first three equations in \cref{eq:sc-diagram-Gamma,eq:sc-diagram-M,eq:sc-diagram-mu,eq:sc-diagram-B,eq:sc-diagram-S,eq:sc-diagram-K}, which are the core equations underlying our diagrammatic method. In the last part of this section, we will discuss the remaining three \cref{eq:sc-diagram-B,eq:sc-diagram-S,eq:sc-diagram-K} that are related to the dressed resonant excitation \!\!\tikzpic{-9}{figsv6_small/notation/B_small.pdf}. Specifically, we will first comment on \cref{eq:sc-diagram-B,eq:sc-diagram-S} that formally define the dressed resonant excitation. Then we will discuss \cref{eq:sc-diagram-K}, which relate the effective Hamiltonian \tikzpic{-11}{figsv6_small/notation/K_small_h.pdf} ($\cl K$) to \!\!\tikzpic{-10.5}{figsv6_small/notation/Gamma_small.pdf} ($\cl\Gamma$) and \!\!\tikzpic{-9}{figsv6_small/notation/B_small.pdf} ($\clu Be^{-i\omega_o^\prime t}$). 

\cref{eq:sc-diagram-B,eq:sc-diagram-S} define the dressed resonant excitation in a self-consistent manner as 
\begin{subequations}\label{eq:dressed-osc-23}
\begin{align}\label{eq:dressed-osc-2}
\tikzpic{-21}{figsv6_small/notation/B.pdf} \;&=  \; \tikzpic{-11}{figsv6_small/notation/band_pass.pdf}\exp\Bigg(\tikzpic{-23}{figsv6_small/canonicity/Husimi_zeta.pdf}\Bigg)\tikzpic{-21}{figsv6_small/notation/A.pdf}\\\label{eq:dressed-osc-3}
&= \tikzpic{-20}{figsv6_small/notation/A.pdf}\; +\; \tikzpic{-30}{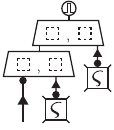} \;+\!  \tikzpic{-35}{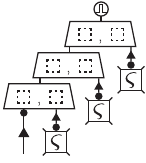} \;+\; \cdots\raisetag{1.2\baselineskip}
\end{align}
\end{subequations}
\begin{subequations}\label{eq:zeta}
\begin{align}\label{eq:zeta-1}
\tikzpic{-13}{figsv6_small/notation/zeta.pdf} &= \;\tikzpic{-23}{figsv6_small/notation/eta.pdf} \;-\; \tikzpic{-11}{figsv6_small/notation/band_stop.pdf}\exp\Bigg(\tikzpic{-23}{figsv6_small/canonicity/Husimi_zeta.pdf}\Bigg)\tikzpic{-21}{figsv6_small/notation/A.pdf}\;+\;\tikzpic{-28}{figsv6_small/canonicity/zeta_k1.pdf}\\\label{eq:zeta-2}
& = \;\tikzpic{-23}{figsv6_small/notation/eta.pdf}  \;-\; \tikzpic{-30}{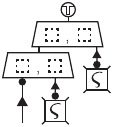}\;-\!  \tikzpic{-35}{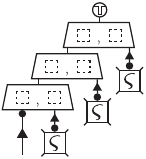} \;-\; \cdots. \raisetag{1.2\baselineskip} 
\end{align}
\end{subequations}
Here, \cref{eq:dressed-osc-2,eq:zeta-1} simply restate \cref{eq:sc-diagram-B,eq:sc-diagram-S}, while in \cref{eq:dressed-osc-3,eq:zeta-2}, we expand the exponentiation operation from the initial equations and obtain two infinite series of similar structure. We note that in \cref{eq:dressed-osc-3} the term
\begin{align*}
\begin{split}
\tikzpic{-28}{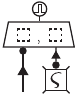} = e^{-i\omega_o^\prime t}\text{Sta}(e^{i\omega_o^\prime t}i\moyal{e^{-i\omega_o^\prime t}\clu A}{i\cl S})
\end{split}
\end{align*}
is not explicitly displayed. This is because $\cl S$ is constructed to be purely oscillating (c.f. \cref{eq:self-consistent-S}) and thus the above term is zero. 

In \cref{eq:dressed-osc-23}, \!\tikzpic{-9}{figsv6_small/notation/B_small.pdf} is defined in terms of the intermediate quantity \!\!\tikzpic{-10}{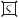}, which is further defined by a self-consistent relation \cref{eq:zeta}. These two coupled equations should be carried out by iteratively substituting \!\!\tikzpic{-10}{figsv6_small/canonicity/zeta_dAs_small.pdf} in \cref{eq:dressed-osc-23,eq:zeta} with \cref{eq:zeta}. This yields
\begin{align}\label{eq:dressed-osc-4}
\begin{split}
\tikzpic{-21}{figsv6_small/notation/B.pdf} = &\;\tikzpic{-20}{figsv6_small/notation/A.pdf}\; +\; \tikzpic{-30}{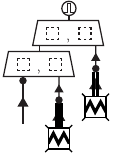} \;+\!  \tikzpic{-35}{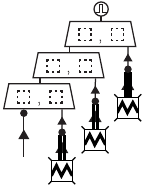}\\
& -  \tikzpic{-65}{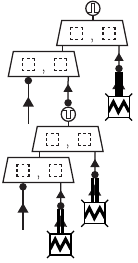} \;\;-  \tikzpic{-65}{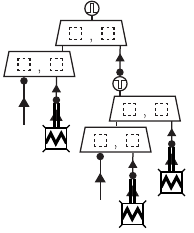} \;+ \cdots.
\end{split}\raisetag{1.2\baselineskip}
\end{align}
Remarkably, \cref{eq:dressed-osc-4} defines the dressed oscillator \!\tikzpic{-9}{figsv6_small/notation/B_small.pdf} in terms the dressed off-resonant excitation \tikzpic{-11}{figsv6_small/notation/eta_small.pdf}, which is a known quantity from \cref{subsec:order-1,subsec:bare-virtual,subsec:dressing}. By substituting the specific diagrams constituting \tikzpic{-11}{figsv6_small/notation/eta_small.pdf}, one can then construct \!\tikzpic{-9}{figsv6_small/notation/B_small.pdf} to any desired order. It is important to note that the leading order contribution of \tikzpic{-11}{figsv6_small/notation/eta_small.pdf} is at order $q_\text{zps}^1$ (see \cref{eq:pert-structure}). Therefore, for a dressed resonant excitation defined by \cref{eq:dressed-osc-4}, the correction to the bare resonant excitation is $\mathcal O(q_\text{zps}^2)$. 

When evaluating a diagram in \cref{eq:dressed-osc-4}, one follows the same rules introduced earlier in this section but with an additional one: for each bare resonant excitation \!\tikzpic{-9}{figsv6_small/notation/A_small.pdf} that goes through $k$ number of \tikzpic{-11}{figsv6_small/notation/husimi_small.pdf} before being terminated by a \tikzpic{-11}{figsv6_small/notation/band_pass.pdf} or \tikzpic{-11}{figsv6_small/notation/band_stop.pdf}, an extra factor of $1/k!$ is associated with it. This factor stems from expanding the exponential functions in \cref{eq:dressed-osc-23,eq:zeta}. For instance, the five diagrams in the right-hand side of \cref{eq:dressed-osc-4} are associated with the overcounting factors of $1/0!$, $1/2!$, $1/3!$, $1/(2!\times2!)$, and $1/(2!\times2!)$, respectively.

Lastly, we discuss \cref{eq:sc-diagram-K},
\begin{align*}
        \tikzpic{-18}{figsv6_small/notation/K.pdf} \;&=\;\tikzpic{-18}{figsv6_small/notation/K_bare.pdf} \;+\; \tikzpic{-36}{figsv6_small/canonicity/B_K_dAs.pdf} \;-\; \tikzpic{-36}{figsv6_small/canonicity/A_K_dAs.pdf},
\end{align*}
that relates the effective Hamiltonian \tikzpic{-11}{figsv6_small/notation/K_small_h.pdf}~($\cl K$) to \tikzpic{-10.5}{figsv6_small/notation/Gamma_small.pdf}~($\cl \Gamma$) and \!\tikzpic{-9}{figsv6_small/notation/B_small.pdf}~($\clu B e^{-i\omega_o^\prime t}$). To see their relationship more explicitly, we can manipulate \cref{eq:sc-diagram-K} (or directly translate \cref{eq:composite-var-Gamma} to diagrammatic representation) and get 
\begin{subequations}\label{eq:gamma-k-b}
    \begin{align}
    \tikzpic{-13}{figsv6_small/notation/Gamma.pdf} \;\,&=\;\, -\tikzpic{-26}{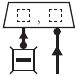},
    \end{align}
    which corresponds to 
    \begin{align}
    \cl\Gamma \;&= -i\moyal{\cl K}{\clu Be^{-i\omega_o^\prime t}}.
\end{align}
\end{subequations}
Physically, this equation indicates that  \tikzpic{-10.5}{figsv6_small/notation/Gamma_small.pdf} ($\cl \Gamma$) is a quantity describing the slow dynamics of motion for \!\tikzpic{-9}{figsv6_small/notation/B_small.pdf} ($\clu B e^{-i\omega_o^\prime t}$). In other words, the equation of motion for \!\tikzpic{-9}{figsv6_small/notation/B_small.pdf} is just 
\begin{align}\label{eq:eom-B}
d_t \tikzpic{-20}{figsv6_small/notation/B.pdf} = \partial_t \tikzpic{-20}{figsv6_small/notation/B.pdf} - i\tikzpic{-13}{figsv6_small/notation/Gamma.pdf}.
\end{align}
Mathematically, constructing  \tikzpic{-11}{figsv6_small/notation/K_small_h.pdf} ($\cl K$) from \tikzpic{-10.5}{figsv6_small/notation/Gamma_small.pdf} ($\cl \Gamma$) and \!\tikzpic{-9}{figsv6_small/notation/B_small.pdf} ($\clu B e^{-i\omega_o^\prime t}$) involves inverting the Husimi bracket in \cref{eq:gamma-k-b}, the operation of which is formally defined as \cref{eq:sc-diagram-K} (and algebraically as \cref{eq:self-consistent-K}). 

To explicitly compute \tikzpic{-11}{figsv6_small/notation/K_small_h.pdf} ($\cl K$), we take \cref{eq:sc-diagram-K} as a self-consistent relation and substituting each \tikzpic{-11}{figsv6_small/notation/K_small_h.pdf} component in the right-hand side of the equation with the equation itself. Carrying this out iteratively yields an infinite series
\begin{align}\label{eq:dressed-K-2}
\begin{split}
    \tikzpic{-19}{figsv6_small/notation/K.pdf} \;=&\;\tikzpic{-20}{figsv6_small/notation/K_bare.pdf} \;+\; \tikzpic{-40}{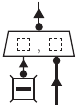} \;-\; \tikzpic{-40}{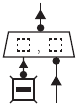}\\
    &+\; \tikzpic{-40}{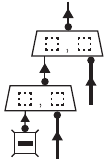} \;-\;\tikzpic{-40}{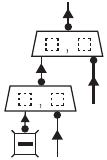} \quad+\quad \cdots.
\end{split}
\end{align}
One can then compute  \!\tikzpic{-11}{figsv6_small/notation/K_small_h.pdf} to the desired order by evaluating the relevant diagrams above provided that \!\tikzpic{-10.5}{figsv6_small/notation/Gamma_small.pdf} and \tikzpic{-9}{figsv6_small/notation/B_small.pdf} are specified.\footnote{In order for the series \cref{eq:dressed-K-2} to converge, a diagram involving \!\tikzpic{-8}{figsv6_small/notation/A_small.pdf} should be evaluated together with those diagrams that look alike but with \!\tikzpic{-8}{figsv6_small/notation/A_small.pdf} replaced by \!\tikzpic{-8}{figsv6_small/notation/B_small.pdf}.} 

We have now discussed all the equations necessary for implementing our diagrammatic method, specifically \cref{eq:sc-diagram-Gamma,eq:sc-diagram-M,eq:sc-diagram-mu,eq:sc-diagram-B,eq:sc-diagram-S,eq:sc-diagram-K}. To conclude this section, we concisely recapitulate this new method. The general goal of our method is to describe the effective dynamics of a driven nonlinear oscillator governed by Hamiltonian \cref{eq:op-H-a-ad}. To achieve this, we seek for a frame transformation $\cl a\rightarrow \clu A$, discussed in \cref{eq:QHB-ansatz}, so that in the transformed frame the dynamics of the bosonic coordinate $\clu A$ is governed by a time-independent Hamiltonian $\cl K$, which we call the effective Hamiltonian. Finding this frame then boils down to solving the equation of motion \cref{eq:eom-phase-expand}, which we further breaks down into 6 coupled self-consistent equations \cref{eq:self-consistent-Gamma,eq:self-consistent-M,eq:self-consistent-mu,eq:self-consistent-B,eq:self-consistent-S,eq:self-consistent-K} by introducing the change of variables in \cref{eq:composite-var}. Due to the structural similarity between \cref{eq:self-consistent} and the equations underlying the classical harmonic balance method, we refer to this algebraic procedure as the quantum harmonic balance (QHB) method. 

Our diagrammatic method is developed through solving \cref{eq:self-consistent-Gamma,eq:self-consistent-M,eq:self-consistent-mu,eq:self-consistent-B,eq:self-consistent-S,eq:self-consistent-K} in their corresponding diagrammatic form \cref{eq:sc-diagram-Gamma,eq:sc-diagram-M,eq:sc-diagram-mu,eq:sc-diagram-B,eq:sc-diagram-S,eq:sc-diagram-K}. Specifically, solving \cref{eq:sc-diagram-Gamma,eq:sc-diagram-M} elucidates the cascaded mixing networks underlying the effective Hamiltonian and the frame transformation. These mixing networks, which are termed as dressed diagrams, represent mixing processes involving the dressed oscillator excitation  \!\tikzpic{-9}{figsv6_small/notation/B_small.pdf}, \!\tikzpic{-9}{figsv6_small/notation/Bs_small.pdf}, the dressed oscillator excitation \!\tikzpic{-11}{figsv6_small/notation/eta_small.pdf}, \!\tikzpic{-11}{figsv6_small/notation/etas_small.pdf}, and the drive excitation \!\tikzpic{-9}{figsv6_small/notation/xi_small.pdf}, \!\tikzpic{-9}{figsv6_small/notation/xis_small.pdf}. Like Feynman diagrams, a dressed diagram to leading order is a bare diagram involving only bare excitations. In \cref{subsec:order-1,subsec:bare-virtual}, we solve \cref{eq:sc-diagram-Gamma,eq:sc-diagram-M} to leading order to introduce the bare diagram and its structure, i.e. the cascaded mixing networks, shared by the its dressed counterpart. In \cref{subsec:dressing}, we then introduce the specific construction of the dressed off-resonant excitation by solving \cref{eq:sc-diagram-mu}. In \cref{subsec:dressing-osc}, we introduce the specific construction of the dressed resonant excitation by solving \cref{eq:sc-diagram-B,eq:sc-diagram-S,eq:sc-diagram-K} This formalizes a diagrammatic tools for the computation of nonlinear quantum effects in an oscillator, up to now used only heuristically in the literature.

\section{Axiomatic Presentation of the
Diagrammatic Perturbation Method}
\label{sec:diagrams-pert}

\subsection{Motivation}

In \cref{sec:averaging}, we have introduced the QHB expansion to diagrammatically construct the effective Hamiltonian \!\tikzpic{-11}{figsv6_small/notation/K_small_h.pdf} ($\cl K$) governing the dynamics of a driven nonlinear oscillator. This method provided us with a systematic way to find the effective Hamiltonian as a sum of diagrams involving different kinds of excitations, where the diagrams themselves are of simple graphical structure as rooted trees. However, the QHB expansion involves an iterative process that is cumbersome to be carried out to high orders. In this section, we introduce an axiomatic approach to the diagrammatic method that yields the same effective Hamiltonian as the QHB expansion. As in the case of Feynman diagrams, this axiomatic approach bypasses the iterative procedure and allows the effective Hamiltonian \!\tikzpic{-11}{figsv6_small/notation/K_small_h.pdf}($\cl K$) to be written down directly at any given order by following diagrammatic rules. We present each step of the axiomatic procedure as a well-defined task in graph theory. Each step in this procedure can thus be efficiently implemented by a computer program. 

\subsection{Recipe for evaluating diagrams}

To construct and evaluate of the diagrams for all  $\cl K^{(n)}$, we proceed with the following procedure:
\begin{enumerate}
    \item Write down the forest comprising all possible \textit{unrooted} trees with $k+2$ leaf vertices for  $0 < k \le n$. In graph theory, an unrooted tree is a tree that has no vertex designated as a root and a forest is a disjoint union of trees. 
    \item Write down all unique diagrams\footnote{Two diagrams are said to be equivalent if one can be recovered from the other by an overall rotation.} a tree represents by, 
    \begin{enumerate}
        \item to each internal vertex, assigning a $\tikzpic{-11}{figsv6_small/notation/star_small.pdf}$ representing a mixer,
        \item to each external edge (connecting a leaf and an internal vertex), assigning \!\tikzpic{-9}{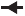} or a \!\tikzpic{-9}{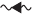} traveling in either direction that are associated with the dressed resonant excitation and the drive excitation, respectively.
        \item and, to each internal edge (connecting internal vertices), assigning a \!\tikzpic{-9}{figsv6_small/notation/prop_dressed_small.pdf} arrow traveling in either direction that represents a dressed off-resonant excitation.
    \end{enumerate}
    \item Eliminate any diagram with nonzero net frequency.
    \item Compute the frequency of the off-resonant excitations by imposing frequency (energy) conservation at each mixer. Eliminate diagrams containing off-resonant excitation at frequency $\omega_o^\prime$.
\end{enumerate}    
The resulting diagrams is a \textit{forest} of  unrooted trees, each an underlying graph shared by a family of rooted trees constituting \!\tikzpic{-11}{figsv6_small/notation/K_small_h.pdf}  $(\cl K)$. This forest is already an unambiguous, succinct\footnote{A more succinct representation is possible by using an underlying \textit{unordered diagram} to represent the ordered ones sharing it --- i.e. for diagrams can be converted to each other through successive permutation of inputs into the mixers, one represents them by one single diagram. We use this representation in the actual code implementation \cite{qhb2022}.}, and intuitive representation of the effective Hamiltonian. Indeed, the diagrammatic heuristics discussed in \cref{sec:intro} are developed at this level but without internal edges elaborated as off-resonant excitations or the resonant excitation being dressed. Evaluating the diagrams generated from the above steps follows the following procedure:
\begin{enumerate}
\setcounter{enumi}{4}
\item For each diagram produced after Step 4, write down all possible unique diagrams it can generate by converting a \tikzpic{-10}{figsv6_small/notation/straight_arrow_dressed_small_r.pdf} traveling away from a mixer to a  \tikzpic{-9}{figsv6_small/notation/Bs_small_r.pdf}; eliminate any diagram if such a replacement is not possible. The resulting diagrams are the dressed diagrams constituting \!\tikzpic{-11}{figsv6_small/notation/K_small_h.pdf}~$(\cl K)$.

\item Find the set of off-resonant excitations participating in the diagrams above. The resulting diagrams are the dressed diagrams constituting \tikzpic{-11}{figsv6_small/notation/eta_small.pdf}~$(\cl\mu e^{-i\omega_o^\prime t})$. 

\item Consider only the bare off-resonant excitations and bare resonant excitation in each diagram constituting \!\tikzpic{-11}{figsv6_small/notation/K_small_h.pdf}. Evaluate the resulting bare diagram following the prescription in \cref{subsec:order-1,subsec:bare-virtual} and example of \cref{eq:order2-term1,eq:order2-term1-unabridged}.

\item Construct the dressed diagrams in \!\tikzpic{-11}{figsv6_small/notation/K_small_h.pdf} and \tikzpic{-11}{figsv6_small/notation/eta_small.pdf} stemming from the diagrams already constructed. Specifically, the dressed off-resonant excitation, expanded by \cref{eq:eta-dressed-3}, depends on \!\tikzpic{-11}{figsv6_small/notation/K_small_h.pdf} while the dressed resonant excitation, expanded by \cref{eq:dressed-osc-4}, depends on \tikzpic{-11}{figsv6_small/notation/eta_small.pdf}. When computing these dressed excitations in a dressed diagram, the diagrams constituting \!\tikzpic{-11}{figsv6_small/notation/K_small_h.pdf} and  \tikzpic{-11}{figsv6_small/notation/eta_small.pdf} should be considered as those already constructed from the previous steps. Evaluate the resulting dressed diagrams following the prescription in \cref{subsec:dressing,subsec:dressing-osc}.

\item Repeat Step 8 until all unique diagrams of the desired order are constructed and evaluated.
\end{enumerate}

We remark that Steps 7-9 contains iterative elaborations of the dressed off-resonant excitation and resonant excitation with the set of constructed diagrams constituting \!\tikzpic{-11}{figsv6_small/notation/K_small_h.pdf} and \tikzpic{-11}{figsv6_small/notation/eta_small.pdf}, while each round of elaboration updates the constructed set itself. It is also possible to avoid the iterative procedure completely by further exploiting the graphical structure behind dressed diagrams. We discuss this and other tricks facilitating code implementation in \cite{qhb2022}, an open-source program to compute the effective Hamiltonian based on the diagrammatic method developed in this work. 

We also note that if one is only interested in specific driven processes, e.g. the terms involving $\xi^2\clu A^{\dagger 3}$ responsible for the three-legged cat discussed in \cref{sec:intro}, in the Steps 2 and 3 above one can directly draw the relevant diagrams, e.g. those involving three \!\tikzpic{-9}{figsv6_small/notation/As_small.pdf} and two  \!\tikzpic{-9}{figsv6_small/notation/xi_small.pdf} as external edges, instead all diagrams of zero net frequency. This provides significant computational advantage of the diagrammatic method over other high-frequency expansions. We illustrate this in more detail in the examples treated in the rest of the work, where we obtain analytical descriptions for high-order processes that were not possible before.

\section{Remarks and Extensions}\label{sec:discussion}

\subsection{General form of the effective Hamiltonian} \label{subsec:eff-Ham}

The axiomatic procedure explained in \cref{sec:diagrams-pert} provides a simple way to construct the diagrams that constitute $\cl K$ as shown in \!\tikzpic{-11}{figsv6_small/notation/K_small_h.pdf}. Furthermore, this procedure also reveals the general structure of these Hamiltonian diagrams, which we elaborate on in this subsection. We would like to remind the reader that the driven oscillator of interest usually has a drive with frequency $\omega_d$ in the vicinity of $q\omega_o/p$, where $\omega_o$ is the natural frequency of the oscillator and $q$ and $p$ are positive integers. When the frequency configuration is such, the effective Hamiltonian is constructed in the rotating frame at $\omega_o^\prime = p\omega_d/q$, which is in the vicinity of $\omega_o$. In this frame, the resonant excitation \!\tikzpic{-9}{figsv6_small/notation/A_small.pdf} is associated with $\clu Ae^{-i\omega_o^\prime t}$, and the drive excitation \!\tikzpic{-9}{figsv6_small/notation/xi_small.pdf} is associated with $\xi e^{-i\omega_d t}$. Since a diagram that constitutes the effective Hamiltonian has zero net frequency, it can be expressed as
\begin{align}\label{eq:discussion-K-general-diagram}
\tikzpic{-48}{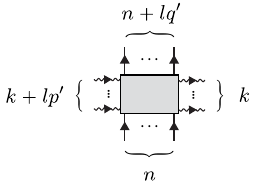}
\end{align}
where $n, k, l$ are natural numbers and $q^\prime, p^\prime$ are coprime such that $p^\prime/q^\prime$ is the simplest form of $p/q$. Note that in later text, different pairs of $q, p$ associated with the same lowest terms $p^\prime/q^\prime$ can represent different nonlinear processes. In \cref{eq:discussion-K-general-diagram}, the gray box should be understood as some diagram whose content can be found by following the procedure in \cref{sec:diagrams-pert}. Collecting all relevant diagrams, we obtain the effective Hamiltonian of the form
\begin{align}\label{eq:discussion-K-general}
    \cl K = \sum_{n>0} K_n\cl{\cu A}^{\dagger n} \cl{\cu A}^{n} + \sum_{l>0}\Omega_{lq^\prime,lp^\prime}\xi^{lp^\prime}\op{\cu A}^{\dagger lq^\prime} + \mathrm{h.c.},
\end{align}
where the first summation groups the diagrams with $l=0$, while each summand itself is computed by summing the relevant diagrams over the index $k$ in \cref{eq:discussion-K-general-diagram}, and the second summation groups the terms with $l>0$, while each summand is computed by summing over the indices $n$ and $k$.

In \cref{eq:discussion-K-general}, we refer to $K_n\cl{\cu A}^{\dagger n} \cl{\cu A}^{n}$ as the energy renormalization terms because they renormalize the energy spacing between Fock states, which are the eigenstates of these terms, from the bare detuning $\delta$ (or $\omega_o$ when the rotating frame transformation at $\omega_o^\prime$ is undone). Terms associated with the ac-Stark shift, Lamb shift, and Kerr nonlinearity belong to this category and we discuss them in more detail in \cref{sec:3lcat,sec:renormalization}. Correspondingly, we refer to $\Omega_{lq^\prime,lp^\prime}\xi^{lp^\prime}\cl{\cu A}^{\dagger lq^\prime} + \mathrm{h.c.}$ as the coupling terms, since they induce a  coupling between Fock states whose indices differ by $l q^\prime$. These terms are responsible for the nontrivial nonlinear processes such as the Kerr-cat and three-legged Schr\"odinger cat states previewed in \cref{sec:intro} and we discuss other processes derived from them in \cref{sec:mpnr,sec:USH}. We also note that $\Omega_{lq^\prime,lp^\prime}$, to leading order, is a complex number whose value is determined by diagrams like that in \cref{eq:discussion-K-general-diagram} with $k=n=0$. At higher order, it is corrected by terms $\propto |\xi|^{2k}\cl{\cu A}^{\dagger n} \cl{\cu A}^{n}$ and as written they can be understood as the renormalization of the coupling amplitude $\Omega_{lq^\prime,lp^\prime}$.

\subsection{Extension to multi-mode and multi-tone systems}
\label{subsec:multi}
The diagrammatic procedure prescribed above can be extended to multi-mode and multi-tone problems, keeping the same conceptual and computational foundations. Specifically, we may be interested in modeling a system in which one nonlinear mode, like the one described by $\op H$ in \cref{eq:op-H-q-p}, is linearly coupled with multiple linear modes and driven by multiple sinusoidal drives. Following the procedure in \cref{app:multi}, we obtain the Hamiltonian in the displaced/rotated frames in the normal modes:
\begin{align}\label{eq:multi-mode}
\begin{split}
    \op H= \sum_k{\delta_k \op a_k\op a_k^\dagger}+ &\sum_{m\ge3}
    \frac{g_m}{m}\Big(\sum_k( {\lambda_k}\op a^\dagger_ke^{i\omega_k^\prime t}+{\lambda_k}\op a_ke^{-i\omega_k^\prime t}) \\
    &\qquad+ \sum_l (\xi^*_le^{i\omega_{d,l}t}+ \xi_l e^{-i\omega_{d,l}t})\Big)^m,
\end{split}
\end{align}
where $\op a_k$ and $\op a_k^\dagger$ are the conjugate bosonic operators for the normal mode $k$ satisfying $[\op a, \op a^\dagger]=\hbar$, $\delta_k =\omega_k-\omega_k^\prime$ is the detuning between the natural frequency and the rotating frame frequency of mode $k$ and $\lambda_k$ is a dimensionless factor that measures the hybridization strength between the mode $k$ and the nonlinear mode. For readers familiar with the energy participation ratio (EPR) method \cite{minev2021} of analyzing Josephson circuits, it is related to $P_k$,  the energy participation ratio of mode $k$ in the nonlinear element, by $\lambda_k^2 = P_k\omega_k/\omega_o$ and $\sum_kP_k =1$. The factor $\xi_l$ are the effective drive amplitude of a drive tone at frequency $\omega_{d,l}$. This Hamiltonian resembles the single mode one \cref{eq:H-tran} and is the general form of the ones describing many cQED experiments \cite{Mundhada2016,leghtas2015}. 

Carrying out the averaging procedure over the Hamiltonian similar to the single mode one, we find that the \textit{only} modifications to the diagrammatic procedure stated in \cref{sec:diagrams-pert} are: 
\begin{itemize}
    \item In Step 2(b), the straight arrows and wavy arrows have different colors, representing different oscillator modes (indexed by $k$) and drive tones (indexed by $l$) at their corresponding frequencies.
    \item Every straight arrow, including the one linked with another straight arrow, is associated with an additional factor $\lambda_k$.
    \item Every double-line arrow \tikzpic{-8}{figsv6_small/notation/prop_small.pdf}, representing a bare off-resonant excitation at frequency $\omega_\text{out}$, is associated with a propagator $ \sum_k \frac{\lambda_k^2}{\omega_\text{out}-\omega_k^\prime}(1-\delta_{\omega_\text{out}, \omega_k^\prime})$. It can be intuited as the off-resonant excitation propagates in the medium of each oscillator mode with a weight factor $\lambda_k$. For a propagator successively dressed by $n$ Kamiltonian terms, the associated compounded propagation factor is $\sum_k \frac{\lambda_k^2}{(\omega_\text{out}-\omega_k^\prime)^{n+1}}(1-\delta_{\omega_\text{out}, \omega_k^\prime})$ (c.f. \cref{eq:eta-dressing_unabridged} for $n=1$ in the single mode case). 
    \item The Husimi Q product is defined over the multiple dimensional phase space as $\cl f\varstar \cl g =  \cl f \left(\sum_k\hbar\overleftarrow{\partial}_{\!\!\clu A_k} \overrightarrow \partial_{\!\!\clu {A}_k^*}\right) \cl g$.
\end{itemize}
We illustrate this procedure further in \cref{sec:renormalization} with a concrete example involving a system with two modes.

\subsection{Extension to open quantum systems}\label{sec:dissipator}

It is important to reiterate that in the perturbation method we have presented in this work, we have constructed a frame in which the slow dynamics of interest is described by a time-independent effective Hamiltonian $\op K$. Besides the Hamiltonian, the passage to this frame also transforms other physical objects, for example, the coupling between the driven nonlinear oscillator and a thermal bath. It turns out that this transformation leads to nontrivial dissipative dynamics such as the heating of the oscillator at zero temperature. In this section, we briefly discuss a diagrammatic way to describe the effective dissipative dynamics of the driven nonlinear oscillator, which can be done by computing some \textit{effective Lindbladian} governing the oscillator. This calculation is a natural extension of that of the effective Hamiltonian we have so far discussed. 

We now consider an extended Hamiltonian describing not only the driven nonlinear oscillator but also its coupling to a thermal bath 
\begin{align}\label{eq:discussion-H-total}
&\op{\cu H}_\text{tot} = \hat{\cu H}_\text{s}(t) + \hat{\cu H}_\text{sb} + \op {\cu H}_\text{b},
\end{align}
where $\hat{\cu H}_\text{s}(t)$ is the oscillator Hamiltonian defined in \cref{eq:op-H-a-ad}, $\hat{\cu{H}}_{\mathrm{b}} = \int_0^\infty d\omega \op b_\omega^\dagger \op b_\omega$ is the thermal bath Hamiltonian containing a continuum of modes each at frequency $\omega$ with bosonic operator $\op b_\omega$, and $\hat{\cu{H}}_{\mathrm{sb}}=-(\hat{a} -\hat{a}^{\dagger})\times \int_0^\infty d\omega h_\omega\left(\hat{b}_\omega-\hat{b}_\omega^{\dagger}\right)$ defines the coupling between the oscillator mode $\op a$ and each bath mode $\op b_\omega$ with strength $h_\omega$. We note that here we impose the commutator $[\op b_\omega, \op b_\omega^\dagger] = \hbar$ to be consistent with that of the $\op a$ mode (cf. \cref{eq:op-H-a-ad}). 

In the diagrammatic method, which is carried out in the Husimi Q space, the effective Hamiltonian of the oscillator $\cl K_\text{s}$ is constructed in a frame defined by $\cl{\cu A}$ whose relationship to the lab frame amounts to $\cl a \rightarrow\xi e^{-i\omega_d t} + e^{-i\omega_o^\prime t}(\cl{\cu B} + \cl\mu)$. In this frame, \cref{eq:discussion-H-total} transforms to 
\begin{subequations}\label{eq:discussion-K-total}
\begin{align}
&\cl{K}_\text{tot} = \cl{K}_\text{s} + \cl{ K}_\text{sb}(t)\\[3pt]\label{eq:discussion-Ksb}
&\cl{{K}}_{\mathrm{sb}}(t)=-\cl C(t)\times\int_0^\infty d\omega h_\omega\left(\cl{b}_\omega e^{-i\omega t}-\cl{b}_\omega^{*}e^{i\omega t}\right)\\
\begin{split}\label{eq:discussion-collaspe}
&\cl C(t) =\cl{\cu B}e^{-i\omega_o^\prime t} -\cl{\cu B}^{*}e^{i\omega_o^\prime t}
+\cl{\mu}e^{-i\omega_o^\prime t} -\cl{\mu}^{*}e^{i\omega_o^\prime t} \\
&\qquad\qquad +\xi e^{-i\omega_d t} +  \xi^* e^{i\omega_d t},
\end{split}
\end{align}
\end{subequations}
where we have also gone to the rotating frame of each $\op b_\omega$ mode $\cl b_\omega\rightarrow \cl b_\omega e^{-i\omega t}$. With $\cl{K}_\text{tot}$ defined  as \cref{eq:discussion-K-total}, one can obtain the effective Lindbladian under the standard Born-Markov approximation\footnote{The standard treatment also involves going to the interaction picture and performing Born-Markov approximation on $e^{L_{\cl K_\text{s}t}}\cl{{K}}_{\mathrm{sb}}(t)$. Yet this treatment can be avoided under the assumption that the thermal bath is white within the neighbourhood around some frequency $\omega_j$ with the width covering the relevant portion of the spectrum of $\cl K_\text{s}$. See supplement of \cite{petrescu2020} for more discussion.} as 
\begin{align}\label{eq:discussion-lindbladian}
\begin{split}
    &\partial_t\cl Q_s =\moyal{\cl{K}_s}{\cl Q_s}+2\pi\sum_j \Big(\mathcal D\left[\sqrt{S_{f\!f}[-\omega_j]}\cl C_{\omega_j}\right] \cl Q_s\\
    &\qquad\qquad\qquad\qquad\;+ \mathcal D\left[\sqrt{S_{f\!f}[\omega_j]}\cl C_{\omega_j}^\dagger\right] \cl Q_s\Big),\\[3pt]
    &\mathcal D [\cl O]\cl f := \cl O \varstar\cl f\varstar \cl O^* -(\cl O^*\varstar\cl O\varstar \cl f + \cl f\varstar\cl O^*\varstar \cl O)/2\\
    &S_{f\!f}[-\omega_j] := (1+\bar n_{\omega_j})h_{\omega_j}, \quad S_{f\!f}[\omega_j] := \bar n_{\omega_j}h_{\omega_j}
\end{split}
\end{align}
where $\cl Q_s$ is the Husimi Q quasiprobability distribution of the system, $\bar{n}_{\omega_j}$ is the thermal photon number at frequency $\omega_j$ in the bath, and $\cl C_{\omega_j}$ is the Fourier component of $\cl C(t)$ defined by \cref{eq:discussion-collaspe} such that $\cl C(t) = \sum_j \cl C_{\omega_j}e^{-i\omega_jt}+\cl C_{\omega_j}^* e^{i\omega_jt}$ with $\omega_j\ge0$. The effect of the bath under the Markov approximation is equivalent to a stochastic force coupled to the system by $if(t)(\cl a -\cl a^*)$ with amplitude spectral density $S_{f\!f}[\pm \omega_j]$.

We note that, remarkably, $\cl C(t)$ defined by \cref{eq:discussion-collaspe} is just the sum of the resonant, off-resonant, and drive excitations participating in diagrams generated by \cref{eq:eom-phase-expand}. Therefore, the transformed collapse operators in \cref{eq:discussion-lindbladian} $\sqrt{S_{f\!f}[-\omega_j]}\cl C_{\omega_j}$ and $\sqrt{S_{f\!f}[\omega_j]}\cl C_{\omega_j}^*$ can be readily expressed as and computed through diagrams. For example, the order $q_\text{zps}^0$ contribution of the collapse operators, which are merely the untransformed ones $\sqrt{S_{f\!f}[-\omega_o^\prime]}\cl{\cu A}$ and $\sqrt{S_{f\!f}[\omega_o^\prime]}\cl{\cu A}^*$, can be expressed as 
\begin{align}\label{eq:discussion-dissipator0}
\tikzpic{-28}{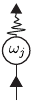},\quad
\tikzpic{-28}{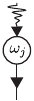},
\end{align}
where 
\begin{align*}
\tikzpic{-20}{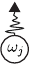} = e^{i\omega_jt}\sqrt{S_{f\!f}[-\omega_j]}\;,\;
\tikzpic{-20}{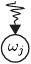} = e^{-i\omega_jt}\sqrt{S_{f\!f}[\omega_j]}
\end{align*}
The two diagrams in \cref{eq:discussion-dissipator0} are understood as the resonant conversion of a bare resonant excitation \!\tikzpic{-8}{figsv6_small/notation/A_small.pdf} (or \!\tikzpic{-8}{figsv6_small/notation/As_small.pdf}) at frequency $\omega_o^\prime$ into a bath excitation  \!\tikzpic{-8.5}{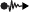} (or \!\tikzpic{-8.5}{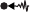}) at frequency $\omega_j$ with strength $\sqrt{S_{f\!f}[-\omega_j]}\propto\sqrt{\bar n_{\omega_j}+ 1}$ and the reverse process with strength $\sqrt{S_{f\!f}[\omega_j]}\propto\sqrt{\bar n_{\omega_j}}$. This association makes physical sense since
the creation of a bath excitation, as in the first diagram of \cref{eq:discussion-dissipator0}, is viable even for a thermal bath at zero temperature, i.e. $\bar n_{\omega_j} =0$. Correspondingly, the absorption of a bath excitation, as in the second diagram of \cref{eq:discussion-dissipator0}, occurs only at finite temperature $\sqrt{S_{f\!f}[\omega_j]}\propto\sqrt{\bar  n_{\omega_j}}$. The resonance condition of the conversion process determines the bath excitation frequency, which, in this case, is $\omega_j = \omega_o^\prime$.

Beyond the leading order, other terms in the transformed collapse operator can be similarly computed through diagrams like \cref{eq:discussion-dissipator0}. Specifically, those corresponding to $\sqrt{S_{f\!f}[-\omega_j]}\cl C_{\omega_j}$ comprise of all possible two-wave mixing processes of net frequency zero involving one bath excitation of frequency $\omega_j>0$ traveling away from one dressed off-resonant excitation or dressed resonant excitation\footnote{The two-wave mixer with one input as the drive excitation also constitutes a valid diagram. Yet its resulting collapse operator $\sqrt{S_{f\!f}[\omega_d]}\xi$ does not alter the evolution of the oscillator state and we thus omit these diagrams here.} of frequency $\pm \omega_j$ traveling into or away from \!\tikzpic{-11}{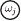}. For example, in the general case of $\omega_d>\omega_o^\prime$ and $\omega_d \ne 2\omega_o^\prime$, the transformed collapse operator at order $q_\text{zps}^1$ involves the off-resonant excitations generated by the 3-wave mixers in \cref{eq:order-1-sum} and reads
\begin{align}\label{eq:discussion-dissipator1}
\begin{split}
&\underbrace{\tikzpic{-5}{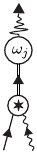}\qquad
\tikzpic{-5}{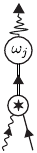}\qquad
\tikzpic{-5}{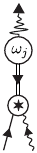}\qquad
\tikzpic{-5}{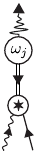}}_{\omega_j = \omega_o^\prime + \omega_d}\qquad
\underbrace{\tikzpic{-5}{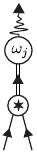}\qquad
\tikzpic{-5}{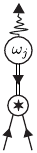}}_{\omega_j = 2\omega_o^\prime}\\
&\underbrace{\tikzpic{-5}{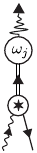}\quad
\tikzpic{-5}{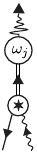}\quad
\tikzpic{-5}{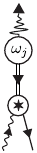}\quad
\tikzpic{-5}{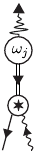}}_{\omega_j =  \omega_d-\omega_o^\prime}\quad
\underbrace{\tikzpic{-5}{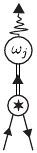}\quad
\tikzpic{-5}{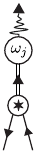}\quad
\tikzpic{-5}{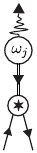}\quad
\tikzpic{-5}{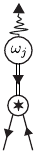}}_{\omega_j = 0}
\end{split}
\end{align}
Following the same evaluation rules introduced in \cref{sec:averaging}, the diagrams in \cref{eq:discussion-dissipator1} are evaluated as $\sum_{\omega_j}\sqrt{S_{f\!f}[-\omega_j]} \cl C_{\omega_j}$ with  
\begin{align}\label{eq:discussion-dissipator1-alg}
\begin{split}
        &\cl C_{\omega_o^\prime + \omega_d} = \frac{2g_3}{\omega_d}\xi\cl{\cu A} - \frac{2g_3}{\omega_d+2\omega_o^\prime}\xi\cl{\cu A},\quad
        \cl C_{2\omega_o^\prime } = \frac{4g_3}{3\omega_o^\prime}\cl{\cu A}^2,\\
        &\cl C_{\omega_d-\omega_o^\prime } = \frac{2g_3}{\omega_d-2\omega_o^\prime }\xi\cl{\cu A}^* - \frac{2g_3}{\omega_d}\xi\cl{\cu A}^*,\quad
        \cl C_{0} = 0.
\end{split}\raisetag{1.2\baselineskip}
\end{align}
Accordingly, the diagrams associated with $\sqrt{S_{f\!f}[\omega_j]}\cl C_{\omega_j}^*$ at these order are just the conjugate diagrams (c.f. \cref{eq:order-2-eta,eq:order-2-etas,eq:discussion-dissipator0}) of the diagrams in \cref{eq:discussion-dissipator1}. 

As suggested by the diagrams, the transformed collapse operators can be understood as the nonlinear mixing between the resonant excitation, drive excitation and the incoherent fluctuation in the thermal bath. Besides a trivial renormalization, these transformed collapse operators can lead to nontrivial physics absent in the nondriven system. For example, in the $\omega_j=\omega_d-\omega_o^\prime$ terms in \cref{eq:discussion-dissipator1,eq:discussion-dissipator1-alg} a drive excitation is simultaneously converted to a resonant excitation and a bath excitation, suggesting a non-zero population in the driven excited states even when the oscillator is coupled to a bath at zero temperature \cite{petrescu2020}. Nevertheless, in this section we content ourselves with extending the diagrammatic representation of the effective Hamiltonian to the effective Lindbladian. We refer the readers to a few recent works for more in-depth discussion regarding the effective dissipative dynamics in the driven nonlinear system, namely \cite{petrescu2020} on the life-time renormalization of Josephson circuit when dispersively readout, and \cite{venkatraman2022_2} on the effective Lindbladian of the Kerr-cat system. In these works the frames hosting the effective Lindbladians are constructed via other high-frequency expansions different from the one used in the current work (see \cref{subsec:other-pert} on how they are related); yet one can easily translate the results into the diagrammatic language. The advantage of our diagrammatic method is to directly capture the diagrams that contribute to the oscillator's decoherence dynamics bypassing the need for high-frequency expansions which get cumbersome at high orders.

\subsection{Choice of rotating frame and the general slow-evolving effective Hamiltonian} \label{subsec:frame}

When computing the effective Hamiltonian, an a priori knowledge we have injected into the analysis is some pertinent rotating frame at $\omega_o^\prime$ in which diagrams are constructed. The choice of rotating frame defines the frequency of the resonant excitation \tikzpic{-8}{figsv6_small/notation/A_small.pdf} $ =\clu Ae^{-i\omega_o^\prime t}$ and plays a crucial role in capturing the relevant nonlinear processes --- in \cref{eq:discussion-K-general-diagram,eq:discussion-K-general} the coupling term $\Omega_{lq^\prime,lp^\prime}\xi^{lp^\prime}\cl{\cu A}^{\dagger lq^\prime} + \mathrm{h.c.}$ exists only in the frame rotating at $\omega_o^\prime = p^\prime\omega_d/q^\prime$ but not in any other frame in the vicinity of $\omega_o^\prime$. It is natural to then pose the question: in the absence of an a priori knowledge concerning this rotating frame, how should one perform QHB to capture the concerned effective dynamics of the oscillator?

In this section, we address this question by introducing the notion of a ``slow-evolving effective Hamiltonian'', which captures the near-resonant dynamics of the oscillator. This is a generalization of the static effective Hamiltonian, which captures the resonant dynamics. To illustrate the necessity of this generalized notion, let us first take the example of the Kerr-cat as a concrete exercise and examine the concerned dynamics in two different frames.

The first frame, as discussed in \cref{sec:intro}, corresponds to a frame rotating at $\omega_o^\prime = \omega_d/2$. This choice entails the knowledge that a Kerr-cat system is the quantum manifestation of the classical period doubling bifurcation (thus $4\pi/\omega_o^\prime = 2\pi/\omega_d$) when the drive is near the vicinity of twice the oscillator frequency. The second frame does not entail any intuition and, for convenience, we choose it to be $\omega_o^{\prime\prime} = \omega_o$ that is close to the first choice $\omega_o^\prime$. We consider the following diagram, which stems from the multinomial expansion in \cref{eq:sc-diagram-Gamma,eq:sc-diagram-M}, that is responsible for the squeezing action generating the Kerr-cat:
\begin{align}\label{eq:discussion-squeezing}
\tikzpic{-48}{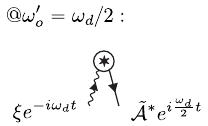},\quad
\tikzpic{-48}{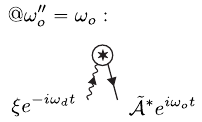}.
\end{align}
In these two frames, this diagram corresponds to $g_3\xi\clu A^*e^{-i\omega_o^\prime t}$ and $g_3\xi\clu A^*e^{-i(\omega_d-\omega_o^{\prime\prime})t}$, respectively, which are resonant and off-resonant with their corresponding rotating frames at $\omega_o^\prime$ and $\omega_o^{\prime\prime}$. Solving the \cref{eq:sc-diagram-Gamma,eq:sc-diagram-M}, the left resonant diagram in \cref{eq:discussion-squeezing} contributes to \tikzpic{-10.5}{figsv6_small/notation/Gamma_small.pdf} while the right off-resonant diagram contributes to \tikzpic{-10}{figsv6_small/notation/M_small.pdf}. Consequently, \tikzpic{-10.5}{figsv6_small/notation/Gamma_small.pdf} constructed in the rotating frame $\omega_o^{\prime\prime}$ will \textit{not} collect this three-wave mixer, causing its corresponding effective Hamiltonian to \textit{fail} in capturing the squeezing action intrinsic to the Kerr-cat system.

To capture the Kerr-cat dynamics constructed in the rotating frame $\omega_o^{\prime\prime}$, we need to relax the requirement for \!\tikzpic{-10.5}{figsv6_small/notation/Gamma_small.pdf} to be strictly resonant with the excitation at $\omega_o$ and allow it to be ``near-resonant'' instead. The rationale for this modification will be discussed shortly. Note that in the driving condition of a Kerr-cat, where $\delta = \omega_o-\omega_d/2\ll\omega_o$, the right diagram in \cref{eq:discussion-squeezing}, which corresponds to $g_3\xi\clu A^*e^{-i(\omega_d-\omega_o)t} = g_3\xi\clu A^*e^{-i(\omega_o+2\delta)t}$, is indeed near-resonant with the oscillator and should now be collected by \!\tikzpic{-10.5}{figsv6_small/notation/Gamma_small.pdf}. Following this revised procedure, we can compute the effective Hamiltonians \!\tikzpic{-11}{figsv6_small/notation/K_small_h.pdf} in each frame as:
\begin{align*}
    \cl K_{\omega_o^{\prime}} &=  \delta \clu A^*\clu A + g_3\xi\clu A^{*2} + \mathrm{c.c.} + \mathcal O(q_\text{zps}^2)\\
    \cl K_{\omega_o^{\prime\prime}} &= g_3\xi\clu A^{*2} e^{2i\delta t} + \mathrm{c.c.} + \mathcal O(q_\text{zps}^2),
\end{align*}
where the subscripts in $\cl K_{\omega_o^{\prime}}$ and $\cl K_{\omega_o^{\prime\prime}}$ denote the rotating frame in which the corresponding Hamiltonian lives. Here $\cl K_{\omega_o^\prime}$ is the same static effective Hamiltonian as \cref{eq:Kerr-cat-RWA}, whose ground state manifold comprises of Schr\"odinger cat states shown in \cref{fig:cat-wigner} (a). In addition, $\cl K_{\omega_o^{\prime\prime}}$ is a slow-evolving Hamiltonian also capturing the Kerr-cat dynamics --- with numerical method, e.g. the Floquet numerical diagonalization which we will introduce in \cref{subsec:other-pert}, one can find that the ground state manifold of $\cl K_{\omega_o^{\prime\prime}}$ are also Shr\"odinger cat states but with slow-evolving phase at frequency $\delta$. Indeed, the slow-evolving $K_{\omega_o^{\prime\prime}}$ can be further converted to a static effective Hamiltonian by going to another rotating frame $\clu A\rightarrow \clu Ae^{i\delta t}$, which yields
\begin{align*}
    \cl K_{\omega_o^{\prime\prime}-\delta} = \delta\clu A^*\clu A + g_3\xi\clu A^{*2} + \mathrm{c.c.}+ \mathcal O(q_\text{zps}^2).
\end{align*}
Note that $\cl K_{\omega_o^{\prime\prime}-\delta}$ is just the effective Hamiltonian $\cl K_{\omega_o^{\prime}}$ constructed in the rotating frame at $\omega_o^\prime = \omega_d/2 = \omega_o^{\prime\prime} - \delta$.

To better understand the motivation behind the concept of the slow-evolving Hamiltonian, let us revisit the general goal of QHB that underlies the diagrams. When analyzing a driven nonlinear oscillator described by a bosonic coordinate $\cl a$, the QHB seeks to find a frame transformation so that the dynamics of $\cl a$ is separated into \tikzpic{-9}{figsv6_small/notation/B_small.pdf} ($\clu B e^{-i\omega_o^\prime t}$) and \tikzpic{-11}{figsv6_small/notation/eta_small.pdf} ($\cl\mu (t) e^{-i\omega_o^\prime t}$). The former, termed as resonant-excitation, is to capture the effective dynamics of interest near-resonant with the oscillator, and the latter, termed as the off-resonant excitation, is to capture the micromotion, which is of less physical importance. It should be recognize that \tikzpic{-9}{figsv6_small/notation/B_small.pdf} ($\clu B e^{-i\omega_o^\prime t}$) and \tikzpic{-11}{figsv6_small/notation/eta_small.pdf} ($\cl\mu (t) e^{-i\omega_o^\prime t}$) both contain an explicit time-dependence as a rotating phase, which occurs at frequency $\omega_o^\prime$ and $m\omega_o^\prime + n\omega_d$ (with $m,n\in \mathbb Z$) for the two excitations, respectively, as well as an implicit time-dependence generated by the effective Hamiltonian \!\tikzpic{-11}{figsv6_small/notation/K_small_h.pdf}($\cl K$) and assumed to be much slower than the rotating phase. The terms  ``near-resonant", ``off-resonant", ``slow'', and ``fast'' here are used in a perturbative sense --- we assume that the detuning between full dynamics of \tikzpic{-9}{figsv6_small/notation/B_small.pdf} ($\clu B e^{-i\omega_o^\prime t}$) and the oscillator frequency $\omega_o$ is much smaller than $\omega_o$ while the detuning associated with \tikzpic{-11}{figsv6_small/notation/eta_small.pdf} ($\cl\mu(t) e^{-i\omega_o^\prime t}$) is of the same order as $\omega_o$. Reinforcing this perturbative hierarchy leads to a separation of time-scale for the dynamics of the oscillator.

In the process of constructing a strictly static Hamiltonian, \tikzpic{-11}{figsv6_small/notation/eta_small.pdf}($\cl\mu (t) e^{-i\omega_o^\prime t}$) is designed to include all terms with explicit frequency $m\omega_o^\prime + n\omega_d\ne\omega_o^\prime$. However, this could lead to the failure to separate the time scales if $$|(m\omega_o^\prime + n\omega_d)-\omega_o^\prime|\ll\omega_o^\prime$$ for some $m,n$, causing \tikzpic{-11}{figsv6_small/notation/eta_small.pdf} ($\cl\mu (t) e^{-i\omega_o^\prime t}$) to contain near-resonant dynamics that is supposed to be captured by \tikzpic{-9}{figsv6_small/notation/B_small.pdf} ($\clu B e^{-i\omega_o^\prime t}$) and \!\tikzpic{-11}{figsv6_small/notation/K_small_h.pdf}($\cl K$). This is the failure mode for the Kerr cat analysis at rotating frame $\omega_o^{\prime\prime}=\omega_o$, as illustrated above. To maintain the perturbative hierarchy in the dynamics of the oscillator, it is neccessary to include the terms with frequency $m\omega_o^\prime + n\omega_d\approx\omega_o^\prime$ in \tikzpic{-9}{figsv6_small/notation/B_small.pdf} ($\clu B e^{-i\omega_o^\prime t}$) instead. This adjustment is equivalent to relaxing the requirement that the effective Hamiltonian $\cl K$ be strictly static, allowing it to be ``slow-evolving'' in the perturbative sense.

In some simpler cases, a well-chosen rotating frame $\omega_o^\prime$ can still lead to time-independent Hamiltonian, while this is not possible in general cases. In particular, if the oscillator only exhibits energy renormalization effects (captured by diagrams with $l=0$ in \cref{eq:discussion-K-general-diagram}) but not coupling effects (captured by diagrams with $l\ne0$) up to the perturbative order of interest, then any rotating frame in the vicinity of $\omega_o$ can result in a static effective Hamiltonian. This can be seen from \cref{eq:discussion-K-general-diagram} by noting that  the diagrams with $l=0$ are always static, independent of the resonant excitation frequency $\omega_o^\prime$, which is determined by the choice of rotating frame. In the case that the oscillator only exhibits one dominant nonlinear process labeled as $(q:p)$, in which $q$ number of oscillator quanta near-resonantly couple to $p$ number of drive photons, there exists a special rotating frame $\omega_o^\prime = p\omega_d/q$ in the vicinity of $\omega_o$, e.g. $(q:p)=(2:1)$ in the Kerr-cat, that renders the coupling diagram in the effective Hamiltonian to be static. Other choices of rotating frame will yield a slow-varying effective Hamiltonian. 

However, in more complicated cases where multiple nonlinear processes coexist, it is impossible to find a rotating frame that can render all these processes static simultaneously. For example, if $(q_1:p_1)$ and $(q_2:p_2)$ processes coexist and $p_1/q_1\ne p_2/q_2$, there exists no rotating frame that can render two diagrams, each involving $q_1$ or $q_2$ incoming resonant excitations and $p_1$ or $p_2$ drive outgoing excitations, respectively, to be static simultaneously. In these cases, the generalized procedure of constructing a slow-evolving effective Hamiltonian is essential to capture the full effective dynamics. We will illustrate this generalized procedure and the power of it in a concrete system in \cref{sec:mpnr} (see \cref{eq:mnr-K-10-4}).

\subsection{Relation to Floquet formalism}\label{subsec:other-pert}

In this subsection, we briefly discuss the relationship between our diagrammatic perturbation method and the widely-used Floquet-based methods \cite{kuchment1993,rahav2003,eckardt2015,casas2001,zhang2017} for analyzing periodically driven nonlinear systems. As will be demonstrated, our diagrammatic approach can be construed as a non-traditional variant of Floquet perturbation theory. The effective Hamiltonian derived through our diagrammatic procedure concurs with specific instances of the Floquet methodologies, notably the Schrieffer-Wolff expansion \cite{rahav2003,eckardt2015,venkatraman2021}.

Floquet theory states that the solution of a time-dependent Schr\"{o}dinger equation $\op{H}(t)|\Psi(t)\rangle = i\hbar \frac{\partial}{\partial t}|\Psi(t)\rangle$ for some periodic Hamiltonian $\op{H}(t+T) = \op{H}(t)$ takes the form of 
\begin{align}\label{eq:floquet-state-2}
|\Psi_m(t)\rangle = e^{-\op \Lambda(t)/i\hbar}e^{\op H_\text{F}t/i\hbar}|\tilde m\rangle,
\end{align}
where $\op\Lambda(t)$, called the micromotion generator, is a periodic function with periodicity $T$, and $\op H_\text{F}$, called the Floquet Hamiltonian, is a time independent function with eigenstates $|\tilde m\rangle$ and eigenvalues $\epsilon^\text{F}_m$. The Floquet state $|\Psi_m(t)\rangle$, written in the form of \cref{eq:floquet-state-2}, conveys a physical picture similar to that in the QHB: the dynamics of a driven state is decomposed into a slow dynamics generated by the static Hamiltonian $\op H_\text{F}$ and a micromotion generated by the periodic function $\op \Lambda(t)$. It is worth noting that the construction of $\op\Lambda$ and $\op H_\text{F}$ is not unique. The static part of $\op\Lambda$, denoted as $\text{Sta}(\op \Lambda)$, can vary as a gauge transformation, and as a result, the Floquet Hamiltonian $\op H_\text{F}$ is uniquely determined up to a gauge transformation. However, the quasienergy spectrum of the system remains invariant across different gauge choices.

To find the solution of the Floquet state, one class of methods, known as Floquet perturbation methods \cite{rahav2003,eckardt2015,casas2001}, takes \cref{eq:floquet-state-2} as an ansatz and plugs it into the Schr\"{o}dinger equation. With some rearrangement, one finds that  the Floquet Hamiltonian $\op H_\text{F}$ is related to $\op H(t)$ by $\op H_\text{F}= e^{\op \Lambda/i\hbar}(\op H(t)+i\hbar\partial_t) e^{-\op \Lambda/i\hbar}$.  Floquet perturbation methods assume that $\op H(t)$, $\op\Lambda(t)$, and $\op H_\text{F}$ each is a perturbative series starting at order one in some relevant perturbative parameter. These methods construct $\op\Lambda(t)$ order by order using specific iterative procedures so that the Floquet Hamiltonian $\op H_\text{F}$ is time-independent at each order. The Schrieffer-Wolff expansion (also known as the van Vleck expansion) \cite{rahav2003,eckardt2015} and the Floquet-Magnus expansion \cite{casas2001} are two notable examples of Floquet perturbation methods each prescribing a particular iterative procedure mentioned above. In addition, different Floquet perturbation methods are associated with different gauges $\text{Sta}(\op \Lambda)$ \cite{venkatraman2021}.

To relate the Floquet formalism and our diagrammatic method, we consider the periodic Hamiltonian $\op{H}(t)$ defined in \cref{eq:H-tran}. This Hamiltonian describes a driven nonlinear oscillator in a displaced and rotated frame at a frequency $\omega_o^\prime = p\omega_d/q$. The period of the Hamiltonian is $T =2\pi q/\omega_d$. As discussed in \cref{sec:averaging}, the effective Hamiltonian $\hat K$ can be computed using diagrams, and it is related to $\op{H}(t)$ by a unitary transformation generated by $\hat S (t)$, which can be found diagrammatically from \cref{eq:zeta}. Therefore, the Floquet states associated with $\op{H}(t)$ are related to the eigenstates of $\hat K$ by
\begin{align}
|\Psi_m(t)\rangle &= e^{-\op S(t)/i\hbar}e^{\op Kt/i\hbar}|\tilde m\rangle\label{eq:floquet-state-decom},
\end{align}
where $|\tilde m\rangle$ denotes the eigenstates of $\op K$. Upon comparison of \cref{eq:floquet-state-decom} and \cref{eq:floquet-state-2}, it becomes evident that the effective Hamiltonian $\op K$ and the generator $\op S$ constructed from our diagrammatic method are respectively a Floquet Hamiltonian $\op H_\text{F}$ and a micromotion generator $\op\Lambda$. Furthermore, since $\op S$ is constructed as a purely oscillating function, the diagrammatic method corresponds to the Floquet perturbation method with the gauge of $\text{Sta}(\op\Lambda)=0$, which is the choice made in the Schrieffer-Wolff expansion \cite{rahav2003,eckardt2015}. 

In the discussion above, we can see that the diagrammatic method serves as an unconventional Floquet perturbation method. It begins with an additive ansatz (see \cref{eq:QHB-ansatz}) rather than an exponential one (see \cref{eq:floquet-state-2}), yet it produces the same micromotion generator $\hat S$ and effective Hamiltonian $\hat K$ as the Schrieffer-Wolff expansion. This equivalence, already analytically demonstrated, will be further illustrated in a concrete example in \cref{sec:renormalization}, where we apply the diagrammatic method to an experimental system previously analyzed using a Schrieffer-Wolff-like expansion in our recent work (\cite{venkatraman2021}).  We will show that both methods indeed yield the same effective Hamiltonian to a very high order in the perturbative parameter.

In comparison to the conventional Floquet perturbation method, a significant advantage of our diagrammatic method is its ability to independently compute each diagram. This feature not only makes the diagrammatic method compatible with parallel computing but also enables the calculation of specific effective Hamiltonian terms of interest, such as the coupling term in \cref{eq:discussion-K-general}, without computing the others. In contrast, the iterative nature of the conventional Floquet methods requires the knowledge of all previous terms in the Floquet Hamiltonian $\op H_\text{F}$ and the micromotion generator $\op \Lambda$ at each order. As a result, the diagrammatic method can be executed more efficiently to a perturbative order that is significantly higher than that achievable by conventional Floquet perturbation methods. This advantage is showcased in \cref{sec:mpnr}, where we analyze the multiphoton resonance process in driven superconducting circuits to an order that was previously unattainable.

Besides the Floquet perturbation methods, a Floquet problem can be solved exactly through Floquet numerical diagonalization, which we will review in more detail in \cref{app:floquet}. Here we consider the time-evolution operator associated with $\op H(t)$ as $\op U(0,t)$. The evolution of a Floquet state over one period of time reads
\begin{subequations}
\begin{align}
\op{U}(0,T)|\Psi_m(0)\rangle &= |\Psi_m(T)\rangle\\
\op{U}(0,T)e^{-\op \Lambda(0)/i\hbar}|\tilde m\rangle &=e^{-\op \Lambda(T)/i\hbar}e^{\epsilon_m^\text{F}T/i\hbar}|\tilde m\rangle\\
&= e^{\epsilon_m^\text{F}T/i\hbar}e^{-\op \Lambda(0)/i\hbar}|\tilde m\rangle.\label{eq:floquet-prop}
\end{align}
\end{subequations}
As written in \cref{eq:floquet-prop}, we see that the eigenvalues and eigenvectors of $\op{U}(0,T)$ are respectively $e^{\epsilon_m^\text{F} T/i\hbar}$ and $e^{-\op \Lambda(0)/i\hbar}|\tilde m\rangle$, while $|\Phi(t)\rangle = e^{-\op \Lambda(t)/i\hbar}|\tilde m\rangle$ is also known as the Floquet mode. Floquet numerical diagonalization exploits this relation and finds the eigenenergies and Floquet modes by diagonalizing $\op{U}(0,T)$, which is obtained by numerically integrating $\op H(t)$ to one period. 

In our work, we perform Floquet numerical diagonalization, in order to capture the exact dynamics, as a benchmark of our diagrammatic method. However, for a strongly driven nonlinear system with around 10 or more states, the quasienergy spectrum and Floquet states solved from Floquet numerical diagonalization can be challenging to decipher, especially under strong driving conditions where many states hybridize with each other. In contrast, our analytical method provides the advantage of mapping the Floquet states into the Fock state basis analytically. This mapping not only reveals the analytical structure of the driven dynamics and but can also serve as a useful tool when employed in conjunction with Floquet numerical diagonalization to decipher the quasienergy spectrum and Floquet states. This will become more clear in the concrete examples treated in \cref{sec:mpnr,sec:IST}.

\section{Application Examples}\label{sec:examples}

\subsection{Stabilization of three-legged Schr\"odinger cats}\label{sec:3lcat}

In this section, we exemplify the axiomatic procedures outlined in \cref{sec:diagrams-pert} by providing a detailed calculation of the three-legged Schr\"odinger cat states, as mentioned in \cref{sec:intro}. This example is particularly instructive as it encompasses all the diagrammatic elements and rules introduced thus far.

In particular, we consider a driven SNAIL superconducting circuit (see \cite{frattini2017}) described by the Hamiltonian
\begin{align}\label{eq:snail-H}
\begin{split}
    \frac{\op{\mathcal{H}}(t)}{\hbar} = \omega_o\op a^\dagger \op a &+ \sum_{m\ge3}\frac{g_m}{m}(\op a + \op a^\dagger)^m \\
    &- i\Omega_d (\op a - \op a^\dagger)\cos\omega_d t.
\end{split}
\end{align}
In accordance with the standard quantum optics convention, here and in the following text, we take $\op a$ and $\op a^\dagger$ to be the usual annihilation and creation ladder operators; in other words, we take their commutator to be 1. In this case, all the diagrammatic rules in \cref{sec:diagrams-pert} remain valid but with three modifications: (1) the effective Hamiltonian diagram \!\tikzpic{-11}{figsv6_small/notation/K_small_h.pdf} is associated with $\op K/\hbar$ instead of $\op K$, (2) each $\hbar$ factor in $\varstar$ is replaced by a factor of 1, and (3) each quantum bond is associated with a factor of 1 instead of $\hbar$ (c.f. \cref{eq:unabridged-order1}). Moreover, when analyzing a Josephson circuit, we take $\varphi_\text{zps}$ in lieu of $q_\text{zps}$ as the perturbative parameter underlying the diagrams. Here $\varphi_\text{zps} \ll 1$ is the zero point spread of the phase $\varphi$ across the Josephson junction, which is the source of the native nonlinearity. In the Hamiltonian \cref{eq:snail-H}, this perturbative structure is manifested as $g_m/\omega_o = \mathcal{O}(\varphi_\text{zps}^{m-2})$.
 
The three-legged Schr\"odinger cat states are obtained when the drive frequency $\omega_d$ is chosen to be in the vicinity of $3\omega_o/2$. Provided that the nonlinearities $g_m$'s are correctly engineered, in a semiclassical picture, the $2/3$rd ultra-subharmonic of the drive, at frequency $2\omega_d/3\approx\omega_o$, can consequently be resonantly stablized through the oscillator. We refer to this process as the $(3:2)$ ultra-subharmonic bifurcation, the general class of which will be discussed in more detail in \cref{sec:USH}. In the quantum regime, this bifurcation manifests when the system adopts the three-legged Schr\"odinger cat states as its ground states. We capture this bifurcation process by transforming $\mathcal{H}(t)$ in \cref{eq:snail-H} to a new frame amounting to $\op a \rightarrow  \op a e^{-i\omega_o^\prime t} + \xi  e^{-i\omega_dt}$, where $\xi =\frac{i\Omega_d\omega_d}{\omega_d^2-\omega_o^2}$ and $\omega_o^\prime = 2\omega_d/3$. The transformed Hamiltonian reads 
\begin{align}\label{eq:snail-H-tran}
\begin{split}
    \frac{\op H(t)}{\hbar} = \delta\op a^\dagger \op a+  \sum_{\substack{m \ge 3 \\ m \in \mathbb{N}}} \frac{g_m}{m}  &(\hat{a} e^{- i \omega_o^\prime t} + \hat{a}^{\dagger} e^{i \omega_o^\prime t} \\[-8pt]
    & \;+ \xi e^{-i \omega_d t} +  \xi^* e^{i \omega_d t} )^m
\end{split}\raisetag{0.4\baselineskip}
\end{align}
with $\delta = \omega_o - \omega_o^\prime$. Now we can use the diagrammatic method to directly write down the effective Hamiltonian associated with \cref{eq:snail-H-tran}. 

The most important effective Hamiltonian terms are those associated with the diagrams of 5-wave composite mixers with two incoming drive excitations and three outgoing resonant excitations. Such diagrams, in algebraic form, read $\Omega_{3,2}\xi^2\op{ \cu A}^{\dagger 3}+\mathrm{h.c.}$ with some prefactor $\Omega_{3,2} = \mathcal O(\varphi_\text{zps}^3)$ to be determined. These terms obey three-fold rotational symmetry in phase space and are directly responsible for the bifurcation. We are interested in finding the leading order contribution of $\Omega_{3,2}$ diagrammatically. To this end, we first follow Step 1 in \cref{sec:diagrams-pert} and write down the forest of unrooted trees involving 5 leaves as 
\begin{align}\label{eq:3leg-tree}
\tikzpic{-28}{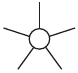} \qquad \tikzpic{-30}{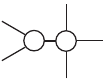}\quad
\tikzpic{-35}{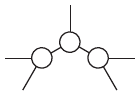}.
\end{align}
Following Steps 2-4, for the five external edges in each tree above, we assign two incoming drive excitations and three outgoing dressed resonant excitations, and for each internal edge we assign a dressed off-resonant excitation traveling in either direction. The first tree in \cref{eq:3leg-tree} consequently yields 2 unique diagrams 
\begin{subequations}\label{eq:3leg-K-12}
\begin{align}\label{eq:3leg-K-1}
\tikzpic{-28}{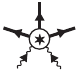} \quad \tikzpic{-30}{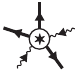},
\end{align}
the second tree yields 14 unique diagrams 
\begin{align}\label{eq:3leg-K-2}
\begin{split}
\tikzpic{-28}{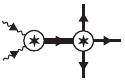} \,
&\tikzpic{-28}{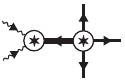} \,
\tikzpic{-28}{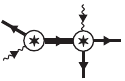} \,
\tikzpic{-28}{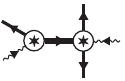} \\
\tikzpic{-28}{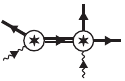} \,
&\tikzpic{-28}{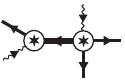} \,
\tikzpic{-28}{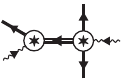} \,
\tikzpic{-28}{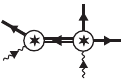} \\
\tikzpic{-28}{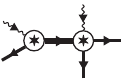} \,
&\tikzpic{-28}{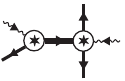} \,
\tikzpic{-28}{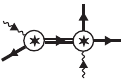} \,
\tikzpic{-28}{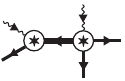} \\
&\tikzpic{-28}{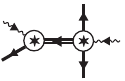} \,
\tikzpic{-28}{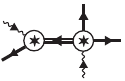},
\end{split}
\end{align}
and the third tree yields 40 unique diagrams, whose explicit can be form written down in a straightforward way. 
\end{subequations}

Each diagram above is to be further transformed following Steps 5-7, which we concretely illustrate with the last diagram in \cref{eq:3leg-K-2} as an example. First, at Step 5, we convert a \tikzpic{-10}{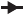} traveling away from a mixer to a  \tikzpic{-9}{figsv6_small/notation/Bs_small.pdf}. This yields three diagrams each a rooted tree:
\begin{align}\label{eq:3leg-K-rooted}
\tikzpic{-28}{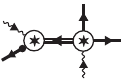} \quad
\tikzpic{-28}{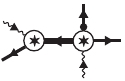} \quad
\tikzpic{-28}{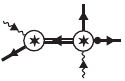} \quad
\end{align}
These diagrams are at the level of Hamiltonian \!\tikzpic{-11}{figsv6_small/notation/K_small_h.pdf}~($\cl K/\hbar$) and, to evaluate them at Step 7 (we discuss Step 6 later), they are further converted to the level of  \!\tikzpic{-10.5}{figsv6_small/notation/Gamma_small.pdf} by removing \!\tikzpic{-9}{figsv6_small/notation/Bs_small.pdf}:
\begin{align}\label{eq:3leg-gamma}
\tikzpic{-28}{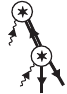} \qquad
\tikzpic{-28}{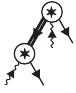} \qquad\;
\tikzpic{-28}{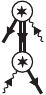}\,,
\end{align}
where we have re-orientated the diagrams in a way similar to those in \cref{sec:averaging}. The three diagrams in \cref{eq:3leg-gamma} to leading order (i.e. considering only the bare propagator contribution of \!\tikzpic{-9}{figsv6_small/notation/prop_dressed_small.pdf} and the bare resonant excitation of \!\tikzpic{-9}{figsv6_small/notation/Bs_small.pdf}) correspond to  
\begin{align}\label{eq:3leg-Gamma-algebra}
\begin{split}
&\quad g_3\xi \varstar\Big(\frac{g_4}{\Delta_{-\omega_d/3}}\xi\varstar \clu A^* \varstar \clu A^*\!\Big)e^{-i\frac{2}{3}\omega_dt} \!\\
&\quad + g_4\Big( \frac{g_3}{\Delta_{-\omega_d/3}}\xi \varstar\clu A^*\!\Big)\varstar \xi \varstar \clu A^*e^{-i\frac{2}{3}\omega_dt}\\
&\quad  + g_4\clu A^* \varstar \Big(\frac{g_3}{\Delta_{-\omega_d/3}}\xi\varstar \clu A^*\Big)\varstar\xi e^{-i\frac{2}{3}\omega_dt} \\
&= -\frac{3g_3g_4}{\omega_d} \xi^2\clu A^{*2}e^{-i\frac{2}{3}\omega_dt},
\end{split}\raisetag{1\baselineskip}
\end{align}
where $\Delta_{-\omega_d/3} = -\frac{1}{3}\omega_d - \frac{2}{3}\omega_d$ is the bare detuning between the off-resonant excitation at $-\omega_d/3$ and the resonant excitation at $\omega_o^\prime = 2\omega_d/3$. The leading order effective Hamiltonian terms associated with \cref{eq:3leg-K-rooted} are computed by integrating \cref{eq:3leg-Gamma-algebra} over $\clu A^*e^{i\frac{2}{3}\omega_dt}$. This results in $-g_3g_4\xi^2\clu A^{*3}/\omega_d -g_3g_4\xi^{*2}\clu A^{3}/\omega_d$, which in the Hilbert space representation reads $-g_3g_4\xi^2\op {\cu A}^{\dagger 3}/\omega_d-g_3g_4\xi^{*2}\op {\cu A}^{3}/\omega_d$.

With the assistance of a computer program \cite{qhb2022}, we construct all the bare diagrams derived from \cref{eq:3leg-tree} and evaluate them as
\begin{align}\label{eq:3leg-coupling}
\begin{split}
    (2g_5 -\frac{165g_3g_4}{8\omega_d} + \frac{195g_3^3}{4\omega_d^2})\xi^2\op{\cu A}^3+ \mathrm{h.c.},
\end{split}\raisetag{1\baselineskip}
\end{align}
which are just the terms displayed in the second line of \cref{eq:3-legged-cat-H}. These terms are directly responsible for the three-legged cat formation as they endorse a 3-fold rotational symmetry in phase space. Since the sought-after terms are already captured, we truncate our perturbative expansion at this order, i.e. $\varphi_\text{zps}^3$, while higher order terms will only perturbatively correct the effective dynamics captured at this order. 

Within order $\varphi_\text{zps}^3$, a different type of diagrams contributing to the effective Hamiltonian are those containing equal number of incoming and outgoing waves for each type of excitation. Following Steps 1-4 in \cref{sec:diagrams-pert}, one constructs 21 such diagrams as 
\begin{align}\label{eq:3leg-renorm-diagram}
\underbrace{\tikzpic{-26}{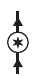}}_\text{Type I} \;
\underbrace{
    \tikzpic{-26}{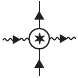} \;
    \tikzpic{-18}{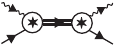} 
    }_\text{Type II}\;
    \underbrace{
    \tikzpic{-26}{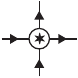}\;
    \tikzpic{-18}{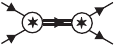}}_\text{Type III}
    \;\cdots\raisetag{0.4\baselineskip}
\end{align}
where the diagrams not explicitly displayed are those similar to the displayed ones with the external edges permuted or the internal edges inverted. Note that in \cref{eq:3leg-renorm-diagram}, we represent the resonant excitations in each diagram in their bare form, with the exception of the first one. This is because the dressed resonant excitation includes corrections to the bare one at order $\mathcal O(\varphi_\text{zps}^2)$ (see \cref{eq:dressed-osc-4}). Consequently, only the first diagram in \cref{eq:3leg-renorm-diagram} with these corrections is $\mathcal O(\varphi_\text{zps}^3)$, while the others are $\mathcal O(\varphi_\text{zps}^4)$, which is beyond the perturbative order of interest. The off-resonant excitations in \cref{eq:3leg-renorm-diagram} are depicted in their dressed form, as they involve $\mathcal O(\varphi_\text{zps})$ corrections beyond the bare ones (see \cref{eq:eta-dressed-3}) and are thus relevant in these diagrams. In addition, based on the leaf vertices, the diagrams in \cref{eq:3leg-renorm-diagram} are further categorized into three types as indicated by the explicit label in \cref{eq:3leg-renorm-diagram} containing both the displayed diagrams and undisplayed diagrams similar to the displayed ones. We will discuss the role of each type in the effective Hamiltonian soon. 

Each diagram in \cref{eq:3leg-renorm-diagram} is evaluated following Step 5-8. Specifically, the bare diagram contribution to \cref{eq:3leg-renorm-diagram} should be evaluated first following Step 5 and 7, in a manner similar to evaluating the diagrams in \cref{eq:3leg-K-12}. We refrain from detailing these steps once more for the sake of brevity.  Furthermore, since the diagrams in \cref{eq:3leg-renorm-diagram} involve dressed oscillator and dressed off-resonant excitations, the relevant Steps 6 and 8, which were omitted in the evaluations of \cref{eq:3leg-K-12}, should also be performed. We now illustrate the procedures of evaluating these diagrams involving dressed components. 

We first consider diagrams involving dressed propagator of the off-resonant excitation by taking the third diagram in \cref{eq:3leg-renorm-diagram} as an example. Following Steps 5 \& 8, we can convert this diagram to the level of  \!\tikzpic{-10.5}{figsv6_small/notation/Gamma_small.pdf} as 
\begin{align}\label{eq:3leg-gamma-4}
\tikzpic{-26}{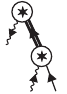} = 
\tikzpic{-26}{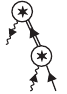} \!\!+\,\;
\tikzpic{-26}{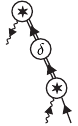} \!\!+\,\; \tikzpic{-26}{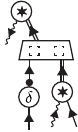} + \mathcal O(\varphi_\text{zps}^4)\raisetag{\baselineskip},
\end{align}
where we have expanded the dressed off-resonant excitation in the left-hand side following \cref{eq:eta-dressed-3}. In algebraic form, the series in the right-hand side of \cref{eq:3leg-gamma-4} reads
\begin{align}
\begin{split}
&\;g_3\xi\varstar\left(\frac{g_3}{\Delta_{5\omega_d/3}}\xi\varstar\clu A\right)e^{-i\frac{2}{3}\omega_dt} \\
&+ g_3\xi\varstar\left(\frac{\delta g_3}{\Delta^2_{5\omega_d/3}}\xi\varstar\clu A\right)e^{-i\frac{2}{3}\omega_dt}\\
&+g_3\xi\!\varstar\!\left(\!\frac{1}{\Delta_{5\omega_d/3}}\moyal{\delta\clu A^*\!\clu A}{\frac{g_3}{\Delta_{5\omega_d/3}}\xi\!\varstar\!\clu A}\!\right)\!e^{-i\frac{2}{3}\omega_dt}\!+\! \mathcal O(\varphi_\text{zps}^4)
\end{split}\nonumber\\
&=\frac{g_3^2}{\omega_d}\xi^2\clu A e^{-i\frac{2}{3}\omega_dt}+ \mathcal O(\varphi_\text{zps}^4)\label{eq:3leg-gamma-4-al}
\end{align}
where $\Delta_{5\omega_d/3} = \frac{5}{3}\omega_d - \frac{2}{3}\omega_d$ is the bare detuning between the off-resonant excitation at $5\omega_d/3$ and the resonant excitation at $\omega_o^\prime = 2\omega_d/3$. Integrating \cref{eq:3leg-gamma-4-al} over $\clu A^*$, one then obtains the leading order contribution of the third diagram \cref{eq:3leg-renorm-diagram} in to the effective Hamiltonian. 

To evaluate the first diagram in \cref{eq:3leg-renorm-diagram}, which involves the dressed resonant excitation, we follow Step 5 and 8 in \cref{sec:diagrams-pert} and convert this diagram to the level of  \!\tikzpic{-10.5}{figsv6_small/notation/Gamma_small.pdf} as 
\begin{align}\label{eq:3leg-gamma-5}
    \tikzpic{-25}{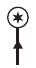} \;=\; \tikzpic{-25}{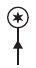} \;+\; \tikzpic{-40}{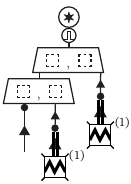} \;+\; \mathcal O(\varphi_\text{zps}^4),
\end{align}
where we have expanded the dressed resonant excitation in the left-hand side following \cref{eq:dressed-osc-4}. Evaluating \cref{eq:3leg-gamma-5} to order $\varphi_\text{zps}^3$ requires plugging in the leading order contribution of the off-resonant excitation \tikzpic{-11}{figsv6_small/notation/eta_small.pdf}. Following Step 6 in \cref{sec:diagrams-pert}, we collect all the off-resonant excitations present in the diagrams in \cref{eq:3leg-K-12,eq:3leg-renorm-diagram}. This yields:
\begin{align}\label{eq:3leg-eta}
    \tikzpic{-23}{figsv6_small/notation/eta.pdf} =\; \tikzpic{-23}{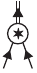} + \; \tikzpic{-23}{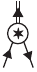} + \; \tikzpic{-23}{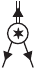} + \; \tikzpic{-23}{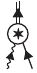} + \; \tikzpic{-23}{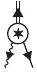} + \; \tikzpic{-23}{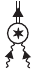} + \cdots,
\end{align}
where we have explicitly displayed some representational terms at order $\varphi_\text{zps}^1$. By plugging \cref{eq:3leg-eta} into \cref{eq:3leg-gamma-5}, we obtain
\begin{align}\label{eq:3leg-gamma-5-2}
    \tikzpic{-25}{figsv6_small/3leg/Gamma5.pdf} \;=\; \tikzpic{-25}{figsv6_small/3leg/Gamma5_1.pdf} \;+\; \tikzpic{-43}{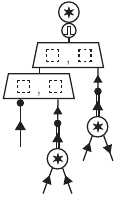}\;+\; \tikzpic{-43}{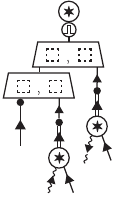} \;+\; \cdots,
\end{align}
where each diagram can be readily evaluated following the rules introduced in \cref{subsec:dressing,subsec:dressing-osc}. For brevity, we only explicate the evaluation of the second diagram in the right-hand side of \cref{eq:3leg-gamma-5} as follows: 
\begin{align*}
\begin{split}
    &\frac{\delta}{2!}\{\!\!\{\moyal{\clu A e^{-i\omega_o^\prime t}}{\int d\clu A^* \, \frac{g_3}{\Delta_{4\omega_d/3}}\clu A\varstar\clu A e^{-i2\omega_o^\prime t}},\\
    &\qquad\qquad\quad\int d\clu A^* \, \frac{g_3}{\Delta_{0}}\clu A\varstar\clu A^*\}\!\!\}
\end{split}\\
& = -\frac{9\delta g_3^2}{4\omega_d^2}\clu A^{*}\clu A^2e^{-i\omega_o^\prime t} -\frac{27\delta g_3^2}{8\omega_d^2}\clu Ae^{-i\omega_o^\prime t},
\end{align*}
where $\Delta_{4\omega_d/3} = \frac{4}{3}\omega_d - \frac{3}{2}\omega_d$ is the detuning between the off-resonant excitation at $4\omega_d/2$ and the resonant excitation at $\omega_o^\prime = 2\omega_d/3$ and $\Delta_{0} = 0 - \frac{3}{2}\omega_d$ is the detuning between the off-resonant excitation at $0$ frequency and the resonant excitation.

Each diagram in \cref{eq:3leg-gamma-5-2} is constructed at the level of \!\tikzpic{-10.5}{figsv6_small/notation/Gamma_small.pdf}. To compute its corresponding contribution to the effective Hamiltonian \!\tikzpic{-11}{figsv6_small/notation/K_small_h.pdf} $(\cl K/\hbar)$, we perform an ``integration" over \!\tikzpic{-9}{figsv6_small/notation/Bs_small.pdf}, as defined by the series in \cref{eq:dressed-K-2}. This operation, to leading order, corresponds to an integration over \!\tikzpic{-9}{figsv6_small/notation/As_small.pdf} $(\clu A^* e^{i\omega_o^\prime t})$, with higher-order corrections being at order $\mathcal O(\varphi_\text{zps}^2)$. When computing the Hamiltonian term up to $\varphi_\text{zps}^3$, these corrections should be taken into account for the first diagram in the right-hand side of \cref{eq:3leg-gamma-5-2}, which itself is at order $\varphi_\text{zps}^1$. In particular, the Hamiltonian term associated with this diagram can be expanded as: 
\begin{align}\label{eq:3leg-K-5-1}
    \tikzpic{-25}{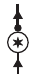} \;&=\; \tikzpic{-25}{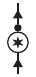}\;+\; \tikzpic{-30}{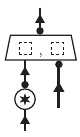}\;-\; \tikzpic{-30}{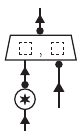} +\; \cdots \\
    &=\; \tikzpic{-25}{figsv6_small/3leg/K5_1.pdf}\;+\; \tikzpic{-54}{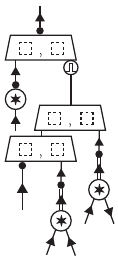}+\quad \tikzpic{-54}{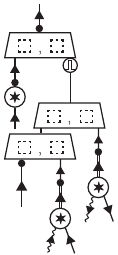} +\; \cdots,\label{eq:3leg-K-5-2}
\end{align}
where \cref{eq:3leg-K-5-1} is just \cref{eq:dressed-K-2} with \!\tikzpic{-10.5}{figsv6_small/notation/Gamma_small.pdf} specified as the first diagram in the right-hand side of \cref{eq:3leg-gamma-5-2}, and in \cref{eq:3leg-K-5-2} we have further plugged in \cref{eq:dressed-osc-4} for  \!\tikzpic{-9}{figsv6_small/notation/B_small.pdf} and \cref{eq:3leg-eta} for \!\tikzpic{-11}{figsv6_small/notation/eta_small.pdf}. The first diagram in \cref{eq:3leg-K-5-2} is simply evaluated as $\int d\clu A^*\delta\clu A = \delta \clu A^*\clu A$, and other diagrams can be evaluated similar to those in \cref{eq:3leg-gamma-5-2}. For example, the algebraic expression associated with the second diagram in \cref{eq:3leg-K-5-2} reads
\begin{align*}
\begin{split}
    &\int d\clu A^* e^{i\omega_o^\prime t} \\
    &\times \{\!\!\{ \delta \clu A^*\clu A \,, \frac{1}{2!}\{\!\!\{\moyal{\clu A e^{-i\omega_o^\prime t}}{\int d\clu A^* \, \frac{g_3}{\Delta_{4\omega_d/3}}\clu A\varstar\clu A e^{-i\omega_o^\prime t}},\\
    &\qquad\qquad\qquad\qquad\qquad\qquad\int d\clu A^* \, \frac{g_3}{\Delta_0}\clu A\varstar\clu A^* e^{i\omega_o^\prime t}\}\!\!\}\}\!\!\}
\end{split}\\
& = \frac{9\delta g_3^2}{8\omega_d}\clu A^{*2}\clu A^2 +\frac{27\delta g_3^2}{8\omega_d}\clu A^*\clu A
\end{align*}

Up to now, we have demonstrated all the diagrammatic procedures required for computing the effective Hamiltonian of the three-legged Schr\"odinger cats in a driven superconducting circuit. To obtain such effective Hamiltonian to order $\varphi_\text{zps}^3$, which is 2 order beyond rotating wave approximation, we have utilized all of the diagrammatic elements and rules established in \cref{sec:averaging}. Computing higher-order Hamiltonian terms in this example, as well as in other examples covered later, involves more complex expansions using the same set of the diagrammatic ingredients. In the remainder of this section, rather than manually performing these intricate diagrammatic calculations, we will employ a computer program \cite{qhb2022} that implements all of the diagrammatic rules we have developed.

With the assistance of the computer program, we compute the algebraic expressions associated with all diagrams in \cref{eq:3leg-renorm-diagram} and, up to order $\varphi_\text{zps}^3$, display it below:
\begin{align}\label{eq:3leg-renorm}
\begin{split}
    &\!\Big(\overset{\substack{\text{detuning}\\[3pt]\uparrow\\[4pt]}}{\delta}
     +\overbrace{\big(6g_4-\frac{180g_3^2}{7\omega_d}+\frac{1494\delta g_3^2}{49\omega_d^2}\big)|\xi|^2}^\text{ac-Stark shift}
     \\
     &+ \underbrace{3g_4 \!- \!\frac{10g_3^2}{\omega_d}+\frac{5\delta g_3^2}{\omega_d^2}}_\text{Lamb shift}\Big)\op{\cu{A}}^\dagger \!\op{\cu{A}}\!+\! \underbrace{\Big( \frac{3}{2}g_4 \!-\! \frac{5g_3^2}{\omega_d} + \frac{5\delta g_3^2}{2\omega_d^2}\Big)}_\text{Kerr nonlinearity}\!\op{\cu{A}}^{\dagger2} \!\op{\cu{A}}^2.
     \end{split}
\end{align}
There exist four different type of terms in \cref{eq:3leg-renorm} and they are related to the three types of diagrams in \cref{eq:3leg-renorm-diagram} as follows: the Type I diagram in \cref{eq:3leg-renorm-diagram} is associated with the detuning term and any other term involving $\delta$ in \cref{eq:3leg-renorm}, Type II diagrams are associated with the ac-Stark shift terms, and Type III diagrams with the the Lamb shift terms and the Kerr nonlinearity terms --- the latter correspond to the Type III diagrams with one quantum bond when represented in the expanded form. We note that the Kerr nonlinearity, at order $\varphi_\text{zps}^2$, vanishes for $g_3/\omega_d = \sqrt{3g_4/10\omega_d}$ (or $g_3/\omega_o^\prime = \sqrt{9g_4/20\omega_o^\prime}$). This is also known as the Kerr-free point in a SNAIL circuit \cite{frattini2018}, where the value of $g_3$ and $g_4$ are tunable with external magnetic flux threaded in the circuit. 

Adding together the effective Hamiltonian terms  from \cref{eq:3leg-coupling,eq:3leg-renorm}, we obtain the full effective Hamiltonian up to the order of $\varphi_\text{zps}^3$ as 
\begin{align}\label{eq:3leg-Keff}
    \frac{\op K}{\hbar} = \sum_{n = 1,2} K_n\op{\cu A}^{\dagger n} \op{\cu A}^{n} + \Omega_{3,2}\xi^2\op{\cu A}^{\dagger 3} + \mathrm{h.c.} + \mathcal O(\varphi_\text{zps}^4),
\end{align}
where the coefficients of $K_1, K_2$ and $\Omega_{3,2}$ are defined by the corresponding ones in \cref{eq:3leg-coupling,eq:3leg-renorm}. The emergence of three-legged cat states from \cref{eq:3leg-Keff} is intricately controlled by the effective Hamiltonian coefficients and its precise description is left for future work. Qualitatively, three-legged cat states are stabilized in the regime $K_2\lesssim\Omega_{3,2}\xi^2$, which is similar to the Kerr-cat states that has been intensively studied recently \cite{puri2017,grimm2020,frattini2022,venkatraman2022}. For the parameters chosen in \cref{fig:cat-wigner} (b), we indeed choose $g_3$ and $g_4$ near the Kerr-free point to find this regime. 

\subsection{Modeling the energy renormalization effects of a driven superconducting circuit and comparison with experimental results}
\label{sec:renormalization}

\begin{figure}
\includegraphics{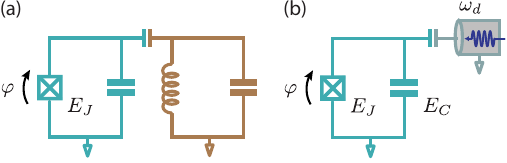}
\caption{Schematic of the circuits of interest. (a) A transmon (in turquoise color) capacitively coupled to a cavity (in brown color). (b) A transmon capacitively coupled to a transmission line with a periodic drive at frequency $\omega_d$.}
\label{fig:renorm-circuit}
\end{figure}

Having provided an illustrative example in the last section, we consider in this section a system of interest in cQED experiments, a transmon coupled to a high-Q cavity as shown in \cref{fig:renorm-circuit} (a). Our goal is to find the energy renormalization effects on both of the transmon on the cavity modes when the latter is prepared in a coherent state with large photon occupation. This system has been employed to engineer GKP states in a recent experiment \cite{eickbusch2022} and modeling its dynamics is essential to achieve high-fidelity quantum control. We also use this example to demonstrate the diagrammatic method in multi-mode systems, the general procedure of which has been introduced in \cref{subsec:multi}.

Specifically, we write the Hamiltonian of the system in normal modes (see \cref{app:multi}) as
\begin{align}
\label{eq:H-sc}
    \frac{\op{H}}{\hbar} = \omega_a \op{a}^{\dagger} \op{a} + \omega_c \op{c}^{\dagger} \op{c} - \frac{E_J}{\hbar}\left( \cos   \hat{\varphi} + \frac{\hat{\varphi}^2}{2} \right),
\end{align}
where $\op a$ and $\op c$ denote annihilation operators associated with the qubit-like and cavity-like modes and $\omega_a$ and $\omega_c$ denote their respective natural frequencies. The operator
$\hat \varphi = \varphi_{\text{zps}, a} (\op a + \op{a}^{\dagger}) + \varphi_{\text{zps}, c} (\op c + \op{c}^{\dagger})$ is the phase operator across the Josephson junction, where $\varphi_{\text{zps}, a}, \varphi_{\text{zps}, c}$ represent the participation of the corresponding modes in the zero point spread of the junction phase. These parameters are determined as $\varphi_{\text{zps}, i}^2 = P_i \hbar \omega_i / 2 E_J$ for $i = a, c$, where $P_i$ represents the ratio between the inductive energy stored in the Josephson junction and the total inductive energy stored in mode $i$ and is a circuit design parameter (see \cite{minev2021} and \cref{app:multi}). The Josephson energy is denoted $E_J$ and the function $\cos(\hat{\varphi}) + \hat{\varphi}^2/2$, up to constants, contains the anharmonic part of the Josephson potential under a Taylor series expansion.  Going into a rotating frame induced by the Hamiltonian $\hbar \omega_a \op{a}^{\dagger} \op{a} + \hbar \omega_c \op{c}^{\dagger} \op{c}$ and with $\cos\hat\varphi$ expanded as a Taylor series around its potential minimum at zero, $\hat H$ in \cref{eq:H-sc} transforms to
\begin{align}
\label{eq:renorm-tra-cav}
\begin{split}
    \frac{\op{H}}{\hbar} &= \sum_{m = 4}^{\infty} \frac{g_m}{m} (\lambda_a (\op a e^{-i \omega_a t} + \op{a}^{\dagger} e^{i \omega_a t})  \\ & \qquad \qquad\qquad +\lambda_c (\op c e^{-i \omega_c t}+ \op{c}^{\dagger} e^{i \omega_c t}))^m,
\end{split}
\end{align}
where $g_m = (-1)^{1+m/2}\omega_a \varphi_{\mathrm{zps}}^{m-2} /2(m-1)!$ for even $m$ and $0$ for odd $m$, with $\varphi_{\mathrm{zps}} = \sqrt{\hbar \omega_a/2 E_{\mathrm{J}}}$ and $\lambda_i^2 = P_i \omega_i/ \omega_a$ for $i = a, c$. In the dispersive regime \cite{Blais2020} we consider in this section, $\lambda_a \sim 1$ and $\lambda_c\ll1$. 

\Cref{eq:renorm-tra-cav} is in the form of \cref{eq:multi-mode}, and the diagrammatic prescription for multi-mode problem introduced in \cref{subsec:multi} can be readily employed. In particular, the 2-mode diagrams are similar to single-mode ones, but with two types of resonant excitations participating. For example, to leading order the diagrams emerged from \cref{eq:renorm-tra-cav} are
\begin{align}\label{eq:renorm-K1-diagram}
\tikzpic{-30}{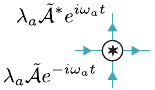} \quad
\tikzpic{-30}{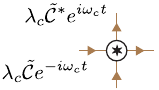} \quad
\tikzpic{-24}{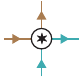} \;\cdots\raisetag{0.8\baselineskip}
\end{align}
where the straight arrows of turquoise and brown color respectively correspond to resonant excitations of the transmon and cavity. As annotated in the equation above, each straight arrow is associated with an algebraic expression $\lambda_a\clu Ae^{-i\omega_at}, \lambda_a\clu A^*e^{i\omega_at}, \lambda_q\clu Ce^{i\omega_qt}$ or $\lambda_q\clu Ce^{i\omega_ct}$, where $(\clu A, \clu A^*)$ and $(\clu C, \clu C^*)$ are respectively the bosonic coordinates of the transmon and the cavity modes in the frame of the effective Hamiltonian. Note that the Husimi Q product $\varstar$ in \cref{eq:renorm-K1-diagram} is now defined over the 2-dimensional phase space as $\cl f\varstar \cl g= \cl f\exp\left(\hbar\overleftarrow{\partial}_{\clu A} \overrightarrow \partial_{\clu {A}^*}+\hbar\overleftarrow{\partial}_{\clu C} \overrightarrow \partial_{\clu {C}^*}\right) \cl g$. The algebraic expression associated with the first order diagrams in \cref{eq:renorm-K1-diagram} reads
\begin{align}\label{eq:renorm-K1}
\begin{split}
   & 3g_4(\lambda_a^2+\lambda_c^2)\lambda_a^2\clu A^*\clu A +     \frac{3g_4}{2}\lambda_a^4\clu A^{*2}\clu A^{2}\\
    &\quad+3g_4(\lambda_a^2+\lambda_c^2)\lambda_c^2\clu C^*\clu C + \frac{3g_4}{2}\lambda_c^4\,\clu C^{*2}\clu C^{2}\\
    &\quad+6g_4\lambda_a^2\lambda_c^2\clu A^*\clu A \,\clu C^*\clu C.
\end{split}
\end{align}
We remark that the first term in \cref{eq:renorm-K1} corresponds to two types of diagrams in the expanded form: 
\begin{align}\label{eq:renorm-K1-qm}
\tikzpic{-27}{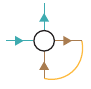} \qquad\;
\tikzpic{-27}{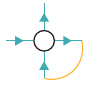}\;,
\end{align}
where each type, which also contains those diagrams look similar to the displayed diagram but with the excitations permuted, is associated with $3g_4\lambda_a^4\clu A^*\clu A$ and $3g_4\lambda_a^2\lambda_c^2\clu A^*\clu A$, respectively. In addition, since $\lambda_c \ll \lambda_a\sim1$, the contribution of the second type diagram is much smaller than that of the first type. In general, due to the smallness of $\lambda_c$,\footnote{We note that $\lambda_c$ is independent of the perturbative parameter $\varphi_\text{zps}$ and usually much smaller than the latter when the cavity has a small participation in the nonlinear mode. This is often the case of a cavity dispersively coupled to the transmon. In the particular system we treat here $\lambda_c = 0.0073$ and $\varphi_\text{zps} = 0.33$.} diagrams involving a quantum bond between cavity excitations are much smaller than those containing the same set of external excitations but not involving such a quantum bond.

At the next order, diagrams involving off-resonant excitations start to emerge, which are evaluated in a way different from the single-mode case and we discuss it now. We consider a specific term
\begin{align}\label{eq:renorm-K2-diagram}
    \tikzpic{-26}{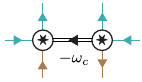} \,=\,
    \tikzpic{-26}{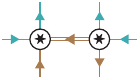}\,+\,   \tikzpic{-26}{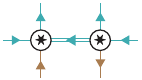}
\end{align}
as an example here. Each diagram above contains an off-resonant excitation at frequency $\omega_\text{out} = -\omega_c$ output by the 4-wave mixer in the right of the diagram. According to the rules discussed in \cref{subsec:multi}, the propagator of the off-resonant excitation in the first diagram is evaluated as \tikzpic{-8}{figsv6_small/notation/prop_small.pdf}$=\lambda_a^2/(\omega_\text{out}-\omega_a)+\lambda_c^2/(\omega_\text{out}-\omega_c)$. This expression motivates expanding the diagram in the left-hand side of \cref{eq:renorm-K2-diagram} as the two diagrams in the right-hand side --- the two colored off-resonant excitations are associated with the propagator $\lambda_a^2/(\omega_\text{out}-\omega_a)$ and $\lambda_c^2/(\omega_\text{out}-\omega_c)$ and we interpret them as the propagation of off-resonant excitation through the cavity mode and the transmon mode, respectively. Similar to \cref{eq:renorm-K1-qm}, the propagation through cavity mode is much smaller than the transmon one due to the smallness of $\lambda_c\ll\varphi_\text{zps}\ll\lambda_a\sim1$. For the sake of simplifying the diagrammatic computation, in the rest of analysis in this section we ignore the diagrams involving cavity-mode quantum bonds or off-resonant excitation propagating through cavity mode. 

\begin{figure}
\includegraphics{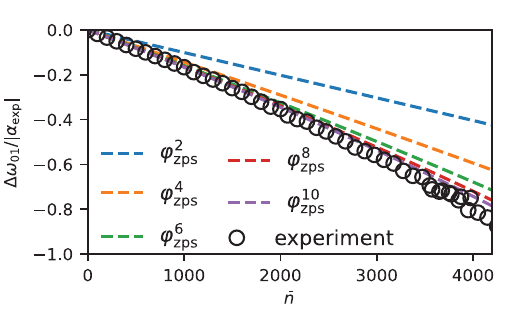}
\caption{Parameter-free model of the ac-Stark shift data of the qubit mode in a transmon-cavity superconducting circuit as a function of cavity occupation number $\bar{n}$.  Open black circles represent experimental data taken from Fig. S2 in \cite{venkatraman2021}. Dashed lines represent the theoretical prediction from the diagrammatic method for different orders of $\varphi_{\mathrm{zps}}$. Theory prediction converges to the experimental data at sufficiently high order of expansion.}
\label{fig:ss-exp}
\end{figure}

With the diagrammatic construction in the 2-mode system illustrated above, we compute the effective Hamiltonian of \cref{eq:renorm-tra-cav} to the order of $\varphi_\text{zps}^{10}$ as
\begin{align}
\label{eq:renorm-Heff}
\frac{\hat {K}}{\hbar} = \sum_{\substack{m, n\ge0,\\ m+n\le6}} K_{m, n} \hat{\cu A}^{\dagger m} \hat{\cu A}^m \hat{\cu C}^{\dagger n} \hat{\cu C}^n  + \mathcal{O}\left(\varphi_{\mathrm{zps}}^{12}\right),
\end{align}
where each $K_{m, n}$ is a function of $\varphi_{\mathrm{zps}}, \lambda_a, \lambda_c, \omega_a, \omega_c$, and we have explicitly suppressed the functional form for brevity. We note that here we assume that $\omega_a$ and $\omega_c$ are incommensurable, and thus, \cref{eq:renorm-Heff} only contains energy renormalization terms.

We test our model by comparing the measured effective Hamiltonian parameters found in a recent experiment \cite{eickbusch2022} consisting of a transmon qubit coupled to a high-Q superconducting cavity.\footnote{Note that in a previous work \cite{venkatraman2021} we developed a different perturbation method to explain the same set of experimental data.} The coefficients of \cref{eq:renorm-tra-cav} are fully determined by the following independently calibrated parameters measured experimentally in the absence of drives:
\begin{align}
\label{eq:exp-paras}
\begin{split}
    \frac{\omega_{01, \mathrm{exp}}}{2 \pi} &= \frac{\omega_a + K_{1,0,\text{exp}}}{2\pi} =6.657\, \mathrm{GHz}, \\ \frac{\omega_{c, \mathrm{exp}}}{2 \pi} &=  5.261\, \mathrm{GHz},\\
     \frac{\chi_{\mathrm{exp}}}{2\pi} &= \frac{K_{1, 1, \mathrm{exp}}}{2\pi} =  -31.2\, \mathrm{kHz}, \\
    \frac{\alpha_{\mathrm{exp}}}{2\pi} &=  \frac{2 K_{2, 0, \mathrm{exp}}}{2\pi} = -193.29\, \mathrm{MHz},
\end{split}
\end{align}
where $\omega_{01, \mathrm{exp}}, \chi_{\mathrm{exp}}$ and $\alpha_{\mathrm{exp}}$ respectively correspond to the measured qubit frequency, qubit-cavity cross-Kerr coefficient, and the qubit anharmonicity, and the subscript ``$\mathrm{exp}$'' indicates an experimentally measurable quantity; see Supplement of \cite{eickbusch2022} for characterization details.
By a numerical diagonalization of \cref{eq:H-sc}, we map \cref{eq:exp-paras} to $E_J/ 2\pi = 32.33~ \mathrm{GHz}$, $\omega_a/ 2\pi = 6.843~\mathrm{GHz}$,  $\omega_c/ 2\pi = 5.261 ~\mathrm{GHz}$,  $\lambda_a = -0.99996$ and $\lambda_c =  0.0073$. Here, we refer two different measurements detailed in \cite{eickbusch2022} and show agreement between the experimental data and the perturbative results.

\begin{figure}[t!]
\includegraphics{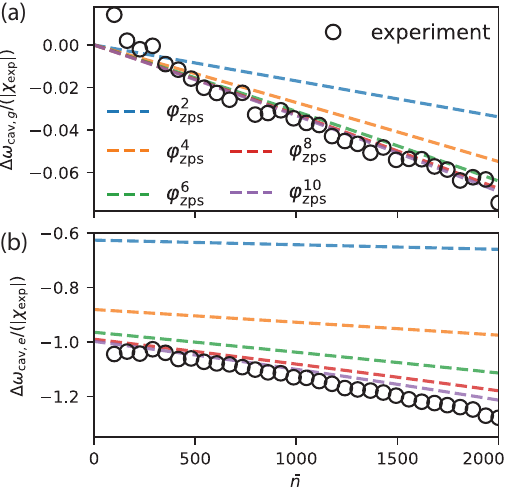}
\caption{Parameter-free model of the cavity frequency data as a function of cavity photon number $\bar{n}$ with the qubit prepared in (a) $|0\rangle$ and (b) $|1\rangle$. Open black circles represent experimentally measured data taken from Fig. S1 in \cite{eickbusch2022}. Dashed lines represent theoretical prediction from the diagrammatic method with zero fit parameters. Theory prediction converges to the measured data at sufficiently high order in the perturbative parameter $\varphi_{\mathrm{zps}}$.} 
\label{fig:ss-exp-cav}
\end{figure}

The first is a measurement of the transmon ac-Stark shift, the photon-number dependent frequency renormalization, as a function of the mean photon number $\bar{n} = |\alpha|^2$ for a coherent state $|\alpha\rangle$ prepared in the cavity. In the effective Hamiltonian $\hat{K}$ in \cref{eq:renorm-Heff}, the measurement represents $\langle 1, \alpha| \hat{K} | 1, \alpha \rangle - \langle 0, \alpha| \hat{K} | 0, \alpha \rangle = \sum_{n = 0}^6 K_{1,n} \bar{n}^n/ \hbar$ as a function of $\bar{n} = |\alpha|^2$, where $|i,\alpha\rangle$ denotes a state with the transmon mode in the $i$-th state and the cavity mode in a coherent state of size $\alpha$. Open black circles in \cref{fig:ss-exp} represent experimentally measured data. Dashed lines represent the theoretical prediction, with different colors representing the perturbative results to different orders. They converge to the experimental data as the expansion is performed to higher orders.

The second is a measurement of the cavity frequency as a function of the cavity photon number $\bar{n}$ with the qubit prepared in (a) $|0\rangle$ and (b) $|1\rangle$. In the effective Hamiltonian $\hat{K}$ in \cref{eq:renorm-Heff}, we model this measurement as the renormalized cross-Kerr coefficient with the cavity in a Fock state $|n\rangle$, i.e., $\langle i, n+1| \hat{K} | i, n+1 \rangle - \langle i, n| \hat{K} | i, n \rangle $ versus $\bar n = n$ where in (a) $|i\rangle = |0\rangle$ and (b) $|i\rangle = |1\rangle$. The expectation value taken over the Fock state $|n\rangle$ serves as a faithful proxy to the measurement performed over the coherent state $|\alpha\rangle$ with $|\alpha|^2 = \bar n$, because the measured frequency is independent of the phase of the cavity field. Open black circles in \cref{fig:ss-exp-cav} are the measured data. Again, dashed lines represent the theoretical prediction, with different colors representing the perturbative result to different orders. They converge to the measured data as the order becomes higher.

With these two experiments, we have demonstrated the accuracy and convergence of the effective Hamiltonian description constructed through diagrams by comparing with the experimental data. Note that, for this experimental system consisting of a strongly anharmonic transmon, the perturbative parameter\footnote{Because the system contains no odd rank nonlinearity, the actual perturbative parameter is actually $\varphi_\text{zps}^2$.} $\varphi_\text{zps} = 0.33$ is only moderately small and thus the perturbation expansion has to be carried out to high order to explain the experiments with large cavity photon occupation. Moreover, we remark that the starting point \cref{eq:H-sc} of this example is an undriven time-independent system. Yet with proper transformation, we obtain a time-dependent Hamiltonian \cref{eq:renorm-tra-cav} that is suitable for the diagrammatic analysis. 

Lastly, we note that in a previous work \cite{venkatraman2021}, we have analyzed the same experimental system using a Schrieffer-Wolff-like expansion, which is a Floquet perturbation method. Here, we confirm that the analytical results obtained from this method and our diagrammatic method are the same for the two measurement schemes discussed above. This equivalence of results in this specific system supports the claim made in \cref{subsec:other-pert} that our diagrammatic method is equivalent to the Floquet perturbation method.

\subsection{Multiphoton resonances in a driven superconducting circuit}\label{sec:mpnr}
As discussed in \cref{subsec:eff-Ham}, when the drive frequency is in the vicinity of $p\omega_o/q$ for some integers $p, q$, the effective Hamiltonian of a driven nonlinear oscillator is of the form 
\begin{align}\label{eq:mnr-H}
    \frac{\op K}{\hbar} = \sum_{n>0} K_n\op{\cu A}^{\dagger n} \op{\cu A}^{n} + \sum_{l>0}\Omega_{lq^\prime,lp^\prime}\xi^{lp^\prime}\op{\cu A}^{\dagger lq^\prime} + \mathrm{h.c.}
\end{align}
where $p^\prime/q^\prime$ equals to $p/q$ in its factored form; in other words, $q^\prime$ and $p^\prime$ are coprime.
From the Hamiltonian in \cref{eq:mnr-H} a variety of nonlinear processes emerge ---  we have demonstrated  in \cref{sec:3lcat} the three-legged Schr\"odinger cat states in the case of $(q:p) = (3:2)$, and in \cref{sec:renormalization} the energy renormalization of the oscillator states when the coupling terms $\Omega_{lq^\prime,lp^\prime}\xi^{lp^\prime}\op{\cu A}^{\dagger lq^\prime} + \mathrm{h.c.}$ are absent in \cref{eq:mnr-H}.  In this section, we discuss another class of processes, multiphoton resonances, due to the coupling terms.

In a nutshell, a multiphoton resonance labeled $(q:p)$ is the coherent coupling between the $i$-th state and $(i+q)$-th state when they are (near-)resonant in the effective frame that $\op K$ lives in. The uncoupled energies of the states are determined by the terms $K_n\op{\cu A}^{\dagger n} \op{\cu A}^{n}$ in \cref{eq:mnr-H} and the coupling, to leading order, is created by $\Omega_{q,p}\xi^{p}\op{\cu A}^{\dagger q} + \mathrm{h.c.}$. In the lab frame, the resonance condition of such a multiphoton resonance translates to $\tilde{E}_{i+q} - \tilde{E}_i = p\hbar\omega_d$, where $\tilde{E}_i$ is the renormalized energy of the $i$-th state under the drive. The coupling between the $i$-th and $(i+q)$-th state can also be understood as a Raman transition between the states mediated by $p$ drive photons. We also note that the $(q:p)$ multiphoton resonances associated with the same simplified fraction $p^\prime/q^\prime$ form a family of processes sharing the same form of the effective Hamiltonian \cref{eq:mnr-H}. Yet these processes each happens at a different resonance condition controlled by $K_n\op{\cu A}^{\dagger n} \op{\cu A}^{n}$, which are determined by the drive parameters.

Multiphoton resonances have been observed in driven nonlinear systems for several decades. In Josephson circuits, recent works have shown experimental \cite{sank2016,zhang2019} and numerical evidence \cite{sank2016,zhang2019,shillito2022} that multiphoton resonances are responsible for anamolous state transitions. It has been demonstrated that these resonances play a major role in qubit readout limiting fidelities and constraining the regime of operation \cite{reed2010,Blais2020}. However, a general analytical description of these processes is missing so far. In this section, we employ the diagrammatic method to characterize multiphoton resonances in Josephson circuits and discuss design principles for their mitigation.

\subsubsection{Example: $(5:3)$ multiphoton resonance in a transmon}\label{subsec:mnr-53}

For the sake of concreteness, in this section we consider a transmon periodically driven through its charge degree of freedom as shown in \cref{fig:renorm-circuit} (b). The Hamiltonian of the system reads
\begin{align}\label{eq:mnr-transmon-H}
    \op{\mathcal{H}}(t) = 4E_C(\op N-N_g)^2 -E_J\cos\op\varphi + E_d\op N\cos\omega_dt,
\end{align}
where the canonically conjugate operators $\op\varphi$ and $\op N$ are the phase and charge operators across the junction in the unit of magnetic flux quantum and Cooper pair respectively, and they satisfy the commutation relation $[\op\varphi, \op N] = i$. The transmon is characterized by the tunneling energy of the Josephson junction $E_J$, the effective charging energy of the shunting capacitance $E_C$, and the charge offset across the junction $N_g$. The drive is characterized by its energy $E_d$ and frequency $\omega_d$. In the rest of this work, we take $N_g=0$ for simplicity.

Expanding the cosine Josephson potential around its minimum at $\op\varphi =0$, the Hamiltonian in \cref{eq:mnr-transmon-H} can be re-expressed in the form describing a driven nonlinear oscillator
\begin{align}\label{eq:mnr-transmon-H-bosonic}
\begin{split}
    \frac{\op{\mathcal{H}}(t)}{\hbar} = \omega_o\op a^\dagger \op a &+ \sum_{\substack{m \ge 4 \\ m \in 2\mathbb Z}}\frac{g_m}{m}(\op a + \op a^\dagger)^m \\
    &- i\Omega_d (\op a - \op a^\dagger)\cos\omega_d t,
\end{split}
\end{align}
where $\omega_o = \sqrt{8E_JE_C}$ is the natural frequency of the oscillator and $g_m = (-1)^{(m-2)/2}\omega_o\varphi_\text{zps}^{m-2}/2(m-1)!$ are the even-rank nonlinearities while the odd-rank ones are zero due to the symmetry of the cosine potential. Here $\op a = (\op\varphi/\varphi_\text{zps} + i\op N/N_\text{zps})/2$ is the annihilation operator and $\varphi_\text{zps}=(2E_C/E_J)^{1/4}$ and $N_\text{zps} = 1/2\varphi_\text{zps}$ are respectively the zero-point spreads of the phase and charge across the junction. The parameter $\Omega_d = N_\text{zps}E_d$ is the drive amplitude.

\begin{figure}
\centering
    \includegraphics{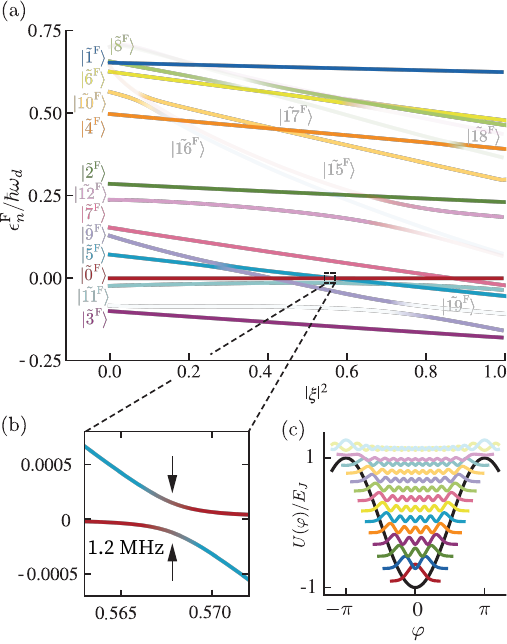}
    \caption{ Energy spectrum of a driven transmon. The transmon ($E_J/h =30$ GHz, $E_C/h= 0.15$ GHz, $N_g$ = 0) and drive parameters ($\omega_d/2\pi = 8.97$ GHz) are chosen to be typical ones in cQED experiments. 
    (a) Quasienegies of Floquet states in the 1st Brillouin zone of the driven transmon defined by \cref{eq:mnr-transmon-H} as a function of drive strength $|\xi|^2$. Under the periodic drive, quasi-energies live in Brillouin zones with width of $\hbar\omega_d$, where $\omega_d/2\pi = 8.97$ GHz. States with large energy difference in lab frame can have near-degenerate quasi-energies and interact with each other, indicated as the anti-crossing in the plot. Solid color lines and transparent lines are intra-well states and running states respectively. The running states interacting weakly with intra-well states are not shown. (b) A zoom-in of (a) which shows an anti-crossing of size $2\Omega^R_{0\leftrightarrow5,3}=1.2~$MHz between $|\tilde{0}\rangle$ and $|\tilde{5}\rangle$ bridged by 3 drive photons, where $\Omega^R_{0\leftrightarrow5,3}$ is the Rabi rate between them. (c) The undriven spectrum of the transmon with the shape of the each wavefunctions depicted and plotted over the Josephson potential $U(\varphi) = -E_J\cos(\varphi)$. The first 13 states are intra-well and others are running states. }
    \label{fig:quasispectrum}
\end{figure}

We first analyze a particular multiphoton resonance $(5:3)$ that is manifested when the drive frequency $\omega_d$ is in the vicinity of $5\omega_o/3$. Similar to the previous examples, we perform the frame transformation on \cref{eq:mnr-transmon-H-bosonic} amounting to $\op a \rightarrow  \op a e^{-i\omega_o^\prime t} + \xi  e^{-i\omega_dt}$, where $\xi =\frac{i\Omega_d\omega_d}{\omega_d^2-\omega_o^2}$ and $\omega_o^\prime = 3\omega_d/5$. The transformed Hamiltonian is of the form 
\begin{align}\label{eq:mnr-transmon-H-dr}
\begin{split}
    \frac{\op H(t)}{\hbar} = \delta\op a^\dagger \op a+  \sum_{\substack{m \ge 4 \\ m \in 2\mathbb Z}} \frac{g_m}{m}  &(\hat{a} e^{- i \omega_o^\prime t} + \hat{a}^{\dagger} e^{i \omega_o^\prime t} \\[-8pt]
    & \;+ \xi e^{-i \omega_d t} +  \xi^* e^{i \omega_d t} )^m,
\end{split}\raisetag{0.4\baselineskip}
\end{align}
with $\delta = \omega_o - \omega_o^\prime$. 

In the effective Hamiltonian computed from \cref{eq:mnr-transmon-H-dr}, the coupling terms to leading order are born from the bare diagrams
\begin{align}\label{eq:5:3_diagram}
\tikzpic{-25}{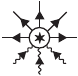} \quad \tikzpic{-25}{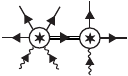}\quad
\tikzpic{-25}{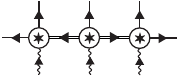}\;\cdots,
\end{align}
where we suppress the diagrams similar to the displayed ones but with the external edges permuted or with the internal ones inverted. These terms are associated with effective Hamiltonian terms $\Omega_{5,3}\xi^3\op{\cu A}^{\dagger 5} + \mathrm{h.c.}$, with
\begin{align}\label{eq:mnr-53-coulping}
    &\Omega_{5,3}^{(6)} = 7g_8- \frac{1745g_4g_6}{18\omega_d}+\frac{21275g_4^3}{72\omega_d^2},
\end{align}
where the superscript $(6)$ indicates the order of $\varphi_\text{zps}$ the coupling term is computed to. The three terms in the right-hand side of \cref{eq:mnr-53-coulping} correspond to diagrams comprising of an 8-wave mixer, a cascade of a 4-wave mixer and a 6-wave mixer, and a cascade of three 4-wave mixers, respectively. We remark that even though these three terms are of the same order ($\varphi_\text{zps}^6$), their prefactors form an ordered list ranging from 7 to $21275/72$, resulting in a ratio of 40.\footnote{If one also considers the $1/(m-1)!$ factor in the definition of $g_m$ (c.f. \cref{eq:mnr-transmon-H-bosonic}), the strength of the $g_4^3$ term is 985 times larger than that of the $g_8$ term!}) This is because the prefactor of an effective Hamiltonian term is related to the number of diagrams associated with it. For the same set of external excitations, the cascaded mixing diagrams contain more permutation-generated configurations than a simple mixer. This makes $7g_8$, an RWA term from \cref{eq:mnr-transmon-H-dr}, a poor approximation for the full coupling term and renders the high order description, and thus our diagrammatic method, necessary for even a qualitative description of the multiphoton resonance process. 

\cref{eq:5:3_diagram} corresponds to the particular coupling between the ground state of the transmon and the 5th excited state when they are near-resonant with 3 drive photons. The relevant dynamics, to leading order, is captured by the effective Hamiltonian \cref{eq:mnr-H} reduced to the submanifold spanned by 0th and 5th Fock states as
\begin{align}\label{eq:mnr-53-K}
    \frac{\hat K}{\hbar} = \begin{bmatrix}
    \mathcal{E}_0 & \Omega_{0\leftrightarrow5,3}^*\\
    \Omega_{0\leftrightarrow5,3}& \mathcal{E}_5
    \end{bmatrix}.
\end{align}
Here $\mathcal{E}_0, \mathcal{E}_5$ are the renormalized energy of the two Fock states and determined by the energy renormalizing terms in \cref{eq:mnr-H} as $\mathcal{E}_i = \langle i|\sum_nK_n\op{\cu A}^{\dagger n}\op{\cu A}^{n}|i\rangle $, where $K_n$'s are functions of $\omega_d$ and $\xi$. The off-diagonal term $\Omega_{0\leftrightarrow5, 3}=\langle5|\op{\cu A}^{\dagger 5}|0\rangle\xi^3\Omega_{5,3}$ is the multiphoton Rabi rate between the two states when resonant, i.e. $\mathcal{E}_0 = \mathcal{E}_5 = 0$, and $\Omega_{5,3}$ to leading order is given by \cref{eq:mnr-53-coulping}.

To verify the effective description given by \cref{eq:mnr-53-K} of the $(5:3)$ resonance, we compare it with an exact calculation, which we find by solving \cref{eq:mnr-transmon-H} via Floquet numerical diagonalization (see \cite{zhang2019} and \cref{app:floquet}). Specifically, we fix $\omega_d$ to be a particular value near $5\omega_o/3$ and, in \cref{fig:quasispectrum}, we plot the Floquet quasienergy spectrum of the transmon as a function of $|\xi|^2$. In the regime where the dominant nonlinear processes are multiphoton resonances, the quasienergies $\epsilon^\text{F}_n$ of the Floquet states $|\tilde n^\text{F}\rangle$ can be understood as $\epsilon^\text{F}_n = \tilde{E}_n\mod \hbar\omega_d$, where $\tilde{E}_n$ is the Stark-shifted energy of the $n$-th state of the transmon in lab frame. For this reason, the quasienergies of the lower lying excited states (those with small non-zero index $n$) in \cref{fig:quasispectrum} (a) decrease as $|\xi|^2$ increases, and to leading order the Stark-shift is $6g_4|\xi|^2 n!$ (c.f. \cref{eq:3leg-renorm}). In this example, the drive frequency $\omega_d$ is chosen in a way that $\epsilon_5^\text{F}$ is slightly larger than $\epsilon_0^\text{F}$ at $\xi=0$ (or, in the lab frame, the undriven energy of the 5th excited state $E_5$ is slightly larger than $E_0 + 3\hbar\omega_d$). As $|\xi|^2$ increases to 0.568, the quasienergy of 5th Floquet state Stark-shifts into resonance with the 0th Floquet state, and one observes an 1.2~MHz anti-crossing predicting a Rabi coupling between the states with strength given by $\Omega_{0\leftrightarrow5, 3}/2\pi = 0.6~$MHz. By comparing this value with the analytical Rabi coupling prediction of 0.67~MHz given by \cref{eq:mnr-53-K}, we find excellent agreement between Floquet numerics and our analytics, which we obtained via our diagrammatic perturbation method.

\subsubsection{Characterization of general multiphoton resonance in transmons}

In general, for a transmon with $N_g=0$, the ground state can couple to the $q$-th excited state if the resonance condition $\tilde E_{q} - \tilde E_0 = p\hbar\omega_d$ is met and $q+p$ has even parity. This leads to a dense and complicated landscape of multiphoton resonances in the drive and oscillator parameter space, which we characterize now. Specifically, we will analyze the lowest order processes $(q:p)$ with $q+p\le 16$ and those occurring inside an experimentally relevant window $\omega_d\in[1.1\omega_o,2\omega_o]$. The methods discussed below can be applied to treat processes outside this window.

\begin{figure}
\centering
    \includegraphics{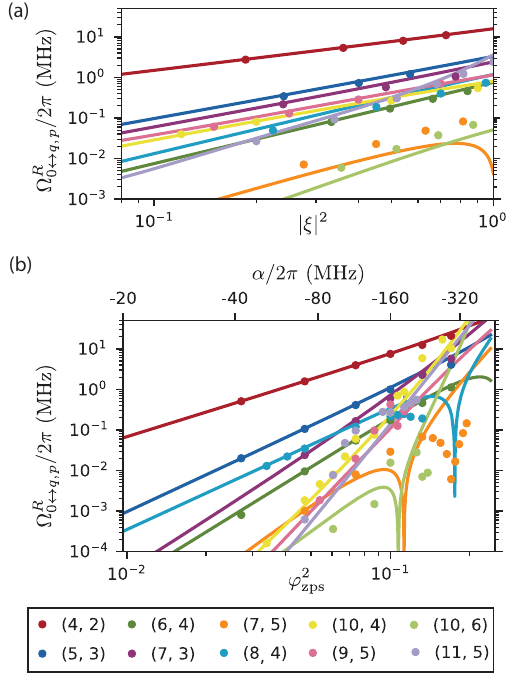}    \caption{Rabi strength $\Omega^R_{0\leftrightarrow q, p}$ between the ground state and $q$-th excited state of a transmon in a $(q:p)$ multiphoton resonance process. In (a), the transmon is of the same parameters ($\varphi_\text{zps}^2 = 0.1, \omega_{01}/2\pi = 5.85$ GHz) as the one in \cref{fig:quasispectrum} and the drive strength $|\xi|^2$ is varied and plotted in log scale. In (b), the transmon frequency $\omega_{01}$ is fixed to be the same as that in (a), the drive strength is fixed to be $|\xi|^2 = 0.5$, and the zero-point spread $\varphi_\text{zps}^2$ is varied and plotted in log scale. The anharmonicities of the transmons $\alpha$ are labeled on the top $x$ axis; to leading order $\alpha=-\omega_o\varphi_\text{zps}^2/4$.  The solid lines in both plots are obtained from the diagrammatic method (c.f. \cref{app:coupling}). For each $(q:p)$ process the perturbative result is computed to $\varphi_\text{zps}^{q+p}$ except for $(10:6)$ and $(11:5)$ processes, 
   for which the results are only computed to the leading order, i.e. $\varphi_\text{zps}^{q+p-2}$, for numerical efficiency. The solid dots are obtained by Floquet simulation in the same way as \cref{fig:quasispectrum} where $\Omega^R_{0\leftrightarrow5,3}$ at $|\xi|^2 = 0.568$ is given as an example.}
    \label{fig:coupling-fit}
\end{figure}

First, in \cref{fig:coupling-fit} (a), we consider the same transmon as the one in \cref{fig:quasispectrum} and plot the Rabi strength of different processes as a function of drive strength $|\xi|^2$ in a range where anomalous transitions are observed in experiments \cite{sank2016,minev2019}. Each dot in the plot, corresponding to a resonance occurring at a particular configuration of drive frequency and amplitude, is obtained from the exact Floquet analysis similar to the one in \cref{fig:quasispectrum}~(a). The solid lines correspond to the Rabi strengths that are obtained from the diagrammatically constructed effective Hamiltonian and they quantitatively agree with the exact values from Floquet simulation.

To compute the Rabi strength from an effective Hamiltonian, different resonance processes require different treatments. Specifically, similar to \cref{eq:mnr-53-K}, for $(q:p) = (4:2), (5:3), (6:4), (7:3),$ or $(7:5)$ the effective Hamiltonian \cref{eq:mnr-H} can be reduced to the submanifold spanned by 0th and $q$-th Fock state. The resulting Hamiltonian reads
\begin{align}\label{eq:mnr-K-1}
    \frac{\hat K}{\hbar} = \begin{bmatrix}
    \mathcal{E}_0 & \Omega_{0\leftrightarrow q,p}^*\\
    \Omega_{0\leftrightarrow q,p}& \mathcal{E}_q
    \end{bmatrix},
\end{align}
where  
\begin{align}\label{eq:mnr-reduced-rel}
\begin{split}
        &\mathcal{E}_i = \langle i|\sum_nK_n\op{\cu A}^{\dagger n}\op{\cu A}^{n}|i\rangle \\
    &\Omega_{i\leftrightarrow i+q, p}=\langle i+q|\op{\cu A}^{\dagger q}|i\rangle\xi^{p}\Omega_{q, p}
\end{split}
\end{align}
are the dressed energy of the $i$-th Fock state and the direct coupling term between $i$-th and $(i+q)$-th states, respectively, and $K_n$ and $\Omega_{q, p}$ are the parameters in the effective Hamiltonian \cref{eq:mnr-H}. In these cases where the relevant states only involve the coupled two, the Rabi strength is simply the off-diagonal term in \cref{eq:mnr-K-1} $\Omega_{0\leftrightarrow q, p}^R = |\Omega_{0\leftrightarrow q, p}|$.

For the processes $(q:p) = (8:4)$ and $(10:6)$, their effective Hamiltonians in the form of \cref{eq:mnr-H} involve not only the coupling term of interest $\Omega_{q,p}\xi^{p}\op{\cu A}^{\dagger q} + \mathrm{h.c.}$ but also $\Omega_{q/2,p/2}\xi^{p/2}\op{\cu A}^{\dagger q/2} + \mathrm{h.c.}$ while the latter is of lower order in $\varphi_\text{zps}$ than the former. As a result, the reduced effective Hamiltonian, to leading order, involves three states and reads
\begin{align}\label{eq:mnr-K-2}
    \frac{\hat K}{\hbar} = \begin{bmatrix}
    \mathcal{E}_0 & \Omega_{0\leftrightarrow q/2,p/2}^*&\Omega_{0\leftrightarrow q, p}^*\\[4pt]
    \Omega_{0\leftrightarrow q/2,p/2} & \mathcal{E}_{q/2}&\Omega_{q/2 \leftrightarrow q, p/2}^*\\[4pt]
    \Omega_{0\leftrightarrow q, p}&\Omega_{q/2 \leftrightarrow q, p/2}&\mathcal{E}_0
    \end{bmatrix}.
\end{align}
The resonant condition for the $(q:p)$ process is $\mathcal{E}_0\approx\mathcal{E}_0$, under which  $\mathcal{E}_{q/2}$ is detuned from both $\mathcal{E}_0$ and $\mathcal{E}_0$ due to the anharmonicity. For these processes, the Rabi strength between ground and $q$-th excited states is computed by performing numerical diagonalization of \cref{eq:mnr-K-2}, which we discuss in more detail in \cref{app:coupling}.

Lastly, we note that in the transmon of the particular design parameters we choose in this example, the resonant processes $(q:p) = (10:4), (9:5)$, and $(11:5)$ each coexists with some other lower-order processes. To compute the Rabi strength, it is necessary to construct \textit{a slow-varying effective Hamiltonian}, instead of a static one, that \textit{captures different classes of multiphoton resonances simultaneously}. The principle behind this procedure has been discussed in \cref{subsec:frame}. Here we take the $(10:4)$ process as an example and leave the detail analysis in \cref{app:coupling}. When the drive strength frequency are chosen so that the ground state is coupled to 10th excited state mediated by 4 drive photons, two other resonant processes are near-resonant: the 10th and 7th excited states are coupled through a $(3:1)$ process, and the ground state and the 7th excited state are coupled through a $(7:3)$ processes. While each of these two processes is hundreds of megahertz detuned from the resonance condition, the coupling strength between the 7th and 10th state is comparable to the detuning (when $|\xi|^2 = 0.2$, the coupling is $\sim 590~$MHz; see footnote\footnote{This coupling strength is much larger than those characterized in \cref{fig:coupling-fit} because (1) the $(3:1)$ process is of lower order in $\varphi_\text{zps}$ and (2) for the same coupling Hamiltonian terms, the matrix element between the excited Fock states is larger than the one between ground state and the corresponding excited state.}) and renders it necessary to include the dynamics of the 7th excited state in the effective description as well. To capture all three processes $(10:4), (3:1),$ and $(7:3)$, we construct diagrams in the rotating frame at $\omega_o^\prime = 2\omega_d/5$ so that the diagrams associated with the coupling term $\xi^4\op{\cu A}^{\dagger 10}$ is static. In addition, we consider the diagrams associated with $\xi\op{\cu A}^{\dagger 3}\exp[i(3\omega_o^\prime-\omega_d)t] + \text{h.c.}$ and $\xi^3\op{\cu A}^{\dagger 7}\exp[i(7\omega_o^\prime-3\omega_d)t] + \text{h.c.}$ to be slow-varying and absorb them to the effective Hamiltonian as well. Noting that $\pm(3\omega_o^\prime-\omega_d) = \mp(7\omega_o^\prime-3\omega_d)=\pm \omega_d/5$, the effective Hamiltonian to leading order reads
\begin{align}\label{eq:mnr-K-10-4}
\begin{split}
        \frac{\op K}{\hbar} = \sum_{n>0} K_n\op{\cu A}^{\dagger n} \op{\cu A}^{n} &+\Omega_{10,4}\xi^{4}\op{\cu A}^{\dagger 10} + \Omega_{3,1}\xi\op{\cu A}^{\dagger 3}e^{i\frac{\omega_d}{5}t} \\
    &+\Omega_{7,3}\xi^{3}\op{\cu A}^{\dagger 7}e^{-i\frac{\omega_d}{5}t}+\mathrm{h.c.},
\end{split}
\end{align}
where each Hamiltonian parameter above is associated with the diagrams constructed following the ordinary rules but excluding those containing near-resonant off-resonant excitations, i.e. those of frequency $(-2\omega_o^\prime +\omega_d)$ or $(-6\omega_o^\prime +3\omega_d)$. This effective Hamiltonian can be reduced to the subspace of relevant oscillator states, i.e.  0th, 3rd, 7th, and 10th Fock states, and reads
\begin{widetext}
\begin{align}\label{eq:mnr-K-3}
\begin{split}
    \frac{\hat K}{\hbar} =\begin{bmatrix}
    \mathcal{E}_0 &\Omega^{*}_{0\leftrightarrow3,1} e^{-i\frac{\omega_d}{5}t}  &\Omega^{*}_{0\leftrightarrow7,3} e^{i\frac{\omega_d}{5}t} & \Omega^{*}_{0\leftrightarrow10,4}\\[5pt]
    \Omega_{0\leftrightarrow3,1} e^{i\frac{\omega_d}{5}t}&\mathcal{E}_{3}&0&\Omega^{*}_{3\leftrightarrow10,3} e^{i\frac{\omega_d}{5}t}\\[5pt]
    \Omega_{0\leftrightarrow7,3}e^{-i\frac{\omega_d}{5}t} &0& \mathcal{E}_{7} & \Omega^{*}_{7\leftrightarrow10,1}e^{-i\frac{\omega_d}{5}t}\\[5pt]
    \Omega_{0\leftrightarrow10,4} & \Omega_{3\leftrightarrow10,3} e^{-i\frac{\omega_d}{5}t}& \Omega_{7\leftrightarrow10,1}e^{i\frac{\omega_d}{5}t} & \mathcal{E}_{10}
    \end{bmatrix},
\end{split}
\end{align}
\end{widetext}
where each entry is related to the parameters in \cref{eq:mnr-K-10-4} by the relation in \cref{eq:mnr-reduced-rel}. We further transform the slow-varying effective Hamiltonian \cref{eq:mnr-K-3} into a static one by the unitary transformation $\op U = \exp[i\frac{\omega_d}{5}(|3\rangle\langle3|-|7\rangle\langle7|)t]$. The resulting effective Hamiltonian reads
\begin{align}\label{eq:mnr-K-4}
    \frac{\hat K^\prime}{\hbar} = \begin{bmatrix}
    \mathcal{E}_0 & \Omega^{*}_{0\leftrightarrow3,1} &\Omega^{*}_{0\leftrightarrow7,3} & \Omega^{*}_{0\leftrightarrow10,4}\\[5pt]
    \Omega_{0\leftrightarrow3,1} & \mathcal{E}_{3} + \omega_d/5 & 0 & \Omega^*_{3\leftrightarrow10,3}\\[5pt]
    \Omega_{0\leftrightarrow7,3}& 0&  \mathcal{E}_{7}-\omega_d/5 & \Omega^{*}_{7\leftrightarrow10,1}\\[5pt]
    \Omega_{0\leftrightarrow10,4} & \Omega_{3\leftrightarrow10,3} & \Omega_{7\leftrightarrow10,1} & \mathcal{E}_{10}
    \end{bmatrix},
\end{align}
and the Rabi strength between the 0th and 10th states $\Omega^R_{0\leftrightarrow10,4}$ is found by diagonalizing \cref{eq:mnr-K-4}. We note that, for the interested range of $|\xi|^2$ in \cref{fig:coupling-fit} (a), \cref{eq:mnr-K-4} is in the domain where $|\mathcal{E}_{10} - (\mathcal{E}_{7}-\omega_d/5)|\sim |\Omega_{7\leftrightarrow10,1}|$. This implies that the 7th and 10th Fock states are strongly hybridized through the $(3:1)$ process when the latter is coupled to the ground state through $(10:4)$. Due to such a hybridization, the $(10:4)$ process, which are of higher order in $\varphi_\text{zps}$, appears stronger than some of its lower-order counterparts in  \cref{fig:coupling-fit}~(a). This is also the case for the $(9:5)$ and $(11:5)$ processes, in which the 9th and 11th states are strongly hybridized with 5th and 8th excited states, respectively. We discuss this in more detail in \cref{app:coupling}. 

We also remark that in the log-log plot  \cref{fig:coupling-fit}~(a), most of the Rabi strengths $\Omega^R_{0\leftrightarrow q, p}$ of different $(q:p)$ processes appear as straight lines with different slopes. This reflects the relation $\Omega^R_{0\leftrightarrow q, p}\propto |\xi|^{p}$ to leading order and the slope is proportional to the number of drive photons $p$ involved in the multiphoton resonance. An important implication is that the strengths of processes involving more drive photons, which are usually of higher order in $\varphi_\text{zps}$, increase faster than that of processes involving fewer drive photons. Therefore when $\xi$ is sufficiently large, some higher-order processes are as strong as the lower orders.

\begin{figure*}
\centering
    \includegraphics{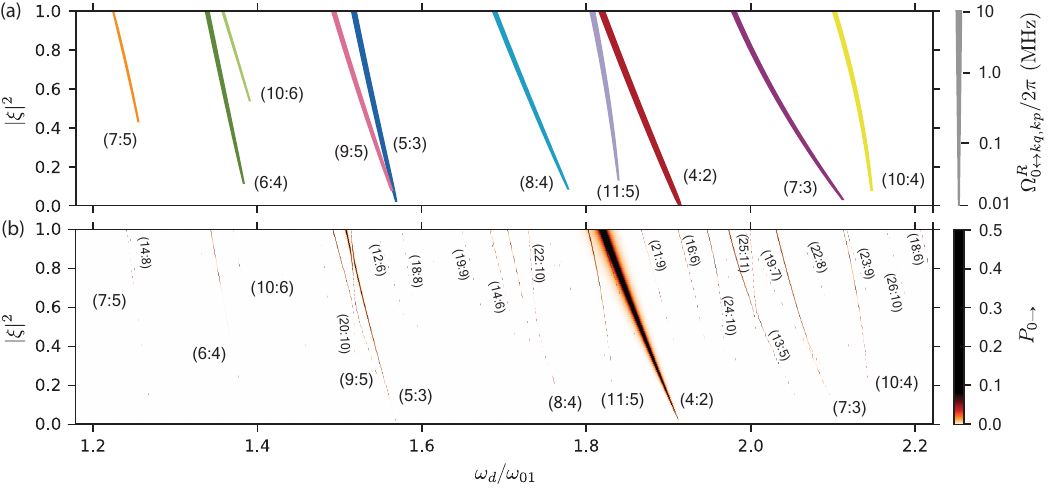}
    \caption{Landscape of $(q:p)$ multiphoton resonances in a driven transmon in the drive parameter space $(\omega_d, |\xi|^2)$. The transmon's parameters are identical to those in \cref{fig:quasispectrum,fig:coupling-fit}, with qubit frequency $\omega_{01}/2\pi = 5.85$~GHz and anharmoncity $\alpha/2\pi = -160$~MHz. (a) Analytical result from our diagrammatic method. Each line marks a $(q:p)$ process with the color matching that in \cref{fig:coupling-fit}. The center of each line corresponds to the location of each resonance process in $(\omega_d, |\xi|^2)$ space and the width corresponds to the Rabi strength $\Omega^R_{0\leftrightarrow q, p}$ between the ground state and $q$-th excited state. (b) Numerical result from Floquet simulation. The excitation possibility out of the ground state $P_{0\rightarrow}$ (see \cref{app:transmon-state} for detail) is plotted in the same drive parameter space as in (a). Each line emerging in the plot is labeled as a $(q:p)$ process, which is inferred from the quasienergy spectrum like \cref{fig:quasispectrum}.}
    \label{fig:spaghetti}
\end{figure*}

Besides the drive strength, the coupling strengths of the multiphoton resonances also depend on the nonlinearities of the oscillator, which in the case of a transmon is controlled by a single parameter $\varphi_\text{zps}$. In \cref{fig:coupling-fit} (b), we plot the Rabi strength as a function of $\varphi_\text{zps}^2$ and, for simplicity, we assume a resonant process does not coexist with others. In this plot, most of the predicted strengths also appear as straight lines in the log-log plot. This reflects the relationship $\Omega^R_{0\leftrightarrow q, p}=\mathcal O( \varphi_\text{zps}^{p+q-2})$ and the slope is thus proportional to $p+q-2$. We remark that for the chosen drive strength $|\xi|^2=0.5$, all the examined processes independent of their perturbative order are $\sim 10$~MHz in Rabi strength when $\varphi_\text{zps}^2 = 0.17$. It implies that in a transmon of $\omega_{01}/2\pi = 5.85 $ GHz and anharmonicity $\alpha/2\pi = 290$~MHz, a design choice close to many experimental implementations \cite{sank2016,eickbusch2021}, the notion of ``higher order process" breaks down as these processes, conventionally believed to be weak, are as strong as the lower order ones.

It is worth noting that, in \cref{fig:coupling-fit} (b), the predicted Rabi strengths of $(7:5), (8:4)$ and $(10:6)$ are completely suppressed at $\varphi_\text{zps}^2 = 0.113, 0.107$, and 0.176, respectively. This is because propagators of some off-resonant excitations involved in these processes sensitively depend on the drive and oscillator frequencies, which vary for different choices of $\varphi_\text{zps}$. At certain value of $\varphi_\text{zps}$, the diagrams containing different off-resonant excitations perfectly cancel out each other. For the $(7:5)$ process, the Floquet simulation gives the actual suppression point at $\varphi_\text{zps}^2 = 0.172$ which is close to the predicted one, a remarkable agreement considering the high order of the $(7:5)$ process in $\varphi_\text{zps}$. For the $(8:4)$ and $(10:6)$ processes, the 8th and 10th states are already running states at the suppression points and thus a Floquet numerical characterization is difficult. Yet one can still observe the precursor of the suppression for both processes around $\varphi_\text{zps}^2= 1.35$, which are also close to the diagrammatic prediction. 

We also remark that, to the best of our knowledge, the existing high-frequency perturbation methods, such as Schrieffer-Wolff expansion, cannot compute the analytical expression for the higher order processes  ($q+p>12$). Even with the knowledge of our previous result \cite{venkatraman2021} which provides symbolic software to compute the correction recursively, the enormous number of terms involved in the full effective Hamiltonian would cause memory overflow. In contrast, the diagram method book-keeps different types of terms in the effective Hamiltonian, allowing for the construction of only the relevant terms and greatly simplifying computations. For example, for the $(11:5)$ process, the full Hamiltonian including energy renormalization terms comprises approximately 200 million unordered diagrams up to the order of $\varphi_\text{zps}^{14}$, while the relevant coupling term only constitutes approximately 0.7 million of these diagrams.  This bookkeeping feature also enables more involved operations in constructing the effective Hamiltonian, for example, as in \cref{eq:mnr-K-10-4}, identifying the slow-varying dynamics that contain particular off-resonant excitations and capturing them in a slow-varying effective Hamiltonian. 

\subsubsection{Landscape of multiphoton resonances in a transmon}\label{subsec:mnr-landscape}

In addition to the Rabi strength, it is also of experimental interest to characterize the location of multiphoton resonances in the drive parameter space $(\omega_d, |\xi|^2)$. We recall that for a $(q:p)$ process coupling the ground state and $q$-th excited state, the resonance condition in lab frame is $\tilde{E}_{q} - \tilde{E}_0 = p\hbar\omega_d$, where $\tilde{E}_n$ is the Stark-shifted energy of the $n$-th state of the oscillator. The Stark-shifted spectrum in the lab frame can be obtained from the spectrum of the effective Hamiltonian, i.e. $\mathcal{E}_n$'s in \cref{eq:mnr-reduced-rel}, by undoing the rotating frame transformation, i.e. $\tilde{E}_n = \mathcal{E}_n + n\omega_o^\prime$. With this, in \cref{fig:spaghetti} (a) we consider the same transmon as the one in \cref{fig:quasispectrum} and plot the landscape of multiphoton resonances involving the ground state in the drive parameter space. In this plot each colored line represents a multiphoton process discussed in \cref{fig:coupling-fit} (a): the center of each line\footnote{For the processes that coexist with others, we obtain the resonant line by diagonalizing the effective Hamiltonian such as \cref{eq:mnr-K-4}.} represents the location, i.e. $\omega_d = (\tilde{E}_{q} - \tilde{E}_0)/p\hbar$, of the corresponding $(q:p)$ process and its width represents the Rabi strength. Here we choose to show the multiphoton resonance only when its strength is greater than 0.01~MHz, which is experimentally relevant in a transmon with typical coherence lifetime $T_1 = 100\;\mu$s. 

We remark that due to the ac-Stark shift term $ (2\alpha |\xi|^2\op{\cu A}^\dagger\op{\cu A} + \mathcal{O}(\varphi_\text{zps}^4))$ in the effective Hamiltonian, most of the processes in \cref{fig:spaghetti} (a) appear as straight lines tilting to the left --- they become resonant at a smaller drive frequency with increasing drive strength. Due to this feature, for any fixed drive frequency $\omega_d$, some multiphoton resonance process will come into resonance as $|\xi|^2$ increases. In this particular transmon ($\omega_{01}/2\pi = 5.85$~GHz, $\alpha/2\pi = -160$~MHz), for most values of drive frequency $\omega_d$ one encounters a strong enough resonance process when $|\xi|^2<1$. This is in agreement with experimental observations \cite{Blais2020,minev2019,sank2016} on transmons with similar parameters, where ``anomalous state transitions" occur when the ac-Stark shift $(\Delta_{ac} = 2|\xi|^2\alpha)$ is of the same order of the transmon anharmonicity $\alpha$. 

To verify the phase diagram of \cref{fig:spaghetti} (a), in \cref{fig:spaghetti} (b) we plot the numerically computed the weight of the Floquet ground state out of the displaced transmon ground state. This quantity, which we denote as $P_{0\rightarrow}$, is obtained from the exact Floquet numerical diagonalization and serves as a proxy to measure the excitation probability out of the ground state. Its definition amounts to $P_{0\rightarrow}= 1 - |\langle\tilde{0}^\text{F} | \tilde {0}_t\rangle|^2$, where $|\tilde{0}^\text{F}\rangle$ is the Floquet ground state and $|\tilde {0}_t\rangle$ is the 0th transmon state with a displacement induced by the drive (see \cref{app:transmon-state} for detail). When away from any resonance process we therefore have $|\tilde{0}^\text{F}\rangle = |\tilde{0}_t\rangle$ and $P_{0\rightarrow}= 0$, and when a $(q:p)$ multiphoton resonance occurs (c.f. \cref{fig:quasispectrum}), we have $|\tilde{0}\rangle = (|\tilde{0}_t\rangle + |\tilde{q}_t\rangle)/\sqrt{2}$ and $P_{0\rightarrow} = 0.5$.\footnote{In the presence of multiple simultaneous multiphoton resonances, as discussed in the example of \cref{eq:mnr-K-4}, $P_{0\rightarrow}$ may exceed 0.5 due to the hybridization of the ground state and multiple excited states.} Under this definition, each colored line in the heat map of \cref{fig:quasispectrum} (b) indicates the hybridization between the driven ground state and an excited state and the width of the line is associated with the Rabi strength of the corresponding process. Besides the ten $(q:p)$ processes with $q+p\le16$ captured in \cref{fig:quasispectrum} (a), in (b) one also observes higher order processes with $q+p>16$ when the drive is sufficiently strong. Indeed, since the rational number $\{q/p|q,p \in \mathbb{Z}^+\}$ forms a dense set, the number of $(q, p)$ resonance processes is also dense in parameter space.  We also note that for $(q:p)$ processes involving highly excited states such as $q =$ 10 and 11, the perturbative results in (a) does does not give the precise location of the process. This is related to general difficulty of perturbatively computing the (dressed) energy of large Fock state where the perturbative condition of \cref{eq:pert-condition} breaks down. 

\begin{figure}
\centering
\includegraphics{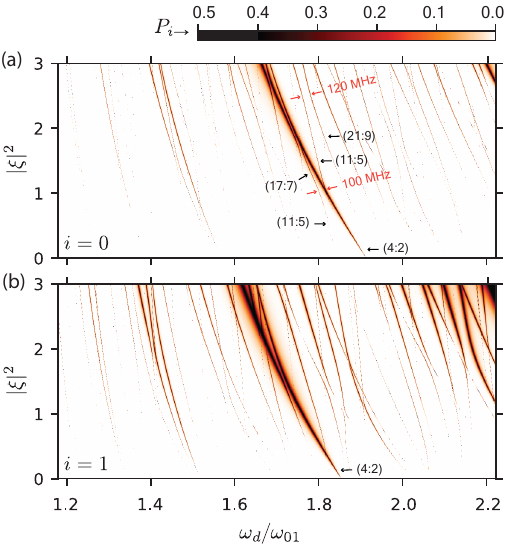}    \caption{Landscape of multiphoton resonances in a transmon in the drive parameter space larger than that in \cref{fig:spaghetti}. The excitation possibility out of the ground state $P_{0\rightarrow}$ is plotted in (a) and that out of the first excited state $P_{1\rightarrow}$ is plotted in (b). Note that the normalization of the color map is different from \cref{fig:spaghetti} for visual compatibility. The tuples $(q:p)$ in black text label the processes the selected resonant lines correspond to. The red arrow and text labels the anti-crossing between relevant resonant lines. When $|\xi|^2<1$, most of the resonances involving the qubit manifold are isolated from each other and we call the transmon in the multiphoton resonance regime. When $|\xi|^2>1$, there are more and more anticrossings among resonance lines, indicating simultaneous existence of multiple resonance processes; we call the transmon gradually phase into quantum diffusion regime as $|\xi|^2$ increases.}
    \label{fig:spaghetti-nonpert}
\end{figure}

For the sake of completeness, in \cref{fig:spaghetti-nonpert} we survey a larger drive parameter space and plot the landscape of  multiphoton resonances involving both the ground and first excited state in the subplots (a) and (b), respectively. In most experiments a multiphoton resonance is consequential if it involves any state in the qubit manifold. We observe that the landscapes of the resonances involving the ground state and the first excited state share a similar structure but with a frequency offset. For example, for a fixed value of $|\xi|^2$ the $(4:2)$ process in (a), which becomes resonant at $\omega_d = (\tilde{E}_4 -\tilde{E}_0)/2\hbar$, occurs at a higher frequency than the same process in (b), which becomes resonant when $\omega_d = (\tilde{E}_5 -\tilde{E}_1)/2\hbar$. This frequency offset is due to the anharmonicity of the transmon and makes the landscape of the multiphoton resonances involving the qubit manifold twice as dense as that involving only the ground state. In addition, processes involving the first excited state are stronger than their counterparts involving the ground state in (a). This is due to the difference in the matrix element between the relevant states as $\langle1|\op {\cu A}^q|q+1\rangle =\sqrt{q+1}\langle0|\op {\cu A}^q|q\rangle$.

Another important feature in the landscape of multiphoton resonance \cref{fig:spaghetti-nonpert} is the anti-crossing between the resonant lines. For example, as labeled in  \cref{fig:spaghetti-nonpert} (a), the $(q_1:p_1) = (11:5)$ resonant line anti-crosses with the $(q_2:p_2) = (17:7)$ line with an energy gap of 100~MHz and with the $(q_3:p_3) = (21:9)$ line with an energy gap of 120~MHz. We recall that each resonant line indexed $(q:p)$ in \cref{fig:spaghetti-nonpert} (a) corresponds to the hybridization between the driven ground and the $q$-th excited state (c.f. \cref{fig:quasispectrum}). Therefore an anti-crossing between two resonant lines indicates the hybridization between two excited states coupled through some multiphoton resonance process, while each state also couples to the ground state of the transmon and thus is perceivable in the $P_{0\rightarrow}$ heat map. In the aforementioned two examples, the anti-crossings respectively imply the coupling between the 11th excited state and 17st excited state through the $(q_2-q_1:p_2-p_1) = (6:2)$ process and between the 11th excited state and 21st excited state through the $(q_3-q_1:p_3-p_1) = (10:4)$ process. The energy gaps in the two anti-crossings only give a proxy that is correlated to the coupling strength in the corresponding multiphoton resonance. From the quasienergy spectrum like the one in \cref{fig:quasispectrum}, we found that, at $|\xi|^2 = 1$, the actual coupling strength between the 11th and 17th excited state is 500~MHz, and, at $|\xi|^2=1$, the coupling strength between the 11th and 21st excited states is 600~MHz.

With a coupling as strong as 500~MHz, the 11th and 17th excited states are significantly hybridized even if the drive is a few GHz detuned from the resonant condition. As a result, even though two resonant lines in \cref{fig:spaghetti-nonpert} (a) are separately labeled as $(17:7)$ and $(11:5)$ processes, for strong drive $|\xi|^2\gtrsim1$, each of the two resonant lines actually implies the concurrence of  $(17:7)$, $(11:5)$ and $(6:2)$ processes in the system; as the drive becomes stronger, the $(21:9)$ and $(10:4)$ processes also become concurrent. From \cref{fig:spaghetti} and \cref{app:coupling}, we also know that 11th state is strongly hybridized with 8th state through a $(3:1)$ process. Therefore in this region of the drive parameter space, the 8th, 11th, 17th, and 21st states form a ``hybridized island'' comprised of the Fock states isolated in the undriven system.

As the drive becomes stronger, in \cref{fig:spaghetti-nonpert} one observes more anti-crossings among the resonant lines implying more states forming more and larger ``hybridized islands'' in the Fock state space.\footnote{Note that the coupling between two excited states are perceivable in \cref{fig:spaghetti-nonpert} only if each state is also coupled to the ground or the excited state. Therefore the actual number of coupling among excited states in a transmon is more dense than shown in \cref{fig:spaghetti-nonpert}.} We therefore refer to the weak drive regime (in this case $|\xi|^2\lesssim1$) in \cref{fig:spaghetti-nonpert} as \textit{multiphoton resonance} regime and the strong drive one as the \textit{quantum diffusion} regime. The latter is named so because, when the excited states in the transmon are sufficiently hybridized, the population in a transmon can efficiently diffuse to all the excited states in the presence of dissipation (see \cref{sec:dissipator}) or with non-zero temperature.\footnote{In the multiphoton resonance regime if the qubit manifold is directly coupled to a running state outside the transmon cosine potential (e.g. through the $(13:5)$ process in \cref{fig:spaghetti} (b)), the transmon will ``ionize'' \cite{reed2010,lescanne2019}. This process is extensively investigated in recent works \cite{shillito2022,cohen2023} through numerical simulation. In the quantum diffusion regime, however, ionization can happen in the absence of such a direct coupling.} We remark these two regimes are similarly identified in the driven Rydberg atom \cite{wang1989}. In classical systems, they respectively correspond to the ultra-subharmonic bifurcation regime (\cref{sec:USH}) and the chaos regime.

We foresee that this regime of extremely strong driving will be reached in search for faster gates, larger bosonic codes and more efficient readout schemes. Accounting for these drive-induced effects is critical for the development of quantum control in superconducting circuits or other platforms modeled by driven nonlinear oscillators. 

\subsubsection{Design principles to mitigate multiphoton resonances}\label{subsec:mnr-design}

We have so far characterized the multiphoton resonances in a transmon of typical design parameters and charted their dense landscape in the drive parameter space. These processes, which are inevitable in a transmon when the drive is sufficiently strong, induce anomalous state transitions or ionization and limit the state-of-art readout scheme and parametric gates. In the following text we comment on a few design principles implied from the discussion above to mitigate multiphoton resonances in driven Josephson circuits. The application of these principles should be examined with specific experimental goals and constraints. 

\textit{1. Choosing the drive frequency informedly.} Since the distribution of multiphoton resonances in the drive parameter space is dense but not uniform, one apparent strategy is to get informed about their landscape as in \cref{fig:spaghetti-nonpert} and design the drive to be at a frequency with a larger range in strength that is resonance free.\footnote{This strategy, along with the strategy of engineering selection rules, are also identified in a recent work \cite{cohen2023}.} In the specific case of \cref{fig:spaghetti-nonpert}, for example, choosing the drive frequency $\omega_d/\omega_{01}\in [1.2,1.3]$ would be a more favorable operational regime than $[1.8,1.9]$. 

Yet, it should be noted that the two plots in \cref{fig:spaghetti-nonpert} assume zero offset charge between the superconducting island in a transmon, i.e. $N_g=0$ in \cref{eq:mnr-transmon-H}. In this case the circuit only contain even order of nonlinearity $g_m$ thus a $(q:p)$ process occurs only if $q+p$ has even parity. This selection rule does not apply to a transmons without control on $N_g$, where the odd order process could be as strong as the even order ones while the resonance condition involving highly excited states significantly disperse with $N_g$. These make the actual landscape of multiphoton resonances in an ordinary transmon with $N_g\ne0$ more dense and complicated than depicted in \cref{fig:spaghetti-nonpert}.

\textit{2. Engineering selection rules.} Another strategy that naturally follows from the discussion above is to impose selection rules, such as $q+p\in2\mathbb Z$, to a Josephson circuit. In a transmon this can be achieved by adding an offset charge gate to enforce $N_g=0$ or shunting the superconducting islands across the junction with an inductive element such that no offset charge accumulates across the junction. We discuss the latter implementation in more detail in \cref{sec:IST}. Other selection rules can be engineered with more sophisticated circuit designs and driving schemes. For example, a recent work \cite{Lu2023} develops of a scheme of pumping a novel Josephson circuit by modulating the magnetic flux threaded in the circuit. With this, different selection rules, such as both $q, p$ being even number for a $(q:p)$ process, can be achieved. 

\textit{3. Suppressing ac-Stark shift.} In \cref{fig:spaghetti,fig:spaghetti-nonpert}, most of the resonant lines, whose location corresponds to the resonant condition of the $(q:p)$ process $\omega_p = (\tilde E_{q+i} - \tilde E_i)/p\hbar$ for $i=0,1$, tilt to the left due to ac-Stark shift. It is clear that if each resonant line has a smaller slope, the resonance-free range of the drive strength for a given drive frequency is larger. Therefore suppressing the ac- Stark shift is another strategy to mitigate multiphoton resonance. This can be achieved in a SNAIL circuit \cite{frattini2017} by operating with the external magnetic flux near a sweet spot. Specifically, with the diagrammatic method we compute the leading order ac-Stark shift term in the effective Hamiltonian of a driven SNAIL as 
\begin{align} \label{eq:mnr-ac-stark}
    \big(6g_4-\frac{4g_3^2}{2\omega_a-\omega_d}-\frac{4g_3^2}{2\omega_a+\omega_d}-\frac{8g_3^2}{2\omega_a}\big)|\xi|^2\op{\cu A}^\dagger\op{\cu A},
\end{align}
where the nonlinearity $g_3$, $g_4$ is controlled by the external magnetic flux (see ref \cite{frattini2017}), and drive frequency is assumed to be away from $2\omega_a$. For any $\omega_d$ there exists a magnetic flux, such that  $g_3,g_4$ are tuned to fully suppress \cref{eq:mnr-ac-stark}. This sweet spot is experimentally illustrated in \cite{sivak2019}.

\textit{4. Reducing $\varphi_\text{zps}$.} To engineering parametric processes in Josephson circuits a central task is to exploit the desired process, e.g. the coupling term in Kerr-cat and three-legged cat Hamiltonian \cref{eq:Kerr-cat-RWA,eq:3-legged-cat-H}, and curb the spurious ones, e.g. the multiphoton resonance processes. Importantly, both the desired and undesired processes are controlled by the circuit design parameter $\varphi_\text{zps}$ and the drive parameter $\xi$ but with different dependencies. In most driving schemes, on one hand, the exploited parametric process are 3-wave or 4-wave mixing with one or two drive photons involved; the parametric strength is therefore usually of the form $\Omega_\text{para}\propto\varphi_\text{zps}^m\xi^n$ with $m=1,2, n=1,2$. For an undesired $(q:p)$ multiphoton resonance, on the other hand, the coupling strength is of the form $\Omega_{q, p}\propto\varphi_\text{zps}^{q+p-2}\xi^{p}$. The goal of curbing the multiphoton resonance without compromising the parametric strength then translates to suppressing the ratio
\begin{align}\label{eq:mnr-ratio}
    \frac{\Omega_{q, p}}{\Omega_\text{para}} \propto\varphi_\text{zps}^{q+p-2-m}\xi^{p-n}.
\end{align}
With the configuration of $k, q,p,m,n$ in experiments, a strategy that usually works is to lower $\varphi_\text{zps}$ while increase $\xi$ --- this will lead to suppression of $\Omega_{q, p}/\Omega_\text{para}$ at the same time 
the parametric strength $\Omega_\text{para}$ is kept constant.

For example, in the Kerr-cat system described by \cref{eq:Kerr-cat-RWA} the interested term is $g_3\xi\op{\cu A}^2$ where $g_3 \propto \varphi_\text{zps}$. Under the constraint of the squeezing strength $\Omega_\text{sqz}=g_3\xi$ being constant, the drive strength $\xi = \Omega_\text{sqz}/g_3\propto \varphi_\text{zps}^{-1}$ is then inversely related to $\varphi_\text{zps}$. In this case, \cref{eq:mnr-ratio} becomes $\Omega_{q, p}/\Omega_\text{sqz} \propto \varphi_\text{zps}^{q-2}$, which implies that with the squeezing strength kept constant, lowering $\varphi_\text{zps}$ by a factor of $m$ will suppress the $(q:p)$ process by a factor of $m^{q-2}$.  

\subsection{Mitigating nonlinear resonance with inductively-shunted transmon}\label{sec:IST}

\begin{figure}
	\includegraphics{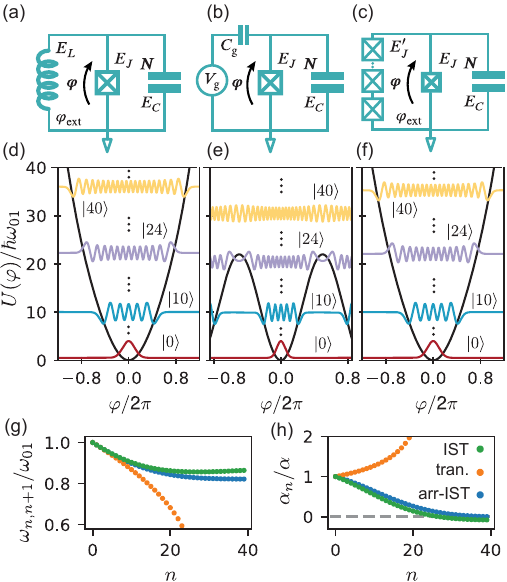}
	\caption{ (a) - (c) Circuit diagrams of an inductively shunted transmon (IST), transmon, and array IST. In (a) and (c), the offset charge represented by $V_\text{g}$ in (b) is nulled by the inductor or the array. (d) - (f) Potentials of the corresponding circuits with selected wavefunctions overlaying on the top. The parameters of each circuits are chosen such that the qubit frequency $\omega_{01}=5.85$~GHz and anharmonicity $ \alpha/2\pi= -70$~MHz are the same across all circuits. The extra knob $r$ in the IST and array-IST allows tunability on $\varphi_\text{zps}$ --- in this case we choose $r=3$, the resulting  $\varphi_\text{zps}= 0.446$ for IST and arr-IST is different from $\varphi_\text{zps}= 0.215$ for the transmon in (e). This can be inferred from the larger spread of ground state in (d) and (f) comparing with (e). (g) transition frequency $\omega_{n,n+1}$ between the $n$th and $(n+1)$-th state as a function of state index. (f) Higher-order nonlinearity $\alpha_n = \omega_{n+1,n+2} - \omega_{n,n+1}$ as a function of state index. For an IST and array-IST, $\omega_{n,n+1}$ and $\alpha_n$ behaves like a Bessel function \cite{verney2019}. Importantly, $\alpha_n$ start to oscillate around 0 when the spread of state $|n>20\rangle$ over $\varphi$ is greater than $2\pi$, the width of the nonlinear part $-E_J\cos\varphi$ in $U_\text{IST}$. This saturation behavior of nonlinearity is unique to ISTs.}
	\label{fig:IST-circuits}
\end{figure}

Based on the design principles discussed above, in this section we introduce a circuit design, the inductively-shunted transmon (IST) \cite{smith2022}, to mitigate multiphoton resonances when the circuit is driven. As shown in \cref{fig:IST-circuits} (a), an IST is an ordinary transmon of tunneling energy $E_J$ and charging energy $E_C$ shunted by a linear inductance of inductive energy $E_L$. The Hamiltonian of an IST is just that of a transmon (the static part in \cref{eq:mnr-transmon-H}) with an additional inductive term
\begin{equation}\label{eq:IST-H}
	\begin{split}
		&\op{H}_\text{IST}  = 4E_C\op{N}^2 + \op{U}_\text{IST}(\op{\varphi})\\
		& \op{U}_\text{IST}(\op{\varphi}) = -E_J\cos\op{{\varphi}} + \frac{1}{2}E_L(\op{{\varphi}}-\varphi_\text{ext})^2
	\end{split}
\end{equation}
with $\op{N}$ and $\op{\varphi}$ being the Cooper-pair number operator and phase operator across the junction. The function  $\op{U}_\text{IST}(\op{\varphi})$ is the nonlinear potential specified by the junction and the inductive shunt, and $\varphi_\text{ext}$ is the reduced magnetic flux threaded in the inductive loop. To imitate the symmetric potential of the transmon, an IST is designed to operate at zero flux bias $\varphi_\text{ext} = 0$. We also demand the ratio between the shunting inductive energy and the junction tunneling energy to be sufficiently large, i.e. $r=E_L/E_J>0.22$; in this regime the potential $U_\text{IST}(\varphi)$ is single-welled and confining such that the wavefunctions are localized around the single minimum of the potential at $\varphi = 0$. 

Expanding the potential $\op U_\text{IST}$ as a Taylor series,  the Hamiltonian \cref{eq:IST-H} can be rewritten in the bosonic basis as 
\begin{subequations}\label{eq:IST-H-bosonic}
\begin{align}
\begin{split}
    \frac{\op{H}_\text{IST}}{\hbar} = \omega_o\op a^\dagger \op a &+ \sum_{\substack{m \ge 4 \\ m \in 2\mathbb Z}}\frac{g_m}{m}(\op a + \op a^\dagger)^m,
\end{split}
\end{align}
where 
\begin{align}\label{eq:IST-H-bosonic-b}
\begin{split}
&\omega_o=\sqrt{8E_CE_J(1+r)}/\hbar\,, \;
g_m = \frac{(-1)^{\frac{m}{2}+1}\omega_o}{2(m-1)!(1+r)}\varphi_\text{zps}^{m-2},\\
&\varphi_\text{zps} = \Big(\frac{2E_C}{E_J(1+r)}\Big)^\frac{1}{4},\; \op a = \frac{1}{2}\Big(\frac{\op\varphi}{\varphi_\text{zps}}+2i\varphi_\text{zps}\op N\Big).
\end{split}\raisetag{0.8\baselineskip}
\end{align}
\end{subequations}
We note that an IST deforms to an ordinary transmon when $E_L\rightarrow0$; similarly, the parameters in  \cref{eq:IST-H-bosonic-b} defining the nonlinear oscillator become those corresponding to a transmon in \cref{eq:mnr-transmon-H-bosonic} when $r\rightarrow0$.

In addition, we remark that fabricating a lossless linear inductance with $E_L$ in the range of interest is difficult despite a few recent pioneering works achieving it with granular aluminum \cite{pop2018,pop2020}. An IST can be alternatively realized by shunting a transmon with an array of $M$ Josephson junctions each with Josephson energy $E_J^\prime$, as shown in \cref{fig:IST-circuits} (c). This array-IST design is described by the same Hamiltonian as \cref{eq:IST-H} but with a modified potential energy $\op U_\text{IST}\rightarrow \op U_\text{array-IST}$ as
\begin{subequations}
    \begin{align}\label{eq:array-IST-H-a}
         \op U_\text{array-IST}(\op\varphi) &= -E_J\cos\op\varphi -ME_J^\prime\cos\frac{\op\varphi-\varphi_\text{ext}}{M}\\\label{eq:array-IST-H-b}
         &=-E_J\cos\op\varphi + \frac{E_J^\prime}{2M}\op\varphi^2 + \mathcal{O}(\frac{\op\varphi^4}{M^3}),
    \end{align}
\end{subequations}
where in \cref{eq:array-IST-H-b} we expand the last term in \cref{eq:array-IST-H-a} as a Taylor series. As written it is clear that $\op U_\text{array-IST}(\op\varphi)$ approximates $\op U_\text{IST}(\op\varphi)$ in \cref{eq:IST-H} to leading order with the effective shunting inductive energy as $E_J^\prime/M$. This can also be observed in the close resemblance of the spectrum of an IST and an array-IST shown in \cref{fig:IST-circuits} (d), (f), (e), and (h). 

The inductive shunt in an IST modifies an ordinary transmon in a few important ways which we now discuss. Firstly, the shunt connects the superconducting islands in a transmon and no charge can accumulate across them. The offset charge $N_g$ in \cref{eq:mnr-transmon-H} is thus warranted to be zero in the IST Hamiltonian $\op H_\text{IST}$ in \cref{eq:IST-H}. As a result,  $\op H_\text{IST}$ by construction is an even function that endorses the selection rule $q+p\in2\mathbb Z$ for $(q, p)$ multiphoton resonance processes --- this reduces the density of the resonances in the drive parameter space, which is only the case for $N_g=0$ in an ordinary transmon. Also due to the suppression of $N_g$ dependence, the energy spectrum of an IST does not disperse as the static charge fluctuates in the environment the circuit 
lives in; in the context of multiphoton resonance, the location of each process in the drive parameter space (c.f. \cref{fig:spaghetti}) is invariant under the static charge fluctuation and provides more experimental controllability.

Secondly, besides the circuit parameters $E_C$ and $E_J$ of a transmon, an IST involves an extra parameter $r=E_L/E_J$ that also controls the configuration, specifically $\omega_o$, $\varphi_\text{zps}$, and $g_m$'s, of the nonlinear oscillator the circuit corresponds to. As evident in \cref{eq:IST-H-bosonic-b}, through tuning $E_J$ and $E_C$ an IST can then be constructed such that its $\omega_o=\sqrt{8E_CE_J(1+r)}/\hbar$ and $g_4=-E_C/3(1+r)$ match those of any given transmon (with $r=0$), while $r$ is left as a free knob to tune $\varphi_\text{zps}=\sqrt{-12(1+r)g_4/\omega_o}$. Here we choose to match $g_4$ the lowest order nonlinearity 
since it usually generates the interested parametric processes and determines the gate and readout speed in experiments. To illustrate the tunability on $\varphi_\text{zps}$, in \cref{fig:IST-circuits} (d)-(h), we plot the spectrum of an IST, a transmon, and an array-IST with circuit parameters that have the same qubit frequency\footnote{Here the qubit parameters $\omega_{01}$ and $\alpha$ are chosen over Hamiltonian parameters $\omega_o$ and $g_4$ since the former pair is directly measurable in experiments.} $\omega_{01}\approx \omega_o -\alpha$ and anharmonicity $\alpha \approx 3g_4$ but different $\varphi_\text{zps}$.

\begin{figure}
	\includegraphics{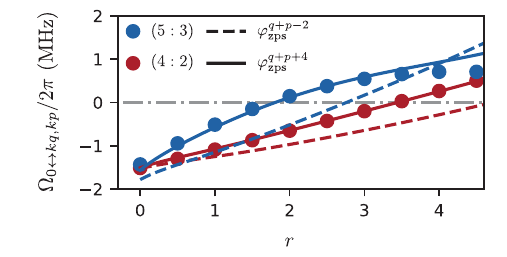}
	\caption{Coupling strength $\Omega_{0\leftrightarrow q, p}$ between the 0th state and $q$-th Fock state as a function of $r = E_L/E_J$ for $(q:p)= (4:2)$ and $(5:3)$ processes in ISTs with $\omega_{01}/2\pi = 5.85$ GHz, $\alpha/2 \pi = -70$~MHz. For better visual comparability, $(4:2)$ is computed for the drive strength at $|\xi|^2= 0.5$, and (5:3) is at $|\xi|^2 = 3$. For $r = 0$ and $3$, the circuits correspond to the transmon and the IST with the same parameters as shown in \cref{fig:IST-circuits} (b) and (a), respectively. Dots correspond to the value inferred from the exact numerical Floquet simulations (c.f. \cref{fig:quasispectrum,eq:mnr-K-1}). Dashed curves are extracted from the leading order result of $\Omega_{q, p}$ in \cref{eq:IST-53,eq:IST-42}. Solid lines are extracted from the analytical results when computed to order $\varphi_\text{zps}^{q+p+4}$. For increasing $r$, the discrepancy between the numerical and perturbative description arises since $\varphi_{\rm{zps}}$ is no longer small.}
	\label{fig:IST-r-knob}
\end{figure}

The independent knob $r$ that revises $\varphi_\text{zps}$ is an asset to control particular multiphoton resonance processes. This is already evident in \cref{fig:coupling-fit} (b) where the $(7:5)$ process in a transmon is fully suppressed at $\varphi_\text{zps}^2 = 0.176$ (yet in this case $\varphi_\text{zps}$ is fixed by specifying $g_4$). Another example is the $(5:3)$ process which we have extensively analyzed for a transmon in \cref{subsec:mnr-53}. In an IST, the corresponding coupling amplitude can be obtained by plugging \cref{eq:IST-H-bosonic-b} into \cref{eq:mnr-53-coulping}, which yields
\begin{align}\label{eq:IST-53}
    \Omega_{5,3} =\frac{(240r^2-6500r+14535)g_4^3}{72\omega_d^2} + \mathcal{O}(\varphi_\text{zps}^8).
\end{align}
To leading order $\varphi_{zps}^{6}$, the coupling amplitude $\Omega_{5,3}$ is a quadratic function in $r$ and is fully suppressed at $r = 2.46$ and 26.42. When the ground and 5th excited states are resonant under this process, it yields a coupling strength between the two states of strength $\Omega_{0\leftrightarrow5,3} = \langle5|\op{\cu A}^{\dagger 5}|0\rangle\xi^3$.  In \cref{fig:IST-r-knob}, we plot the coupling strength $\Omega_{0\leftrightarrow5,3}$ as a function of $r$; the Floquet numerical simulation gives a suppression point at $r = 1.79$, which is in good agreement with the analytical result when computed to $\varphi_\text{zps}^{12}$. In the same figure we also plot the coupling strength of $(4:2)$ process, stems from
\begin{align}\label{eq:IST-42}
    \Omega_{4,2} =\frac{(6r-27)g_4^2}{2\omega_d}+ \mathcal{O}(\varphi_\text{zps}^6).
\end{align}
We note that the suppression point of $ \Omega_{4,2}$ to leading order is at $r=4.5$ (and at order $\varphi_\text{zps}^{10}$ is at $r=3.45$), which is different from that of the $(5:3)$ process. In general, different parametric processes have different suppression points due to the different off-resonant excitations constituting their effective Hamiltonians. Therefore an IST can only be designed to suppress one particular multiphoton resonance. Yet this could still be a valuable tool for experiments that require special drive configurations where some specific multiphoton resonance is inevitable. 

Another benefit stemming from the inductive shunt in an IST is the confinement of the potential. As shown in \cref{fig:IST-circuits}, a transmon with a cosine potential only contains $\sim\sqrt{E_J/E_C}$ number of states inside the potential well while all other states are ``running-states" \cite{koch2007}. When directly coupled to the latter through some multiphoton resonance process, a transmon in the qubit manifold will ``ionize'' \cite{shillito2022} and, in the presence of dissipation, heat up to some highly mixed states \cite{verney2019}. In an IST (array-IST), however, the potential is (largely) confined and all states (a large number of states) are inside the potential well. Consequently, when the drive strength $|\xi|$ is sufficiently small such that the qubit is within the multiphoton resonance regime (c.f. \cref{fig:spaghetti-nonpert}), the ionization of an IST is fully suppressed. 

We remark that, similar to the transmon discussed in \cref{subsec:mnr-landscape}, an IST in general can still diffuse to all driven states through ``quantum diffusion'' when $|\xi|^2$ is sufficiently large, which also leads to heating up to highly mixed states. Yet \cite{verney2019} has shown numerical evidence suggesting that this instability in an IST can be fully suppressed when $r$ is sufficiently large. This observation is further supported by a recent work \cite{burgelman2022}, which proves that when $r>1.89$, a driven IST in the classical limit cannot undergo ``period-doubling cascading'', a process believed to be necessary for entering chaotic regime (the classical counterpart of quantum diffusion). While the deep quantum diffusion regime is beyond the perturbative analysis concerned in this work, it is useful to comment on this suppression in the context of multiphoton resonance and relate it to the diagrammatic language, which we briefly discuss on now. 

In an IST, the cosine part $-E_J\cos\op\varphi$ of the potential $\op U_\text{IST}$ in \cref{eq:IST-H} is responsible for the nonlinearity, which is diluted by the linear part $E_L\op\varphi^2/2$. The ratio $1/(1+r) = E_J/(E_J+E_L)$ can be viewed as a dimensionless measure of the available nonlinearity in the system. Heuristically, when a wavefunction explores a larger domain of  $(-\pi,\pi]$, the more nonlinearity it inherits from the potential. For example, for the IST with smaller $\varphi_\text{zps}$, the span of wavefunctions over $\varphi$ is smaller and so is the nonlinearity $g_m$'s in \cref{eq:IST-H-bosonic-b}.  In the presence of drive, the oscillator gains energy since it is displaced by $\sim2\varphi_\text{zps}|\xi|$ away from the potential minimum at $\varphi = 0$. This lets the oscillator states explore a larger range of the nonlinear potential and thus the strength of the relevant nonlinear processes increases. Moreover, in an IST, the amount of nonlinearity, measured by $1/(1+r)$, is limited and it is expected that the nonlinearity inherited by a state is also finite\footnote{One can intuit is in the limiting case $r\rightarrow\infty$, in which in cosine nonlinear potential is infinitely diluted by the linear one. In this case, the IST becomes an harmonic oscillator and no nonlinear phenomena should be expected.}. Phenomenologically, this corresponds to an observation that when a state explores more than one period of the cosine potential, the strengths of associated nonlinear processes are bounded. 

\begin{figure}
	\includegraphics{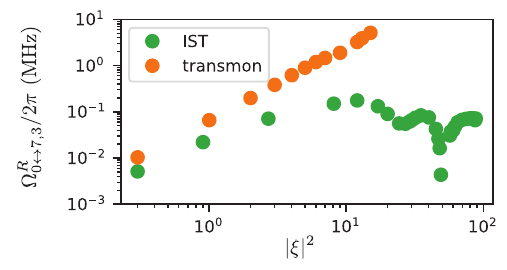}
	\caption{Rabi strength between the ground and 7th excited states under a $(7:3)$ resonance as a function of drive strength. Green and orange dots correspond to the resonance in the transmon and IST with parameters the same as \cref{fig:IST-circuits}~(d) and (e), respectively. Each data point is obtained from Floquet simulation similar to those in \cref{fig:coupling-fit,fig:IST-r-knob}. In the transmon, the Rabi strength increases monotonically in $|\xi|^2$. In the IST, however, the Rabi strength saturates around $|\xi|^2 = 12$ and then oscillate around 0 like a Bessel function for a stronger drive. }
	\label{fig:IST-bessel}
\end{figure}

This boundedness can be already observed in \cref{fig:IST-circuits}~(h), where the higher order anharmonicity diminishes and has a Bessel-function-like behavior for highly excited states\footnote{In this case, the energy is inserted  through exciting the transmon to its high-energy eigenstates instead of through displacing the transmon via drive. Both of these ways can be seen as exploiting the nonlinearity of the system through letting the wavefunction interacting with a larger range of nonlinear potential.} that span over one period of cosine potential. In the case of multiphoton resonance, we illustrate this boundedness in \cref{fig:IST-bessel}, where we compare the Rabi strength of $(7:3)$ process in a transmon and an IST of $r=3$ and with frequency and anharmonicity of the two circuits matched. While the Rabi strength of this process monotonically increase with $|\xi|$ in the transmon, in the IST it is bounded to $0.2~$MHz and has a Bessel-function-like behavior. Phenomenologically, the saturation drive strength corresponds to the driven ground state being displaced by $2\varphi_\text{zps}|\xi|\approx \pi$ in $\varphi$ coordinate. When all multiphoton resonances are sufficiently bounded with large enough $r$, the size of ``hybridized islands'', which are formed by coexistence of many strong multiphoton resonances, will remain finite in the Fock state space and connect to the qubit manifold in a confined region in the drive parameter space. This confinement leads to the suppression of quantum diffusion.

In the context of the diagrammatic method, this boundedness can be understood as the cancellation between the lower order process and higher order processes. Specifically, as discussed in \cref{eq:discussion-K-general}, the leading order diagram responsible for a nonlinear process is perturbatively corrected by some more complicated mixing diagram of higher order in $\varphi_\text{zps}$ and involving more pairs of drive or resonant excitations. Diagrams at different order in general have different signs and could (partially) cancel out each other (e,g. the $(7:5)$ process in \cref{fig:coupling-fit}~(b)), while  contribution of the higher order diagrams become larger as $|\xi|$ and $\langle \op{\cu A}^\dagger \op{\cu A}\rangle$ increase. In an IST, the perturbative parameter of the diagrammatic expansion is $\varphi_\text{zps} = \sqrt{-12(1+r)g_4/\omega_o}$, implying that a larger $r$ yields a larger $\varphi_\text{zps}$ when $g_4/\omega_o$ (or $\alpha/\omega_{01}$) is kept constant.\footnote{We insist on keeping $g_4/\omega_o$ or $\alpha/\omega_{01}$ constant instead of keeping $E_C$, $E_J$, or $\varphi_\text{zps}$ constant. In the latter cases, increasing $r=E_L/E_J$ leads to trivial decreasing of all $g_m$'s, and thus all nonlinear processes, including the desired parametric process, will be suppressed.} Consequently, when $r$ becomes larger, the contribution of diagrams of higher order in $\varphi_\text{zps}$ will become more comparable with the lower order ones and lead to saturation of the strength of the nonlinear effects as observed in \cref{fig:IST-bessel}. In general, the larger $r$ is, the earlier such saturation point will be reached in the drive strength $|\xi|$. We note that the discussion here is only meant to provide a heuristic account on the suppression of quantum diffusion in ISTs, while a systematic study is left to a future work.

\subsection{Ultrasubharmonic bifurcation in classical driven nonlinear oscillator}\label{sec:USH}

In this last example, we apply the diagrammatic method to a classical driven Duffing oscillator and characterize the ultra-subharmonic (USH) bifurcation, a general bifurcation class exhibited in this system. These processes beyond having attracted intense theoretical interest in the nonlinear dynamics community for their intricate topological structure and close relationship to chaos, also constitute the essential physics underlying many quantum devices, such as parametric amplifiers \cite{siddiqi2004,vijay2009,frattini2017,sivak2019}, which are Josephson circuits operating in the weakly-nonlinear and strongly-dissipative regime. Also, importantly, treating this classical example showcases that our diagrammatic method provides a unified framework to analyze both classical and quantum driven dynamics, allowing intuition from the former to be applied to the latter.

Specifically, we consider a Duffing equation
\begin{align}\label{eq:ush-duffing}
    d_t^2{\cl x} + \gamma d_t{\cl x} + \cl x + c_3\cl x^2+c_4\cl x^3 = e^{-i\nu t}+e^{i\nu t},
\end{align}
where $\cl x$ is the position coordinate\footnote{Here we use tilded symbol $\cl x$ to denote that it is a variable in phase space and contrast it with the Hilbert space operator such as $\op x$. This is consistent with the notation in \cref{eq:eom-phase} and onward.} of a Duffing oscillator, $c_3$ and $c_4$ are its third- and fourth-rank nonlinearities, $\gamma$ is the damping rate, and $\nu$ is the dimensionless frequency of a drive coupled to the oscillator. Here we take the drive to have unit amplitude (i.e. the prefactor of $e^{\pm i\nu t}$ to be one), yet for a general drive amplitude one can recover \cref{eq:ush-duffing} by re-scaling $x$, $c_3$ and $c_4$ (see \cref{app:duffing}). In the following analysis we also assume the perturbative structure $c_3^2\sim c_4\ll1$ and $\gamma\ll1$. 

\begin{figure}
	\includegraphics{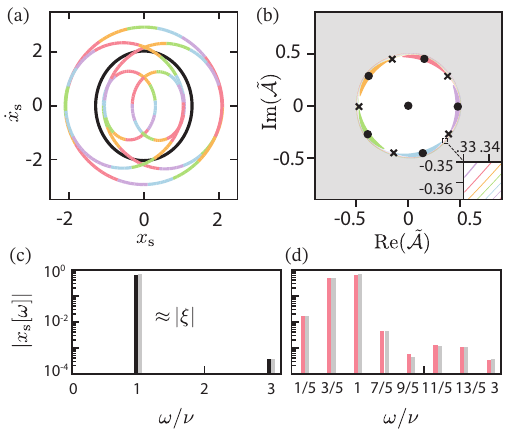}
	\caption{$(5:3)$ ultra-subharmonic bifurcation in  a driven Duffing oscillator. Panel (a) phase portrait of the steady state of \cref{eq:ush-duffing} with $\gamma = 10^{-5}, c_3 = 0,  c_4 = -0.03$. The black line corresponds to the trivial harmonic solution of period $2\pi/\nu$ and the colored lines corresponds to 5-degenerate bifurcated solution of period $10\pi/\nu$ and equally spaced in phase. Panel (b) Phase portrait of the dynamics governed by the effective EOM \cref{eq:ush-eom-eff}. Black dots marks the nodes (steady states) of the EOM and crosses marks the saddles. Each colored region labels the initial states that will evolve to the node enclosed by the region. Grey color corresponds to the region where different types of initial states densely interleaved, as illustrated in the inset. Panel (c), (d) Fourier components of the steady state solution, with the harmonic solution plotted in (c) and  the bifurcated solutions in (d). Black and red bars are obtained by Fourier decomposition of the numerical solution, as shown in (a), to the Duffing equation. Grey bars are predicted from the diagrammatic method (c.f. \cref{eq:ush-frame}).}
	\label{fig:ush-ss}
\end{figure}

\subsubsection{Example: $(5:3)$ USH bifurcation}

As shown in \cref{fig:ush-ss} (a) and (c), a Duffing equation usually admits a trivial harmonic solution where $\cl x(t)$ has the same periodicity $2\pi/\nu$ as the drive and contains a main response at frequency $\nu$ and small ones at the overtones. The main response is approximately 
\begin{align}\label{eq:ush-linear}
x_\text{lin}(t) =\xi e^{-i\nu t} +  \xi^* e^{i\nu t},
\end{align}
which is the solution of the linearized Duffing equation $d_t^2{\cl x} + \gamma d_t{\cl x} + \cl x= \cos\nu t$ with $\xi = 1/(1-\nu^2-i\gamma\nu)$, and the perturbative correction to it at the overtones are due to the mixing of the linear response through the oscillator nonlinearity. 

Besides the harmonic solution, when $\nu$ is in the vicinity of $q/p$ with $q, p$ being coprimes, the Duffing equation \cref{eq:ush-duffing} could admit a set of $q$-fold degenerate solutions that are of period $2q\pi/v$ and evenly spaced in phase. As shown in \cref{fig:ush-ss} (a) and (d), where we take the particular case $(q:p) = (5:3)$ as an example, these solutions contain a major response at frequency $p\nu/q$ in addition to the one at $\nu$. Physically it can be understood as the $p/q$-th ultra-subharmonic of the drive is generated and resonantly stabilized by the nonlinear oscillator. Therefore we call this process $(q:p)$ ultra-subharmonic bifurcation.  

We want to apply the diagrammatic method to analytically characterize the general class of ultra-subharmonic bifurcation. To this end, we first apply a two-step frame transformation to \cref{eq:ush-duffing} similar to those applied to \cref{eq:op-H-a-ad} to arrive at \cref{eq:H-tran} --- the first step is to a displaced frame $\cl x \rightarrow \cl x + x_\text{lin}(t)$ and the second is to a rotating frame $\cl x \rightarrow \cl x\cos(p\nu t/q) + d_t{\cl x}\sin(p\nu t/q)$. In the new frame, \cref{eq:ush-duffing} transforms to 
\begin{align}\label{eq:ush-eom-tran}
\begin{split}
    d_t{\cl a} = -i(\delta - \frac{i\gamma}{2})\cl a -ie^{i\frac{p}{q}\nu t}&\sum_{m=3,4} g_m\Big(\cl a e^{-i\frac{p}{q}\nu t}+\cl a^* e^{i\frac{p}{q}\nu t}\\
    &\quad + \xi e^{-i\nu t}+ \xi^* e^{i\nu t}\Big)^{m-1}, 
\end{split}
\end{align}
where  $\cl a =(\cl x+id_t{\cl x})/2$ is the complex coordinate satisfying the canonical relation $\{\cl a, \cl a^*\}_{\cl a, \cl a^*}=1$, and $\{\cl f, \cl g\}_{\cl a, \cl a^*} =-i\partial_{\cl a}\cl f\partial_{\cl a^*}\cl g+i\partial_{\cl a}\cl f\partial_{\cl a^*}\cl g $ is the Poisson bracket (also the $\hbar^0$ order of Husimi bracket in \cref{eq:Husimi-transform}) defined over the phase space $(\cl a, \cl a^*)$. The parameter $\delta = 1-p\nu/q$ is the detuning between the natural frequency of the oscillator and $p/q$-th ultra-subharmonic of the drive and $g_m=c_m/2$ is the rescaled nonlinearity. Here we also symmetrized the damping to both the position and momentum coordinate, which is a valid approximation for $\gamma\ll 1$. 

Written as it is, \cref{eq:ush-eom-tran} can be identified as a Hamilton equation of motion $d_t{\cl a} = -\{\cl H, \cl a\}_{\cl a, \cl a^*}$ with the corresponding Hamiltonian being 
\begin{align}\label{eq:ush-H-tran}
\begin{split}
    \cl H(t) = (\delta-i\frac{\gamma}{2})\cl a^* \cl a+  \sum_{m=3,4} \frac{g_m}{m}  &\Big(\cl{a} e^{- i \frac{p}{q}\nu t} + \cl{a}^{*} e^{i\frac{p}{q}\nu t} \\[-8pt]
    & \;+ \xi e^{-i \omega_d t} +  \xi^* e^{i \omega_d t}\Big)^m.
\end{split}\raisetag{-0.5\baselineskip}
\end{align}
\Cref{eq:ush-H-tran} is just the classical counterpart of \cref{eq:H-tran}, the starting point of the quantum harmonic balance (QHB) detailed in \cref{sec:averaging}, but with the modification $\delta\rightarrow \delta + i\gamma/2$ incorporating the damping in a complex Hamiltonian description. We note that the Husimi Q phase-space, in which QHB operates, deforms to the phase space of classical mechanics when $\hbar \rightarrow 0$. Consequently, \cref{eq:ush-H-tran} can be described by an effective Hamiltonian whose diagrammatic representation is just the $\hbar^0$ order contribution of the diagrams, i.e. those without any quantum bond, born from \cref{eq:H-tran}. 

For example, for the $(5:3)$ USH bifurcation shown in \cref{fig:ush-ss}, the Duffing equation to leading orders can be described by an effective Hamiltonian 
\begin{align}\label{eq:ush-H-53}
    \cl K = \sum_n K_n\clu A^{*n}\clu A^n + \Omega_{5,3}\xi^{3}\clu A^{*5} + \mathrm{c.c.}
\end{align}
where $(\clu A, \clu A^*)$ are the complex coordinates of the transformed frame that $\cl K$ lives in. In this specific example, we take $g_3=0$ for simplicity and, from the diagram, get the coefficient $K_1 = (\delta -i\gamma/2 + 6g_4|\xi|^2) + \mathcal O(g_4^2)$. Note that $K_1$ here contains the ac-Stark shift term $6g_4|\xi|^2$ but not the Lamb shift term as that in \cref{eq:3leg-renorm} does. In the diagrammatic representation, these two terms correspond expanded diagrams like 
\begin{align}\label{eq:ush-K1-diagram}
\qquad\underbrace{\tikzpic{-26}{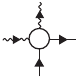}}_\text{ac-Stark shift} \qquad
\underbrace{
    \tikzpic{-26}{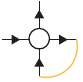} 
    }_\text{Lamb shift (absent in \cref{eq:ush-H-53})}
\end{align}
and it is clear that the Lamb shift, due to its quantum nature, is absent in the classical Duffing system. 
For the coupling term $\Omega_{5,3}$, the corresponding diagrams are of the type 
\begin{align}\label{eq:ush-couple-diagram}
\tikzpic{-26}{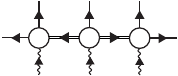}\quad
\tikzpic{-26}{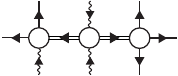}\quad\cdots
\end{align}
These diagrams are also present in the $(5:3)$ multiphoton process in a transmon as shown in \cref{eq:5:3_diagram}. Therefore to leading order the coupling amplitude $\Omega_{5,3} = 21275g_4^3/72\nu^2+ \mathcal O(g_4^4)$ in the Duffing system has the same prefactor as those involving $g_4^3$ in \cref{eq:mnr-53-coulping}. For other coefficients in \cref{eq:ush-H-53} one can be similarly compute them through the diagrammatic method and for simplicity we do not display them here explicitly.

With the effective Hamiltonian $\cl K$ governing the $(5:3)$ USH bifurcation computed as in \cref{eq:ush-H-53}, we can numerically solve its equation of motion
\begin{align}\label{eq:ush-eom-eff}
d_t\clu A = -\{\cl K, \clu A\}
\end{align}
and obtain the time evolution of any classical state with a given initial condition. This leads to \cref{fig:ush-ss}~(b), in which we plot the phase portrait of the $(5:3)$ USH bifurcation in the $(\clu A, \clu A^*)$ phase space. The stationary oscillatory states, corresponding to those with $\frac{d\cl A}{dt}=0$, are labeled as the black dots and crosses. In particular, the dots, on the one hand, are the nodes of the effective EOM, and all the initial states in a colored (or white) region will evolve to a node enclosed by the corresponding region. In this sense, each node represents a steady state, which is also called an attractor, of the driven Duffing oscillator. The crosses in \cref{fig:ush-ss} (b), on the other hand, are the saddles of the effective EOM lying on the separatrix of the phase space regions of different colors. They are unstable stationary points in phase space, and the slightest deviation from them results in attraction to a node instead. There are six nodes in total, with one at the origin and the other five evenly spaced in a circle centered around the origin. Since the phase space $(\clu A, \clu A^*)$ is in a rotating frame at frequency $3\nu/5$, the node at the origin, denoting $\clu A = 0$, corresponds to the harmonic steady state (the black lines in \cref{fig:ush-ss} (a) and (c)) that has no response at the $3/5$-th USH of the drive. Accordingly, the other five nodes correspond to the set of 5-fold degenerate USH steady states (the colored lines in \cref{fig:ush-ss} (a) and (d)).

In \cref{fig:ush-ss} (c) and (d), we have also computed the Fourier components for the steady states of $\cl x$ using the diagrammatic method and represented them in grey color. To obtain these, we first observe that the transformed bosonic coordinate $(\clu A, \clu A^*)$ in the effective frame, where $\cl K$ lives in, is related to $\cl x$ in the lab frame, where the Duffing equation \cref{eq:ush-duffing} lives in, by the successive frame transformations:
\begin{subequations}\label{eq:ush-frame}
\begin{align}\label{eq:ush-frame-a}
\begin{split}
    \cl x =&\;\clu Ae^{-i\frac{3}{5}\nu t}+\clu A^*e^{i\frac{3}{5}\nu t} +\xi e^{-i\nu t} +  \xi^* e^{i\nu t} \\
    &+ (\cl\sigma+\cl\mu )e^{-i\frac{3}{5}\nu t}+ (\cl\sigma^*+\cl\mu^*) e^{i\frac{3}{5}\nu t},
\end{split}\\
\begin{split}
    =&\;\clu Be^{-i\frac{3}{5}\nu t}+\clu B^*e^{i\frac{3}{5}\nu t} +\xi e^{-i\nu t} +  \xi^* e^{i\nu t} \\
    &+ \cl\mu e^{-i\frac{3}{5}\nu t}+ \cl\mu^*e^{i\frac{3}{5}\nu t},
\end{split}\label{eq:ush-frame-b}
\end{align}
\end{subequations}
where $\cl x$ is the position coordinate in the lab frame, $\xi$ is the linear response of the oscillator to the drive (c.f. \cref{eq:ush-linear}),  and $\cl\sigma = \cl\sigma(\clu A, \clu A^*)$ and $\cl\mu = \cl\mu(\clu A, \clu A^*, t)$ are the static and oscillating parts of the frame constructed by the QHB expansion in \cref{sec:averaging}. In \cref{eq:ush-frame-b}, we further introduce the change of variable $\clu B=\clu A+\cl\sigma$ as done in \cref{eq:composite-var}. With this, each pair of conjugate terms in the right-hand side of \cref{eq:ush-frame-b} diagrammatically corresponds to the dressed resonant excitations, the drive excitations, and the dressed off-resonant excitations, respectively. Remarkably, different type of excitations in the diagrammatic method, which were constructed to facilitate the perturbation expansion, now can be mapped to physical observables, i.e. the  Fourier components of the steady state of $\cl x$ shown in \cref{fig:ush-ss} (c). For example, the Fourier component at $3\nu/5$ corresponds to 
\begin{align}\label{eq:ush-res}
     \tikzpic{-25}{figsv6_small/notation/B.pdf} \quad  \tikzpic{-25}{figsv6_small/notation/Bs.pdf}\quad
     \tikzpic{-30}{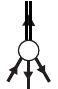}\quad
     \tikzpic{-30}{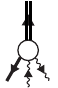}\quad
     \tikzpic{-30}{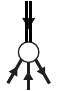}\quad
      \tikzpic{-30}{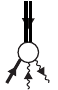}\quad \cdots
\end{align}
where each diagram above has a rotating phase of $e^{\pm i 3\nu t/5}$. Likewise, the Fourier component at frequency $7\nu/5$  correspond to the diagrams 
\begin{align}\label{eq:ush-off-res}
     \tikzpic{-30}{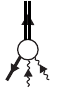}\quad
     \tikzpic{-30}{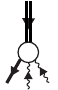}\quad
     \tikzpic{-30}{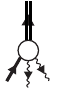}\quad
      \tikzpic{-30}{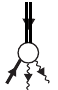}\quad \cdots
\end{align}
Each Fourier component in $\cl x$ can be expressed as a series of diagrams similar to those in \cref{eq:ush-res,eq:ush-off-res}. By summing their algebraic expressions, which are functions of $\clu A$, $\clu A^*$, and $t$, and plugging in the steady states for $\clu A$ and $\clu A^*$ obtained from \cref{eq:ush-eom-eff}, we can obtain the perturbatively computed steady states of $\cl x$. These steady states are shown as grey bars in \cref{fig:ush-ss} (c) and (d) and are in good agreement with the exact values obtained from numerical integration.
Such agreement provides compelling evidence that our diagrammatic method accurately captures classical bifurcation processes.

While the discussion above suffices for computing the steady state of the oscillator's position $\cl x$, we further note that the steady state can be obtained only from the bare diagrams. This simplification will shed light on the connection between the diagrammatic method and the classical harmonic balance method. To derive this, we make two observations. First, we note that for any diagram evaluated with the bosonic coordinate $\clu A$ in the steady state, the dressed propagator in it, formally defined by \cref{eq:eta-dressed-4} in the quantum case, can be simplified by omitting the Poisson (Husimi) bracket term corresponding to $i\{\cl K,\,\tikzpic{-9}{figsv6_small/notation/dash_box.pdf}\,\}$ in the classical case. This can be seen from the definition of the Poisson bracket
\begin{align}
   \{\cl K, \cl f\}_{\clu A, \clu A^*} = -i(\partial_{\clu A}\cl K\partial_{\clu A^*}\cl f - \partial_{\clu A^*}\cl K\partial_{\clu A}\cl f) 
\end{align}  
for some generic function $\cl f$. In a steady state with $d_t{\clu A} = i\partial_{\clu A^*}\cl K=0$, the Poisson bracket $i\{\cl K,\,\tikzpic{-9}{figsv6_small/notation/dash_box.pdf}\,\}$ in \cref{eq:eta-dressed-4} therefore vanishes. As a result, the dressed propagator, which is evaluated as a infinite series \cref{eq:eta-dressed-3} for the full Hamiltonian, can be directly evaluated as $1/(\omega_\text{out} - 1 + i\gamma/2)$ for steady states, where $\omega_\text{out}$ is the frequency of the off-resonant excitation associated with the propagator.

Our second observation is that the resonant excitations in diagrams representing the steady states of $\cl x$ (e.g., those in \cref{eq:ush-res,eq:ush-off-res}) take the dressed form \!\tikzpic{-9}{figsv6_small/notation/B_small.pdf} $(\clu Be^{-i\frac{3}{5}\nu t})$ rather than the bare form \!\tikzpic{-9}{figsv6_small/notation/A_small.pdf} $(\clu Ae^{-i\frac{3}{5}\nu t})$. By directly computing the steady state of $\clu B$, we can avoid the cumbersome process of expanding $\clu B$ in terms of $\clu A$ and $\clu A^*$ (see \cref{eq:dressed-osc-4}) and then plugging in the steady state of $\clu A$. From \cref{eq:eom-B}, we know that the equation of motion for  \!\tikzpic{-9}{figsv6_small/notation/B_small.pdf} $(\clu Be^{-i\frac{3}{5}\nu t})$  is simply
\begin{align*}
d_t \tikzpic{-20}{figsv6_small/notation/B.pdf} = \partial_t \tikzpic{-20}{figsv6_small/notation/B.pdf} - i\tikzpic{-13}{figsv6_small/notation/Gamma.pdf}.
\end{align*}
Thus, to find the steady state of \!\tikzpic{-9}{figsv6_small/notation/B_small.pdf}, we need only find its corresponding value that solves $\tikzpic{-10.5}{figsv6_small/notation/Gamma_small.pdf} =0$.

Based on our two observations, we can conclude that computing the steady states of $\cl x$ does not require knowledge of the dressing of the resonant and off-resonant excitations through the infinite series in \cref{eq:dressed-osc-4} and \cref{eq:eta-dressed-3}. Rather, the dressed resonant excitation can be treated as a standalone quantity, and the dressed propagator can be evaluated in its bare form $1/(\omega_\text{out} - 1 + i\gamma/2)$. In this sense, the bare diagrams contain all the information of the steady states of a classical driven nonlinear oscillator. Indeed, the classical harmonic balance method, which focuses only on the steady states of the oscillator, can be seen as solving the equation of motion \cref{eq:eom-phase-expand} subject to the constraint that $d_t\clu A = -i\partial_{\clu A^*}\cl K = 0$, $\{\cl K, \cl \eta\} =0$, and $\cl\sigma=0$. Solving this modified equation of motion diagrammatically yields the bare diagrams introduced in \cref{subsec:diagram-prep,subsec:order-1,subsec:bare-virtual}.

\begin{figure*}
	\includegraphics{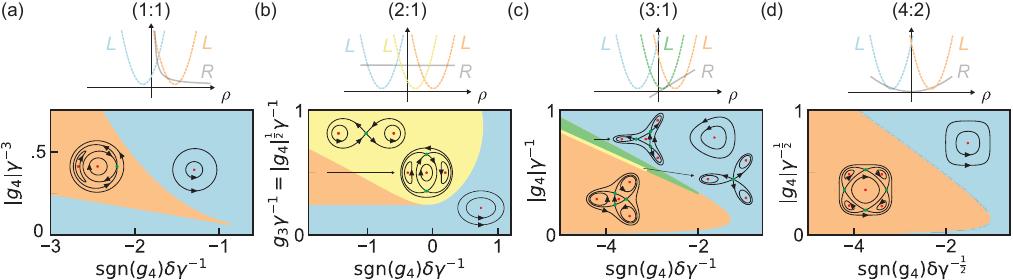}
	\caption{Geometric illustration and bifurcation domain diagrams for lowest order USH processes in classical driven Duffing oscillator, speciﬁcally  $(q : p) =$ (a) $(1 : 1)$, (b) $(2 : 1)$, (c) $(3 : 1)$, and (d) $(4 : 2)$ processes. (a)-(d) top row: the left- and right-hand side of \cref{eq:ush-eom-rho} are represented by the two curves $L$, a parabola, and $R$, a $q-2$ degree curve, in $\rho$. Blue, yellow and orange parabolas represents different choice of $\delta$ such that there exist zero, one, or two bifurcation solutions, i.e. $\rho>0$ points that the two curves intersect. The green parabola in $(3:1)$ represents a special case that two bifurcation solutions both locate on the right half of the parabola, in contrast with the orange parabola that the two solutions locate on different halves. (a)-(d) bottom row: Domain diagrams for the corresponding process in the $\delta$-$g_4$ plane with the axes rescaled by $\gamma$ to pertinent power. For the sake of concrete comparison, we take $|\xi|=1$ for the processes other than $(1:1)$ and $g_3 = \sqrt{|g_4|}$ for $(2:1)$ and 0 otherwise. Distinct domains are marked by different colors, each corresponding to a parabola location in the top row plots. Also plotted within each domain is its characteristic non-dissipative phase diagram governed by \cref{eq:ush-eom-eff-expand} with $\gamma =0$. The blue domain marks the absence of bifurcation and the corresponding phase diagram has a single trivial node at the origin, marked in red, representing the trivial harmonic response solution. Stationary orbits are marked around the node, with arrows indicating direction of motion. In the orange domain, other than the trivial node emerge two additional $q$-fold multiplicities of nontrivial nodes and saddles, marked in red and green respectively. The positions of the nodes and saddles are obtained by solving analytically \cref{eq:ush-eom-r}. Another nontrivial domain colored in yellow, where the $q$-saddle points coalesce at the origin, occurs only in lower-order processes such as in (b) where it is a 2-D area in parameter space and is further reduced to a line in (c) if $\gamma =0$.  Another nontrivial domain colored in green is unique to (c), or $q=3$ cases in general, where the each saddle shares the same phase $\theta$ as a bifurcation node. Any higher-order process contains only the blue and orange domains like in (d).}
	\label{fig:ush-phase}
\end{figure*}

\subsubsection{Structure of general $(q:p)$ USH bifurcation}

In the preceding text, we have so far assumed to find the steady states in the phase space $(\clu A, \clu A^*)$ by numerically solving the effective EOM \cref{eq:ush-eom-eff} for $d_t{\clu A} = 0$. Now we discuss the analytical structure of the steady state solutions in the parameter space of the driven Duffing oscillator. For a $(q:p)$ USH bifurcation process, the steady states of the EOM \cref{eq:ush-eom-eff} satisfies
\begin{align}\label{eq:ush-eom-eff-expand}
\begin{split}
    (\Delta - \frac{i\gamma}{2})\clu A + 2K_2\clu A^*\clu A^2 +\mathcal O(g_4^2)= - q\Omega_{q,p}\xi^p\mathcal A^{*q-1},
\end{split}
\end{align}
where we have plugged the effective Hamiltonian of the general form  
\begin{align}\label{eq:ush-K-qp}
\cl K = \sum_n K_n\clu A^{*n}\clu A^{n} + \Omega_{p,q}\xi^{*p}\clu A^q + \mathrm{c.c.},
\end{align}
and the renormalized detuning $\Delta = \delta + 6g_4|\xi|^2 + 4g_3^2|\xi|^2/(\nu-2)-4g_3^2|\xi|^2/(\nu+2) + 8g_3^2|\xi|^2 +\mathcal O(g_4^2)$ and the second order renormalization coefficient $K_2 = \frac{3g_4}{2} - \frac{10g_3^2}{3} +\mathcal O(g_4^2)$ are computed through diagrams. 

To find the steady state $\clu A =  A_\text{s}$ satisfying \cref{eq:ush-eom-eff-expand}, we further express the equation in the polar coordinate $ A_\text{s} = r e^{i\theta}$ and it reads
\begin{align}\label{eq:ush-eom-r}
     2K_2r^3  + \Delta r - \frac{i\gamma}{2}r + \mathcal{O}(g_4^2) =- q\Omega_{q,p}\xi^pr^{q-1}e^{-iq\theta}.
\end{align}
This equation always admits a trivial solution $r=0$ corresponding to the harmonic response of the oscillator to the drive. We note that the phase factor in the right-hand side of \cref{eq:ush-eom-r} satisfies $e^{-iq\theta} = e^{-iq(\theta +2k\pi/q)}$ for $k\in\mathbb Z$. Therefore, each nontrivial solution in $r\ne0$ has a $q$-fold multiplicity in phase $\theta$; these nontrivial solutions, if they exist, represent the $q$-fold degenerate bifurcated steady states that are equispaced by an angle $2\pi/q$.

To solve for these nontrivial solutions, we further multiply \cref{eq:ush-eom-r} with its complex-conjugate and introduce $\rho = r^2$ to arrive at
\begin{align}\label{eq:ush-eom-rho}
    (2K_2\rho + \Delta)^2 + \frac{\gamma^2}{4} +  \mathcal{O}(g_4^2) = q^2\Omega_{q,p}^2|\xi|^{2p}\rho^{q-2}.
\end{align}
Geometrically, solving \cref{eq:ush-eom-rho} is to find the intersection between two curves $L$ and $R$, corresponding to the left- and right-hand side of the equation, with $\rho>0$. To leading order, $L = (2K_2 \rho + \Delta )^2 + \gamma^2/4$ represents a parabola and $R = q^2\Omega_{q,p}^2 |\xi|^{2p} \rho^{q-2}$ represents a $q-2$ degree curve. In the top row of \cref{fig:ush-phase}, we illustrate this geometric picture for the representational cases of $(1 : 1), (2 : 1), (3 : 1), (4 : 2)$ USH processes\footnote{We note that the $(1:1)$ instability, which is well known as Duffing bifurcation \cite{nayfeh1995,vijay2009}, is analyzed using a modified USH method that we review in \cref{app:duffing}. Yet it follows the same structure as \cref{eq:ush-eom-rho} and for the sake of completeness we compare it with other USH processes together.}. They respectively correspond to the intersection of the parabola with a hyperbola $R \propto \rho^{-1}$, a constant line $R \propto \rho^{0}$, a straight line $R \propto \rho^{1}$, and a parabola $R \propto \rho^{2}$. The intersection between the two curves is controlled on three types of perturbative parameters: the damping rate $\gamma$, the nonlinearity strengths $g_3 = \mathcal O(g_4^\frac{1}{2})$ and $g_4$, and the bare detuning $\delta = 1- p\nu/q$. We now discuss this dependence in more detail.

First, the damping rate $\gamma$ determines the bottom of the parabola $L$ and the bifurcation strength $\Omega_{q,p} = \mathcal O(g_4^{\frac{p+q}{2}-1})$ determines the height of the curve $R$. As we are interested in finding the solution within the perturbative regime, i.e. $\rho = \mathcal O(g_4^0)$ (c.f. \cref{eq:pert-condition}),  $\gamma$ needs to be of the same order as or smaller than the bifurcation strength $\Omega_{q,p}$. This condition\footnote{For $(1:1)$ process this condition is $g_4\gtrsim \gamma^3$, which we derive in \cref{app:duffing}.} translates to $g_4 \gtrsim \gamma^{2/(p+q-2)}$ for $p+q\ne 2$, which brings the two different perturbative parameters $\gamma$ and $g_4$ on equal footing. 

In addition, the detuning $\delta$ controls the center of the parabola $R$ and, when the damping is sufficiently weak, it determines the number of nontrivial solutions in $\rho$. Different solutions, based on their stability properties, can be classified as either nodes or saddles (c.f. \cref{fig:ush-ss}). With this, in the bottom row of \cref{fig:ush-phase} we plot the domain diagram for different USH processes in the parameter space $(\delta, g_4)$ while each axis is rescaled by $\gamma$ to some pertinent order. Different colors represent different configuration of solutions and the insets give the representational phase diagram of each of these regimes.

We remark that the domain diagrams of some lower-order USH processes, such as the $(1 : 1)$ and the $(2 : 1)$ USH process have been computed in previous works \cite{wustmann2019}. However, the perturbative structure among the USH processes and an accurate domain diagram taking into account all the frequency-renormalizing terms for higher order processes has not been constructed in previous literature to the best of our knowledge. This is partially because with increasing $q, p$ the complexity of the analysis increases exponentially. Moreover, historically there was no practical incentive to construct domain diagrams for higher-order USH processes: the regime of weak-damping was practically inaccessible. But this has changed with the advent of technologies that can simultaneously access the parameter regimes of weak damping and ﬁnite nonlinearity, for instance with the superconducting Josephson junction element. Here, we have constructed the domain diagrams of some novel USH processes in  \cref{fig:ush-phase} and for processes higher than the $(4 : 2)$ process, with the help of the diagrammatic method, one can use the computer programs to analytically or numerically compute the domain diagrams.

\subsubsection{Unification of USH bifurcation, multiphoton resonance, and Schr\"odinger cat states in a semiclassical picture}

Lastly, we remark that the effective Hamiltonian \cref{eq:ush-K-qp} governing classical USH bifurcations has the same form as \cref{eq:mnr-H}, from which two drastically different quantum processes could emerge: Schr\"odinger cat state and multiphoton resonance. In the rest of this section we discuss the unification of these two processes in a semiclassical picture. 

For illustrative purpose, we consider a simple example of a $(2:1)$ process taking place in a driven quantum nonlinear oscillator described by \cref{eq:nl-osc-H}. With the diagrammatic method, the $(2:1)$ process to order $\mathcal{O}(\varphi_\text{zps}^2)$ can be described by a quantum effective Hamiltonian 
\begin{align} \label{eq:ush-K-21}
    \op K/\hbar = (\Delta_\text{cl} + 2K_2)\op{\cu A}^\dagger \op{\cu A} +K_2\op{\cu A}^{\dagger2} \op{\cu A}^2+ \Omega_\text{sqz}\op{\cu A}^{\dagger 2}+\mathrm{h.c.},
\end{align}
where $\Delta_\text{cl} = \omega_o-\omega_d/2 + 6g_4|\xi|^2-18g_3^2|\xi|^2/\omega_d$ is the renormalized detuning due to the classical diagrams, $K_2 = 3g_4/2 - 20g_3^2/3\omega_d$, $\Omega_\text{sqz} = \frac{2i\Omega_dg_3}{\omega_d}$ is the squeezing strength, and the $2K_2$ term in the above equation is the Lamb shift resulting from the quantum diagrams (c.f. \cref{eq:ush-K1-diagram}). We note that \cref{eq:ush-K-21} is just \cref{eq:Kerr-cat-RWA} with beyond RWA terms explicitly computed.  

\begin{figure}
	\includegraphics{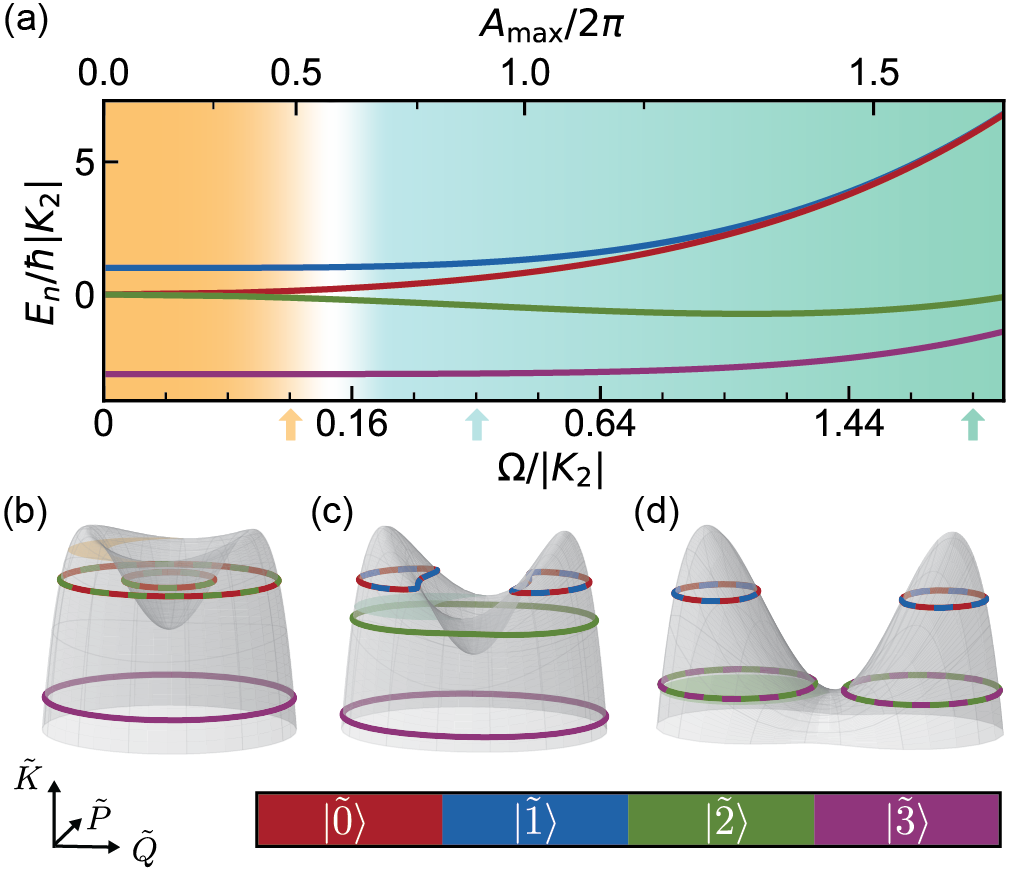}
	\caption{Demonstration of the $(2:1)$ process with two quantum manifestations: multiphoton resonance and Schr\"odinger cat states. (a) Eigenenergy spectrum of \cref{eq:ush-K-21} with $\Delta_\text{cl} = -3K_2$ as a function of reduced squeezing strength $|\Omega_\text{sqz}|/K_2$ in square root scale. Eigenstate $|\tilde n\rangle$ is indexed by the Fock state $|n\rangle$ that it adiabatically connects to at $\Omega_\text{sqz} = 0$. The top $x$-axis measures the number of action quanta enclosed by each metapotential well away from the origin, as illustrated in (b) - (d). Resonance and Cat domains, colored as yellow and turquoise, correspond to regions that live in the two quantum manifestations. (b) - (d) semiclassical configurations of quantum states at $|\Omega_\text{sqz}/K_2| = 0.09, 0.36, 1.96$. The metapotential surface is given by \cref{eq:ush-Kcl-21} with states shown as quantized orbits enclosing $k+1/2$ action quanta (with $k\in \mathbb N$). Shaded areas represent $A_\text{max}$. }
	\label{fig:ush-mnr2cat}
\end{figure}

In \cref{fig:ush-mnr2cat} (a), we plot the eigenspectrum of $\op K$ with parameters $K_2 < 0$ (to resemble the negative anharmonicity of a usual Josephson circuit), $\Delta_\text{cl} = -3K_2$ and $\Omega_\text{sqz}$ varying. The parameters are chosen so that the Fock states $|0\rangle$ and $|2\rangle$ are degenerate eigenstates of $\op K$ at $\Omega_\text{sqz} =0$. With moderately non-zero $\Omega_\text{sqz}$,  $|0\rangle$ and $|2\rangle$ are thus resonantly coupled by the squeezing term $\Omega_\text{sqz}\op{\cu A}^{\dagger 2} + \mathrm{h.c.}$, a process we refer as the multiphoton resonance and has been extensively examined in \cref{sec:mpnr}. In \cref{fig:ush-mnr2cat}~(a) this resonance is characterized by the energy repulsion between the eigenstates $|\tilde0\rangle, |\tilde2\rangle\approx|0\rangle\pm |2\rangle$ in the yellow-color resonance domain and the energy gap corresponds to twice the Rabi rate between the two Fock states. When $\Omega_\text{sqz}$ becomes larger, \cref{eq:ush-K-21} enters the Schr\"odinger cat domain and admits a two-fold near-degenerate ground state comprising Schr\"odinger cat-like states (c.f. \cref{fig:cat-wigner} (a) for a $\Delta_\text{cl}\approx -2K_2$ case). This is characterized by the ``pairwise kissing'' \cite{frattini2022} in the spectrum  \cref{fig:ush-mnr2cat} (a), specifically the closing of the eigenenergies of $|\tilde 0\rangle, |\tilde 1\rangle$ and those of $|\tilde 2\rangle, |\tilde 3\rangle$ in the tourquoise-color regime.

To shed light on the continuous transition between resonance and cat domains, we introduce a semicalssical object, \textit{metapotential}, which is the energy surface defined by
\begin{align} \label{eq:ush-Kcl-21}
    \cl K/\hbar = \Delta_\text{cl}\cl{\cu A}^* \cl{\cu A} +K_2\cl{\cu A}^{* 2} \cl{\cu A}^2+ \Omega_\text{sqz}\op{\cu A}^{* 2}+\mathrm{c.c.}.
\end{align}
Note that \cref{eq:ush-Kcl-21} is just the classical counterpart of \cref{eq:ush-K-21}; in the diagrammatic representation, the former comprises of the subset of classical diagrams that constituting the latter.

In \cref{fig:ush-mnr2cat} (b) - (d), we plot the metapotential \cref{eq:ush-Kcl-21} in the phase space $(\cl Q, \cl P)$, where $\cl Q, \cl P$ are the reduced position and momentum coordinates and related to the complex coordinate by $\clu A = (\cl Q+i\cl P)/\sqrt2$. Three particular values of $\Omega_\text{sqz}$ are chosen with each corresponding to the system in the resonance regime, early in the cat regime, and deep in the cat regime. An equi-energy line in the metapotential corresponds to an orbit of the classical state and a local extreme, or node, corresponds to a stationary state of the classical nonlinear oscillator governed by \cref{eq:ush-Kcl-21}. All three metapotentials in \cref{fig:ush-mnr2cat} contain two nodes away from the origin, each sitting at the top of a hill whose footprint grows as $\Omega_\text{sqz}$ increasing. We denote the area of the footprint by $A_\text{max}$ and it measures the maximum action enclosed by each hill. In  \cref{fig:ush-mnr2cat} (b) and (c), the origin is another node sitting at the bottom of a valley, while in (d) the origin becomes a saddle point. One can immediately identify that the metapotential in (b) and (c) corresponds to the orange domain of the $(2:1)$ bifurcation domain diagram \cref{fig:ush-phase} (b) where there are three nodes and two saddles while metapotential in \cref{fig:ush-mnr2cat} (d) corresponds to the yellow domain that contains two nodes and one saddle.

In the semiclassical picture, a metapotential can accommodate quantized orbits satisfying the Einstein–Brillouin–Keller (EBK) quantization condition ---  the action function enclosed by a quantize orbit is half-integer action quanta, for any integer $n$. Written mathematically, this well-known quantization condition reads
\begin{align}
    \oint d\cl P \cl Q = (n+\frac{1}{2})2\pi,
\end{align}
where $2\pi$ is the action quantum\footnote{For an unscaled phase space with the canonical commutation relation $[\op q,\op p]=i\hbar$, the action quantum is the Planck constant $h$. For the phase space with the canonical commutation relation $[\op Q,\op P]=i$ the reduced action quantum is $h/\hbar = 2\pi$.} in the phase space $(\cl Q, \cl P)$. With this, the resonance and cat domains in \cref{fig:ush-mnr2cat} become clear. In particular, subplot (b) exemplifies the resonance domain with $A_\text{max}/2\pi < 0.5$, where $|0\rangle,|2\rangle$ are resonantly coupled by $\Omega_\text{sqz}\op{\cu A}^{\dagger 2} + \mathrm{h.c.}$. Semiclassically, eigenstates $|\tilde0\rangle, |\tilde2\rangle\approx |0\rangle\pm|2\rangle$ are superpositions of two orbits that enclose 0.5 and 2.5 action quanta, which live in the inside and outside metapotential wells respectively and correspond to the Fock state $|0\rangle, |2\rangle$. \cref{fig:ush-mnr2cat} (c) exemplifies the emergence of cat domain with $0.5<A_\text{max}<1.5$ such that each metapotential hill away from the origin is big enough to host one quantized orbit. Eigenstates $|\tilde0\rangle,|\tilde1\rangle$ are Schr\"odinger cat-like states constituted by the two orbits living equidistant to the origin with opposite phase. With $\Omega_\text{sqz}$ increasing, the energy barrier separated the two orbits increases and suppresses the tunneling between the two orbits as characterized by the closing of energy gap between the two. Lastly, the subplot (d)  exemplifies a region deep in the cat domain with $A_\text{max}>1.5$ so that more than one orbits live in each metapotential hill away from the origin and thus exist multiple pairs of cat-like states.

Remarkably, here we have identified a shared object, the metapotential, that governs both the classical USH bifurcation and the quantum processes, i.e. the multiphoton resonance and stabilization of general Schr\"odinger cat-like states, in driven nonlinear oscillators. The shape of metapotential determines the configurations of classical stationary states (or steady states in presence of damping) and is characterized by the domain diagrams in \cref{fig:ush-phase}. In the quantum regime, the characteristic area $A_\text{max}$ of the metapotential relative to the action quantum further determines configuration of the quantized orbits and thus the quantum processes. The metapotential, together with the diagrammatic method transcending classical and quantum regimes, unifies these nonlinear dynamical processes that were seemingly unrelated. Finally, the semiclassical picture in \cref{fig:ush-mnr2cat} has been successful in explaining the squeeze-driven Kerr oscillator spectrum and the well-flip lifetime of the Kerr-cat qubit measured in recent experiments \cite{frattini2022,venkatraman2022}, which showcases its potential in developing future qubit encodings.

\section{Conclusion}\label{sec:conclusion}

In this work, we have developed a novel diagrammatic method to compute the effective Hamiltonian of a nonlinear oscillator driven by an oscillatory force. The core principle of our method involves constructing perturbatively a canonical frame transformation that decouples the near-resonant dynamics of the oscillator from the off-resonant one. In a rotating frame at a frequency near-resonant with the oscillator, the effective Hamiltonian is then captured through mixing products of zero net frequency. The cascade of these mixing products is represented by a diagram, a special graph where the internal vertices represent frequency-mixing factors, the external edges correspond to dressed resonant excitations or drive excitations, and the internal edges correspond to dressed propagators associated with off-resonant excitations. Similar to Feynman diagrams, mixing diagrams are evaluated by multiplying the algebraic expressions associated with each component in the diagram. The diagram's pictorial structure serves to organize the associated Hamiltonian term according to perturbative order and the types of excitations involved. This bookkeeping feature allows for the computation of specific types of effective Hamiltonian terms directly at arbitrary orders. This feature is also parallel computation friendly. 

At the foundation of these diagrams lies a novel self-consistent perturbation expansion which we called Quantum Harmonic Balance (QHB). This expansion is based on the phase-space formulation of quantum mechanics, a choice justified by the natural connection between diagram counting rules and the Husimi product, which involves taking partial derivative of phase-space variables. Additionally, the phase-space formulation establishes a correspondence between the space of quantum Hilbert space operators and the space of the classical phase-space functions, enabling the treatment of quantum and classical nonlinear dynamics on equal footing. In developing QHB, we have also devised a new procedure to ensure the canonicity of the frame transformation underlying the diagrammatic method. This rigorous procedure is applicable to the family of averaging methods whose canonicity is often not ensured, further enhancing its utility.

In the field of quantum computing, our diagrammatic method provides three valuable features. First, it facilitates the analytical description of effective dynamics for driven nonlinear oscillators, which model numerous quantum computing devices and their control schemes, arbitrarily to high order. Second, it captures and reveals the structure of nonlinear processes through the intricate design of the diagrams. Lastly, it treats quantum and classical driven processes on equal footing, allowing knowledge from classical nonlinear dynamics to be applied to quantum systems. These features are essential for characterizing beyond-RWA phenomena that constrain many state-of-the-art technologies, for achieving precise control of quantum systems via parametric drives, and for designing innovative qubit encoding, readout, and control schemes through Floquet engineering. To demonstrate the power of our diagrammatic method, we have explored several pertinent examples in the superconducting circuits platform, such as a proposal to realize a three-legged Schrödinger cat state, modeling energy-renormalization effects in a recent superconducting experiment in the strongly driven regime, a comprehensive characterization of multiphoton resonances in a driven transmon, a proposal for an innovative inductively shunted transmon circuit, and a characterization of classical ultra-subharmonic bifurcation in driven oscillators. Our method can be applied and extended to various other relevant systems, including those involving multiple drive tones and quantum modes, open quantum systems coupled to a dissipative bath, or superconducting circuits driven through magnetic flux. In summary, the results presented in this article provide the groundwork for a systematic perturbation theory of the control of non-linear bosonic quantum systems through an oscillating force of variable frequency and strength. 

\section*{Acknowledgements}\label{sec:acknowledgement}

XX and JV thank Yaxing Zhang for the tutoring on the Schrieffer-Wolff expansion and on his pioneering analysis of multiphoton resonance using effective Hamiltonian theory during the initial stages of this work. RGC thanks Brice Ravon for bringing to his attention Groenewold’s theorem. The authors thank Steven M. Girvin, Alexandru Petrescu, Leonid Glazeman and Shruti Puri for the valuable discussions and Alec Eickbusch for the useful discussion on the experimental data presented in \cref{sec:renormalization}. This research was sponsored by the Air Force Office of Scientific Research under award number FA9550-19-1-0399, and by the National Science Foundation (NSF) under award numbers 1941583 and 2124511. The views and conclusions contained in this document belong to the authors and should not be interpreted as representing the official policies, either expressed or implied, of the grant agencies, or the U.S. Government. The U.S. Government is authorized to reproduce and distribute reprints for Government purposes notwithstanding any copyright notation herein.

\apptoc
\begin{appendices}

\appendix

\section{Glossary}\label{app:gloss}
\centerline{\underline{Graph theory}}
\;\\[3pt]
\noindent\textbf{loop}: a closed sequence of connected vertices that starts and ends at the same vertex without revisiting any vertex in the sequence.\\[5pt]
\noindent\textbf{tree}: an undirected graph in which any two vertices are connected by exactly one path; i.e. a tree has no loop. \\[5pt]
\noindent\textbf{leaf vertex}: a vertex with degree one, i.e. only connecting to one other vertex.\\[5pt]
\noindent\textbf{external edge}: an edge connecting to a leaf vertex.\\[5pt]
\noindent\textbf{internal edge}: an edge not connecting to any leaf vertex.\\[5pt]
\noindent\textbf{rooted tree}: a tree in which one leaf vertex has been designated the root.\\[5pt]
\noindent\textbf{forest}: an undirected graph in which any two vertices are connected by at most one path; in other words, the graph consists of a disjoint union of trees.\\[3pt]

\centerline{\underline{Diagrammatic method}}
\;\\[3pt]
\noindent\textbf{diagram}: a special graph in which vertices correspond to interactions, internal edges correspond to propagators, and external edges correspond to excitations.\\[5pt]
\noindent\textbf{resonant excitation}: an excitation at the oscillator frequency represented by a straight edge. \\[5pt] 
\noindent\textbf{drive excitation}: an excitation at the drive frequency represented by a wiggly edge. \\[5pt] 
\noindent\textbf{off-resonant excitation}: an excitation off resonant from the oscillator frequency represented by a diagram whose output is a double-line edge. \\[5pt] 
\noindent\textbf{bare diagram}: a diagram in which the excitations and propagators are in the bare form. \\[5pt]
\noindent\textbf{dressed diagram}: a diagram in which the excitations and propagators are in the dressed form.\\[5pt]
\noindent\textbf{unexpanded diagram}: a diagram in which each internal vertex is a \tikzpic{-11}{figsv6_small/notation/star_small.pdf} and the dressed excitations involve \tikzpic{-11}{figsv6_small/notation/husimi_small.pdf}. The operations underlying the algebraic expressions associated with unexpanded diagrams are Husimi product $\varstar$ and Husimi bracket $\moyal{\tikzpic{-9}{figsv6_small/notation/dash_box.pdf}\,}{\tikzpic{-9}{figsv6_small/notation/dash_box.pdf}}$. See \cref{eq:order-1-sum,eq:order2-term1,eq:eta-dressed-3} for examples.  One could say that the unexpanded diagram has dressed vertices. \\[5pt]
\noindent\textbf{expanded diagram}: a fully evaluated representation of the unexpanded diagram in which  \tikzpic{-11}{figsv6_small/notation/star_small.pdf} and \tikzpic{-11}{figsv6_small/notation/husimi_small.pdf} are algebraically evaluated and represented in diagrammatic form. See \cref{eq:order2-term1-unabridged,eq:order2-term1-unabridged,eq:eta-dressing_unabridged} for examples. The product underlying the algebraic expressions associated with expanded diagrams is the ordinary commutative product. \\[5pt]
\noindent\textbf{quantum bond}: an edge in an expanded diagram linking two bare resonant excitation. The algebraic expression associated with a quantum bond is $\hbar$. A quantum bond is colored in orange.\\[5pt] 
\noindent\textbf{classical diagram}: an expanded diagram in which there is no quantum bond.\\[5pt]
\noindent\textbf{quantum diagram}: an expanded diagram in which there is one or more quantum bonds.\\[5pt]
\noindent\textbf{ordered diagram}: a diagram where the subdiagrams of each vertex has a definite order around that vertex.\\[5pt]
\noindent\textbf{unordered diagram}: the equivalence class of topologically equivalent ordered diagrams.\\[5pt]

\section{List of Symbols}

\renewcommand*{\arraystretch}{1.4}
\begin{longtable}[h]{@{\hskip7pt}p{0.7in}p{2.5in}}
 \multicolumn{2}{c}{\normalsize{\underline{Driven nonlinear oscillator}}}\\[3pt]
$\omega_o$ & natural oscillator frequency\\
$\omega_{01}$& transition frequency between ground and first excited state\\
 $g_m$ & $m$-th rank nonlinearity\\
    $\omega_d$ & drive frequency\\
    $\Omega_d$ & drive amplitude\\
    $\xi$ & dimensionless drive amplitude\\
    $\omega_o^\prime$ & frequency of the rotating frame\\
    $\delta$ & bare detuning ($=\omega_o-\omega_o^\prime$)\\
    $\op q$ & position operator\\
    $\op p$ & momentum operator\\
    $q_\text{zps}, p_\text{zps}$ & zero-point spread of the position and momentum coordinates\\
    $\op U$ & potential operator\\
    $c_m$ & $m$-th rank coefficient of the potential $\op U$\\
    $\op a, \op a^\dagger$ & bosonic operators\\
    $\op{\mathcal H}$ & Hamiltonian in lab frame\\[5pt]
     \multicolumn{2}{c}{\normalsize{\underline{Phase space representation of quantum mechanics}}}\\[3pt]
     $\mathfrak H$&Husimi Q map\\
    $\cl O$ & phase space variable/function corresponding
to some operator $\op O$ in Hilbert space \\
    $\varstar$ &  Huisimi Q product\\
    $\star$ & Groenewold/Moyal product \\
    $\cl f\cl g\;$ & ordinary product between $\cl f$ and $\cl g$\\
    $\moyal{\,}{}$ & Husimi bracket, based on the Husimi Q product\\
    $\{\,,\}$ & Poisson bracket\\
    $(\cl f+ \cl g)^m_\varstar$ & polynomial expansion of $\cl f$ and $\cl g$ with $\varstar$ as the underlying product\\[5pt]
     \multicolumn{2}{c}{\normalsize{\underline{Quantum harmonic balance in phase space representation}}}\\[3pt]
$\clu A, \clu A^*$ &  bosonic variables in the transformed frame\\
$\cl\mu$ & off-resonant part of the micromotion captured by additive map ($\cl a = \clu A + \cl \sigma + \cl\mu$) \\
$\cl\sigma$ & resonant part of the micromotion captured by additive map ($\cl a = \clu A + \cl \sigma + \cl\mu$) \\
$\clu B, \clu B^*$ &  dressed bosonic variables in the transformed frame ($ = \clu A+\cl\sigma$)\\
$\cl K$ & effective Hamiltonian\\
$\cl K_{\omega_o^\prime}$ & effective Hamiltonian in the rotating frame at frequency $\omega_o^\prime$\\
$\omega_\text{out}$ &  output frequency of a diagram producing off-resonant excitation\\
$\tilde\omega_\text{out}$ &  renormalized output frequency of a diagram producing off-resonant excitation\\
$\Delta$ & detuning between the oscillator renormalized frequency and the natural oscillator frequency\\
$\Delta_\text{cl}$ &  classical contribution of $\Delta$\\
$\Delta_\omega$ &  detuning between the off-resonant excitation at frequency $\omega$ and the resonant excitation\\
$(q:p)$ & characterizes a process involving conversion between $p$ drive excitations and $q$ resonant excitations\\
$K_n$ &  $n$-th order coefficient of Kerr-like term in the effective Hamiltonian\\
$\Omega_{q,p}$ & coupling strength in a $(q:p)$ process\\
$\Omega_{q,p}^{(k)}$ & coupling strength in a $(q:p)$ process under perturbative expansion to $k$-th order\\
$\Omega_\text{sqz}$ & squeezing strength ($ = \Omega_{2,1}$)\\
$A_\text{max}$ & maximum action enclosed by an energy well in the semiclasscial metapotential \\
$Z$ & a general variable in the additive map of frame transformation\\
$\cu Z, \zeta$ & slow and fast dynamics of the variable $Z$\\
$\op S$ & generator of an exponential map of frame transformation\\
$L_{\cl S} = \moyal{\cl S}{\cdot}$ & Lie derivative with respective to phase-space function $\cl S$\\[5pt]
     \multicolumn{2}{c}{\normalsize{\underline{Multi-tone, multi-mode system}}}\\[3pt]
$\omega_k$ & natural frequency of mode $k$\\
$\omega_k^\prime$ & frequency of rotating frame for mode $k$\\
$\delta_k$ & bare detuning for mode $k$ ($=\omega_k-\omega_k^\prime$)\\
$\lambda_k$ & phase participation ratio of $k$ mode $k$\\
$p_k$ & energy participation ratio of mode $k$\\
$\xi_l$ & effective drive amplitude of drive tone $l$\\
$\omega_{d,l}$ & frequency of drive tone $l$\\
$\varphi_k$ & zero-point spread of phase contributed by mode $k$\\
$\op c, \op c^\dagger$ & bosonic operators of cavity-like mode\\
$\op{\cu C}, \op{\cu C}$ & bosonic operators of cavity-like mode in the transformed frame\\
$K_{m,n}$ & energy renormalization coefficient of the $\op{\cu A}^n\op{\cu A}^{\dagger n}\op{\cu C}^m\op{\cu C}^{\dagger m}$ term in a 2-mode problem\\
$\chi$ & cross-Kerr between qubit-like and cavity-like mode in a 2-mode problem\\
$\alpha$ & coherent state complex amplitude in \cref{sec:renormalization}\\
$\bar{n}$ & coherent state photon number $(=|\alpha|^2)$\\[5pt]
     \multicolumn{2}{c}{\normalsize{\underline{Dissipative dynamics}}}\\[3pt]
$\op{\mathcal H}_\text{tot}$ & Hamiltonian of the driven oscillator and the thermal bath\\
$\op{\mathcal H}_\text{s}$ & Hamiltonian of the driven oscillator \\
$\op{\mathcal H}_\text{b}$ & Hamiltonian of the thermal bath\\
$\op{\mathcal H}_\text{sb}$ & Hamiltonian of the oscillator-bath coupling\\
$\op b_\omega, \op b_\omega^\dagger$ & bosonic operators for frequency-$\omega$ bath mode\\
$\op \rho_s$ & density matrix operator of the oscillator\\
$\op K_\text{tot}$ & effective Hamiltonian of the oscillator and thermal bath\\
$\op K_\text{s}$ & effective Hamiltonian of the oscillator\\
$\op K_\text{sb}$ & effective oscillator-bath coupling\\
$h_\omega$ & oscillator-bath coupling strength\\
$\op C(t)$ & effective oscillator coupling operator\\
$\op C_{\omega_j}$ & Fourier component of $\op C(t)$ at frequency $\omega_j$\\
$\bar n_{\omega_j}$ & thermal photon number at $\omega_j$\\
$\mathcal D [\op O]$ & Lindblad superoperator over loss operator $\op O$\\[5pt]
 \multicolumn{2}{c}{\normalsize{\underline{Josephson circuits}}}\\[3pt]
$E_C$ & charging energy\\
$E_J$ & Josephson energy\\
$E_L$ & inductive energy\\
$N_g$ & gate offset charge\\
$\varphi_\text{ext}$ & reduced external flux\\
$E_J^\prime$ & Josephson energy of each junction in
an array shunt\\
$M$ & junction number in an array shunt\\
$R$ & ratio between $E_L$ and $E_J$\\
$\alpha$ & anharmonicity in \cref{sec:mpnr,sec:IST}\\
$\op N$ & Cooper pair number operator\\
$\op \varphi$ & phase operator\\
$N_\text{zps}, \varphi_\text{zps}$ & zero-point spread of $\op N$ and $\op\varphi$\\
$E_d$ & drive energy\\
$\op U_\text{IST}$ & potential operator of an IST circuit
\\[5pt]
     \multicolumn{2}{c}{\normalsize{\underline{Floquet analysis of multiphoton resonance}}}\\[3pt]
$|i\rangle$ & $i$-th Fock state\\
$\mathcal{E}_{i}$ & energy of $i$-th Fock state dressed by the renormalization term\\
$\epsilon_i$ & eigenenergy of $i$-th state in the effective Hamiltonian\\
$|\tilde{i}\rangle$ & the eigenstate of the effective Hamiltonian that is associated with the $i$-th Fock state\\
$\tilde{E}_i$ & renormalized energy of $i$-th driven state in lab frame\\
$|\tilde{i}^\text{F}\rangle$ & Floquet state associated with the  $i$-th Fock state\\
$\epsilon_i^\text{F}$ & quasienergy of $i$-th Floquet state\\
$\Omega_{k\leftrightarrow k+q, p}$ & Hamiltonian coupling strength between $k$-th and $k+q$-th states mediated by $q$ drive photons\\
$\Omega^R_{k\leftrightarrow k+q, p}$ & Rabi rate between $k$-th and $k+q$-th states mediated by $q$ drive photons\\
$\Omega_\text{para}$ & strength of some parametric process\\
$P_{i\rightarrow}$ & weight of leaving $i$-th Fock state\\[5pt]
     \multicolumn{2}{c}{\normalsize{\underline{Classical Duffing oscillator}}}\\[3pt]
$\nu$ & dimensionless drive frequency\\
$\gamma$ & damping rate\\
$x_\text{lin}$ & linear response of the position coordinate to the drive\\
$x_\text{s}$ & steady state of position coordinate $\cl x$\\
$\dot{x}_\text{s}$ & steady state of the velocity $d_t\cl x$\\
$A_\text{s}$ & steady state of complex coordinate $\clu A$\\
$(r,\theta)$ & polar coordinate of steady state in phase space\\
$\rho = r^2$ & square of the amplitude of the steady state coordinate\\
$L, R$ & left- and right-hand side of the steady state equation
\end{longtable}

\section{Frame Transformation}\label{app:frame}

To prepare the Hamiltonian \cref{eq:op-H-a-ad} for the perturbative analysis, we perform two-step frame transformations. The first one is a displacement transformation generated by  $\op U_D = \exp(\alpha^*_\text{lin}\op a - \alpha_\text{lin}\op a^\dagger)$ where
\begin{align}\label{app:eq:a-lin}
    \alpha_\text{lin}(t) = \frac{\Omega_d}{2}\Big(\frac{1}{\omega_d-\omega_o}e^{-i\omega_dt} - \frac{1}{\omega_d+\omega_o}e^{i\omega_dt}\Big)
\end{align}
is the linear response of the oscillator to the drive. Under this transformation, the Hamiltonian transforms to 
\begin{align}
\op H^\prime &= \op U_D\op H\op U_D^\dagger -i\hbar \op U_D \partial_t\op U_D^\dagger\\
&=\omega_o \op a^\dagger\op a +  \sum_{\substack{m \ge 3 \\ m \in \mathbb{N}}} \frac{g_m}{m} (\hat{a} + \hat{a}^{\dagger} + \xi e^{-i \omega_d t} +  \xi^* e^{i \omega_d t} )^m,\label{app:frame-H-D}
\end{align}
where $\xi = \frac{\Omega_d\omega_a}{\omega_o^2-\omega_d^2}$. If we focus on the nonlinear terms in \cref{app:frame-H-D}, this transformation, which is defined as $\op a \rightarrow \op a + \alpha_\text{lin}(t)$, amounts to $\op a \rightarrow \op a + \xi e^{-i\omega_d t}$. Note that, in the case that the oscillator is driven through momentum degree of freedom by $-i\Omega_d(\op a-\op a^\dagger)\cos\omega_dt$, such as in \cref{eq:mnr-transmon-H-bosonic}, the linear response should be modified as 
\begin{align}\label{app:eq:a-lin-mom}
    \alpha_\text{lin}(t) = \frac{i\Omega_d}{2}\Big(\frac{1}{\omega_d-\omega_o}e^{-i\omega_dt} + \frac{1}{\omega_d+\omega_o}e^{i\omega_dt}\Big),
\end{align}
and the Hamiltonian in the displaced frame has the same form as \cref{app:frame-H-D} but with modified $\xi = \frac{i\Omega_d\omega_d}{\omega_d^2-\omega_o^2}$. In addition to the displaced frame transformation, we further perform a rotating frame transformation to \cref{app:frame-H-D} generated by $\op U_R = \exp(-i\omega_o^\prime \op a^\dagger \op a)$ and the resulting Hamiltonian is \cref{eq:H-tran}. 

We also note that the displacement transformation can alternatively be absorbed into the quantum harmonic balance (QHB) procedure developed in \cref{sec:averaging}. Specifically, we consider the Hamiltonian \cref{eq:op-H-a-ad} under the rotating frame transformation generated by $\op U_R$: 
\begin{align}\label{app:H-rotate}
\begin{split}
    \op H^{\prime\prime} = &\delta\op a^\dagger \op a+  \sum_{\substack{m \ge 3 \\ m \in \mathbb{N}}} \frac{g_m}{m} (\hat{a} e^{- i \omega_o^\prime t} + \hat{a}^{\dagger} e^{i \omega_o^\prime t})^m\\
    &+ \Omega_d  (\hat{a} e^{- i \omega_o^\prime t} + \hat{a}^{\dagger} e^{i \omega_o^\prime t})\cos(\omega_d t),
\end{split}
\end{align}
where $\delta = \omega_o - \omega_o^\prime$. Instead of starting with \cref{eq:H-tran}, we can perform  QHB over \cref{app:H-rotate} directly. In particular, in the Husimi Q phase space, we seek for a frame transformation $\cl a = \clu A + \cl\sigma+\cl\mu$ and the equation of motion $d_t\cl a = -\moyal{\cl a}{\cl H^{\prime\prime}}$ correspondingly reads
\begin{align}\label{app:eq:eom-phase-expand}
\begin{split}
     \partial_{\clu A^*}\cl K  + i\partial_t \cl\mu = &\sum_{\substack{m \ge 3}} g_me^{i\omega_ot}  \Big((\cl{\cu A}+\cl\sigma+\cl\mu) e^{- i \omega_o t} \\[-6pt]
     &\qquad\qquad+ (\cl{\cu A}^* + \cl\sigma^*+ \cl\mu^*) e^{i \omega_o t}\Big)^{m-1}_\varstar  \\
     & + e^{i\omega_o^\prime t}\frac{\Omega_d}{2}( e^{-i\omega_d t} +  e^{i\omega_d t})\\
     &+\delta(\clu A + \cl\sigma+\cl\mu) +i\moyal{\cl K}{ \cl\sigma+\cl\mu}_{\cl{\cu A}, \cl{\cu A}^*}.
\end{split}
\end{align}
The QHB is to find $\partial_{\clu A^*}\cl K, \cl\sigma$ and $\cl\mu$ order by order in the perturbative parameter $q_\text{zps}$. Here we assume $\cl\mu$ to be nonzero at order $q_\text{zps}^0$, in stead of being zero in \cref{eq:eom-phase-expand}, and assign order $q_\text{zps}^0$ to $\Omega_d$. As a result, \cref{app:eq:eom-phase-expand} to leading order reads
\begin{align}
\partial_{\clu A^*}\cl K^{(0)}  + i\partial_t \cl\mu^{(0)} =  e^{i\omega_o^\prime t}\frac{\Omega_d}{2}( e^{-i\omega_d t} +  e^{i\omega_d t}).
\end{align}
By construction $\partial_{\clu A^*}\cl K$ is time-independent and $\cl\eta$ is time-dependent, and therefore at this order we have 
\begin{subequations}
    \begin{align}
    &\partial_{\clu A^*}\cl K = 0,\\
    &e^{-i\omega_o^\prime t}\cl\mu^{(0)} = \frac{\Omega_d/2}{\omega_d-\omega_o^\prime}e^{-i\omega_dt} + \frac{\Omega_d/2}{-\omega_d-\omega_o^\prime}e^{i\omega_dt}.\label{app:eq:eta}
\end{align}
\end{subequations}
We recall that, according to \cref{eq:sc-diagram}, the diagrammatic representation of $ e^{-i\omega_o^\prime t}\cl\mu$ is the off-resonant excitation \!\tikzpic{-10}{figsv6_small/notation/eta_small.pdf}. Therefore \cref{app:eq:eta} is just two off-resonant exictations at order $q_\text{zps}^0$: both are of strength $\Omega_d/2$, and each is at frequency $\pm\omega_d$ and with propagator $1/(\pm\omega_d - \omega_o^\prime)$, respectively. It should be noted that the linear response $\alpha_\text{lin}$ in \cref{app:eq:a-lin} is the just $e^{-i\omega_o^\prime t}\cl\eta^{(0)}$ but with dressed propagator $1/(\pm\omega_d - \omega_o)$ defined by \cref{eq:eta-dressed-3}. At next order, both of these two dressed off-resonant excitations can travel into or away from a mixer (c.f. \cref{eq:sc-diagram}). In this sense, $\xi e^{-i\omega_d t}$ in \cref{app:frame-H-D} is just the combination of the off-resonant excitation at $\omega_d$ traveling into the mixer and the one at $-\omega_d$ traveling away (whose algebraic expression is the conjugate of the second term in \cref{app:eq:eta}), which two have the same effect of adding frequency $\omega_d$ to the mixing process. 

\begin{widetext}
\section{Proof of \texorpdfstring{$\partial_{\cl a^*} (\cl a + \cl a^{*})_{\varstar}^m = m (\cl a + \cl a^{*})_{\varstar}^{m-1}$}{*}}\label{app:star-chain}
Our goal is to prove that chain rule applies to the expression $(\cl a + \cl a^{*})_{\varstar}^m$, i.e.
\begin{align}
    \partial_{\cl a^*} (\cl a + \cl a^{*})_{\varstar}^m = m (\cl a + \cl a^{*})_{\varstar}^{m-1}.
\end{align}
To show this, we first invoke the following identity:
\begin{align}
\label{eq:aad}
    (\alpha \cl a + \beta \cl a^{*} )^{m}_{\varstar} = \sum_{k = 0}^{m} \binom{m}{k} \alpha^{m-k} \beta^k \sum_{j = 0}^{\min{(k, m-k)}} \frac{1}{2^j}j!   \binom{k}{j} \binom{m-k}{j} \hbar^j   \cl a^{* k-j} \cl a^{m- k - j} 
\end{align}
From \cref{eq:aad}, we have
\begin{align}
\begin{split}
&\partial_{\cl a^*} (\cl a + \cl a^{*})_{\varstar}^m \\&= \sum_{k = 1}^{m} \frac{m!}{k! (m-k)!} \sum_{j = 0}^{\min{(k-1, m-k)}} \frac{j!}{2^j}   \frac{k!}{j! (k-j-1)!} \frac{(m-k)!}{(m-k-j)! j!} \hbar^j   \cl a^{* k-j-1} \cl a^{m- k - j}.
\end{split}
\end{align}
Redefining the index $k$ as $l = k -1$, we arrive at
\begin{align}
\begin{split}
&\partial_{\cl a^*} (\cl a + \cl a^{*})_{\varstar}^m \\&= m\sum_{l = 0}^{m-1} \frac{(m-1)!}{K! (m-1-l)!} \sum_{j = 0}^{\min{(l, m-1-l)}} \frac{1}{2^j} j!  \frac{l!}{j! (l-j)!} \frac{(m-1-l)!}{(m-1-l-j)! j!} \\ & \qquad \hbar^j   \cl a^{* l-j} \cl a^{m-1- l - j} \\
&= m\sum_{l = 0}^{m-1} \binom{m-1}{l} \sum_{j = 0}^{\min{(l, m-1 - l)}} \frac{1}{2^j} j! \binom{l}{j}\binom{m-1-l}{j} \hbar^j \cl a^{* l-j} \cl a^{m-1- l - j} \\
&= m (\cl a + \cl a^{*})_{\varstar}^{m-1}.
\end{split}
\end{align}
Q.E.D.
\end{widetext}

\section{Quantum Harmonic Balance as a Refined Averaging Method}\label{app:averaging}

The quantum harmonic balance (QHB) expansion developed in \cref{sec:averaging} is inspired by a general class of perturbative theory known as averaging methods, but it also includes important improvement to such approach. In this section, we discuss the relationship between them in more detail.  

Firstly, we note that the shared theme of the averaging methods is to decompose the dynamics of some physical object $Z = \cu Z + \zeta(\cu Z, ^c\!\cu Z, t)$ associated with a driven system into some slow dynamics captured by $\cu Z$ and micromotion captured by $\zeta(\cu Z, ^c\!\cu Z, t)$, where $^c\!\cu Z$ is the conjugate variables to $\cu Z$ \cite{venkatraman2021}. To achieve this separation of time-scales, these methods follow an iterative procedure operating on the EOM over $Z$ similar to that in the specific case of the QHB operating on the Heisenberg EOM \cref{eq:eom-phase} over $Z=\cl a$. Other examples of averaging methods include the well-established Krylov-Bogoliubov (KB) method \cite{landau1976,krylov1937,bogoliubov1961,rahav2003}, which analyzes the Hamilton EOM over the classical position coordinate $Z=\cl q$; the secular averaging theory \cite{buishvili1979}, which analyzes the Liouville EOM over the density matrix $Z=\op\rho$; and the higher-order RWA method \cite{mirrahimi2015}, which analyzes the Schr\"odinger EOM over the wave function $Z=|\psi\rangle$.

It is important to recognize that the QHB expansion provides a novel perspective for treating $Z = \cu Z + \zeta(\cu Z, ^c\!\cu Z, t)$ as a frame transformation, whereas conventional averaging methods merely treat it as an ansatz for the EOM. This modification is crucial in developing a simple diagrammatic description as well as eliminating errors in the QHB. To demonstrate this, we note that when taking the time derivative of the specific function $\zeta(\cu Z, ^c\!\cu Z, t) = \cl\sigma(\clu A,\clu A^*)+\cl\mu(\clu A,\clu A^*, t)$ associated with the QHB, a conventional averaging method, such as the KB method, will simply apply the chain rule and obtain:
\begin{align}\label{eq:eta-eom-kb}
    d_t(\cl\sigma+\cl\mu) =  \partial_t\cl\mu + \partial_{\clu A}(\cl\sigma+\cl\mu) \,d_t\clu A+ \partial_{\clu A^*}(\cl\sigma+\cl\mu) \,d_t\clu A^*.
\end{align}
Although this treatment appears proper and is indeed correct for the classical system, it is incorrect for the quantum phase-space functions $\cl\sigma(\clu A,\clu A^*)$ and $\cl\mu(\clu A,\clu A^*, t)$ in which the bosonic coordinates $\clu A,\clu A^*$ are related by non-commutative Husimi bracket. To correctly derive the time-derivative of $\cl\sigma+\cl\mu$ and, more importantly, to make such an expression diagram-compatible, it is crucial to treat $\cl a = \clu A + \cl\sigma+\cl\mu$ as a canonical frame transformation. The Heisenberg EOM over $\cl\eta(\clu A, \clu A^*, t)$ consequently reads
\begin{align}
     d_t(\cl\sigma+\cl\mu) =  \partial_t\cl\mu - \moyal{\cl K}{\cl\sigma+\cl\mu},
\end{align}
where $\cl K$ is the effective Hamiltonian governing $\clu A$. Noting that $- \moyal{\cl K}{\cl\sigma+\cl\mu} = \partial_{\clu A}(\cl\sigma+\cl\mu) \,d_t\clu A+ \partial_{\clu A^*}(\cl\sigma+\cl\mu)\,d_t\clu A^* + \mathcal O(\hbar)$, we observe that \cref{eq:eta-eom-kb} obtained from conventional averaging methods introduces an error at order $\mathcal O(\hbar)$ and will result in an erroneous EOM, which differs from \cref{eq:eom-phase}, the EOM that QHB aims to solve.

In addition, the conventional averaging procedure does not always leads to a canonical transformation. In particular, the averaging procedure solving the EOM over $Z$ involves iteratively constructing $\zeta$ from $\partial_t\zeta$, which leaves the time-independent part $\text{Sta}(\zeta)$ unconstrained. While $\text{Sta}(\zeta)$ seems to be a gauge free to vary, a generic $\text{Sta}(\zeta)$ does not necessarily ensure that the transformation $Z\rightarrow \mathcal Z+\zeta$ is canonical. Most averaging methods do not address this potential issue on canonicity directly, as their iterative procedures are self-consistent by design. Consequently, some methods, such as the Krylov-Bogoliubov (KB) method \cite{landau1976,krylov1937,bogoliubov1961,rahav2003} and the secular averaging method \cite{buishvili1979}, are canonical without explicit enforcement. In contrast, methods like the higher-order Rotating Wave Approximation (RWA) \cite{mirrahimi2015} and a modified KB method \cite{grozdanov1988} are not canonical.

The QHB method goes a step further by carefully constructing $\cl\sigma$, which corresponds to $\text{Sta}(\zeta)$ in the generic averaging method, to ensure that the resulting frame transformation $\cl a\rightarrow \clu A+\cl\sigma+\cl\mu$ is canonical. This is achieved by demanding the frame transformation $\cl a\rightarrow \clu A+\cl\sigma+\cl\mu$ in the additive form to be re-expressed as an exponential map $\cl a\rightarrow e^{L_{\cl S}} \clu A$ generated by a real function $\cl S$ (c.f. \cref{eq:exponential-map}). In \cref{app:canonicity} we will formally prove the equivalence between the additive and exponential forms of the frame transformation underlying the USH method. This enforcement on canonicity also constitutes as an improvement to the general averaging method. In \cref{app:RS}, we will also adopt this procedure to modify the Brillouin-Wigner perturbation method to make it canonical.

\section{Equivalence between the additive and exponential forms of the frame transformation underlying USH}\label{app:canonicity}

To ensure the canonicity of the USH method, we have assumed that the additive frame transformation $\cl a \rightarrow \cl{\cu A}+\cl\sigma(\clu A, \clu A^*)+\cl\mu(\clu A, \clu A^*, t)$ can be equivalently expressed as an exponential map $\cl a \rightarrow e^{L_{\cl S}}\clu A$, where $L_{\cl S}\cl f = \moyal{\cl S}{\cl f}$ is the Lie derivative with respective to $\cl S$. In this section, we rigorously prove the equivalence of the frame transformation in these two representations. For the ease of understanding, we choose to prove such equivalence in the more widely-used Hilbert space associated with the Lie derivative is $L_{\op S}\op f = [\op S,\op f]/i\hbar$. These two formulations are isomorphic to each other and the following proof will only rely on the properties of their shared Lie algebraic structure \cite{venkatraman2021}. Moreover, we define $\cl\eta(\clu A, \clu A^*, t)=\cl\sigma(\clu A, \clu A^*)+\cl\mu(\clu A, \clu A^*, t)$ for the sake of conciseness.

In Hilbert space, the first representation of the frame transformation is the additive form $\op a \rightarrow \op{\cu A} + \op\eta$. This frame transformation is constructed by solving the equation of motion \cref{eq:A-eom}, which we can rewrite as follows:
\begin{align}
\label{eq:A-eom-app}
\begin{split}
  \frac{1}{i\hbar}\comm{\op K}{\op{\cu{A}}} &- \partial_t \op \eta + \frac{1}{i \hbar} \comm{\op K}{\op \eta} \\
  & \quad\;= \frac{1}{i \hbar} \comm{\op H(\op{\cu A}+\op\eta, \op{\cu A}^\dagger+\op\eta^\dagger, t)}{\op{\cu A}+\op\eta}.
\end{split}
\end{align}
Here $\op H(\op a, \op a^\dagger, t)$ is the time-dependent Hamiltonian describing the driven system of interest and $\op K(\op{\cu A},\op{\cu A}^\dagger)$ is the time-independent effective Hamiltonian we solve for. The second representation of the frame transformation in Hilbert space is an exponential map  $\op a \rightarrow e^{-\op S/i\hbar} \op {\cu A} e^{\op S/i\hbar}$. With the conventional Schrieffer-Wolff expansion\cite{eckardt2015,venkatraman2021}, it is known that one can find a time-independent Hamiltonian 
\begin{align}\label{eq:Kp-exp}
\op K^\prime = e^{\op S/i\hbar}(\op H(\op{\cu A}, \op{\cu A}^\dagger ,t)+i\hbar\partial_t) e^{-\op S/i\hbar}
\end{align}
with a generator $\op S$ that is hermitian and purely rotating. To prove the equivalence of these two representations, it is sufficient to prove the equivalence of $\op K$ that solves \cref{eq:A-eom-app} and $\op K^\prime$ defined by \cref{eq:Kp-exp} provided that $\op\eta$ and $\op S$ are related by 
\begin{align}\label{app:eq:add-exp-rel-2}
    \op{\cu A} + \op \eta(\op{\cu A}, \op{\cu A}^\dagger ,t) =& e^{\op S/i\hbar} \op {\cu A} e^{-\op S/i\hbar},
\end{align}
which is just \cref{eq:exponential-map-phase}.

\noindent\textit{Proof}:

To prove the equivalence of $\op K$ and $\op K^\prime$, we rewrite \cref{eq:A-eom-app} as follows:
\begin{gather}
\begin{split}
  \frac{1}{i\hbar}\comm{\op K}{\op{\cu{A}}} &- \partial_t \op \eta + \frac{1}{i \hbar} \comm{\op K}{\op \eta} \\
  & \quad\quad= \frac{1}{i \hbar} \comm{\op H(\op{\cu A}+\op\eta, \op{\cu A}^\dagger+\op\eta^\dagger, t)}{\op{\cu A}+\op\eta},
\end{split}\\[5pt]
\Updownarrow\nonumber \quad\text{\cref{app:eq:add-exp-rel-2} } \\[5pt]
\begin{split}
  \frac{1}{i\hbar}\comm{\op K}{\op{\cu{A}}} &- \partial_t \op \eta + \frac{1}{i \hbar} \comm{\op K}{\op \eta} \\
  & \quad\quad= \frac{1}{i \hbar} \comm{e^{\op S/i\hbar}\op H(\op{\cu A}, \op{\cu A}^\dagger, t)e^{-\op S/i\hbar}}{\op{\cu A}+\op\eta},
\end{split}\label{eq:app-prove-2}\\[-5pt]
\Updownarrow\nonumber \quad \text{\cref{eq:Kp-exp}}\\[5pt]
\begin{split}
\frac{1}{i\hbar}\comm{\op K}{\op{\cu{A}}+\op\eta} &- \partial_t \op \eta = \frac{1}{i \hbar} \comm{\op K^\prime-i\hbar e^{\op S/i\hbar}\partial_te^{-\op S/i\hbar}}{\op{\cu A}+\op\eta} ,
\end{split}\label{eq:app-prove-3}
\end{gather}
Note that in writing the right-hand side of \cref{eq:app-prove-2}, we have also employed the identity of unitary transformation 
\begin{align*}
    \op f(e^{\op S/i\hbar}\op{\cu A}e^{-\op S/i\hbar},e^{\op S/i\hbar}\op{\cu A}^\dagger e^{-\op S/i\hbar})=e^{\op S/i\hbar}\op f(\op{\cu A},\op{\cu A}^\dagger)e^{-\op S/i\hbar}.
\end{align*}
Now we can express the task of proving $\op K=\op K^\prime$ as proving the equation:
\begin{align}
\partial_t\eta = \comm{e^{\op S/i\hbar}\partial_te^{-\op S/i\hbar}}{\op{\cu A}+\op\eta}, \label{eq:app-prove-4}
\end{align}
which becomes apparent from \cref{eq:app-prove-3}.

To prove this equation, we begin by writing:
\begin{align}
\partial_t\eta &= \partial_t(e^{\op S/i\hbar}\op{\cu A}e^{-\op S/i\hbar} - \op{\cu A})\nonumber\\
&= \partial_t(e^{\op S/i\hbar}\op{\cu A}e^{-\op S/i\hbar})\nonumber\\
&= (\partial_te^{\op S/i\hbar})\op{\cu A}e^{-\op S/i\hbar}+ e^{\op S/i\hbar}\op{\cu A}(\partial_te^{-\op S/i\hbar})\nonumber\\
& = (\partial_te^{\op S/i\hbar})e^{-\op S/i\hbar}e^{\op S/i\hbar}\op{\cu A}e^{-\op S/i\hbar}\nonumber\\
&\qquad+ e^{\op S/i\hbar}\op{\cu A}e^{-\op S/i\hbar}e^{\op S/i\hbar}(\partial_te^{-\op S/i\hbar})\nonumber\\
& = (\partial_te^{\op S/i\hbar})e^{-\op S/i\hbar}(\op{\cu A}+\op\eta)\nonumber\\
&\qquad+ (\op{\cu A}+\op\eta)e^{\op S/i\hbar}(\partial_te^{-\op S/i\hbar})\nonumber\\
& = -e^{\op S/i\hbar}(\partial_te^{-\op S/i\hbar})(\op{\cu A}+\op\eta)\nonumber\\
&\qquad+ (\op{\cu A}+\op\eta)e^{\op S/i\hbar}(\partial_te^{-\op S/i\hbar}),\label{eq:app-last-one}
\end{align}
where we used the product rule and the fact that $\partial_t(e^{\op S/i\hbar}e^{-\op S/i\hbar}) = (\partial_te^{\op S/i\hbar})e^{-\op S/i\hbar} + e^{\op S/i\hbar}\partial_te^{-\op S/i\hbar} =0$.

Equation \cref{eq:app-last-one} can be further simplified to:
\begin{align*}
\partial_t\eta = \comm{e^{\op S/i\hbar}\partial_te^{-\op S/i\hbar}}{\op{\cu A}+\op\eta},
\end{align*}
which is equivalent to \cref{eq:app-prove-4}. This completes the proof.

Lastly, having demonstrated that the frame transformation constructed using QHB constitutes a unitary transformation in the Hilbert space, we can conclude that the effective Hamiltonian $\cl K$ is a real function. Therefore, we have established the validity of all the assumptions made in \cref{sec:averaging} when formulating the QHB expansion.

\section{A self-consistent reformulation of Rayleigh-Schr\"odinger perturbative expansion}\label{app:RS}
In this section, we reformulate the well-established Rayleigh-Schr\"odinger (RS) non-degenerate perturbative expansion in a self-consistent form inspired by the USH method developed in \cref{sec:averaging}. Such reformulation will (1) allow for a more intuitive interpretation of the RS expansion, (2) bridge the RS expansion to another well-established perturbative method, the Brillouin-Wigner expansion, (3) lead to a diagrammatic method to carry out the RS expansion, and (4) lead to a conjecture on the equivalence between the RS expansion and the characteristic equation of a Hamiltonian matrix.

\suppresstocsubsection{A brief review on the Rayleigh-Schr\"odinger non-degenerate perturbation theory}

For the Hamiltonian 
\begin{align}\label{eq:Hamiltonian}
\hat H = \hat H_0 + \epsilon \hat V
\end{align}
where $\epsilon =1$ is there only to keep track of the expansion, the Rayleigh-Schr\"odinger perturbation theory seeks to find the perturbative effect of $\hat V$ on the eigen-spectrum and eigenstates of $\hat H$ relative to those of $\hat H_0$. To achieve this, it assumes that the eigenstate $|n\rangle$ of $\hat H$ can be expressed as 
\begin{align}\label{eq:state-pert}
|n\rangle = |n^{(0)}\rangle + \epsilon|n^{(1)}\rangle + \epsilon^2|n^{(2)}\rangle + \cdots 
\end{align}
where $|n^{(0)}\rangle$ is the eigenstate of $\hat H_0$, i.e.
\begin{align}
\hat H_0|n^{(0)}\rangle = E_n^{(0)}|n^{(0)}\rangle.
\end{align}
Similarly, the eigenenergy $E_n$ associated with $|n\rangle$ is assumed to be 
\begin{align}\label{eq:energy-pert}
E_n = E_n^{(0)}+\epsilon E_n^{(1)}+\epsilon^2 E_n^{(2)}+ \cdots
\end{align}
With this, the time-independent Schro\"odinger equation associated with \cref{eq:Hamiltonian} is
\begin{align}\label{eq:schrodinger-pert}
\begin{split}
&(\hat H_0 + \epsilon \hat V)(|n^{(0)}\rangle + \epsilon|n^{(1)}\rangle + \cdots ) \\
&\qquad\;= ( E_n^{(0)}+\epsilon E_n^{(1)}+  \cdots)(|n^{(0)}\rangle + \epsilon|n^{(1)}\rangle + \cdots )
\end{split}
\end{align}
One then can solve \cref{eq:schrodinger-pert} iteratively at each order with the constraint 
\begin{align}\label{eq:norm}
\langle n|n\rangle = 1,
\end{align}
which ensures the proper normalization of the perturbed states. 

As a result, the leading order corrections of the eigenenergy and eigenstates respectively read
\begin{align}
&E_n^{(1)} = \langle n|V|n\rangle,\\
&E_n^{(2)} = \sum_{k \ne n} \frac{\left |\left \langle k^{(0)} \right |V\left |n^{(0)} \right \rangle \right |^2} {E_n^{(0)} - E_k^{(0)}} ,\\
&|n^{(1)}\rangle =  \sum_{k \ne n} \left |k^{(0)}\right\rangle \frac{\left\langle k^{(0)}\right|V\left|n^{(0)}\right\rangle}{E_n^{(0)}-E_k^{(0)}},\\
\begin{split}
&|n^{(2)}\rangle = \sum_{k\neq n}\sum_{\ell \neq n} \left |k^{(0)}\right\rangle \frac{\left \langle k^{(0)} \right |V \left |\ell^{(0)} \right \rangle \left \langle \ell^{(0)} \right |V \left |n^{(0)} \right \rangle}{\left (E_n^{(0)}-E_k^{(0)}\right ) \left (E_n^{(0)}-E_\ell^{(0)} \right )} \\
&\qquad\qquad -\sum_{k\neq n}\left |k^{(0)}\right\rangle \frac{\left \langle k^{(0)} \right |V\left |n^{(0)} \right \rangle \left \langle n^{(0)} \right |V\left |n^{(0)} \right \rangle}{\left (E_n^{(0)}-E_k^{(0)} \right )^2} \\
&\qquad\qquad- \frac{1}{2} \left |n^{(0)} \right \rangle\sum_{k \ne n} \frac{|\left \langle k^{(0)} \right |V\left |n^{(0)} \right \rangle|^2}{\left (E_n^{(0)}-E_k^{(0)} \right )^2}
\end{split}
\end{align}

\suppresstocsubsection{R-S method as an operator perturbation theory}\label{sec:op-pert}
\subsubsection{Problem setup}
We now reformulate the RS expansion in the language of the operator perturbation method. We first remark that, in the eigenbasis of $\hat H_0$, the Hamiltonian $\hat H$ is a matrix whose off-diagonal entries only perturbatively correct the diagonal ones, which are $E_0^{(0)}, E_1^{(0)}$, .... In this language, the goal of RS expansion can be thought as seeking for a frame transformation 
\begin{align}\label{eq:state-transform}
|\psi\rangle\rightarrow|\varphi\rangle
\end{align}
such that the transformed Hamiltonian 
\begin{align}\label{eq:H-transform}
\hat H\rightarrow \hat K
\end{align}
is a diagonal matrix. Similar to \cref{eq:QHB-ansatz}, we can relate the transformed quantities to their counterparts in the original frame as 
\begin{align}\label{eq:ansatz-state}
&|\psi\rangle = |\varphi\rangle + \hat\sigma|\varphi\rangle + \hat\mu|\varphi\rangle,\\\label{eq:ansatz-H}
&\hat K = \hat H_0 + \hat K_I,
\end{align}
where $\hat\sigma$ is assumed to be a diagonal matrix, $\hat\mu$ is a purely off-diagonal matrix, and $\hat K_I$ --- the rationale for this notation will become apparent shortly -- is a diagonal matrix that perturbatively corrects $\hat H_0$. We note that \cref{eq:ansatz-state,eq:ansatz-H} are just a matrix representation of \cref{eq:state-pert,eq:energy-pert}. Moreover, to ensure the proper normalization of the transformed state, which in turn guarantees the canonicity of the sought-after transformation, \cref{eq:norm} translates to the constraint as
\begin{align}
(1+\hat\sigma^\dagger + \hat\mu^\dagger)(1+ \hat\sigma + \hat\mu) = 1. 
\end{align}
In other words, the change of basis matrix $1+\hat\sigma+\hat\mu$ is unitary. 

To find the the transformation, i.e. $\hat\sigma, \hat\mu$, and $\hat K_I$, and relate the ordinary perturbation theory to the operator-valued one, it is convenient, yet not necessary\footnote{One can derive the results below without going to the interaction picture. Yet such derivation involves inverse operation of commutator, which could be obscure to readers not familiar with such operation. Going into the interaction translates such operation to time-integration.}, to go to the interaction picture\footnote{Here we assume that $\hat V$ is purely off-diagonal. For the case that $\hat V$ contains non-zero diagonal terms, they can be absorbed into $\hat H_0$.} generate by $\hat H_0$. In such picture, \cref{eq:ansatz-state,eq:ansatz-H} translates to 
\begin{align}\label{eq:ansatz-state-I}
&|\psi (t)\rangle \rightarrow |\psi_I\rangle = |\varphi_I\rangle + \hat\sigma |\varphi_I\rangle + \hat\mu_I(t)|\varphi_I\rangle,\\\label{eq:ansatz-H-I}
&\hat K\rightarrow \hat K_I. 
\end{align}
Note that, in \cref{eq:ansatz-state-I}, the diagonal matrix $\hat\sigma$ is left unchanged in the interaction picture while the purely off-diagonal matrix $\hat\mu$ gains time-dependence. This is because the generator of the interaction picture $\hat H_0$ is diagonal. Indeed, in the interaction picture, all the diagonal entries in a matrix, such as those in $\hat\sigma_I$ and $\hat K_I$, remains time-independent while the off-diagonal entries transforms as 
\begin{align}\label{eq:O-I}
 \hat O_{mn}\rightarrow  \hat O_{mn} e^{(E_n^{(0)}-E_m^{(0)})t/i\hbar}
\end{align}
where $\hat O_{mn}$ is the entry at row $m$ and column $n$ of an operator $\hat O$ in the lab frame.

\subsubsection{A self-consistent representation}
In the interaction picture, the system of interest behaves like a time-dependent driven system. We are now ready to carry out the averaging procedure associated with the operator perturbation theory over the time-dependent Schr\"odinger equation governing $|\psi_I\rangle$ and $|\varphi_I\rangle$
\begin{align}\label{eq:schrodinger-psi-I}
&i\hbar\partial_t|\psi_I\rangle = \epsilon\hat V_I|\psi_I\rangle,\\\label{eq:schrodinger-varphi-I}
&i\hbar\partial_t|\varphi_I\rangle = \hat K_I|\varphi_I\rangle.
\end{align}
Plugging \cref{eq:ansatz-state-I} into \cref{eq:schrodinger-psi-I}, we get
\begin{align}
\begin{split}
(1+\hat\sigma)i\hbar\partial_t|\varphi_I\rangle &+ i\hbar(\partial_t\hat\mu_I(t))|\varphi_I\rangle + i\hbar\hat\mu_I(t)\partial_t|\varphi_I\rangle\\
&\qquad\quad=\epsilon\hat V_I(t)(1+\hat\sigma+\hat\mu_I(t))|\varphi_I\rangle.
\end{split}
\end{align}
By further plugging in \cref{eq:schrodinger-varphi-I}, we get
\begin{align}\label{eq:eom}
(1\!+\!\hat\sigma)\hat K_I + i\hbar\partial_t\hat\mu_I(t) + \hat\mu_I(t)\hat K_I = \epsilon\hat V_I(t)(1\!+\!\hat\sigma\!+\!\hat\mu_I(t)).\raisetag{-3pt}
\end{align}
Therefore, the goal of the perturbation treatment now translates to solving \cref{eq:eom} under the constraints that $\hat K_I$ and $\hat\sigma$ being diagonal (time-independent), $\hat\mu_I(t)$ being purely off-diagonal (time-dependent), and that the transformation is canonical, i.e.
\begin{align}\label{eq:canonicity-I}
(1+\hat\sigma^\dagger + \hat\mu_I^\dagger(t))(1+ \hat\sigma + \hat\mu_I(t)) = 1. 
\end{align}

\cref{eq:eom} is tedious to solve as it is. To better illustrate its structure and physical significance, we introduce two composite variables, which allow to break down \cref{eq:eom} into a few simpler self-consistent equations. This is similar to the change of variables in the QHB method discussed in \cref{eq:change-variable}. Specifically, we introduce
\begin{subequations}\label{eq:change-variable}
\begin{align}\label{eq:mu-prime}
&\hat\mu^\prime_I(t) = \hat\mu_I(t)(1+\hat\sigma)^{-1},\\\label{eq:M-prime}
&\hat M^\prime_I(t) = i\hbar\partial_t \hat\mu_I^\prime(t) + \hat\mu_I^\prime(t)\hat K_I
\end{align}
\end{subequations}

With these new variables, \cref{eq:eom} can be rewritten as 
\begin{subequations}\label{eq-app:self-consistent}
\begin{align}\label{eq:eom-comp}
&\hat K_I + \hat M_I^\prime = \epsilon\hat V_I(t) + \epsilon \hat V_I(t)\hat\mu_I^\prime(t)\\
&\hat\mu^\prime_I(t) = \frac{1}{i\hbar}\int dt\left (\hat M_I^\prime(t) - \hat\mu_I^\prime(t)\hat K_I\right)\label{eq:mu-1}
\end{align}
with the constrain that $\hat K_I, \hat\sigma$ being diagonal while $\hat M^\prime_I(t)$ and $\hat\mu_I(t)$ being purely off-diagonal. Note that by writing \cref{eq:eom-comp,eq:mu-1} we have used the fact that $\hat K_I$ and $1+\hat\sigma$, which are both diagonal matrices, commute. This is a special properties for the ordinary perturbation problem where the transformed Hamiltonian $\hat K$ is diagonal, while, for more general cases of applying averaging procedure over driven systems, such simplification is not always possible.  

Besides \cref{eq:eom-comp,eq:mu-1}, the canonicity constraint \cref{eq:canonicity-I} can also be rewritten in the new variables in \cref{eq:change-variable}. With a few lines of algebra, we can get a simple expression
\begin{align}\label{eq:sigma}
1+\hat\sigma &= \left(1+\text{Sta}\left(\hat\mu_I^{\prime\dagger}(t)\hat\mu_I^\prime(t)\right) \right)^{-\frac{1}{2}}
\end{align}
\end{subequations}
where $\text{Sta}(\hat O)$ here means taking the static, thus the diagonal, part of a given matrix $\hat O$. Similarly, another notation that will be used shortly is $\text{Rot}(\hat O)$, which takes the rotating, thus off-diagonal, part of a given matrix $\hat O$. 

\cref{eq:eom-comp,eq:mu-1} are two simple self-consistent equations that express $\hat K_I, \hat M_I^\prime(t),$ and $\hat\mu^\prime_I(t)$ on the left-hand side of the equations as a function of themselves or other quantities on the right-hand side. In addition, \cref{eq:sigma} expresses $\hat\sigma$ in terms of $\hat\mu^\prime$. By iteratively substituting these equations into the right-hand side of themselves, we can obtain the expressions for $\hat K_I, \hat M_I^\prime(t), \hat\mu^\prime_I(t)$ and $\hat\sigma$ at each order of the perturbative parameter $\epsilon$. For example, to leading order we have
\begin{align*}
&\hat K_I^{(1)} = \text{Sta}\left(\epsilon\hat V_I(t)\right) = 0,\\
&\hat M_I^{\prime(1)} (t) = \text{Rot}\left(\epsilon\hat V_I(t)\right),\\
&\hat\mu^{\prime(1)}_I(t) = \frac{1}{i\hbar}\int dt \hat M_I^{\prime(1)} (t),\\
&\hat\sigma^{(1)} = 0,
\end{align*}
and at order 2 we have 
\begin{align*}
&\hat K_I^{(2)} = \text{Sta}(\epsilon\hat V_I(t)\hat\mu^{\prime(1)}_I(t)),\\
&\hat M_I^{\prime(2)} (t) = \text{Rot}(\epsilon\hat V_I(t)\hat\mu^{\prime(1)}_I(t)),\\
&\hat\mu^{\prime(2)}_I(t) = \frac{1}{i\hbar}\int dt \left(\hat M_I^{\prime(2)} (t)-\hat\mu_I^{\prime(1)}(t)\hat K^{(1)}_I\right) \\
&\qquad\quad\,= \frac{1}{i\hbar}\int dt \hat M_I^{\prime(2)} (t),\\
&\hat\sigma^{(2)} = -\frac{1}{2}\text{Sta}\left(\hat\mu_I^{\prime\dagger(1)}(t)\hat\mu_I^{\prime(1)}(t)\right).
\end{align*}

\subsubsection{Physical significance}

The self-consistent equations in \cref{eq-app:self-consistent} are not only simple but also allow for meaningful physical interpretation, which we now discuss.  


We first comment on \cref{eq:mu-1}:
\begin{align*}
    \hat\mu^\prime_I(t) = \frac{1}{i\hbar}\int dt\left (\hat M_I^\prime(t) - \hat\mu_I^\prime(t)\hat K_I\right)
\end{align*}
which defines the relationship between the off-diagonal matrices $\hat M_I^\prime(t)$ and $\hat \mu_I^\prime(t)$. As discussed in \cref{eq:O-I}, we denote the $m$-th row and $n$-th column entry in $\hat M_I^\prime(t)$ as 
\begin{align}
\hat M_{I,mn}^\prime= M_{I,mn}^\prime e^{(E_n^{(0)}-E_m^{(0)})t/i\hbar}.
\end{align}
Noting that $\hat K_I$ is diagonal by definition, we can write each entry in $\hat \mu_I^\prime(t)$ as 
\begin{subequations}\label{eq:mu-2}
\begin{align}\label{eq:mu-2-a}
\hat \mu_{I,mn}^\prime &= \frac{M_{I,mn}^\prime}{E_n^{(0)}-E_m^{(0)}}e^{(E_n^{(0)}-E_m^{(0)})t/i\hbar} - \frac{1}{i\hbar}\int dt \hat \mu_{I,mn}^\prime \hat K_{I, nn}\raisetag{20pt}\\[5pt]
\begin{split}\label{eq:mu-2-b}
&=\Big(\frac{M_{I,mn}^\prime}{E_n^{(0)}-E_m^{(0)}} - \frac{M_{I,mn}^\prime K_{I,nn}}{(E_n^{(0)}-E_m^{(0)})^2} + \frac{M_{I,mn}^\prime K_{I,nn}^2}{(E_n^{(0)}-E_m^{(0)})^3}\\
&\qquad - \frac{M_{I,mn}^\prime K_{I,nn}}{(E_n^{(0)}-E_m^{(0)})^2} + \cdots)e^{(E_n^{(0)}-E_m^{(0)})t/i\hbar}
\end{split}\raisetag{20pt}\\[5pt]\label{eq:mu-2-c}
& = \frac{M_{I,mn}^\prime}{E_n^{(0)}+K_{I,nn}-E_m^{(0)}}e^{(E_n^{(0)}-E_m^{(0)})t/i\hbar}\\\label{eq:mu-2-d}
& = \frac{1}{E_n-E_m^{(0)}}\hat M_{I,mn}^\prime
\end{align}
\end{subequations}
In writing \cref{eq:mu-2-b}, we have iteratively substituted $\hat \mu_{I,mn}^\prime$ on the right-hand side of \cref{eq:mu-2-a} by \cref{eq:mu-2-a} itself. Then, by identifying that \cref{eq:mu-2-b} is a  Taylor series of $a/(1+x)$ type of function, in \cref{eq:mu-2-c} we resum the series to it original form. Finally, in writing \cref{eq:mu-2-d}, we have used the identity that $E_n = E_n^{(0)}+K_{I,nn}$, where $E_n$ is the exact eigenenergy of $\hat H$ (or $\hat K$ defined in \cref{eq:ansatz-H}) associated with state indexed $n$. 

For convenience, we can also introduce a matrix $\hat P_H$ defined as
\begin{align}\label{eq:PH}
\hat P_{H,mn} = \begin{cases}
    \frac{1}{E_n-E_m^{(0)}} &\text{for } n\ne m\\[10pt]
    0&\text{for } n= m
\end{cases}
\end{align}
and thus \cref{eq:mu-2} can be rewritte as 
\begin{align}\label{eq:mu-3}
\hat\mu_I^\prime =\hat P_{H}\circ\hat M_I^\prime,
\end{align}
where $\circ$ is the Hadamard product (element-wise multiplication) between matrices. Note that the subscript in $\hat P_H$ denotes that the specific construction of $\hat P_H$ is determined by $\hat H$. 

\cref{eq:mu-3} gives a very simple relation between $\hat M_I^\prime(t)$ and $\hat \mu_I^\prime(t)$, where each entry $(m,n)$ in the latter differs from its counterpart by a fraction, which can be interpreted as a ``propagator'' associated with a detuned coupling. The denominator of the propagator is the ``dressed" energy difference between the responding states --- specifically, the difference between the \textit{perturbed energy} associated with state $m$ and the \textit{unperturbed energy} associated with state $n$. 


Now we move to discuss \cref{eq:eom-comp}
\begin{align*}
\hat K_I + \hat M_I^\prime(t) = \epsilon\hat V_I(t) + \epsilon \hat V_I(t)\hat \mu_I^\prime(t).
\end{align*}
With \cref{eq:mu-3} that gives a simple relation between $\hat M_I^\prime(t)$ and $\hat \mu_I^\prime(t)$, the above equation should be understood as a self-consistent expression for $\hat K_I$ and $\hat M_I^\prime(t)$ (or $\hat\mu_I^\prime(t)$) under the constraint that the former is diagonal and the latter is purely off-diagonal. In other words, $\hat K_I$ corresponds to the diagonal (time-independent) components of the right-hand-side of \cref{eq:eom-comp}, while $\hat M_I^\prime(t)$ corresponds to the off-diagonal (time-dependent). Therefore, we have
\begin{align}\label{eq:M}
\hat M_I^\prime &= \epsilon\hat V_I + \text{Rot}\Big(\epsilon \hat V_I\hat \mu_I^\prime\Big)
\end{align}
Plugging in \cref{eq:mu-3}, we get the simple self-consistent expression for $\hat\mu^\prime_I$ as
\begin{subequations}\label{eq:mu-4}
\begin{align}\label{eq:mu-4-a}
\hat \mu_I^\prime &= \hat P_{H}\circ\epsilon\hat V_I + \hat P_{H}\circ\Big(\epsilon \hat V_I\hat \mu_I^\prime\Big)\\
\begin{split}\label{eq:mu-4-b}
&=\epsilon\hat P_{H}\circ\hat V_I + \epsilon^2\hat P_{H}\circ\Big( \hat V_I\left(\hat P_{H}\circ\hat V_I\right)\Big)\\
&\quad+ \epsilon^3\hat P_{H}\circ\Big( \hat V_I\left(\hat P_{H}\circ\Big( \hat V_I\left(\hat P_{H}\circ\hat V_I\right)\Big)\right)\Big)\\
&\quad+\cdots
\end{split}\raisetag{8pt}
\end{align}
\end{subequations}
where \cref{eq:mu-4-b} is obtained from \cref{eq:mu-4-a} by iteratively substituting the $\hat \mu^\prime_I$ on the right-hand side by \cref{eq:mu-4-a} itself. Also note that in these expressions we have dropped $\text{Rot}$ because the definition of $\hat P_H$ in \cref{eq:PH} ensures that the static (i.e. diagonal) term is zero in the Hadmard product of $\hat P_H$ and another matrix. Likewise, $\hat K_I$ can be expressed as
\begin{subequations}\label{eq:K}
\begin{align}\label{eq:K-a}
\hat K_I &=  \text{Sta}\Big(\epsilon \hat V_I\hat \mu_I^\prime\Big)\\
\begin{split}\label{eq:K-b}
&= \epsilon^2\text{Sta}\Big( \hat V_I\left(\hat P_{H}\circ\hat V_I\right)\Big)\\
&\quad+ \epsilon^3\text{Sta}\Big( \hat V_I\left(\hat P_{H}\circ\Big( \hat V_I\left(\hat P_{H}\circ\hat V_I\right)\Big)\right)\Big)+\cdots
\end{split}
\end{align}
\end{subequations}

\begin{figure}
    \centering
    \includegraphics{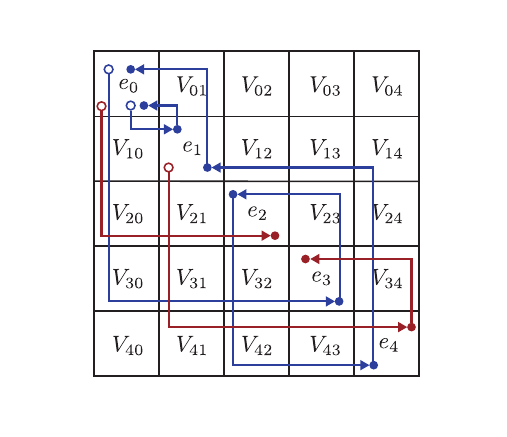}
    \caption{An example of $\hat H$ and the paths representing contributions to $\hat K_I$ and $\hat \mu_I^\prime$. There are two red paths ($e_0\rightarrow e_2$ and $e_1\rightarrow e_4\rightarrow e_3$) and two blue paths ($e_0\rightarrow e_1 \rightarrow e_0$ and $e_0\rightarrow e_3\rightarrow e_2\rightarrow e_4\rightarrow e_1 \rightarrow e_0$).}
    \label{fig:matrix}
\end{figure}

 One can compare \cref{eq:mu-4,eq:K} with the Brillouin-Wigner (BW) perturbative expansion and easily see that  they are equivalent to those in the latter. Therefore, we have shown the equivalence between the RS expansion and the BW expansion. Moreover, a problem associate with BW expansion is that the frame underlying it, which corresponds to $|\psi\rangle\rightarrow(1+\hat\mu^\prime)|\varphi\rangle$, is not canonical \cite{mikami2016}. This is evident from the fact that $|\psi\rangle\rightarrow(1+\hat\sigma)(1+\hat\mu^\prime)|\varphi\rangle$ constitutes a canonical transformation for non-zero $\hat\sigma$ in \cref{eq:sigma}. In other words, in the derivation above, we have also found a transformation $(1+\hat\sigma)$ defined by \cref{eq:sigma} that will make the BW expansion canonical. 

In addition, the expressions for $\hat \mu^\prime_I$ and $\hat K_I$ in \cref{eq:mu-4-b,eq:K-b} allows them to be represented diagrammatically as the ``paths'' in $\hat H$. Intuitively speaking, they can be thought as paths containing multiple segments each stemming from a $\hat V$ factor and associated with some propagator stemming from $\hat P_H$. 

We take a $\hat H$ shown in \cref{fig:matrix} as an example, where the diagonal entries $e_n$'s are unperturbed eigenenergy and off-diagonal terms $V_{mn}$ are the perturbation. On top of the matrix we draw blue and red paths to represent contributions to $\hat K_I$ and $\hat \mu^\prime_I$ in \cref{eq:K-b} and \cref{eq:mu-4-b}, respectively. Each path has a starting point, denoted by an open circle, and one or multiple stops, denoted by solid circles. Both the starting point and the stops are in the diagonal entries. A starting point, or a stop, is connected to another stop by a counter-clockwise directional edge that travels only horizontally or vertically and takes one left turn at the off-diagonal entry that directly couples the diagonal entries associated with the two ends of the edge. This off-diagonal entry is the one associated with the edge while other entries that the edge traverses are not relevant here.  Moreover, while a path can terminate at any stop, it has to terminate when reaching a stop at the same entry as the starting point. We color those paths returning to the starting point by blue color and others by red color. 

Structurally, each path in \cref{fig:matrix}, which constitutes of one or more directional edges associated with some off-diagonal entries cascaded together, resembles a term in  $\hat \mu^\prime_I$ and $\hat K_I$ defined by \cref{eq:mu-4-b,eq:K-b}, which constitutes of one or more $\hat P_H$ and $\hat V_I$ factors cascaded together. Algebraically, a path starting from the $n$-th diagonal entry and end at $m$-th diagonal entry can be evaluated in a simple way so that it represents a term contributing to entry $(m,n)$ in $\hat \mu^\prime_I$, when $m\ne n$ in the red paths, or in $\hat K_I$, when $m=n$ in the blue paths. Specifically, a directional edge connecting the $k$-th diagonal entry to the $l$-th diagonal entry is evaluated as $V_{lk}$, a stop at $k$'s diagonal entry for $k\ne n$ is evaluated as $1/(E_n - E_k^{(0)})$, and the path, evaluated in the interaction picture, is also associated with an overall phase $e^{E_n^{(0)}-E_m^{(0)}t/i\hbar}$. 

For example, the red path $e_0\rightarrow e_2$ in \cref{fig:matrix} is evaluated as 
\begin{align}
\frac{V_{20}}{e_0^\prime-e_2}e^{(e_0-e_2)t/i\hbar},
\end{align}
where $e_m^\prime$ is taken as the perturbed eigenenergy associated with state $m$. This path corresponds to the leading order contribution the $\hat \mu_{I,20}^\prime$. The longer red path in \cref{fig:matrix} correspondingly contributes to $\hat \mu_{I,41}^\prime$ and is evaluated as 
\begin{align}
\quad\frac{V_{41}V_{34}}{(e_1^\prime - e_4)(e_1^\prime - e_3)}e^{(e_1-e_3)t/i\hbar}.
\end{align}
Recalling the relation \cref{eq:ansatz-state}, we recognize that the two terms above are associated with the perturbative correction to the eigenstate of $0$ and $1$, each of which gains a small contribution of the unperturbed state $2$ and $3$, respectively. We will discuss this in more detail shortly. 

Like the red paths, the short blue path in \cref{fig:matrix} is evaluated as 
\begin{align}
\frac{|V_{10}|^2}{e_0^\prime-e_1} 
\end{align}
and the long blue path is evaluated as 
\begin{align}
\frac{V_{30}V_{23}V_{42}V_{14}V_{01}}{(e_0^\prime - e_3)(e_0^\prime - e_2)(e_0^\prime - e_4)(e_0^\prime - e_1)}.
\end{align}
Both of the two terms above contribute to $K_{I, 00}$, which represents the perturbative correction to the unperturbed eigenenergy $e_0$.

\suppresstocsubsection{Examples}

\subsubsection{tridiagonal $3\times3$ matrix}

\begin{figure}
    \centering
    \includegraphics{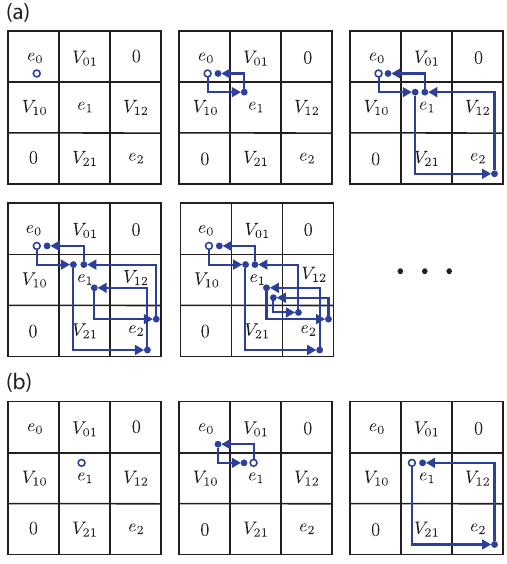}
    \caption{Diagrammatic representation for the perturbed eigenenergy associated with (a) state 0 and (b) state 1. }
    \label{fig:e0}
\end{figure}

To illustrate and verify the results above and to draw more implications from them, we first investigate a simple example of a $3\times3$ matrix
\begin{align}\label{eq:3x3m}
    \hat H = \begin{bmatrix}
        e_0&V_{01}&0\\
        V_{10}&e_1&V_{12}\\
        0&V_{21}&e_2
    \end{bmatrix}
\end{align}
and compute the eigenvalue of it. From the diagrammatic representation, the exact eigenenergy $e_0^\prime$ associated with state 0 can be represented by the infinite number of paths shown in \cref{fig:e0} (a). This series corresponds to the algebraic expression:
\begin{align}\label{eq:e0-1}
\begin{split}
    e_0^\prime &= e_0 + \frac{|V_{01}|^2}{e_0^\prime-e_1}+ \frac{|V_{01}|^2|V_{12}|^2}{(e_0^\prime-e_1)^2(e_0^\prime-e_2)}\\
    &\quad + \frac{|V_{01}|^2|V_{12}|^4}{(e_0^\prime-e_1)^3(e_0^\prime-e_2)^2}+ \frac{|V_{01}|^2|V_{12}|^6}{(e_0^\prime-e_1)^4(e_0^\prime-e_2)^3}+\cdots
\end{split}\\
& = e_0 + \frac{|V_{01}|^2}{e_0^\prime-e_1 - \frac{|V_{12}|^2}{e_0^\prime-e_2}},\label{eq:e0-2}
\end{align}
where in writing \cref{eq:e0-2} we have resummed the series in \cref{eq:e0-1} as a $a/(1-x)$ type of function under the assumption that $|V_{12}|^2/(e^\prime_0-e_2)<1$. Note that the pole in \cref{eq:e0-2} is at $e_0^\prime = e_1 + |V_{12}|^2/(e_0^\prime + e_2)$. Here the bare energy $e_1$ still gets dressed in a way but different from its exact energy, i.e. $e_1 + |V_{12}|^2/(e_0^\prime + e_2)\ne e_1^\prime$. The physical significance and implications of this difference are to be further investigated. 

For the perturbed energy $e_2^\prime$ associated with state $2$, we can write the perturbed energy of the state $2$ in the same form as \cref{eq:e0-2}:
\begin{align}\label{eq:e2}
    e_2^\prime =  e_2 + \frac{|V_{12}|^2}{e_2^\prime-e_1 - \frac{|V_{01}|^2}{e_2^\prime-e_0}}.
\end{align}

For the perturbed energy $e_1^\prime$ of the state $1$, it diagrammatically corresponds to the three paths in \cref{fig:e0} (b), whose algebraic expression is 
\begin{align}\label{eq:e1}
    e_1^\prime = e_1 + \frac{|V_{01}|^2}{e_1^\prime-e_0} + \frac{|V_{12}|^2}{e_1^\prime-e_2}.
\end{align}

There are two remarks to be made about this example. First, \cref{eq:e0-2,eq:e1,eq:e2} are three uncoupled self-consistent equations each only contain one unknown, $e_0^\prime$, $e_1^\prime$, $e_2^\prime$, respectively. If we denote the unknown of each equation as a generic unknown, say $\lambda$, one can easily verify that these three equations are equivalent to each other. Indeed, they are just the re-expression of the characteristic equation $\det(\hat H-\lambda \hat I)=0$ for $\hat H$, i.e.:
\begin{align}
\begin{split}
(\lambda-e_0)(\lambda-e_1)&(\lambda-e_2) - |V_{01}|^2(\lambda-e_2)\\
&\qquad\quad- |V_{12}|^2(\lambda-e_0) =0
\end{split}
\end{align}
 Interestingly, the expansion of each eigenvalue when carried out to all orders contain the information of all other eigenvalues. It can be understood as a reorganization of the characteristic function, which is usually hard to interpret and solve, to a self-consistent form that favors one particular state. 

Secondly, we remark that the expression for $e_0^\prime$ defined by \cref{eq:e0-1} contains infinite number of terms, which correspond to infinite number of paths in \cref{fig:e0} (a), each containing repetitive features (loops  traversing $e_1$ and $e_2$). With resummation of these terms/paths, we arrive at \cref{eq:e0-2} that only contains finite number of terms. By slightly modifying the diagrammatic rules, \cref{eq:e0-2} can similarly be represented by finite number of paths (e.g. only the first two paths in \cref{fig:e0}). This feature is startling from the perspective of a perturbative theory, yet it is trivial considering the observation that the expression for $e_0^\prime$ in \cref{eq:e0-2} is equivalent to the characteristic equation of $\hat H$. 

\subsubsection{generic $3\times3$ matrix}
With the tridiagonal $3\times3$ matrix treated, we now consider the generic $3\times3$ matrix
\begin{align}\label{eq:3x3m-generic}
    \hat H = \begin{bmatrix}
        e_0&V_{01}&V_{02}\\
        V_{10}&e_1&V_{12}\\
        V_{20}&V_{21}&e_2
    \end{bmatrix}
\end{align}

From perturbative method and the diagrammtic computation, we have
\begin{align}
\begin{split}\label{eq:e0-full-1}
     e_0^\prime&= e_0 + \frac{|V_{01}|^2}{e_0^\prime-e_1 - \frac{|V_{12}|^2}{e_0^\prime-e_2}} +\frac{|V_{02}|^2}{e_0^\prime-e_1 - \frac{|V_{12}|^2}{e_0^\prime-e_2}}\\
     &\quad + \frac{V_{10}V_{21}V_{02}}{(e_0^\prime -e_1)(e_0^\prime -e_2)} + \frac{V_{20}V_{12}V_{01}}{(e_0^\prime -e_2)(e_0^\prime -e_1)}
\end{split}\\
\begin{split}\label{eq:e1-full-1}
     e_1^\prime&= e_1 + \frac{|V_{01}|^2}{e_1^\prime-e_0 - \frac{|V_{02}|^2}{e_1^\prime-e_2}} +\frac{|V_{12}|^2}{e_1^\prime-e_2 - \frac{|V_{02}|^2}{e_1^\prime-e_0}}\\
     &\quad + \frac{V_{21}V_{02}V_{10}}{(e_1^\prime -e_2)(e_1^\prime -e_0)} + \frac{V_{01}V_{20}V_{12}}{(e_1^\prime -e_0)(e_0^\prime -e_2)}
\end{split}\\
\begin{split}\label{eq:e2-full-1}
     e_2^\prime&= e_2 + \frac{|V_{12}|^2}{e_2^\prime-e_1 - \frac{|V_{01}|^2}{e_2^\prime-e_0}} +\frac{|V_{02}|^2}{e_2^\prime-e_0 - \frac{|V_{01}|^2}{e_2^\prime-e_1}}\\
     &\quad + \frac{V_{12}V_{01}V_{20}}{(e_2^\prime -e_1)(e_2^\prime -e_0)} + \frac{V_{02}V_{10}V_{21}}{(e_2^\prime -e_0)(e_2^\prime -e_1)}
\end{split}
\end{align}
Multiplied by the first two denominators appearing of the the right-hand side of each equation and with some rearrangement, the three equations above can be written as
\begin{align}
    f(e_0^\prime) - \frac{|V_{12}|^2}{(e_0^\prime-e_1)(e_0^\prime-e_2)}f(e_0^\prime) = 0\\
    f(e_1^\prime) - \frac{|V_{20}|^2}{(e_1^\prime-e_0)(e_1^\prime-e_2)}f(e_1^\prime) = 0\\
    f(e_2^\prime) - \frac{|V_{01}|^2}{(e_2^\prime-e_0)(e_2^\prime-e_1)}f(e_2^\prime) = 0
\end{align}
where 
\begin{align}
\begin{split}
    f(\lambda) = &(\lambda-e_0)(\lambda-e_1)(\lambda-e_2) - (\lambda-e_0)|V_{12}|^2\\[5pt]
    &\quad- (\lambda-e_1)|V_{02}|^2- (\lambda-e_2)|V_{01}|^2 \\[5pt]
    &\quad+ V_{01}V_{12}V_{20} + V_{10}V_{21}V_{02}
\end{split}
\end{align}
Therefore, $e_0^\prime, e_1^\prime, e_2^\prime$ are the solutions for $f(\lambda) = 0$. Moreover, one can also easily verify that $\det(\hat H-\lambda\hat I) = f(\lambda)$ and thus $e_0^\prime, e_1^\prime, e_2^\prime$ are the exact eigenvalues of $\hat H$. In other words, \cref{eq:e0-full-1,eq:e1-full-1,eq:e2-full-1} are three self-consistent forms of the characteristic equation $\det(\hat H-\lambda\hat I) =0$ for a generic $3\times3$ matrix. 


With the example of the $3\times3$ matrix treated, it is very tempting to generalize our observations to any hermitian Hamiltonian. Here we generalize them as conjunctures without providing a rigorous proof.

\textit{Conjecture:} The perturbative expression for the eigenenergy of a state can be expressed as a self-consistent equation. When carried out to all-orders, this equation can be resummed in a ways such that it is equivalent to the characteristic equation of the Hamiltonian. 

\section{Hamiltonian of Multimode Nonlinear Oscillator in Normal Modes}\label{app:multi}

\begin{figure}
\includegraphics{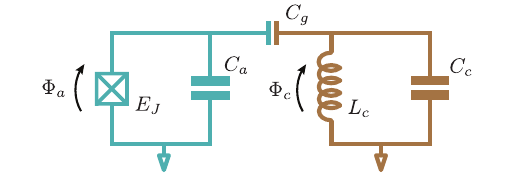}
\caption{Schematic of a transmon (in turquoise color) capacitively coupled to a cavity (in brown color). }
\label{fig:app-circuit}
\end{figure}

In this section, we derive the Hamiltonian for multimode systems where a nonlinear oscillator is coupled to several linear oscillators. For the sake of concreteness, we consider the system treated in \cref{sec:renormalization} where a transmon is capacitively coupled to a cavity mode as shown in \cref{fig:app-circuit}, and the procedure below apply to general multi-mode problem in general.

The Lagrangian of the transmon-cavity system reads
\begin{align}
    \begin{split}
        \mathcal L = &\frac{1}{2}(C_a+C_g)\dot\Phi_a^2 + E_J\cos(\varphi_a) \\
        &+ \frac{1}{2}(C_c+C_g)\dot\Phi_a^2 - \frac{1}{2L_c} \Phi_c^2 - C_g\dot\Phi_a\dot\Phi_c
    \end{split}\\\label{app:eq:L-split}
    & = \mathcal L_\text{lin} + E_J\cos(\varphi_a) + \frac{1}{2L_J} \Phi_a^2
    \intertext{with}
    \begin{split}
    \mathcal L_\text{lin} =& \frac{1}{2}(C_a+C_g)\dot\Phi_a^2 - \frac{1}{2L_J} \Phi_a^2 \\
    &+ \frac{1}{2}(C_c+C_g)\dot\Phi_a^2 - \frac{1}{2L_c} \Phi_c^2 - C_g\dot\Phi_a\dot\Phi_c,
    \end{split}\label{app:eq:L-lin}
\end{align}
where $\Phi_i$ for $i = a, c$ is the flux across the inductive element in the transmon and cavity modes, respectively, $\varphi_i = 2e\Phi_i/\hbar$ is the reduced flux, $L_J = \hbar^2E_J/e^2$ is the effective linear inductance of the Josephson junction, and each capacitance is related to the its capacitive energy, a notion used in \cref{fig:renorm-circuit,fig:IST-circuits}, by $C_i = e^2/2E_{C, i}$. In \cref{app:eq:L-split}, we further separate the Lagrangian into the linear part $\mathcal L_\text{lin}$, defined in \cref{app:eq:L-lin}, and the nonlinear part, i.e. the last two terms in \cref{app:eq:L-split}. Note that $\mathcal L_\text{lin}$ describes two LC-oscillators that are capacitively coupled. Our goal now is to find the normal modes of it following the standard procedure \cite{goldstein2002}. To do this, we first write the equation of motion generated by $\mathcal L_\text{lin}$ as 
\begin{align}
\begin{bmatrix}
    \ddot\Phi_a\\
    \ddot\Phi_c
\end{bmatrix} = \boldsymbol C^{-1}\boldsymbol L\begin{bmatrix}
    \Phi_a\\
    \Phi_c
\end{bmatrix}\label{eq:app:eom-multi}
\end{align}
where
\begin{align}
    \boldsymbol{C} = \begin{bmatrix}
        C_a+C_g&-C_g\\
        -C_g& C_c+C_g
    \end{bmatrix},\quad \boldsymbol{L} = \begin{bmatrix}
        \frac{1}{L_J}&0\\
        0& \frac{1}{L_c}
    \end{bmatrix}\
\end{align}
are the capacitance and inductance matrix. We then can find matrix $\boldsymbol{P}$ that diagonalize $\boldsymbol C^{-1}\boldsymbol L$ and thus \cref{eq:app:eom-multi} can be rewritten as 
\begin{align}
\boldsymbol{P}^{-1}\begin{bmatrix}
    \ddot\Phi_a\\
    \ddot\Phi_c
\end{bmatrix} = \boldsymbol{P}^{-1}\boldsymbol C^{-1}\boldsymbol L\boldsymbol{P}\boldsymbol{P}^{-1}\begin{bmatrix}
    \Phi_a\\
    \Phi_c
\end{bmatrix}, \quad\boldsymbol{P} = \begin{bmatrix}
    \beta_1 &\beta_2\\
    \beta_3 &\beta_4
\end{bmatrix},
\end{align}
where $\boldsymbol{P}^{-1}\boldsymbol C^{-1}\boldsymbol L\boldsymbol{P}$ is a diagonal matrix. Written as it is, it is easy to see that the normal modes of $\mathcal L_\text{lin}$ are 
\begin{align}\label{app:eq:normal-modes}
    \begin{bmatrix}
        \Phi_A\\
        \Phi_C
    \end{bmatrix} = \boldsymbol{P}^{-1} \begin{bmatrix}
        \Phi_a\\
        \Phi_c
    \end{bmatrix}.
\end{align}
Plugging \cref{app:eq:normal-modes} into \cref{app:eq:L-lin}, we get a decoupled linear Lagrangian 
\begin{align}
    \begin{split}
    \mathcal L_\text{lin} =& \frac{1}{2}C_A\dot\Phi_A^2 - \frac{1}{2L_A} \Phi_A^2 
    + \frac{1}{2}C_C\dot\Phi_C^2 - \frac{1}{2L_C} \Phi_C^2
    \end{split},\label{app:eq:L-lin-normal}
\end{align}
with 
\begin{align}
    \begin{split}
        C_A &= (C_a+C_g)\beta_1^2+(C_c+C_g)\beta_3^2 - 2C_g\beta_1\beta_3\\
        C_C &= (C_a+C_g)\beta_2^2+(C_c+C_g)\beta_4^2 - 2C_g\beta_2\beta_4\\
        \frac{1}{L_A} &= \frac{\beta_1^2}{L_J}+\frac{\beta_3^2}{L_c}, \quad      \frac{1}{L_C} = \frac{\beta_2^2}{L_J}+\frac{\beta_4^2}{L_c}.
    \end{split}
\end{align}
The coupling between normal modes $\Phi_A$ and $\Phi_C$ are then completely absorbed into the nonlinear terms $E_J\cos(\varphi_a) + \frac{1}{2L_J} \Phi_a^2$ in $\mathcal L$ in \cref{app:eq:L-split} with $\Phi_a = \beta_1\Phi_A + \beta_2\Phi_C$. With this, we write the Hamiltonian associated with $\mathcal L$ in normal modes and get \cref{eq:H-sc} as\footnote{For notational simplicity, in \cref{eq:H-sc}, we use $\op a$ and $\op c$, in stead of $\op a_A$ and $\op a_C$, to represent the bosonic operators in normal modes, use lower cases $a,c$, instead of $A,C$, in the subscripts to denote the corresponding normal modes, and use $\op\varphi$, instead of $\op\varphi_a$, to denote the phase across the junction.}
\begin{align}
\label{app:eq:H-sc}
    \frac{\op{H}}{\hbar} = \omega_A \op{a}_A^{\dagger} \op{a}_A + \omega_C \op{a}^{\dagger}_C \op{a}_C - \frac{E_J}{\hbar}\left( \cos   \hat{\varphi}_a + \frac{\hat{\varphi}_a^2}{2} \right),
\end{align}
where $\omega_i = \sqrt{1/C_iL_i}$ and the phase operator of each mode is related to the bosonic ladder operators by $\op\Phi_i =(\hbar^2L_i/2C_i)^{1/4} (\op a_i + \op a^\dagger_i)$ for $i = A, C$. The bare phase operator $\op\varphi_a$ can be written as $\op\varphi_a = \varphi_{\text{zps}, A} (\op a_A + \op{a}_A^{\dagger}) + \varphi_{\text{zps}, C} (\op a_C + \op{a}_C^{\dagger})$ where $\varphi_{\text{zps}, A} = \beta_1(4L_A/C_A)^{1/4}\sqrt{e^2/\hbar}$, $\varphi_{\text{zps}, C} = \beta_2(4L_C/C_C)^{1/4}\sqrt{e^2/\hbar}$ are the participation of the corresponding modes in the zero point spread of the junction phase. We note that, for Josephson circuits, $\varphi_{\text{zps}, A}$ and $\varphi_{\text{zps}, C}$ can be alternatively computed \cite{minev2021} as $\varphi_{\text{zps}, i}^2 = P_i \hbar \omega_i / 2 E_J$, where $P_i$, a design parameter of the circuit, is the energy participation ratio of normal mode $i$ in the Josephson junction. The two notations are related to each other by $P_A = \beta_1^2 L_A/L_J$ and $P_C = \beta_2^2 L_C/L_J$.

Expanding the nonlinear terms in \cref{app:eq:H-sc}, we get the expression of the nonlinear oscillator 
\begin{align}
\label{app:eq:H-sc-2}
\begin{split}
        \frac{\op{H}}{\hbar} = \omega_A \op{a}_A^{\dagger} \op{a}_A + \omega_C \op{a}^{\dagger}_C \op{a}_C &+\sum_{m>2}\frac{g_m}{m} \Big(\lambda_A(\op a_A+\op a_A^\dagger)\\
        &\quad+\lambda_A(\op a_A+\op a_A^\dagger)\Big)
\end{split}
\end{align}
where $g_m = (-1)^{1+m/2}\omega_A \varphi_{\mathrm{zps}}^{m-2} /2(m-1)!$ for even $m$ and $0$ for odd $m$, with $\varphi_{\mathrm{zps}} = \sqrt{\hbar \omega_A/2 E_{\mathrm{J}}}$ and $\lambda_i^2 = P_i \omega_i/ \omega_A$ for $i = A, C$. 

When the circuit is capacitively driven with drives of different frequencies, additional terms $-\sum_l\sum_{i = A, C} i\Omega_{i,l}(\op a_i - \op a_i^\dagger)\cos\omega_{d,i,l}t$ should be added to \cref{app:eq:H-sc-2}. Following the frame transformations in \cref{app:frame} for each mode and each drive tone, one will then obtain the Hamiltonian in form of \cref{eq:multi-mode}, which is the starting point of diagrammatic analysis on multimode and multitone problems.

\section{Computing Coupling Strength Through the Effective Hamiltonian}\label{app:coupling}

In this section, we detail the analysis on the cases that several multiphoton resonances coexist. In particular, we consider the transmon in \cref{fig:quasispectrum} and examine the coupling between the ground and 10th excited states through a $(10:4)$ process. As discussed in \cref{sec:mpnr}, when this process is on resonance, nearly resonant are the coupling between the ground and 7th excited state through a $(7:3)$ process and the coupling between the 7th and 10th excited state through a $(3:1)$ process. Specifically, as we will show, the coupling between the 7th and 10th state could be comparable with the energy difference between the two state in the effective frame. Therefore, when $(10:4)$ process is on resonance, both 7th and 10th excited states are effectively hybridized with the ground state. This renders it insufficient to compute the effective Hamiltonian only capturing $(10:4)$ process in the effective dynamics and leaving those involving the 7th excited state to the micromotion. 

To capture all the relevant effective dynamics, we construct the diagrams in the rotating frame at $\omega_o^\prime = 2\omega_d/5$ so that the diagrams associated with the coupling term $\xi^4\op{\cu A}^{\dagger 10}$ is static. Moreover, we also consider the diagrams responsible for $(3:1)$ and $(7:3)$ process to be part of the effective dynamics. For example, when carrying out the quantum harmonic balance, there exists two diagrams
\begin{align}\label{app:eq:slow-diagram}
\begin{split}
        \tikzpic{-21}{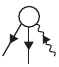}&=g_4\xi\cl{\cu A}^{*2}e^{i(2\omega_o^\prime - \omega_d)t}=g_4\xi\cl{\cu A}^{*2}e^{-i\omega_dt/5}\\
    \tikzpic{-21}{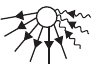}&=g_{10}\xi^3\cl{\cu A}^{*6}e^{i(6\omega_o^\prime -3\omega_d)t} = g_{10}\xi^3\cl{\cu A}^{*6}e^{-i3\omega_dt/5}.
\end{split}
\end{align}
In the ordinary USH procedure discussed in \cref{sec:averaging}, these two terms contribute to \!\tikzpic{-10.5}{figsv6_small/notation/M_small.pdf} ($\cl M$) in \cref{eq:sc-diagram}, i.e. the micromotion, since they are time-dependent. Yet, in the modified treatment here, we consider these two terms slow-varying and pertinent to the effective dynamics. Therefore, we collect these terms to \!\tikzpic{-10.5}{figsv6_small/notation/Gamma_small.pdf} ($\cl\Gamma$) and thus the latter becomes slow-varying. Upon integration over $\clu A^* e^{i\omega_o^\prime t}$, the two terms in \cref{app:eq:slow-diagram} contribute $\frac{1}{3}g_4\xi\cl{\cu A}^{*3}e^{i\omega_dt/5} +\mathrm{c.c.}$ and $\frac{1}{7}g_{10}\xi^3\cl{\cu A}^{*7}e^{-i\omega_dt/5} +\mathrm{c.c.}$ to the effective Hamiltonian, which are responsible for the $(3:1)$ and $(7:3)$ process, respectively. We also note that, porvided that in the rotating frame $\omega_o^\prime=2\omega_d/5$ the $(10:4)$ process is captured by some static effective Hamiltonian terms, if we consider the $(3:1)$ process as slow-varying, then the $(7:3)$ process simultaneously becomes slow-varying. This is because the Hamiltonian terms responsible for the latter two process are of opposite frequency and can cascade together to yield a static process, i.e. $(3+7:1+3) = (10:4)$.

Moreover, since terms like those in \cref{app:eq:slow-diagram} no longer contribute to \!\tikzpic{-10.5}{figsv6_small/notation/M_small.pdf} ($\cl M$), they do not create off-resonant excitations contributing to \!\tikzpic{-11}{figsv6_small/notation/eta_small.pdf} ($\cl \mu e^{-i \omega_o^\prime t}$) that participates in higher order mixing process, either. When constructing diagrams at higher order, one should should eliminate those diagrams involving such off-resonant excitations. For example, in the ordinary treatment, a term reading
\begin{align}\label{app:eq:invalid104-diagram}
\tikzpic{-33}{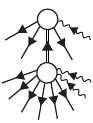} = \frac{g_6g_8}{-\omega_d/5-\omega_o^\prime}\clu A^{*9}\xi^3e^{-i\omega_o^\prime t}
\end{align}
will contribute to \!\tikzpic{-10.5}{figsv6_small/notation/Gamma_small.pdf} ($\cl\Gamma$) and resulting the coupling term in the $(10:4)$ process. In the treatment that includes $(3:1)$ and $(7:3)$ processes in the effective dynamics, however, the term in \cref{app:eq:invalid104-diagram} does not exist either in \!\tikzpic{-10.5}{figsv6_small/notation/Gamma_small.pdf} or \!\tikzpic{-10.5}{figsv6_small/notation/M_small.pdf} because the off-resonant excitation involved in \cref{app:eq:invalid104-diagram} is created by the second term in \cref{app:eq:slow-diagram} and thus spurious.

Following the modified diagrams rules stated above, we obtain the effective Hamiltonian in Hilbert space as:
\begin{align}\label{app:eq:mnr-K-10-4}
\begin{split}
        \frac{\op K}{\hbar} = \sum_{n>0} K_n\op{\cu A}^{\dagger n} \op{\cu A}^{n} &+\Omega_{10,4}\xi^{4}\op{\cu A}^{\dagger 10} + \Omega_{3,1}\xi\op{\cu A}^{\dagger 3}e^{i\frac{\omega_d}{5}t} \\
    &+\Omega_{7,3}\xi^{3}\op{\cu A}^{\dagger 7}e^{-i\frac{\omega_d}{5}t}+\mathrm{h.c.},
\end{split}
\end{align}
where the exact expression of each Hamiltonian parameter above is obtained from the computer program \cite{qhb2022} with the modified procedure. This effective Hamiltonian can be further reduced to the subspace of relevant oscillator states, i.e.  0th, 3rd, 7th, and 10th Fock states, and reads
\begin{widetext}
\begin{align}\label{app:eq:mnr-K-3}
\begin{split}
    \frac{\hat K}{\hbar} =\begin{bmatrix}
    \mathcal{E}_0 &\Omega^{*}_{0\leftrightarrow3,1} e^{-i\frac{\omega_d}{5}t}  &\Omega^{*}_{0\leftrightarrow7,3} e^{i\frac{\omega_d}{5}t} & \Omega^{*}_{0\leftrightarrow10,4}\\[5pt]
    \Omega_{0\leftrightarrow3,1} e^{i\frac{\omega_d}{5}t}&\mathcal{E}_{3}&0&\Omega^{*}_{7\leftrightarrow10,3} e^{i\frac{\omega_d}{5}t}\\[5pt]
    \Omega_{0\leftrightarrow7,3}e^{-i\frac{\omega_d}{5}t} &0& \mathcal{E}_{7} & \Omega^{*}_{7\leftrightarrow10,1}e^{-i\frac{\omega_d}{5}t}\\[5pt]
    \Omega_{0\leftrightarrow10,4} & \Omega_{7\leftrightarrow10,3} e^{-i\frac{\omega_d}{5}t}& \Omega_{7\leftrightarrow10,1}e^{i\frac{\omega_d}{5}t} & \mathcal{E}_{10}
    \end{bmatrix},
\end{split}
\end{align}
\end{widetext}
where each entry is related to the parameters in \cref{app:eq:mnr-K-10-4} by the relation in \cref{eq:mnr-reduced-rel}. We note that, even though the 3rd excited state is not involved in any near-resonant process here, it should still be included in \cref{app:eq:mnr-K-3} for consistency. Specifically, one should understand the 3rd excited state, which is off-resonantly coupled to the ground state and 10th excited state through the $(3:1)$ and $(7:3)$ processes, respectively, as an intermediate state in the coupling between ground and 10th excited state. This is similar to the 7th excited state, which is off-resonantly couple to the ground state through the $(7:3)$ process and near-resonantly couple to the 10th state through the $(3:1)$ process.

\begin{figure}
\includegraphics{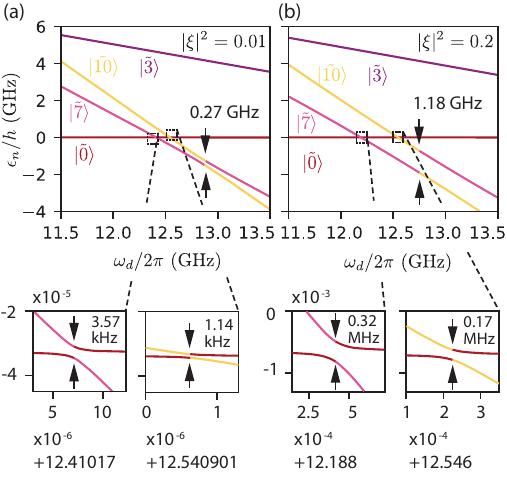}
\caption{Spectrum of \cref{app:eq:mnr-K-4} as a function drive frequency $\omega_d$ for drive strength $|\xi|^2 = 0.01$ in panel (a) and $|\xi|^2 = 0.2$ in panel (b). The drive frequency is \textit{not} rescaled by $\omega_{01}$ as in other plots for easier comparison with the y-axis. Lines of different color corresponds to the eigenenergy $\epsilon_n$ of the eigenstate $|\tilde n\rangle$, where the eigenstate is index by the corresponding Fock state it has the biggest overlap with. Each pair of opposite arrows marks an anti-crossing between two eigenstates. Specifically, the left inset in each panel plots the anticrossing between the ground and 7th excited state through the $(7:3)$ process, the right inset plots the anticrossing between the ground and 10th excited state through the $(10:4)$ process, and the arrows in the main plot of each panel marks the anticrossing between the 7th and 10th excited states through the $(3:1)$ process. The Rabi strength $\Omega_{n\leftrightarrow q, p}^R$ between the corresponding Fock states resulting from the $(q:p)$ process is half of the anti-crossing size.}
\label{app:fig:10-4-spectrum}
\end{figure}

Remarkably, the slow-varying effective Hamiltonian \cref{app:eq:mnr-K-3} can be further transformed into a static one under the unitary transformation $\op U = \exp[i\frac{\omega_d}{5}(|3\rangle\langle3|-|7\rangle\langle7|)t]$. The resulting effective Hamiltonian reads
\begin{align}\label{app:eq:mnr-K-4}
    \frac{\hat K^\prime}{\hbar} = \begin{bmatrix}
    \mathcal{E}_0 & \Omega^{*}_{0\leftrightarrow3,1} &\Omega^{*}_{0\leftrightarrow7,3} & \Omega^{*}_{0\leftrightarrow10,4}\\[5pt]
    \Omega_{0\leftrightarrow3,1} & \mathcal{E}_{3} + \omega_d/5 & 0 & \Omega^*_{3\leftrightarrow10,3}\\[5pt]
    \Omega_{0\leftrightarrow7,3}& 0&  \mathcal{E}_{7}-\omega_d/5 & \Omega^{*}_{7\leftrightarrow10,1}\\[5pt]
    \Omega_{0\leftrightarrow10,4} & \Omega_{3\leftrightarrow10,3} & \Omega_{7\leftrightarrow10,1} & \mathcal{E}_{10}
    \end{bmatrix}.
\end{align}
We find the Rabi strength between the 0th and 10th states by numerically diagonalizing \cref{app:eq:mnr-K-4}. In \cref{app:fig:10-4-spectrum}, we plot the eigenenergies of \cref{app:eq:mnr-K-4} as a function of drive frequency $\omega_d$ for two choices of drive strength, specifically $|\xi|^2=0.01$ in panel (a) and   $|\xi|^2=0.2$ in panel (b). Besides the ground state, the eigenenegies of other states, to leading order, linearly depend on the drive frequency because $\mathcal{E}_n = n\delta + \mathcal{O}(\varphi_\text{zps}^2)$ (c.f. \cref{eq:mnr-reduced-rel}), where $\delta = \omega_o - 2\omega_d/5$. When two states are resonant, the off-diagonal terms in \cref{app:eq:mnr-K-4} couples them and result in anti-crossing features. The Rabi strength between the two states, which equals to half of size of the anti-crossing, can then be read off from the plot directly. Note that when the ground and 10th excited states are resonant, the coupling between the 7th and 10th excited states through the $(3:1)$ process is 0.135 GHz for $|\xi|^2 = 0.01$. This coupling strength is comparable to the energy difference between the 7th and 10th excited states and thus the two states are strongly hybridized. When the drive strength increases to $|\xi|^2=0.2$, as shown in   \cref{app:fig:10-4-spectrum} (b), the coupling between the 7th and 10th excited states is 0.59 GHz, which is comparable to the energy difference between the two states in the drive frequency window $\omega_d/2\pi \in [11.5 ~\text{GHz}, 13.5 ~\text{GHz}]$ considered here. We therefore can interpret that, in this region of the drive parameter space, the 7th and 10th excited states form a hydribized island in the state space, which is the onset of quantum confusion.

\begin{figure}
\includegraphics{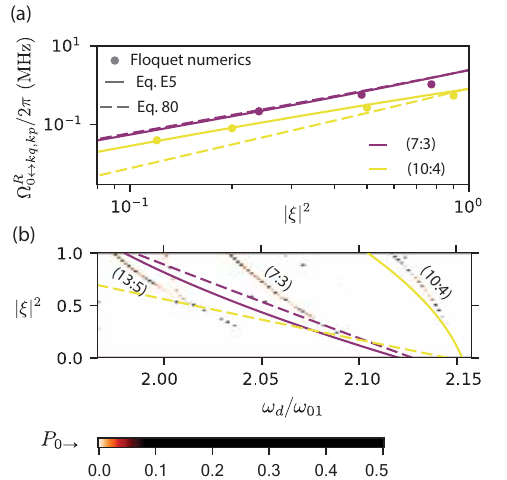}
\caption{Comparison between the exact Floquet numerics with \cref{app:eq:mnr-K-4} (same as \cref{eq:mnr-K-4}), the effective Hamiltonian that captures $(3:1),(7:3),(10:4)$ processes simultaneously, and \cref{eq:mnr-K-1}, the effective Hamiltonian from the ordinary treatment that only captures $(10:4)$ or $(7:3)$ process one at a time. Panle (a): the Rabi strength for the $(7:3)$ and $(10:4)$ process as a function of drive strength $|\xi|^2$. The dots are obtained from the Floquet numerics similar to the one shown in \cref{fig:quasispectrum}. The solid line is obtained from numerical diagonalization of \cref{app:eq:mnr-K-4} similar to the one shown in \cref{app:fig:10-4-spectrum}. The dash line corresponds to the value of $ \Omega_{0\leftrightarrow q, p}$ in \cref{eq:mnr-K-1}. The dots and solid lines are the same as those in \cref{fig:coupling-fit}. Panel (b): the location of the  resonances in the drive parameter space. The heat map, which is a subsection of \cref{fig:spaghetti} (b), plots the excitation possibility out of the ground state $P_{0\rightarrow}$. Each excitation line in the heat map corresponds to a resonance process labeled by $(q:p)$ next to it. The solid lines are obtained from numerical diagonalization of \cref{app:eq:mnr-K-4} like shown in \cref{app:fig:10-4-spectrum}, in which two particular drive strengths are analyzed. The solid lines correspond to the center of the corresponding resonance lines in \cref{fig:spaghetti} (a). The dashed lines are obtained from \cref{eq:mnr-K-1}, where the resonance condition corresponds to the drive parameters that give $\mathcal{E}_0 = \mathcal{E}_0$.}
\label{app:fig:compare}
\end{figure}

In \cref{app:fig:compare}, we compare the Floquet numerical results with the analytical results from \cref{app:eq:mnr-K-4} (same as \cref{eq:mnr-K-4}), the effective Hamiltonian that captures $(3:1),(7:3),(10:4)$ processes simultaneously, and from \cref{eq:mnr-K-1}, the effective Hamiltonian from the ordinary treatment that only captures $(10:4)$ or $(7:3)$ process one at a time. Specifically, panel (a) plots the Rabi strength as a function of drive strength, and panel (b) plots the locations of the multiphoton resonances. For the $(10:4)$ resonance (in yellow), \cref{app:eq:mnr-K-4} gives results closer to the exact Floquet result on both the Rabi strength and the resonance location. For the $(7:3)$ resonance (in purple),  \cref{app:eq:mnr-K-4} only marginally modifies the Rabi strength predicted by \cref{eq:mnr-K-1}. This is because that, on top of the direct coupling between the ground and 7th excited state induced by $\Omega_{0\leftrightarrow7,3}\sim\mathcal O(\varphi_\text{zps}^8)$, \cref{app:eq:mnr-K-4} modifies  \cref{eq:mnr-K-1} by including only a higher order process ---  a cascaded coupling the ground and 7th excited state through the coupling between the ground and 10th excited state, induced by $\Omega_{0\leftrightarrow10,4}\sim\mathcal O(\varphi_\text{zps}^{12})$ in \cref{app:eq:mnr-K-4}, and that between the 10th and 7th excited states, induced by $\Omega_{7\leftrightarrow10,1}\sim\mathcal O(\varphi_\text{zps}^{2})$ in \cref{app:eq:mnr-K-4}. We further note that, in \cref{app:fig:compare} (b), the locations of the $(7:3)$ resonance predicted from \cref{app:eq:mnr-K-4,eq:mnr-K-1} both deviate considerably from the Floquet result when $|\xi|^2$ increases. This is because, at large drive strength, the 7th excited states also near-resonantly couple to the 13th excited state through a $(6:2)$ process. We expect to find better agreement on the resonance location of the $(7:3)$ process if \cref{app:eq:mnr-K-4} is expanded to include the 13th excited states and its associated interaction with other pertinent states. 

Lastly, we note that, in \cref{fig:coupling-fit,fig:spaghetti}, the $(9:5)$ and $(11:5)$ resonances also coexist with other resonances. Specifically, the $(9:5)$ resonance coexists with two other resonances --- the $(5:3)$ resonance, through which the ground and 5th excited state are coupled, and the $(4:2)$ resonance, through which the 5th and 9th excited states are coupled. The $(11:5)$ resonance coexists with four other resonances --- the $(8:4)$ resonance, through which the ground and 8th excited state are coupled, the $(3:1)$ resonance, through which the 8th and 11th excited states are coupled, the $(4:2)$  resonance, through which the ground and 4th excited state are coupled, and the $(7:3)$  resonance, through which the 4th and 11th excited states are coupled. In the results shown in \cref{fig:coupling-fit,fig:spaghetti}, the $(9:5)$ and $(11:5)$ resonances are treated in a similar way as the $(10:4)$ process illustrated above. Moreover, this general procedure of constructing slow-varying effective Hamiltonian can be employed to include more resonances at the same time and thus extend the applicability of the diagrammatic method to deeper quantum diffusion regime.

\section{Floquet Numerical Diagonalization}\label{app:floquet}

In this section, we briefly review the Floquet formalism \cite{shirley1965,floquet1983,grifoni1998}, which we have used to numerically compute dynamics of driven Josephson circuits in \cref{sec:mpnr,sec:IST}. According to Floquet theory, the time-dependent Schr\"{o}dinger equation $\op{H}(t)\Psi(t) = i\hbar \frac{\partial}{\partial t}\Psi(t)$ is satisfied by Floquet states
\begin{align}
    \Psi_m(t) = e^{-i\epsilon_m^{F} t/\hbar}\Phi_m(t)
\end{align}
where $\epsilon_m^{F}$ is called quasienergy, and $\Phi_m(t) = \Phi_m(t+T)$ is called Floquet mode, which is periodic in time with the same periodicity as $\op{H}(t)$. With this relation, the time-dependent Schr\"{o}dinger equation reduces to an eigenvalue equation:
\begin{align}\label{eq:floquet_mode_eigen_H}
    (\op{H}(t) - i\hbar\partial_t) \Phi_m(t) = \epsilon_m\Phi_m(t)
\end{align}
It can be seen that, at any time $t$, the above equation holds and the quasienergy structure of the system is invariant. In this sense, Floquet theorem tranforms a time-dependent problem to a \emph{quasi-stationary} one, i.e. the quasienergies are time-independent while the Floquet modes are oscillating with period $T$. Moreover, Floquet states form a complete basis of the Hilbert space of the driven system, while each Floquet state $\Psi_m(t)$ can be mapped from a set of Floquet modes
 \begin{align}
    \Phi_{m,k}(t) = \Phi_m(t)e^{ik\omega_d t}, \, k\in \mathbb{Z}
 \end{align}
namely,  $\Psi_{m}(t) = e^{-i\epsilon_m^\text{F} t/\hbar}\Phi_m(t)= e^{-i\epsilon_{m,k}^\text{F} t/\hbar} \Phi_{m,k}(t)$ where $\epsilon_{m,k} = \epsilon_m+k\hbar\omega_d$ is the corresponding quasienergy. 

A Floquet state is analogous to a Bloch state \cite{bloch1929}, which sees a periodic potential in space and whose quasimomentum lives in a Brillouin zone. Analogously, a Floquet state sees a periodic potential in time, in whose conjugate space the quasienergy $\epsilon_{m,k}^\text{F}$ lives in a Brillouin zone with width of $\hbar\omega_d$, where $k$ denotes the index of the Brillouin zone. To capture the interaction between different Floquet states, it is sufficient to study the states in the \emph{reduced Brillouin zone}, in which the quasienergy is defined modulo $\hbar\omega_d$, i.e. $\epsilon_m^\text{F} = \epsilon_{m,k}^\text{F} \mod{\hbar\omega_d}$.

Floquet modes and quasienergies are related to the propagator by the relation:
\begin{align}
    \op{U}(0,T)\Phi_m(0) = e^{-i\epsilon_m^\text{F} T/\hbar}\Phi_m(0)
\end{align}
where $\op{U}(0,t)$ is the propagator of $\op{H}(t)$. This relation gives a numerical recipe to implement Floquet theorem by numerically diagonalizing $\op{U}(0,T)$ to find $\epsilon^\text{F}_m$ and $\Phi_m(0)$. This is saying that, by diagonalizing the propagator at one particular time $T$, it is sufficient to obtain the energy structure of the system. If needed, one can further calculate Floquet mode at an arbitrary time $t$ with the relation $ \Phi_m(t) = e^{i\epsilon_m^\text{F} t/\hbar}\op{U}(0,t)\Phi_m(0)$. We use the python package Quantum Toolbox in Python (QuTiP) to implement the above procedures.

\section{Estimated Transmon States Under Displacement and Excitation Possibility}\label{app:transmon-state}

\begin{figure*}
\centering    \includegraphics[width=1.75\columnwidth]
{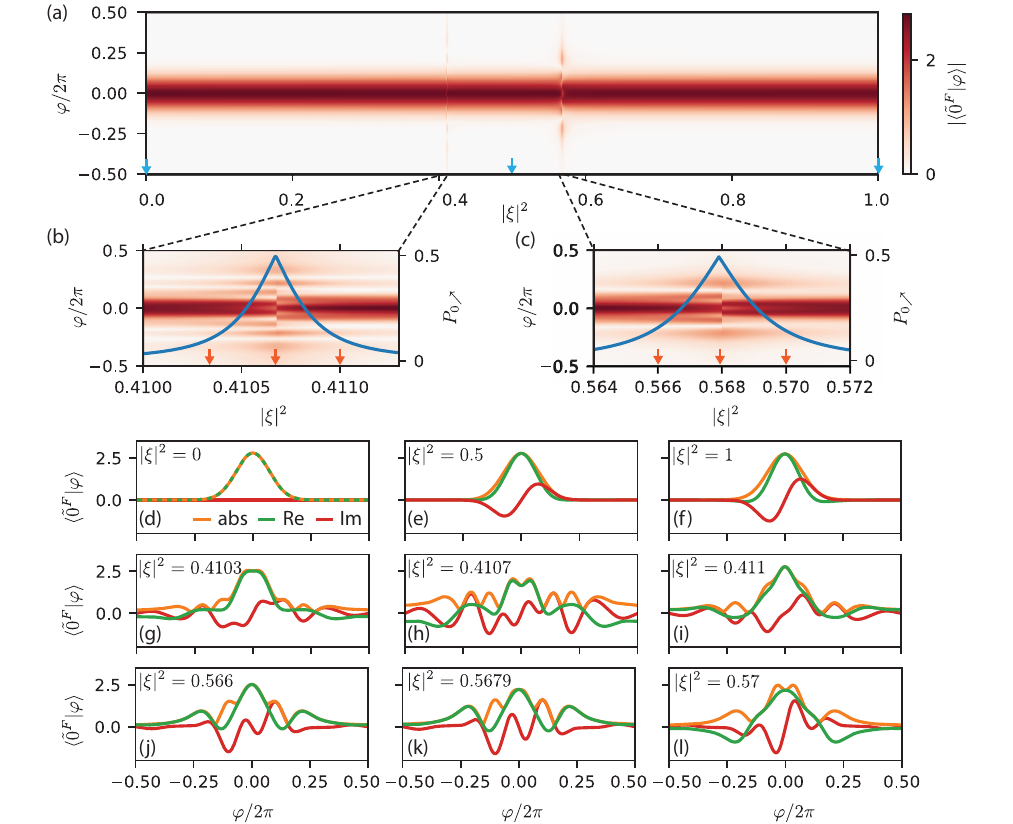}
    \caption{Wave function of Floquet mode $|\tilde 0^\text{F}\rangle$, as defined in \cref{app:eq:0F}, for a driven transmon governed by \cref{eq:mnr-transmon-H}. The transmon and drive are of the same parameters as those in \cref{fig:quasispectrum}, specifically $E_J/h = 30$~GHz, $E_C/h$ = 0.15~GHz, $N_g=0$, $\omega_d/2\pi = 8.97$~GHz, and $|\xi|^2\in[0,1]$. Panel (a)-(c): wave function of $|\tilde 0^\text{F}\rangle$ in $\varphi$ coordinate as a function of $|\xi|^2$. Panel (b) and (c) are zoom-ins of (a) around the $(9:5)$ and $(5:3)$ multiphoton resonances, respectively. Blue lines, labeled by the y-axis on the right, are $P_{0\rightarrow}$ the transition probability out of ground state as defined in \cref{app:eq:p0p1}. Panel (d)-(l): wave function at specific choices of $|\xi|^2$ as marked by the arrows in (a)-(c). Orange, green, and red lines are respectively the absolute value, real part, and imaginary part of the wave function. Panel (d)-(f) correspond to the three blue arrows in (a) at $|\xi| = 0, 0.5$, and 1. The wave functions are Gaussian-like functions and representational for most choices of $|\xi|^2$ in (a). With $|\xi|^2$ increasing, the imaginary parts of these wave functions increase, which correspond to a displacement in the charge coordinate that is conjugate to $\varphi$. Panel (g)-(i) plot the wave function of $|\xi|^2 = 0.4103, 0.4107$, and 0.411 as indicated by the red arrows in (b). These wave functions correspond to hybridization of the ground and 9th excited states due to the $(9:5)$ multiphoton resonance.  Panel (j)-(l) plot the wave function of $|\xi|^2 = 0.566, 0.5679$, and 0.57 as indicated by the red arrows in (c). These wave functions correspond to hybridization of the ground and 5th excited states due to the $(5:3)$ multiphoton resonance. }
    \label{app:fig:wave-function}
\end{figure*}

In \cref{sec:mpnr}, we analyzed multiphoton resonances in a driven transmon. To characterize the landscape of the resonances in the drive parameter space, \cref{fig:spaghetti,fig:spaghetti-nonpert} have plotted the excitation probability from the ground and first excited states respectively defined as 
\begin{align}\label{app:eq:p0p1}
P_{0\rightarrow} = 1-|\langle\tilde 0^\text{F}|\tilde 0_\text{t}\rangle|, \quad P_{1\rightarrow} = 1-|\langle\tilde 1^\text{F}|\tilde 1_t\rangle|,
\end{align}
where $|\tilde 0^\text{F}\rangle, |\tilde 1^\text{F}\rangle$ are the Floquet modes associated with the transmon driven ground and first excited state, and $|\tilde 0_\text{t}\rangle, |\tilde 1_t\rangle$ are the approximate driven ground and first excited state of the transmon. In this section, we detail the computation of $|\tilde 0_\text{t}\rangle, |\tilde 1_t\rangle$ and the state index assignment of $|\tilde 0^\text{F}\rangle, |\tilde 1^\text{F}\rangle$. 

To properly define $|\tilde 0_\text{t}\rangle, |\tilde 1_t\rangle$ and thus $P_{0\rightarrow}, P_{1\rightarrow}$, we assume that the qubit manifold of the transmon is not highly hybridized with other states for most of the interested region in the drive parameter space --- in other words, the driven transmon is not deep in the quantum diffusion regime. Under this assumption, the driven ground state, to leading order, is a displaced ground state with $\alpha_\text{lin}(t) = \frac{i\Omega_d}{2}[e^{-i\omega_dt}/(\omega_d-\omega_o)+e^{i\omega_dt}/(\omega_d+\omega_o)]$ as defined by \cref{app:eq:a-lin-mom}, where drive strength $\xi$ is related to $\Omega_d$ by $\xi = \frac{i\Omega_d\omega_d}{\omega_d^2-\omega_o^2}$. To verify this assumption, in \cref{app:fig:wave-function}, we plot the wavefunction of $|\tilde 0^\text{F}\rangle$ in $\varphi$ as a function of drive strength. For now,  $|\tilde 0^\text{F}\rangle$ at a particular set of drive parameters can be understood as the Floquet mode $\Phi_m(t=0)$ (c.f. \cref{eq:floquet_mode_eigen_H}) of the Hamiltonian \cref{eq:mnr-transmon-H} that has the largest overlap with the undriven transmon ground state $|0_\text{t}\rangle$; its formal definition will be discussed soon. The drive frequency in \cref{app:fig:wave-function} is chosen to be $\omega_d/2\pi = 8.97$~GHz and the drive strength to be $|\xi|^2\in[0,1]$, which are the same as those chosen in \cref{fig:quasispectrum}. As shown in \cref{app:fig:wave-function} (a), for most choices of $|\xi|^2$, the ground Floquet mode $\langle\tilde 0^\text{F}|\varphi\rangle$ is centered around $\varphi=0$ with a Gaussian-like distribution, where $|\varphi\rangle$ is the phase state. In particular, as shown in \cref{app:fig:wave-function} (d), when the drive strength is $|\xi|^2=0$, $|\tilde 0^\text{F}\rangle$ is just the undriven ground state $|0_\text{t}\rangle$. With the drive strength increasing, as examplified in \cref{app:fig:wave-function} (e) and (f) for $|\xi|^2=0.5, 1.0$, the Floquet mode $\langle\tilde 0^\text{F}|\varphi\rangle$  is still centered around $\varphi =0$ but has increasing imaginary component. This indicates that $|\tilde 0^\text{F}\rangle$ is displaced in the charge degree of momentum $\op N$, which is consistent with the phase of the linear response $\alpha_\text{lin}(t)$ that is purely imaginary at $t=0$. 

Our goal is to construct an approximate state $|\tilde 0_\text{t}\rangle$ that captures the displacing effect on the transmon ground state due to the drive but ignore other effects such as multiphoton-resonance. To do this, similar to \cref{app:fig:wave-function} (a), we perform Floquet numerical diagonalization at other 125 drive frequencies evenly spaced in $\omega_d/2\pi \in [6.9~\text{GHz}, 13~\text{GHz}]$, and for each drive frequency we sample 100 drive strength points in $|\xi|^2\in[0, 1]$. This range is the same as that in \cref{fig:spaghetti}. Afterwards, we obtain $|\tilde 0^\text{F}\rangle$, at each given set of drive parameters, by finding the Floquet mode with largest overlap with $|0_\text{t}\rangle$. For those Floquet mode with $|\langle\tilde 0^\text{F}|0_\text{t}\rangle|>0.8$, we fit them with an approximate state $|\tilde{0}_t\rangle$, which can be understood as a displaced transmon ground state as a function of drive frequency $\omega_d$ and drive strength $\xi$. In particular, we decompose $|\tilde{0}_t\rangle$ in the undriven transmon eigenbasis and fit each component as a function of $\omega_d, \xi$:
\begin{align}\label{app:eq:wv-fit}
    \langle\tilde 0_\text{t}(\omega_d, \xi)|k_\text{t}\rangle = \sum_{j\ge0,l\ge0} C_{k, j, l}\omega_d^j\xi^l,
\end{align}
where $|k_\text{t}\rangle$ is the $k$-th state of the undriven transmon and $C_{k,j,l}$'s are the fitted parameters. 

\begin{figure}
\includegraphics{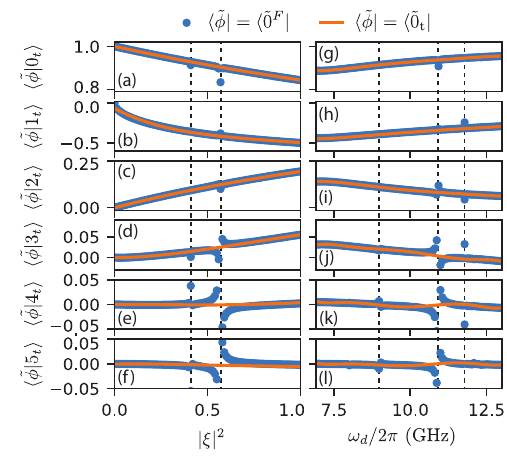}
\caption{ Decomposition of the driven ground states of the transmon analyzed in \cref{app:fig:wave-function} in the basis of eigenstates $\{|k_\text{t}\rangle|t\ge0\}$ of the undriven transmon described by Hamiltonian \cref{eq:mnr-transmon-H} with $E_d=0$. The contribution from $k\ge6$ states are not shown here as they are negligible. Blue dots correspond to the Floquet modes $|\tilde{0}^\text{F}\rangle$, as defined by \cref{app:eq:0F}, associated with the transmon ground state. Here only the Floquet modes with $\langle\tilde 0^\text{F}|0_\text{0}\rangle$ are shown. The orange lines correspond to the approximate driven ground state $|\tilde 0_\text{t}\rangle$ as defined in \cref{app:eq:wv-fit}, which fits the blue dots in the $(\omega_d, \xi)$ drive parameter space. Panel (a)-(f): decomposition of the ground state for $\omega_d/2\pi = 8.97$~GHz and varying $|\xi|^2$. The drive parameters are the same as those in \cref{fig:quasispectrum,app:fig:wave-function}. The dashed lines, from left to right, mark the center of $(9:5)$ and $(5:3)$ multiphoton resonances. Panel (g)-(l):decomposition of the ground state for $|\xi|^2 = 0.5$ and varying $\omega_d$. The dashed lines, from left to right, mark the center of $(5:3), (4:2)$, and $(7:3)$ multiphoton resonances. }
\label{app:fig:wv-decompose}
\end{figure}

In \cref{app:fig:wv-decompose}, we compare the fitted state $|\tilde 0_\text{t}\rangle$ with the actual Floquet mode $|\tilde 0^\text{F}\rangle$. In particular, in panels (a)-(f), we choose drive parameters $\omega_d/2\pi = 8.97$~ GHz and $|\xi|^2\in[0,1]$, which are the same as those in \cref{fig:quasispectrum,app:fig:wave-function}, and plot the overlap between $|\tilde 0_\text{t}\rangle$ and the first six eigenstates of the undriven transmon. The fitted state $|\tilde 0_\text{t}\rangle$ agrees well with the Floquet modes $|\tilde 0^\text{F}\rangle$ for most drive strengths except around the two dashed lines. These two dashed lines corresponds to the $(9:5)$ resonances and $(5:3)$ resonances as shown in \cref{fig:quasispectrum,app:fig:wave-function} (b), and (g)-(i). We comment on them later. In panel (g)-(l), we choose drive parameters $\omega_d/2\pi\in[6.9~\text{GHz}, 13~\text{GHz}]$ and $|\xi|^2=0.5$. The fitted state $|\tilde 0_\text{t}\rangle$ agrees well with the Floquet modes $|\tilde 0^\text{F}\rangle$ for most drive strengths except around the three dashed lines. These lines, from left to right, correspond to the $(5:3), (4:2)$ and $(7:3)$ resonances (see \cref{fig:spaghetti}).

With the approximate ground state $|\tilde 0_\text{t}\rangle$ computed, we formally define $|\tilde 0^\text{F}\rangle$, at a specific choice of drive parameters, as 
\begin{align}\label{app:eq:0F}
    |\tilde 0^\text{F}\rangle = \mathop{\arg \max}\limits_{|\Phi_m(t=0)\rangle}|\langle\Phi_m(t=0)|\tilde 0_\text{t}\rangle|,
\end{align}
where $|\Phi_m(t)\rangle$ is the Floquet mode of the Hamiltonian \cref{eq:mnr-transmon-H}. The excitation probability from the ground state $P_{0\rightarrow} = 1- |\langle \tilde 0^\text{F}|\tilde 0_\text{t}\rangle|^2$ is then inferred by how much the Floquet mode $|\tilde 0^\text{F}\rangle$ is hybridized with states outside the displaced transmon ground state. For instances, in \cref{app:fig:wave-function} (b) and (c), we plot $|\langle\tilde 0^\text{F}|\varphi\rangle|$ in the window of $|\xi|^2$ that the ground state is on resonance with the 9th excited state through a $(9:5)$ process and with the 5th excited state through a $(5:3)$ process, respectively. In stead of a Gaussian-like function, the Floquet mode $|\langle\tilde 0^\text{F}|\varphi\rangle|$ shows more nodes and indicates the hybridization with the higher excited states. The blue lines in \cref{app:fig:wave-function} (b) and (c) shows $P_{0\rightarrow}$ as a function of $|\xi|^2$. It can be seen that, when away from the resonances, the Floquet mode $|\tilde 0^\text{F}\rangle$ is approximately a displaced ground state and thus $P_{0\rightarrow}$ is around zero. When the ground state is exactly on resonances with the 9th or the 5th excited states, $P_{0\rightarrow}=0.5$ indicates the maximum hybridization of the ground state and the corresponding excited state. 

In \cref{fig:spaghetti-nonpert} (a), we also plot $P_{0\rightarrow}$ in the extended drive parameter space with $|\xi|^2$ larger than those considered in \cref{app:fig:wave-function,app:fig:wv-decompose}. This requires to compute $|\tilde 0_\text{t}\rangle$ in this extended parameter space in an iterative manner. In particular, as in \cref{app:fig:wv-decompose}, we first identify a set of Floquet modes $|\tilde{0}^\text{F}\rangle$ associated with the transmon ground state in the smaller range of $|\xi|^2$ by selecting those Floquet modes having overlap with undriven transmon state $|0_\text{t}\rangle$ larger than 0.8. By fitting these Floquet modes as \cref{app:eq:wv-fit}, we obtain $|\tilde 0_\text{t}\rangle$ at the first iteration. Then we consider the $|\xi|^2$ with larger value and, following \cref{app:eq:0F}, identify a larger set of Floquet modes $|\tilde{0}^\text{F}\rangle$ associated with the transmon ground state to be fitted. This iterative procedure allows to compute $|\tilde 0_\text{t}\rangle$ to large drive strength until the driven ground state enters the quantum diffusion regime. The approximate first excited state $|\tilde 1_t\rangle$, which is used to obtained $P_{0\rightarrow} = 1- |\langle \tilde 1^\text{F}|\tilde 1_\text{t}\rangle|^2$ in \cref{fig:spaghetti-nonpert} (b), is computed in a similar manner and for brevity we suppress its detail calculation here.

\section{Duffing Equation and Duffing Bifurcation}\label{app:duffing}

The most general form of Duffing equation containing the 3rd and 4th rank nonlinearities is 
\begin{align}\label{app:eq:ush-duffing}
    d_t^2{\cl x^\prime} + \gamma d_t{\cl x^\prime} + \cl x^\prime + c_3^\prime\cl x^{\prime2}+c_4^\prime\cl x^{\prime3} = f(e^{-i\nu t}+e^{i\nu t}),
\end{align}
where $\cl x^\prime$ is the position coordinate of the Duffing oscillator, $c_3^\prime, c_4^\prime$ are the 3rd and 4th rank nonlinearities, $\gamma$ is the damping rate, and $f$ the drive amplitude. These parameters specifying a Duffing oscillator can be further reduced by introducing $\cl x= \cl x^\prime/f$, $c_3= fc_3^\prime$, and  $c_3\rightarrow f^2c_4^\prime$. With these rescaled quantities, \cref{app:eq:ush-duffing} transforms to
\begin{align} \label{app:eq:ush-duffing-2}
    d_t^2{\cl x} + \gamma d_t{\cl x} + \cl x + c_3\cl x^2+c_4\cl x^3 = e^{-i\nu t}+e^{i\nu t},
\end{align}
which is just \cref{eq:ush-duffing}, the Duffing equation we analyzed in \cref{sec:USH}. In other words, the drive amplitude $f$ in \cref{app:eq:ush-duffing} is effectively an extra knob to control the nonlinearities in the Duffing oscillator and can be absorbed into the latter.

In \cref{sec:USH}, we have discussed the general ultra-subharmonic bifurcation process when the drive is in the vicinity of $q/p$. However, for the specific case of $(q:p) = (1:1)$, or the Duffing bifurcation \cite{nayfeh1995,vijay2009}, we use a modified quantum harmonic balance method, which we outline in this section.   

Similar to the treatment in \cref{app:frame}, we first transform \cref{app:eq:ush-duffing-2} into a rotating frame by $\cl x \rightarrow \cl x\cos(\nu t) + d_t{\cl x}\sin(\nu t)$. In the new frame, \cref{app:eq:ush-duffing-2} can be re-expressed in the bosonic coordinate as 
\begin{align}\label{app:eq:ush-eom-bos}
\begin{split}
    d_t{\cl a} = &-i(\delta - \frac{i\gamma}{2})\cl a -ie^{i\nu t}\sum_{m=3,4} g_m\Big(\cl a e^{-i\nu t}+\cl a^* e^{i\nu t}\Big)^{m-1}\\
    &-\frac{i}{2}e^{i\nu t}(e^{-i\nu t}+e^{i\nu t}), 
\end{split}\raisetag{\baselineskip}
\end{align}
where $\cl a =(\cl x+id_t{\cl x})/2$ is the complex coordinate satisfying the canonical relation $\{\cl a, \cl a^*\}_{\cl a, \cl a^*}=1$, and $\{\cl f, \cl g\}_{\cl a, \cl a^*} =-i\partial_{\cl a}\cl f\partial_{\cl a^*}\cl g+i\partial_{\cl a}\cl f\partial_{\cl a^*}\cl g $ is the Poisson bracket defined over the phase space $(\cl a, \cl a^*)$. The parameter $\delta = 1-\nu$ is the detuning between the natural frequency of the oscillator and the drive frequency, and $g_m=c_m/2$ is the rescaled nonlinearity. Here we also symmetrzied the damping to both the position and momentum coordinate, which is a valid approximation for $\gamma\ll 1$. 

Written as it is, \cref{app:eq:ush-eom-bos} can be identified as a Hamilton equation of motion $\dot{\cl a} = -\{\cl H, \cl a\}_{\cl a, \cl a^*}$ with the corresponding Hamiltonian being 
\begin{align}\label{app:eq:ush-H-tran}
\begin{split}
    \cl H(t) = &(\delta-i\frac{\gamma}{2})\cl a^* \cl a+  \sum_{m=3,4} \frac{g_m}{m}  \left(\cl{a} e^{- i\nu t} + \cl{a}^{*} e^{i\nu t}\right)^m\\
    &+ \frac{1}{2}\left(\cl{a} e^{- i\nu t} + \cl{a}^{*} e^{i\nu t}\right)\left(e^{-i\nu t}+e^{i\nu t}\right).
\end{split}\raisetag{\baselineskip}
\end{align}
In the general case of $(q:p)\ne(1:1)$, as discussed in \cref{app:frame}, we often further transform \cref{app:eq:ush-H-tran} to the linear response of the oscillator to the drive and arrive at \cref{eq:ush-H-tran}. In the case of $(1:1)$ bifurcation here, however, the drive term $\frac{1}{2}\left(\cl{a} e^{- i\nu t} + \cl{a}^{*} e^{i\nu t}\right)\left(e^{-i\nu t}+e^{i\nu t}\right)$ itself contains static component, this transformation should be modified as $\cl a \rightarrow \cl a - \frac{1}{2(\nu+1)}e^{i\nu t}$ and the transformed Hamiltonian reads
\begin{align}\label{app:eq:ush-H-tran-2}
\begin{split}
    \cl H(t) = (\delta-i\frac{\gamma}{2})\cl a^* \cl a&+  \sum_{m=3,4} \frac{g_m}{m}  \Big(\cl{a} e^{- i\nu t} + \cl{a}^{*} e^{i\nu t} \\[-8pt]
    & \;+ \xi e^{-i \nu t} +  \xi^* e^{i \nu t}\Big)^m + \frac{1}{2}(\cl a + \cl a^*),
\end{split}\raisetag{-0.5\baselineskip}
\end{align}
where $\xi = - \frac{1}{2(\nu+1)}e^{-i\nu t}$. \cref{app:eq:ush-H-tran-2}  is of the same form of \cref{eq:ush-H-tran} but with an additional term $(\cl a + \cl a^*)/2$. Taking this equation as the starting point of our diagrammatic analysis, we follow the same treatment discussed in \cref{sec:USH} and obtain the equation governing the steady state of $(1:1)$ bifurcation as:
\begin{align}\label{app:eq:ush-eom-rho}
    (2K_2\rho + \Delta)^2 + \frac{\gamma^2}{4} +  \mathcal{O}(g_4^2) = \rho^{-1},
\end{align}
where $\rho = |\cu A_s|^2$, $|\cu A_s|$ is the steady state of the bosonic coordinate in the effective frame, and $K_2$ is the energy dressing coefficient defined in \cref{eq:ush-K-qp}. We note that \cref{app:eq:ush-eom-rho} shares the same form as \cref{eq:ush-eom-rho}, which governs the steady state of general ultra-subharmonic bifurcation process $(q:p)\ne(1:1)$. The domain diagram in \cref{fig:ush-phase} (a) is also generated by \cref{app:eq:ush-eom-rho}. 
\end{appendices}

\reftoc
\bibliography{main}

\end{document}